\documentclass[%
 reprint,
 superscriptaddress,
 amsmath,
 aps,
 prb
]{revtex4-2}

\usepackage[pdftex]{graphicx, color}
\usepackage[pdftex,
            colorlinks=true,
            citecolor=blue,
            linkcolor=blue]{hyperref}
\usepackage{newtxtext, newtxmath}
\usepackage{siunitx}
\usepackage{bm}
\usepackage{braket, mathtools}
\usepackage{bbm}
\usepackage{ulem}
\usepackage{tabularx}
\usepackage{comment}

\graphicspath{{./images/}}

\begin{document}

\title{Vertex correction for the linear and nonlinear optical responses in superconductors:\\
multiband effect and topological superconductivity}

\author{Hiroto Tanaka}
\email[]{tanaka.hiroto.54z@st.kyoto-u.ac.jp}
\affiliation{%
Department of Physics, Graduate School of Science, Kyoto University, Kyoto 606-8502, Japan
}%

\author{Youichi Yanase}
\affiliation{%
Department of Physics, Graduate School of Science, Kyoto University, Kyoto 606-8502, Japan
}%

\date{\today}

\begin{abstract}
Intensive research has revealed intriguing optical responses in topological materials. This paper focuses on the optical responses in $s$-wave superconductors with a Rashba spin-orbit coupling and a magnetic field, one of the platforms of topological superconductivity. 
On the one hand, to satisfy some conservation laws in superconducting responses, 
it is essential to take into account collective excitation modes.
On the other hand, the optical response is a promising phenomenon for detecting hidden collective modes in superconductors. 
In this paper, we investigate the effect of collective excitation modes on the linear and second-order optical responses 
based on the self-consistent response approximation, which is formulated using the Kadanoff-Baym method. Our main results reveal that the Higgs mode enhances the optical responses when the Fermi level is close to the Dirac point. 
The enhancement is due to the multiband effects characterized by interband pairing.
We also demonstrate the sign reversal of the photocurrent conductivity 
around the topological transition with increasing the Zeeman field.
This finding supports the prediction in our previous work without considering collective excitation modes 
[H. Tanaka, {\it et al.}, Phys. Rev. B $\bm{110}$, 014520 (2024)].
The sign reversal phenomenon is attributed to the magnetic injection current modified by the Higgs mode, and is proposed for a bulk probe of topological superconductors. We also discuss the interplay of quantum geometry and collective modes.
\end{abstract}

\maketitle

\section{Introduction}
Optical measurements have revealed remarkable properties of quantum condensed matter. In the research of superconductors, linear optical measurement is an established technique for measuring the superconducting gap~\cite{Degiorgi1994, Molegraaf2002} and the magnetic penetration length~\cite{Wang1998}.
Theoretically, the linear optical responses in conventional Bardeen-Cooper-Schrieffer (BCS) superconductors are considered to be well understood by the dirty-limit theory of Mattis and Bardeen~\cite{Mattis1958}. 
In this paradigm, impurity-mediated optical transitions are essential.

On the other hand, cumulative studies have revealed that symmetry breaking diversifies the mechanisms of linear optical responses in superconductors. 
Recent theories in the clean limit predict finite linear optical conductivity in the presence of the dc supercurrent, which breaks the time-reversal ($\mathcal{T})$ and the inversion parity ($\mathcal{P}$) symmetry~\cite{Crowley2022, Papaj2022}. 
In the diffusive limit, the supercurrent also activates the Higgs mode in the linear optical phenomena, even though there is no linear coupling between the Higgs mode and the electromagnetic field~\cite{Moor2017}.

The multiple superconducting order parameters in multiband superconductors allow novel contributions to the linear optical conductivity through collective excitations. 
The Leggett mode, the relative phase fluctuation between different bands, is a collective mode unique to multiband superconductors~\cite{Leggett1966}, and 
appears in the linear optical conductivity~\cite{Kamatani2022, Nagashima2024}.
Even an uncondensed pairing channel can be relevant. 
When two Cooper channels are competing, collective excitation due to the uncondensed Cooper pairs, known as the Bardasis-Schriefer (BS) mode~\cite{Bardasis1961}, can cause linear optical conductivity~\cite{Lee2023}.

Recent progress in experimental techniques enables us to access nonlinear optical phenomena in superconductors. The nonlinear optics potentially yield rich information on the degrees of freedom inactive in the linear optical responses.
Indeed, for conventional BCS superconductors without supercurrent, the Higgs mode can be generated by a two-photon process through the nonlinear coupling between the Higgs mode and the electromagnetic field, although a single-photon process for the Higgs mode is prohibited by the particle-hole symmetry~\cite{Tsuji2015, Anderson1958_2}.
Experimentally, the resonant enhancement of third harmonic generation has been observed in $\mathrm{NbN}$~\cite{Matsunaga2014, Shimano2020}.
Furthermore, nonlinear optics can characterize nonreciprocity in superconducting states.
In noncentrosymmetric superconductors, notable nonreciprocal response phenomena occur in the nonlinear regime~\cite{Nagaosa2024}. 
For example, the nonreciprocal superfluid density, a nonreciprocal correction to the superfluid density, leads to the nonreciprocal Meissner effect~\cite{Watanabe2022_2} and the low-frequency divergence of the nonreciprocal optical conductivity~\cite{Watanabe2022}, which are closely related to each other. 

Another intriguing aspect of nonlinear optics is its close relation to quantum geometry in momentum space.
The nonlinear optical effects are sensitive to the quantum-geometric properties such as the Berry curvature and quantum metric, and accordingly topological nature of quantum materials~\cite{Orenstein2021}. 
An example is the photo-induced direct current, that is, the photocurrent generation, in noncentrosymmetric superconductors~\cite{Xu2019, Watanabe2022, Tanaka2023, Huang2023, Tanaka2024, Raj2024, watanabeS2024, WatanabeS2025}. 
Recent theoretical work~\cite{Watanabe2022, Tanaka2024}, using the Kubo formula applied to the Bogoliubov de-Gennes Hamiltonian, revealed that the quantum-geometric quantities give characteristic contributions to the photocurrent conductivity. 
For example, the Berry curvature derivative leads to divergent behavior in the low-frequency regime~\cite{Watanabe2022}, and the quantum geometry around the Dirac point significantly enhances the resonant components of photocurrent generation~\cite{Tanaka2024}.
Moreover, theories predicted that the sign of photocurrent conductivity reverses at the transition to the topological superconducting state~\cite{Tanaka2024, Raj2024}.

In the mean-field formalism, a naive calculation of optical responses may lead to a violation of charge conservation and potentially to unphysical results due to the $U(1)$-gauge symmetry breaking of the mean field~\cite{schrieffer2018}. 
For the linear response, Nambu resolved this issue by incorporating the Nambu-Goldstone (NG) mode through the vertex correction consistently with the gap equation~\cite{Nambu1960}. 
After his study, vertex correction has been adopted in the formulation of linear optical responses for a wide range of superconductors~\cite{Dai2017, Papaj2022, Oh2024}.
Other various methods~\cite{He2017, Guo2013, Kadanoff1961, Kulik1981, Zha1995, Ioan2000, Anderson2016, Boyack2016, Lutchyn2008} beyond the mean-field formalism have also been proposed to take into account the collective mode.
Recent theoretical studies numerically demonstrated non-negligible effects of the vertex correction on the second-order nonlinear responses by the Nambu's method~\cite{Huang2023} and the "Consistent Fluctuations of Order Parameters" method~\cite{watanabeS2024, WatanabeS2025}.

Although the relationship of photocurrent conductivity to the quantum-geometric properties and topological nature has been focused on, the many-body effect on the quantum-geometric contributions to optical responses remains to be uncovered.
In this paper, we investigate the vertex correction for the linear and nonlinear optical responses in topological superconductors.
The linear and second-order nonlinear optical responses are defined as 
\begin{align}
J^{\alpha}_{(1)}(\omega) =& \sigma^{\alpha\beta}(\omega)E^{\beta}(\omega), \\
J^{\alpha}_{(2)}(\omega) =& \int \frac{d\Omega}{2\pi} \sigma^{\alpha;\beta\gamma}(\omega; \Omega, \omega - \Omega)E^{\beta}(\Omega)E^{\gamma}(\omega - \Omega),
\end{align}
with the linear optical conductivity $\sigma^{\alpha\beta}(\omega)$ and the nonlinear optical conductivity $\sigma^{\alpha;\beta\gamma}(\omega; \Omega, \omega - \Omega)$.
Our method for calculating optical responses is based on the self-consistent response approximation. When constructing the self-consistent response approximation, we adopt the Kadanoff-Baym method~\cite{Baym1961, kadanoff1962}, which maintains the total particle number conservation law. 
The formula produces physically reasonable results not only for conventional superconductors but also for various unconventional superconductors.
We focus on 
the $s$-wave Rashba superconductor in the presence of the Zeeman field, which is one of the typical models of topological superconductors~\cite{Sato2009, Sato2010, Sau2010}. 
Our results show that the amplitude (Higgs) mode enhances the optical responses when interband pairing, Cooper pairs of different bands, is significant. This condition is satisfied in the topological superconducting state.
Our previous theory without vertex correction predicted the sign reversal of the photocurrent conductivity, defined as $\sigma^{\alpha;\beta\gamma}_{\mathrm{PC}}(\omega)=\sigma^{\alpha;\beta\gamma}(0;\omega, -\omega)$, at the topological transition~\cite{Tanaka2024}.
We numerically demonstrate that the sign reversal of the photocurrent conductivity is robust to the collective excitation modes. The amplitude mode enhances the resonant component and makes observation of the sign reversal phenomenon more feasible.

The outline of the paper is given below. 
In Sec.~\ref{sec:particle_conservation}, we discuss the particle number conservation law in the effective Hamiltonian with separable attractive interaction. 
In Sec.~\ref{sec:Kadanoff-Baym}, the Kadanoff-Baym method is adopted to derive the equation of motion of Green functions which satisfies the total particle number conservation law.
We formulate the self-consistent response approximation for linear and second-order nonlinear responses in Sec.~\ref{sec:self-consistent}.
In subsequent sections, we study the linear and second-order nonlinear optical conductivity in the Rashba superconductor by using the self-consistent response approximation.
Before the numerical demonstration, Sec.~\ref{sec:numerical_demonstration} outlines topological superconductivity and interband pairing in the Rashba superconductor model.
After that, we present our main results in Secs.~\ref{sec:linear} and \ref{sec:nonlinear}.
In Sec.~\ref{sec:linear}, we show the linear optical conductivity and discuss the effect of the collective modes and interband pairing. 
In Sec.~\ref{sec:nonlinear}, we investigate the photocurrent conductivity around the topological transition, demonstrating sign reversal and the role of the collective modes.
In Sec.~\ref{sec:discussion}, we discuss the effect of the long-range Coulomb interaction and material platforms.
Finally, we summarize our work in Sec.~\ref{sec:summary}.
Throughout this paper, we present formulas with the unit $\hbar=1$ (Dirac constant) and $\mathrm{e}=1$ (electron charge).

\section{Particle number conservation law in effective Hamiltonian}
\label{sec:particle_conservation}
The effective Hamiltonian with separable attractive interaction is often used to study unconventional superconductivity without computational difficulties. We can adopt an effective Hamiltonian of general form to study various classes of unconventional superconductivity, such as anisotropic superconductivity, helical superconductivity, and multiband superconductivity. In this section, we discuss gauge symmetry and the conservation law of the effective Hamiltonian. The effective Hamiltonian does not always preserve the local gauge symmetry, while the total particle number is conserved.

\subsection{Effective Hamiltonian}
We consider a pairing interaction in momentum space,
\begin{align}
  H_{\mathrm{pair}} =-\sum_{\bm{k},\bm{k}^{\prime}, \bm{q}}U_{abcd}(\bm{k},\bm{k}^{\prime}, \bm{q})C_{\bm{k}+\bm{q}a}^{\dagger}C_{-\bm{k}+\bm{q}b}^{\dagger}C_{-\bm{k}^{\prime}+\bm{q}c}C_{\bm{k}^{\prime}+\bm{q}d}, \label{eq:general_pairing_potential}
\end{align}
where $C_{\bm{k}a}$ ($C_{\bm{k}a}^{\dagger}$) is the annihilation (creation) operator of electrons.
We here reduce the pairing interaction to a simple form 
which stabilizes superconductivity that we aim to study. 

If the total momentum of Cooper pairs is zero, we can omit $\bm{q}\neq \bm{0}$ terms in Eq.~(\ref{eq:general_pairing_potential}). The pairing interaction for $\bm{q} =0$ can be represented by bases of the irreducible representation $\Gamma$ with respect to the symmetry of the system~\cite{Sigrist1991},
\begin{align}
    &U_{abcd}(\bm{k},\bm{k}^{\prime}, \bm{q}) = \notag \\
    &\sum_{\Gamma}V(\Gamma)\sum_{m}\varphi_{ab}(\Gamma, m;\bm{k})\varphi^{\dagger}_{cd}(\Gamma, m;\bm{k}^{\prime})\delta_{\bm{q},\bm{0}}.
\end{align}
When we focus on one of the bases of the pair potential $\varphi_{ab}(\tilde{\Gamma}, \tilde{m};\bm{k})\eqqcolon\varphi_{ab}(\bm{k})$, we can reduce the pairing interaction,
\begin{align}
  U_{abcd}(\bm{k},\bm{k}^{\prime}, \bm{q}) = \frac{U}{2V}\varphi_{ab}(\bm{k})\varphi_{cd}^{\dagger}(\bm{k}^{\prime})\delta_{\bm{q},\bm{0}}. 
\end{align}
The basis function $\varphi_{ab}(\bm{k})$ has the following symmetry: $\varphi_{ab}(\bm{k}) = -\varphi_{ba}(-\bm{k})$.
When we consider Cooper pairs which have a finite total momentum $2\bar{\bm{q}}$, we revise the effective pairing interaction,
\begin{align}
  U_{abcd}(\bm{k},\bm{k}^{\prime}, \bm{q}) = \frac{U}{2V}\varphi_{ab}(\bm{k})\varphi_{cd}^{\dagger}(\bm{k}^{\prime})\delta_{\bm{q},\bar{\bm{q}}}.
  \label{eq:effective_int_1}
\end{align}
By this simplification, the pairing interaction Hamiltonian is written by
\begin{align}
  H_{\mathrm{pair}} =-\frac{U}{2V}\sum_{\bm{k},\bm{k}^{\prime}}\varphi_{ab}(\bm{k})\varphi_{cd}^{\dagger}(\bm{k}^{\prime})c_{\bm{k}a}^{\dagger}c_{-\bm{k}b}^{\dagger}c_{-\bm{k}^{\prime}c}c_{\bm{k}^{\prime}d},
\label{eq:pairing_Hamiltonian}
\end{align}
where we denote $c_{\bm{k}a} \equiv C_{\bm{k} + \bar{\bm{q}} a}$ for simplicity.

We can deal with the pairing interaction with the mean-field approximation. The pair potential corresponding to the pairing symmetry $\varphi_{ab}(\bm{k})$ is introduced by
\begin{align}
&\Delta = \frac{U}{V}\sum_{\bm{k}}\varphi_{ab}^{\dagger}(\bm{k})\left<c_{\bm{k}
  b}c_{-\bm{k}
  a}\right>, \\
  &\Delta_{ab}(\bm{k}) = \varphi_{ab}(\bm{k})\Delta,
\end{align}
approximating the interaction term of Hamiltonian as,
\begin{align}
   \frac{1}{2}\sum \left[ c_{-\bm{k}a}\Delta^{\dagger}_{ab}(\bm{k})c_{\bm{k}b} + c_{\bm{k}a}^{\dagger}\Delta_{ab}(\bm{k})c_{-\bm{k}b}^{\dagger}\right] + \frac{1}{2U}\sum_{\bm{k}}|\Delta|^{2}.
\end{align}

\subsection{Total particle number conversation law}
We here discuss the gauge symmetry and the particle number conservation law in the effective Hamiltonian. We consider the effective Hamiltonian with vector potential
\begin{align}
  H(\tau) =& \sum_{\bm{k}}\mathcal{H}_{ab}(\bm{k};\bm{A}(\tau))C_{\bm{k}a}^{\dagger}C_{\bm{k}b} \nonumber \\
  -&\sum_{\bm{k},\bm{k}^{\prime},{\bm q^\prime}}U_{abcd}(\bm{k},\bm{k}^{\prime}, \bm{q}^{\prime})C_{\bm{k}+\bm{q}^{\prime}a}^{\dagger}C_{-\bm{k}+\bm{q}^{\prime}b}^{\dagger}C_{-\bm{k}^{\prime}+\bm{q}^{\prime}c}C_{\bm{k}^{\prime}+\bm{q}^{\prime}d}, \label{eq:effective_Hamiltonian}
\end{align}
where $\tau$ is the (imaginary) time and we adopt the velocity gauge $\mathcal{H}_{ab}(\bm{k};\bm{A}(\tau)) = \mathcal{H}_{ab}(\bm{k} - \bm{A}(\tau))$.
The effective Hamiltonian has the global gauge symmetry as it is invariant under the gauge transformation $C_{\bm{k}a}\rightarrow e^{i\theta}C_{\bm{k}a}$. Thus, the total particle number conservation law is derived from the Heisenberg equation
\begin{align}
  \frac{\partial}{\partial \tau} N(\tau) =0,
\end{align}
where we define the total particle number $N$ as
\begin{align}
  N = \sum_{\bm{k}, i} C^{\dagger}_{\bm{k}i}C_{\bm{k}i} = \sum_{\bm{k}, i} c^{\dagger}_{\bm{k}i}c_{\bm{k}i}.
\end{align}

\subsection{Artificial local gauge symmetry breaking}
The effective Hamiltonian preserves the global gauge symmetry but does not always have the local gauge symmetry. We assume the total momentum of Cooper pairs $2\bar{\bm{q}}$, so the effective pairing interaction is described as
\begin{align}
  &\sum_{\bm{k},\bm{k}^{\prime}} U_{abcd}(\bm{k},\bm{k}^{\prime})c_{\bm{k}a}^{\dagger}c_{-\bm{k}b}^{\dagger}c_{-\bm{k}^{\prime}c}c_{\bm{k}^{\prime}d} \notag \\
  =& \sum_{\Delta\bm{k}, \, \bar{\bm{k}}} \,\, \sum_{\bm{r},\,\bm{r}^{\prime}, \, \Delta\bm{r}, \, \bar{\bm{r}}} \chi_{abcd}(\Delta \bm{k}, \bar{\bm{k}}) e^{2i\Delta\bm{k}\cdot \Delta \bm{r}}e^{2i\bar{\bm{k}}\cdot \bar{\bm{r}}}e^{2i\bar{\bm{q}}\cdot (\bm{r} - \bm{r}^{\prime} +\bar{\bm{r}})} \notag\\
  &\times N^{-4}C_{a}^{\dagger}(\bm{r}) C_{b}^{\dagger}(\bm{r} + \Delta\bm{r} + \bar{\bm{r}}) C_{c}(\bm{r}^{\prime} + \Delta\bm{r}) C_{d}(\bm{r}^{\prime}+ \bar{\bm{r}}),
\end{align}
where we introduce the description as
\begin{align}
    &\bar{\bm{k}} = \frac{\bm{k} + \bm{k}^{\prime}}{2}, \quad \Delta \bm{k} = \frac{\bm{k} - \bm{k}^{\prime}}{2}, \\
    &\chi_{abcd}(\Delta \bm{k}, \bar{\bm{k}})  = U_{abcd}(\bm{k},\bm{k}^{\prime}).
\end{align}
We can rewrite $\chi_{abcd}$ as the real-space function,
\begin{align}
    &\tilde{\chi}_{abcd}(\Delta\bm{r}, \bar{\bm{r}}, \bm{r} - \bm{r}^{\prime}) = \notag\\
    &\sum_{\Delta\bm{k}, \bar{\bm{k}}} \chi_{abcd}(\Delta \bm{k}, \bar{\bm{k}}) e^{2i\Delta\bm{k}\cdot \Delta \bm{r}}e^{2i\bar{\bm{k}}\cdot \bar{\bm{r}}}e^{2i\bar{\bm{q}}\cdot (\bm{r} - \bm{r}^{\prime} +\bar{\bm{r}})}.
\end{align}
Thus, the effective interaction term does not necessarily have local gauge symmetry. Although some gauge-invariant formalisms of responses in superconductors are based on the Ward-Takahashi identity~\cite{Nambu1960,schrieffer2018,Huang2023, watanabeS2024, WatanabeS2025}, these approaches cannot always be applied to the effective Hamiltonian because the Ward-Takahashi identity assumes the local gauge symmetry. Therefore, we formulate the self-consistent response approximation based on the Kadanoff-Baym method instead of the Ward-Takahashi identity.
The Kadanoff-Baym method is a powerful technique for the theoretical analysis of electromagnetic phenomena, even in the nonlinear regime~\cite{Jujo2005, Rostami2021npj}.

The local gauge symmetry breaking is an artificial effect due to the reduction of the microscopic interaction term into the effective pairing interaction. When we restore the local gauge symmetry of the Hamiltonian, the Ward-Takahashi identity can be derived from the corrected Hamiltonian exactly. In the next section, we explain how to restore the local gauge symmetry.

\subsection{Restoration of the local gauge symmetry}
We here show the restoration of the local gauge symmetry in the effective interaction term. When we introduce the effective Hamiltonian, we assume the Cooper pairs' total momentum $2\bar{\bm{q}}$ and ignore the other $\bm{q}\neq\bar{\bm{q}}$ terms. Here, we reintroduce the $\bm{q}\neq\bar{\bm{q}}$ terms. Then, we obtain the interaction Hamiltonian
\begin{align}
  &\sum_{\bm{k},\bm{k}^{\prime}} U_{abcd}(\bm{k},\bm{k}^{\prime}) 
\sum_{\bm{q}} C_{\bm{k}+\bm{q}a}^{\dagger}C_{-\bm{k}+\bm{q}b}^{\dagger}C_{-\bm{k}^{\prime}+\bm{q}c}C_{\bm{k}^{\prime}+\bm{q}d} \label{eq:q_corrected_effective_int} \notag \\
  =&  N^{-3}\sum_{\Delta\bm{k}, \bar{\bm{k}}}\sum_{\bm{r}, \Delta\bm{r}, \bar{\bm{r}}} \chi_{abcd}(\Delta \bm{k}, \bar{\bm{k}}) e^{2i\Delta\bm{k}\cdot \Delta \bm{r}}e^{2i\bar{\bm{k}}\cdot \bar{\bm{r}}} \notag \\
  &\times C_{a}^{\dagger}(\bm{r}) C_{b}^{\dagger}(\bm{r} + \Delta\bm{r} + \bar{\bm{r}}) C_{c}(\bm{r} + \Delta\bm{r}) C_{d}(\bm{r}+ \bar{\bm{r}}).
\end{align}
When $\chi_{abcd}(\Delta \bm{k}, \bar{\bm{k}})$ is decomposed as
\begin{align}
    \chi_{abcd}(\Delta \bm{k}, \bar{\bm{k}}) = \sum_{\alpha} \Delta\chi^{\alpha}_{abcd}(\Delta \bm{k}) + \sum_{\beta} \bar{\chi}^{\beta}_{abcd}(\bar{\bm{k}}), \label{eq:decomp_cond}
\end{align}
we can rewrite Eq.~(\ref{eq:q_corrected_effective_int}) as
\begin{align}
  \sum_{\bm{r}, \Delta\bm{r}} \Delta\chi_{abcd}(\Delta \bm{r})
  C_{a}^{\dagger}(\bm{r}) C_{b}^{\dagger}(\bm{r} + \Delta\bm{r}) C_{c}(\bm{r} + \Delta\bm{r}) C_{d}(\bm{r}) \label{eq:Delta_int} \\
  +\sum_{\bm{r}, \bar{\bm{r}}} \bar{\chi}_{abcd}(\bar{\bm{r}})
  C_{a}^{\dagger}(\bm{r}) C_{b}^{\dagger}(\bm{r} + \bar{\bm{r}}) C_{c}(\bm{r}) C_{d}(\bm{r}+\bar{\bm{r}}) \label{eq:bar_int},
\end{align}
where the Fourier transform $\Delta\chi_{abcd}(\Delta \bm{r})$ and $\bar{\chi}_{abcd}(\bar{\bm{r}})$ are given by
\begin{align}
    \Delta\chi_{abcd}(\Delta \bm{r}) &= N^{-2}\sum_{\Delta\bm{k}}\sum_{\alpha} \Delta\chi^{\alpha}_{abcd}(\Delta \bm{k}) e^{2i\Delta\bm{k}\cdot \Delta \bm{r}}, \\
    \bar{\chi}_{abcd}(\bar{\bm{r}})&=N^{-2}\sum_{\bar{\bm{k}}}\sum_{\beta} \bar{\chi}^\beta_{abcd}(\bar{\bm{k}}) e^{2i\bar{\bm{k}}\cdot \bar{\bm{r}}}.
\end{align}
Equation~\eqref{eq:bar_int} is invariant under the local gauge transformation $C_{a}(\bm{r})\rightarrow e^{i\theta(\bm{r})}C_{a}(\bm{r})$.
Thus, if $\chi_{abcd}(\Delta \bm{k}, \bar{\bm{k}})$ can be decomposed as Eq.~\eqref{eq:decomp_cond}, the local gauge symmetry is restored by reintroducing the summation over $\bm{q}$.
This condition is satisfied in typical interactions such as the Coulomb interaction.

When we focus on a specific pairing symmetry with a basis function $\varphi_{ab}(\bm{k})$, we may assume the separable form 
\begin{align}
    \chi_{abcd}(\Delta \bm{k}, \bar{\bm{k}}) = \frac{U}{2V}\varphi_{ab}(\bm{k})\varphi_{cd}^{\dagger}(\bm{k}^{\prime}).
\end{align}
In the case of the $s$-wave pairing symmetry $\varphi_{ab}(\bm{k}) = \varepsilon_{ab}$ ($\varepsilon_{ab}$ is a constant antisymmetric tensor), Eq.~\eqref{eq:decomp_cond} is satisfied. However, for the basis function $\varphi_{ab}(\bm{k};\mathrm{B_{1g}})=(\cos k_x - \cos k_y)\varepsilon_{ab}$ of the $\mathrm{B_{1g}}$ representation for the tetragonal $\mathrm{D_{4h}}$ point group, the condition is not satisfied because $\chi_{abcd}(\Delta \bm{k}, \bar{\bm{k}})$ has a term proportional to $\cos k_{x}\cos k_{y}^{\prime}$ which cannot be written as $\Delta\chi_{abcd}(\Delta \bm{k})$ or $\bar{\chi}_{abcd}(\bar{\bm{k}})$. We can avoid this problem by introducing an additional interaction term in the A$_{\rm 1g}$ channel with
$\varphi_{ab}(\bm{k};\mathrm{A_{1g}})=(\cos k_x + \cos k_y)\varepsilon_{ab}$, 
\begin{align}
    &\varphi_{ab}(\bm{k};\mathrm{B_{1g}})\varphi^{\dagger}_{cd}(\bm{k}^{\prime};\mathrm{B_{1g}}) + \varphi_{ab}(\bm{k};\mathrm{A_{1g}})\varphi^{\dagger}_{cd}(\bm{k}^{\prime};\mathrm{A_{1g}}) \notag \\
    &=\left(\Delta\chi(\Delta \bm{k}) +  \bar{\chi}(\bar{\bm{k}})\right)\varepsilon_{ab}\varepsilon^{\dagger}_{cd},
\end{align}
where $\Delta\chi(\Delta \bm{k})$ and $\bar{\chi}(\bar{\bm{k}})$ are described as
\begin{align}
    \Delta\chi(\Delta \bm{k}) &= \left(\cos(k_{x} - k_{x}^{\prime}) + \cos(k_{y} -k_{y}^{\prime})\right), \\
    \bar{\chi}(\bar{\bm{k}}) &= \left(\cos(k_{x} + k_{x}^{\prime}) + \cos(k_{y} +k_{y}^{\prime})\right).
\end{align}
Although we can restore the local gauge symmetry by reintroducing the summation over $\bm{q}$, the computational tasks will increase because we need to consider the $\bm{q}\neq \bar{\bm{q}}$ terms.

\section{Equation of motion based on the Kadanoff-Baym method}
\label{sec:Kadanoff-Baym}
We consider the imaginary time evolution of Green's functions based on the Kadanoff-Baym method. We verify that the equation of motion derived from the Kadanoff-Baym method satisfies the total particle number conservation law although the mean-field approximation, where we only consider the time evolution of Bogoliubov quasiparticles, does not satisfy the conservation law. When we assume the effective pairing interaction as Eq.~\eqref{eq:effective_int_1}, the imaginary time-dependent effective Hamiltonian in Eq.~\eqref{eq:effective_Hamiltonian} is approximated as
\begin{align}
  H(\tau) &= \sum_{\bm{k}}\mathcal{H}_{ab}(\bm{k}+\bm{q};\bm{A}(\tau))c_{\bm{k}a}^{\dagger}c_{\bm{k}b} \notag \\
  &+ \frac{1}{2}\sum\left[ c_{-\bm{k}a}\Delta^{+}_{ab}(\bm{k},\tau:\bm{A})c_{\bm{k}b} + c_{\bm{k}a}^{\dagger}\Delta_{ab}(\bm{k},\tau;\bm{A})c_{-\bm{k}b}^{\dagger}\right],
\end{align}
where we ignore the constant term for simplicity and denote the Cooper pairs' total momentum as $2\bm{q}$. We introduce the mean field $\Delta_{ab}$ as it depends on the imaginary time and the vector potential. The 
mean field is determined by the Kadanoff-Baym method as we show later.

We introduce the Green's functions ($0 < \tau, \tau^{\prime} < \beta$) as
\begin{align}
  \mathcal{G}_{ab}(\bm{k}, \tau, \tau^{\prime}) &=-\left<T_{\tau}[c_{\bm{k}a}(\tau)c^{+}_{\bm{k}b}(\tau^{\prime})]\right>_{\mathrm{tot}}, \\
  \mathcal{F}_{ab}(\bm{k}, \tau, \tau^{\prime}) &=-\left<T_{\tau}[c_{\bm{k}a}(\tau)c_{-\bm{k}b}(\tau^{\prime})]\right>_{\mathrm{tot}}, \\
  \mathcal{G}^{+}_{ab}(\bm{k}, \tau, \tau^{\prime}) &=-\left<T_{\tau}[c^{+}_{-\bm{k}a}(\tau)c_{-\bm{k}b}(\tau^{\prime})]\right>_{\mathrm{tot}}, \\
  \mathcal{F}^{+}_{ab}(\bm{k}, \tau, \tau^{\prime}) &=-\left<T_{\tau}[c^{+}_{-\bm{k}a}(\tau)c^{+}_{\bm{k}b}(\tau^{\prime})]\right>_{\mathrm{tot}},
\end{align}
where we adopt the Heisenberg representation
\begin{align}
  c_{\bm{k}a}(\tau) =& S^{+}(0, \tau)c_{\bm{k}a}S(\tau, 0), \label{eq:c_tau_def}\\
  c^{+}_{\bm{k}a}(\tau) =& S^{+}(0, \tau)c^{\dagger}_{\bm{k}a}S(\tau, 0), \\
  S(\tau, 0) =& T_{\tau}\exp\left[-\int^{\tau}_{0}H(\tilde{\tau})d\tilde{\tau}\right],\\
  S^{+}(0, \tau) =& T^{+}_{\tau}\exp\left[\int^{\tau}_{0}H(\tilde{\tau})d\tilde{\tau}\right].
\end{align}
The notation $T_{\tau}$ ($T^{+}_{\tau}$) means the (inverse) imaginary time-ordered product. The expectation value $\left<\cdots\right>_{\mathrm{tot}}$ is defined as
\begin{align}
  \left<\hat{X}\right>_{\mathrm{tot}}=\frac{1}{Z[\bm{A}]}\mathrm{Tr}\left[\hat{X}S(\beta,0)\right], \quad Z[\bm{A}]=\mathrm{Tr}[S(\beta,0)]. \label{eq:tot_expectation}
\end{align}
The imaginary time evolution of the Green's function is determined by the Heisenberg equation
\begin{align}
  \frac{\partial}{\partial \tau} c_{\bm{k}a}(\tau) = [H_{\mathrm{H}}(\tau), c_{\bm{k}a}(\tau)], \\ \frac{\partial}{\partial \tau} c_{\bm{k}a}^{+}(\tau) = [H_{\mathrm{H}}(\tau), c_{\bm{k}a}^{+}(\tau)] ,\\
  H_{\mathrm{H}}(\tau) = S^{+}(0, \tau)H(\tau)S(\tau, 0).
\end{align}
Thus, we derive the equation of motion
\begin{align}
  &\left(-\frac{\partial}{\partial \tau} \delta_{aj} - \overleftrightarrow{\mathcal{H}}_{aj}(\bm{k};\bm{A}(\tau))\right)\overleftrightarrow{\mathcal{G}}_{jb}(\bm{k}, \tau, \tau^{\prime};\bm{A}) \notag \\
  &= \overleftrightarrow{\mathcal{G}}_{aj}(\bm{k}, \tau, \tau^{\prime};\bm{A})\left(\overleftarrow{\frac{\partial}{\partial \tau^{\prime}}} \delta_{aj} - \overleftrightarrow{\mathcal{H}}_{aj}(\bm{k};\bm{A}(\tau^{\prime}))\right) \notag \\
  &= \delta(\tau - \tau^{\prime})\delta_{ab}, \label{eq:EOM}
\end{align}
where we rewrite the Green's function in the Nambu space
\begin{align}
\overleftrightarrow{\mathcal{G}}_{ab}(\bm{k}, \tau, \tau^{\prime};\bm{A}) =&
  \begin{pmatrix}
    \mathcal{G}_{ab} & \mathcal{F}_{ab} \\
    \mathcal{F}^{+}_{ab} & \mathcal{G}^{+}_{ab}
  \end{pmatrix}
  (\bm{k}, \tau, \tau^{\prime};\bm{A}), 
\end{align}
and the Bogoliubov-de Gennes (BdG) Hamiltonian
\begin{align}
&\overleftrightarrow{\mathcal{H}}_{ab}(\bm{k}, \tau;\bm{A}) \notag \\
  &=
  \begin{pmatrix}
    \mathcal{H}_{ab}(\bm{k}+ \bm{q}; \bm{A}(\tau)) & \Delta_{ab}(\bm{k},\tau;\bm{A}) \\
    \Delta^{+}_{ab}(\bm{k}, \tau; \bm{A}) & -\mathcal{H}^{\top}_{ab}(-\bm{k}+ \bm{q}; \bm{A}(\tau))
  \end{pmatrix}.  
\end{align}
%
We can rewrite Eq.~\eqref{eq:EOM} as
\begin{align}
  \int d\tau_{1}\overleftrightarrow{\mathcal{G}}^{-1}_{aj}(\bm{k}, \tau, \tau_{1};\bm{A})\overleftrightarrow{\mathcal{G}}_{jb}(\bm{k}, \tau_{1}, \tau^{\prime};\bm{A}) = \delta(\tau - \tau^{\prime})\delta_{ab}, \label{eq:EOM_3} \\
  \int d\tau_{1}\overleftrightarrow{\mathcal{G}}_{aj}(\bm{k}, \tau, \tau_{1};\bm{A})\overleftrightarrow{\mathcal{G}}^{-1}_{jb}(\bm{k}, \tau_{1}, \tau^{\prime};\bm{A}) = \delta(\tau - \tau^{\prime})\delta_{ab}, \label{eq:EOM4}
\end{align}
where we define $\overleftrightarrow{\mathcal{G}}^{-1}$ as
\begin{align}
  \overleftrightarrow{\mathcal{G}}^{-1}_{ij}(\bm{k}, \tau, \tau^{\prime};\bm{A}) &= \left\{\delta_{ij}\frac{\partial}{\partial \tau^{\prime}} - \overleftrightarrow{\mathcal{H}}_{ij}(\bm{k},\tau^{\prime};\bm{A})\right\}\delta(\tau - \tau^{\prime}) .
\end{align}

Next, we consider the time evolution of the total particle number. The expectation value of the total particle number is given by the normal Green's function as
\begin{align}
  \left<N(\tau)\right>_{\mathrm{tot}} = \sum_{\bm{k}, a} \mathcal{G}_{aa}(\bm{k}, \tau^{-}, \tau^{+};\bm{A}),
  \label{eq:particle_number}
\end{align}
where we define $\tau_{\pm} = \tau \pm \delta$ with an infinitesimal positive number $\delta > 0$. Thus, the imaginary time derivative of $\left<N(\tau)\right>_{\mathrm{tot}}$ is given by
\begin{align}
  &\frac{d}{d\tau}\left<N(\tau)\right>_{\mathrm{tot}} \notag \\
  &=\sum_{\bm{k}, a} \left[\left(\frac{\partial}{\partial \tau} + \frac{\partial}{\partial \tau^{\prime}} \right ) \mathcal{G}_{aa}(\bm{k}, \tau, \tau^{\prime};\bm{A})\right]_{\tau=\tau^{-}, \tau^{\prime}=\tau^{+}}\\
    &=\sum_{a, j, \bm{k}}  \mathcal{G}_{aj}(\bm{k}, \tau^{-}, \tau^{+};\bm{A})\mathcal{H}_{ja}(\bm{k}+\bm{q}, \tau;\bm{A}) \notag \\
    &-\sum_{a, j, \bm{k}}\mathcal{H}_{aj}(\bm{k}+\bm{q}, \tau;\bm{A})\mathcal{G}_{ja}(\bm{k}, \tau^{-}, \tau^{+};\bm{A}) \notag \\
  &+\sum_{a, j, \bm{k}}  \mathcal{F}_{aj}(\bm{k}, \tau^{-}, \tau^{+};\bm{A})\Delta^{+}_{ja}(\bm{k}, \tau;\bm{A}) 
  \notag \\
  &-\sum_{a, j, \bm{k}}\Delta_{aj}(\bm{k}, \tau;\bm{A})\mathcal{F}^{+}_{ja}(\bm{k}, \tau^{-}, \tau^{+};\bm{A}). \label{eq:fourth_term}
\end{align}
We used Eq.~(\ref{eq:EOM}) in the second equation. The first and second terms cancel each other after the summation over the index $a$ and $j$, while the third and fourth terms remain. When we only consider the dynamics of Bogoliubov quasiparticles but ignore the imaginary time dependence of the mean field $\Delta_{aj}(\bm{k})$, the third and fourth terms can be finite and the total particle number is not necessarily conserved. Therefore, the imaginary time dependence of the mean field is essential for the total particle number conservation in the superconducting state.

Here, we assume that the imaginary time dependence of the mean field is determined by the Kadanoff-Baym method,
\begin{align}
  \Delta^{+}_{ab}(\bm{k}, \tau; \bm{A}) &= -\frac{U}{V} \varphi^{\dagger}_{ab}(\bm{k})\sum \varphi_{dc}(\bm{k}^{\prime})\mathcal{F}^{+}_{cd}(\bm{k}^{\prime}, \tau^{+}, \tau^{-};\bm{A}), \label{eq:MF_1}\\
  \Delta_{ab}(\bm{k}, \tau; \bm{A}) &= -\frac{U}{V} \varphi_{ab}(\bm{k})\sum \varphi^{\dagger}_{dc}(\bm{k}^{\prime})\mathcal{F}_{cd}(\bm{k}^{\prime}, \tau^{+}, \tau^{-};\bm{A}). \label{eq:MF_2}
\end{align} 
Then, the third and fourth terms in 
Eq.~\eqref{eq:fourth_term} are rewritten as 
\begin{align}
  \sum \mathcal{F}_{aj}&(\bm{k}, \tau^{-}, \tau^{+};\bm{A})\Delta^{+}_{ja}(\bm{k}, \tau;\bm{A}) \notag\\
  =- \frac{U}{V} &\left[\sum \mathcal{F}_{aj}(\bm{k}, \tau^{-}, \tau^{+};\bm{A})\varphi^{\dagger}_{ja}(\bm{k})\right] \notag \\
  &\times\left[\sum \varphi_{dc}(\bm{k}^{\prime})\mathcal{F}^{+}_{cd}(\bm{k}^{\prime}, \tau^{+}, \tau^{-};\bm{A})\right], \label{eq:F_Delta}
\end{align}
\begin{align}
  \sum  \Delta_{aj}&(\bm{k}, \tau;\bm{A})\mathcal{F}^{+}_{ja}(\bm{k}, \tau^{-}, \tau^{+};\bm{A}) \notag \\
  = - \frac{U}{V}&\left[\sum \varphi^{\dagger}_{dc}(\bm{k}^{\prime})\mathcal{F}_{cd}(\bm{k}^{\prime}, \tau^{+}, \tau^{-};\bm{A})\right] \notag \\
  &\times\left[\sum \varphi_{aj}(\bm{k})\mathcal{F}^{+}_{ja}(\bm{k}, \tau^{-}, \tau^{+};\bm{A})\right]. \label{eq:Delta_F}
\end{align}
Using the relation $\mathcal{F}_{ab}(\bm{k}, \tau^{-}, \tau^{+};\bm{A})=\mathcal{F}_{ab}(\bm{k}, \tau^{+}, \tau^{-};\bm{A})$, we verify that the third and fourth terms in 
Eq.~\eqref{eq:fourth_term} cancel each other and the total particle number conservation law $\frac{d}{d\tau} \left<N(\tau)\right>_{\rm tot}=0$ is satisfied. We can also obtain the standard gap equation for superconductivity from the assumptions (\ref{eq:MF_1}) and (\ref{eq:MF_2}) when the external electric field is absent ($\bm{A}=\bm{0}$). Thus, Eqs.~\eqref{eq:MF_1} and \eqref{eq:MF_2} are expected to be appropriate as the mean-field pair potential under the external electric field.  In the next section, we derive the formula of the self-consistent response approximation from Eqs.~(\ref{eq:EOM_3}) and (\ref{eq:EOM4}).

\section{Self-consistent response approximation}
\label{sec:self-consistent}

We demonstrated that the total particle number conservation law is satisfied when we apply the Kadanoff-Baym method and consider the imaginary time dependence of the mean-field pair potential. Therefore, the self-consistent response approximation formulated below, which is derived from the equation of motion based on the Kadanoff-Baym method, is physically reasonable. Because the Green's function contains the effect of electric fields un-perturbatively, we can derive the formulas of linear and nonlinear responses that satisfy the total number conservation in any order of the electric field expansion. 
This section shows the derivation of the linear and second-order nonlinear response formulas based on the self-consistent response approximation. The vertex correction is taken into account through the electric field dependence of the mean-field pair potential.

\subsection{Derivation of electric current from grand partition function}

The time evolution of the electric current is formulated by the equation of motion based on the Kadanoff-Baym method. 
Following statistical mechanical relations, we derive the electric current from varying a parameter in the grand partition function. 
For consistency between the statistical mechanical relations and the Kadanoff-Baym method, 
the grand partition function should be introduced in an appropriate way. Following Baym's study, we show that the Luttinger-Ward approach satisfies consistency~\cite{Baym1962}. In the following, the particle number and the electric current are derived from the grand partition function. 

In the Luttinger-Ward approach~\cite{Luttinger1960, Baym1962}, $\ln Z$ is evaluated as an integral of the potential energy with respect to the coupling constant parameter $\lambda$,
\begin{align}
    \ln Z = \mathrm{tr} \ln [-\overleftrightarrow{\mathcal{G}}_{0}] + \int^{1}_{0}\frac{d\lambda}{2\lambda}\mathrm{tr}\overleftrightarrow{\Sigma}_{\lambda}\overleftrightarrow{\mathcal{G}}_{\lambda}, \label{eq:partition_appro}
\end{align}
where $\overleftrightarrow{\Sigma}_{\lambda}$ and $\overleftrightarrow{\mathcal{G}}_{\lambda}$ are the self-energy and the Green's function obtained with replacing the coupling constant $U \rightarrow \lambda U$. When we adopt an approximation for $\overleftrightarrow{\Sigma}_{\lambda}$ and $\overleftrightarrow{\mathcal{G}}_{\lambda}$, Eq.~(\ref{eq:partition_appro}) gives the corresponding approximation for $\ln Z$. After integration with respect to the coupling constant parameter $\lambda$, $\ln Z$ is represented as
\begin{align}
    \ln Z = - \left[\Phi - \mathrm{tr}\overleftrightarrow{\Sigma}\overleftrightarrow{\mathcal{G}} +\mathrm{tr} \ln \left(-\overleftrightarrow{\mathcal{G}}\right) \right],
\end{align}
where we introduce a functional $\Phi$ of $\overleftrightarrow{\mathcal{G}}$ such that 
\begin{align}
 \delta \Phi = \mathrm{tr}\overleftrightarrow{\Sigma}\delta\left(\overleftrightarrow{\mathcal{G}}\right ).
\end{align}

First, we show the relation between the total particle number $\left<N(\tau)\right>$ and the approximate $\ln Z$~\cite{Baym1962}. We consider the chemical potential dependent on the imaginary time $\tau$. The derivative of $\ln Z$ with respect to $\mu(\tau)$ leads to 
\begin{align}
    \frac{\partial \ln Z}{\partial \mu(\tau)} = -\left(\mathrm{tr}\overleftrightarrow{\Sigma}\frac{\partial \overleftrightarrow{\mathcal{G}}}{\partial \mu(\tau)} - \frac{\partial}{\partial \mu (\tau)}\mathrm{tr}\overleftrightarrow{\Sigma}\overleftrightarrow{\mathcal{G}} -\mathrm{tr}\frac{\partial \overleftrightarrow{\mathcal{G}}^{-1}}{\partial \mu(\tau)}\overleftrightarrow{\mathcal{G}}\right).
\end{align}
On the other hand, we have the relation
\begin{align}
    \frac{\partial \overleftrightarrow{\mathcal{G}}^{-1}(\tau_{1}, \tau_{2})}{\partial \mu(\tau)} &= \frac{\partial \overleftrightarrow{\mathcal{G}}_{0}^{-1}(\tau_{1}, \tau_{2})}{\partial \mu(\tau)} - \frac{\partial \overleftrightarrow{\Sigma}(\tau_{1}, \tau_{2})}{\partial \mu(\tau)} \\
    &= \delta(\tau_{1} - \tau)\delta(\tau_{2} - \tau) - \frac{\partial \overleftrightarrow{\Sigma}(\tau_{1}, \tau_{2})}{\partial \mu(\tau)},
\end{align}
and thus, 
\begin{align}
    \frac{\partial \ln Z}{\partial \mu(\tau)} = \mathrm{tr}\overleftrightarrow{\mathcal{G}}(\tau, \tau^{+}) =\left<N(\tau)\right>.
\end{align}
Thus, the statistical mechanical definition of the particle number is consistent with Eq.~\eqref{eq:particle_number}.
In the mean-field theory of superconducting states, the self-energy $\Sigma$ is given by the mean-field pair potential $\Delta_{ab}$ and $\Delta^{+}_{ab}$. When we apply the Kadanoff-Baym method to define the mean field, the total particle number $\left<N(\tau)\right>$ is conserved in the self-consistent approximation, as we showed in the previous section.

Next, we derive the electric current by considering the change of $\ln Z$ with varying the vector potential $\bm{A}(\tau)$ in $\mathcal{H}_{ab}(\bm{k}+\bm{q};\bm{A}(\tau))$, 
\begin{align}
  \left<J^{\alpha}(\tau; \bm{A})\right> 
  &=\frac{\partial \ln Z[\bm{A}]}{\partial A^{\alpha}(\tau)} \\
  &= \frac{1}{2} \sum \mathrm{Tr}\left[\overleftrightarrow{J^{\alpha}}_{ab}(\bm{k};\bm{A}(\tau))\overleftrightarrow{\mathcal{G}}_{ba}(\bm{k}, \tau^{-}, \tau^{+};\bm{A})\right].
\end{align}
We introduced the current operator,
\begin{align}
    \overleftrightarrow{J}^{\alpha_{1}\cdots\alpha_{n}}_{ab}(\bm{k};\bm{A}(\tau)) = 
    (-1)^{n}\left.\frac{\partial^{n}\overleftrightarrow{\mathcal{H}}^{(0)}_{ab}(\bm{k};\bm{A})}{\partial A^{\alpha_{1}}\cdots \partial A^{\alpha_{n}}}\right |_{\bm{A}=\bm{A}(\tau)},
\end{align}
with use of the following notation,
\begin{align}
    \overleftrightarrow{\mathcal{H}}^{(0)}_{ab}(\bm{k};\bm{A})=
    \begin{pmatrix}
    \mathcal{H}_{ab}(\bm{k}+ \bm{q}; \bm{A}) & 0 \\
    0 & -\mathcal{H}^{\top}_{ab}(-\bm{k}+ \bm{q}; \bm{A})
  \end{pmatrix}.
\end{align}
It should be noticed that the electric current $\left<J^{\alpha}(\tau; \bm{A})\right>$ depends on the vector potential through the explicit current operator's dependence on the instant vector potential $\bm{A}(\tau)$ and through the implicit dependence of the Green's function on the imaginary-time history of $\bm{A}(\tau)$.

\subsection{Linear response}
First, we formulate the self-consistent response approximation for the linear response functions. 
The vertex correction corresponding to the ladder approximation appears due to the dependence of the mean-field pair potential on the vector potential. 

From the vector potential dependence of the electric current $\left<J^{\alpha}(\tau; \bm{A})\right> $, 
the imaginary-time linear response function $\tilde{\chi}^{\alpha\beta}$ is defined as
\begin{align}
  \tilde{\chi}^{\alpha\beta}(\tau, \tau^{\prime}) = \frac{\delta   \left<J^{\alpha}(\tau; \bm{A})\right> }{\delta A^{\beta}(\tau^{\prime})}.
\end{align}
Here, we omit the index $a, b$ and rewrite the electric current for simplicity,
\begin{align}
  \left<J^{\alpha}(\bm{A}(\tau))\right>=\frac{1}{2}\mathrm{Tr}[J^{\alpha}(\tau)\mathcal{G}(\tau^{-},\tau^{+})].
\end{align}
From Eqs.~(\ref{eq:EOM_3}) and (\ref{eq:EOM4}), we obtain the relation
\begin{align}
  \frac{\delta\mathcal{G}(\tau, \tau^{\prime})}{\delta A^{\alpha}(\bar{\tau}_{1})} =& -\int d\tau_{1}d\tau_{2}\mathcal{G}(\tau, \tau_{1})\frac{\delta\mathcal{G}^{-1}(\tau_{1}, \tau_{2})}{\delta A^{\alpha}(\bar{\tau}_{1})}\mathcal{G}(\tau_{2}, \tau^{\prime}) \notag \\
  =&-\int d\tau_{1}d\tau_{2} \mathcal{G}(\tau, \tau_{1})\Gamma^{\alpha}(\tau_{1}, \tau_{2};\bar{\tau}_{1})\mathcal{G}(\tau_{2}, \tau^{\prime}) \label{eq:1-photon_vertex},
\end{align}
where the vertex function $\Gamma^{\alpha}$ is introduced as
\begin{align}
 \Gamma^{\alpha}(\tau_{1}, \tau_{2};\bar{\tau}_{1})\equiv \frac{\delta\mathcal{G}^{-1}(\tau_{1}, \tau_{2})}{\delta A^{\alpha}(\bar{\tau}_{1})}.
 \label{eq:vertex_def}
\end{align}
Thus, the linear response function $\tilde{\chi}^{\alpha\beta}$ is represented by the Green's function and the vertex function as
\begin{widetext}
\begin{align}
  \tilde{\chi}^{\alpha\beta}(\tau, \tau^{\prime}) 
  =& -\frac{1}{2} \sum \mathrm{Tr}\left[\overleftrightarrow{J}^{\alpha\beta}_{ab}(\bm{k})\overleftrightarrow{\mathcal{G}}_{ba}(\bm{k}^{\prime}, \tau^{-}, \tau^{+})]\delta(\tau - \tau^{\prime})\right] \notag \\
  &-\frac{1}{2} \sum \int d\tau_{1} \mathrm{Tr}\left[\overleftrightarrow{J}^{\alpha}_{ab}(\bm{k})\overleftrightarrow{\mathcal{G}}_{bi}(\bm{k}, \tau, \tau_{1})\overleftrightarrow{\Gamma}^{\beta}_{ij}(\bm{k}, \tau_{1};\tau^{\prime})\overleftrightarrow{\mathcal{G}}_{ja}(\bm{k}, \tau_{1}, \tau)\right].
  \label{eq:linear_chi_tau}
\end{align}
\end{widetext}

Here, we show 
a self-consistent equation that determines the vertex function.
By the definition of the vertex function Eq.~\eqref{eq:vertex_def}, we have 
\begin{align}
  &\overleftrightarrow{\Gamma}^{\alpha}_{ab}(\bm{k}, \tau_{1}, \tau_{2};\tau^{\prime}) \notag \\
  =&\left[\overleftrightarrow{J}^{\alpha}_{ab}(\bm{k})\delta(\tau_{2}- \tau^{\prime})-\frac{\delta\overleftrightarrow{\Delta}_{ab}(\bm{k}, \tau_{2};\bm{A})}{\delta A^{\alpha}(\tau^{\prime})}\right]\delta(\tau_{1} - \tau_{2}) 
  \label{eq:vertex_self}
  \\
  =&\overleftrightarrow{\Gamma}^{\alpha}_{ab}(\bm{k}, \tau_{2};\tau^{\prime}) \delta(\tau_{1} - \tau_{2}) \label{eq:vertex_Delta_relation}.
\end{align}
In the last line, a simplified notation of the vertex function is introduced and the $\tau_1$-dependence is omitted.
We define $\overleftrightarrow{\Delta}$ as
\begin{align}
  \overleftrightarrow{\Delta}_{ab}(\bm{k}, \tau;\bm{A}) = 
  \begin{pmatrix}
    0 & \Delta_{ab} \\
    \Delta^{+}_{ab} & 0
  \end{pmatrix}
  (\bm{k}, \tau; \bm{A}).
\end{align}
It can be rewritten as 
\begin{align}
  &\overleftrightarrow{\Delta}_{ab}(\bm{k}, \tau;\bm{A}) \notag \\
  =& -\frac{U}{V}\varphi_{ab}(\bm{k})\sum \varphi^{\dagger}_{dc}(\bm{k}^{\prime})\overleftrightarrow{\mathrm{P}_{+}}\overleftrightarrow{\mathcal{G}}_{cd}(\bm{k}^{\prime}, \tau^{+}, \tau^{-}; \bm{A})\overleftrightarrow{\mathrm{P}_{-}} \notag \\
  &-\frac{U}{V}\varphi_{ab}^{\dagger}(\bm{k})\sum \varphi_{dc}(\bm{k}^{\prime})\overleftrightarrow{\mathrm{P}_{-}}\overleftrightarrow{\mathcal{G}}_{cd}(\bm{k}^{\prime}, \tau^{+}, \tau^{-}; \bm{A})\overleftrightarrow{\mathrm{P}_{+}} \label{eq:Delta_Green},
\end{align}
where we use the matrices in the Nambu space such that
\begin{align}
  \overleftrightarrow{\mathrm{P}_{+}} =
  \begin{pmatrix}
    1 & 0 \\
    0 & 0
  \end{pmatrix},\quad   
  \overleftrightarrow{\mathrm{P}_{-}} =
  \begin{pmatrix}
    0 & 0 \\
    0 & 1
  \end{pmatrix}.
\end{align}
The mean-field pair potential $\overleftrightarrow{\Delta}$ depends on the vector potential only through the Green's function. 
We see that the vector potential derivative of the mean-field pair potential contributes to the vertex function in Eq.~\eqref{eq:vertex_self}. 
The self-consistent equation of the vertex function is obtained from Eqs.~(\ref{eq:1-photon_vertex}), (\ref{eq:vertex_Delta_relation}), and (\ref{eq:Delta_Green}).

\begin{figure}[htbp]
 \includegraphics[width=0.9\linewidth]{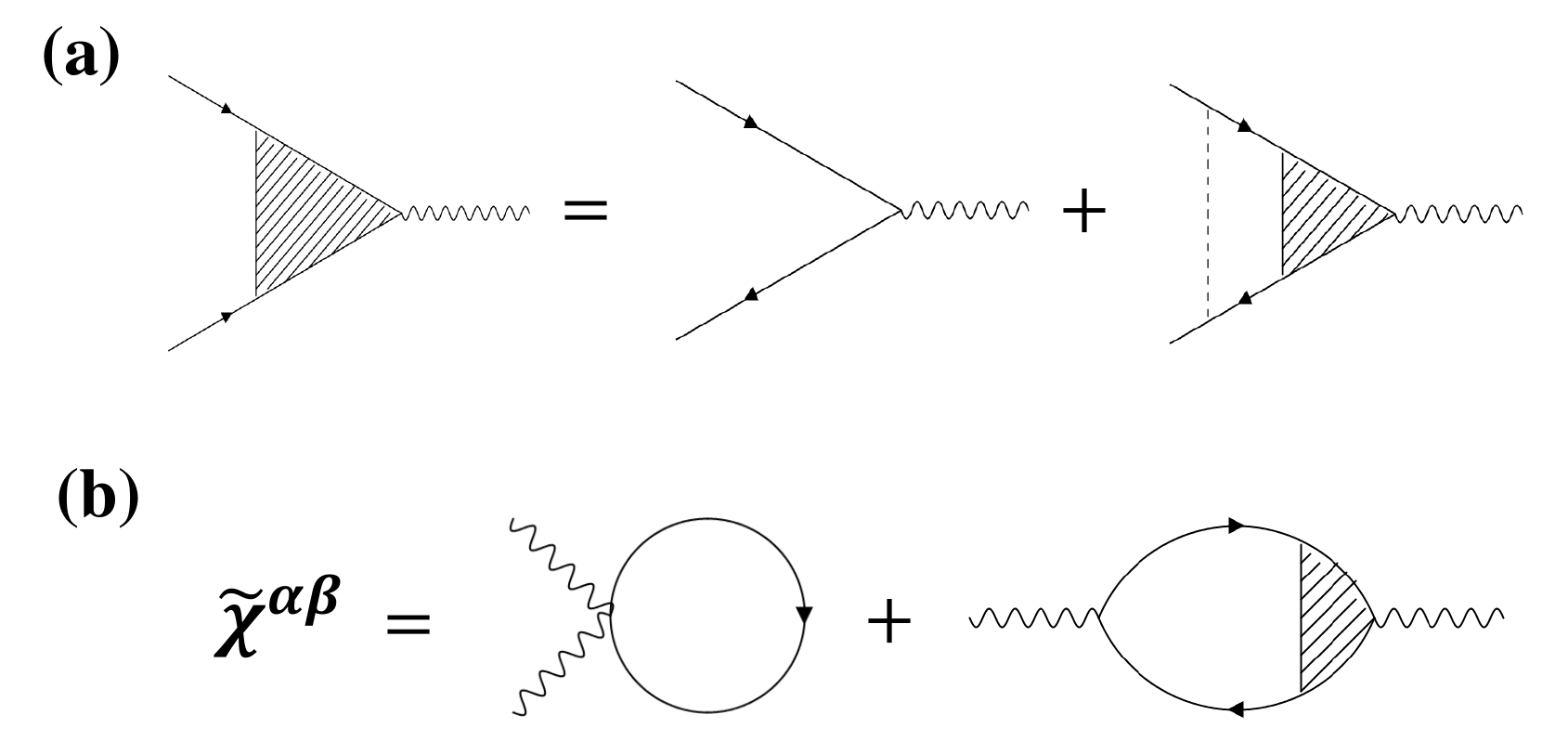}
 \caption{Feynman diagrams 
 of
 (a) the self-consistent equation for the vertex function and (b) the linear response function $\tilde{\chi}^{\alpha\beta}$.}
 \label{fig:sefl_eq_vertex}
\end{figure}

We can simply describe the self-consistent equation by introducing the superoperator $\overleftrightarrow{U}_{ab}(\bm{k}, \tau_{1})[X]$ acting on $X\equiv \overleftrightarrow{X}_{ab}(\bm{k},\tau)$ as, 
\begin{align}
  \overleftrightarrow{\Gamma}^{\alpha}_{ab}(\bm{k}, \tau_{1};\tau^{\prime}) = \overleftrightarrow{J}^{\alpha}_{ab}(\bm{k})\delta(\tau_{1}- \tau^{\prime}) + \overleftrightarrow{U}_{ab}(\bm{k}, \tau_{1})[\Gamma^{\alpha}(\tau^{\prime})]. \label{eq:self_eq_vertex}
\end{align}
The expression of the  superoperator $\overleftrightarrow{U}_{ab}(\bm{k}, \tau_{1})[X]$ is described in Appendix~\ref{App:def_of_superop}.
The Feynman diagram representing the self-consistent equation is depicted in Fig.~\ref{fig:sefl_eq_vertex}(a).  

Here, we introduce another superoperator that transforms the bare electric current operator to the vertex function.
We consider a self-consistent equation of an operator $Z$,
\begin{align}
  \overleftrightarrow{Z}_{ab}(\bm{k},\tau) = \overleftrightarrow{X}_{ab}(\bm{k}, \tau) +  \overleftrightarrow{U}_{ab}(\bm{k}, \tau)[Z]. \label{eq:Z_def}
\end{align}
When the solution $Z$ of the equation is uniquely determined, the $Z$ can be regarded as a function of $X$ such that
\begin{align}
  \overleftrightarrow{Z}_{ab}(\bm{k}, \tau) = \overleftrightarrow{Z}_{ab}(\bm{k}, \tau)[X].
\end{align}
Thus, the function $Z[X]$ is a superoperator, and the self-consistent equation of the vertex function is rewritten as
\begin{align}
  \overleftrightarrow{\Gamma}^{\alpha}_{ab}(\bm{k}, \tau_{1};\tau^{\prime}) = \overleftrightarrow{Z}_{ab}(\bm{k}, \tau_{1})[J^{\alpha}\delta(\tau_{1} - \tau^{\prime})]. \label{eq:self-consistent_by_Z}
\end{align}

Finally, we summarize the self-consistent response approximation for the linear response function.
The imaginary-time linear response function $\tilde{\chi}^{\alpha\beta}$ is given by Eq.~\eqref{eq:linear_chi_tau}.
The diagrammatic illustration of this formula is shown in Fig.~\ref{fig:sefl_eq_vertex}(b).
The vertex function $\Gamma$ is determined by the self-consistent equation [Fig.~\ref{fig:sefl_eq_vertex}(a)], which corresponds to the ladder approximation.
We can get the linear response function in the Matsubara frequency presentation $\tilde{\chi}^{\alpha\beta}(i\omega, i\omega^{\prime})$ after the Fourier transformation with respect to the imaginary time. We introduce a simplified notation of the linear response function,
\begin{align}
    \tilde{\chi}^{\alpha\beta}(i\omega, i\omega^{\prime}) = \tilde{\chi}^{\alpha\beta}(i\omega)\beta\delta(\omega - \omega^{\prime}),
\end{align} 
where we use the imaginary time translation symmetry to omit the $\omega^{\prime}$-dependence.

\subsection{Nonlinear response}

\begin{figure*}[htbp]
 \includegraphics[width=0.9\linewidth]{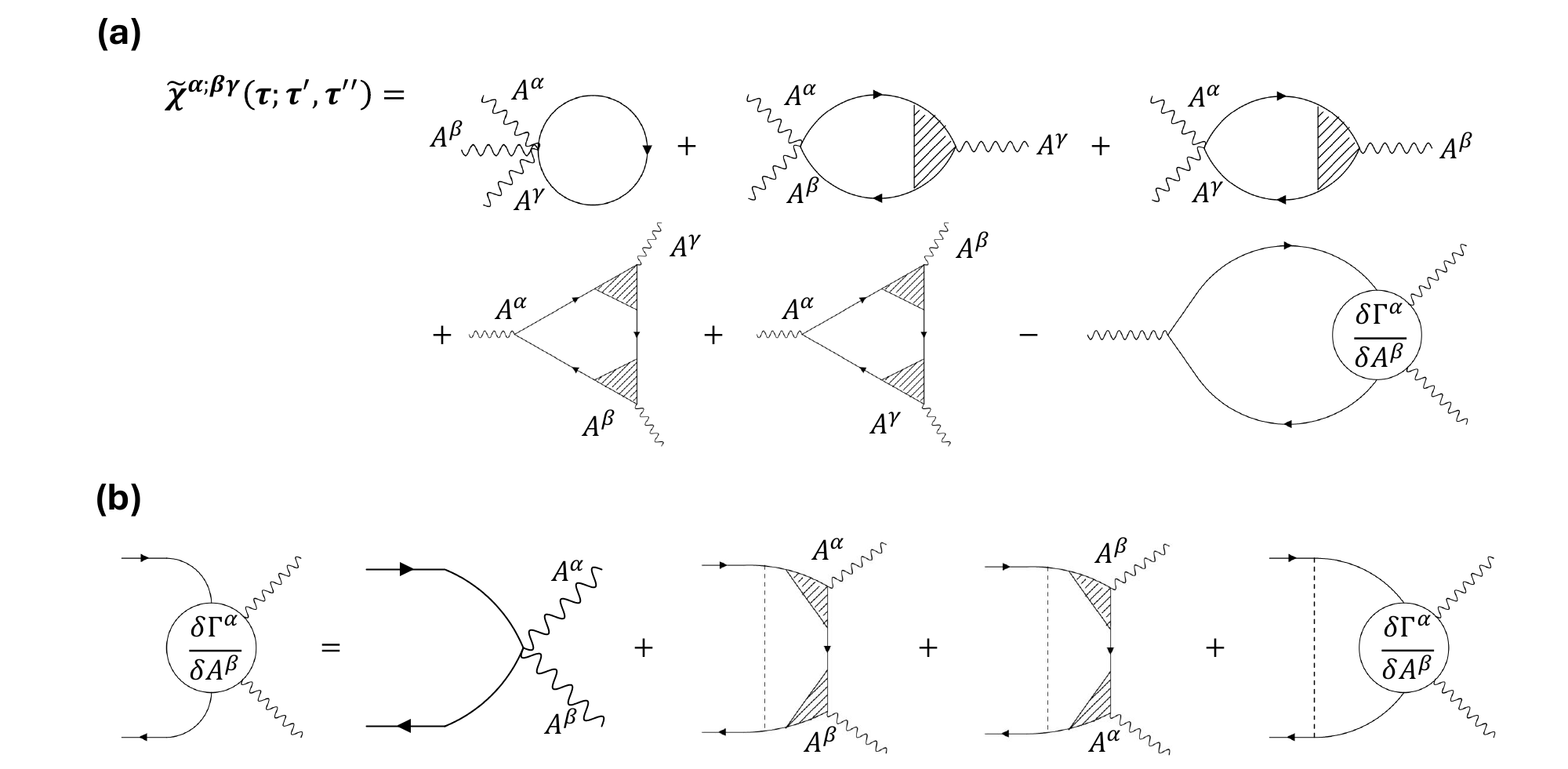}
 \caption{Feynman diagrams 
 of (a) the second-order nonlinear response function $\tilde{\chi}^{\alpha;\beta\gamma}$ and (b) the self-consistent equation for the two-photon vertex. }
 \label{fig:nonlinear_chi}
\end{figure*}

Next, we derive the self-consistent response approximation for the second-order nonlinear responses. The second-order nonlinear response function $\tilde{\chi}^{\alpha;\beta\gamma}$ is defined as 
\begin{align}
  \tilde{\chi}^{\alpha;\beta\gamma}(\tau;\tau^{\prime},\tau^{\prime\prime})\equiv \frac{1}{2!} \frac{\delta^{2}\left<J^{\alpha}(\tau)\right>}{\delta A^{\beta}(\tau^{\prime})\delta A^{\gamma}(\tau^{\prime\prime})}.
\end{align}
We differentiate Eq.~(\ref{eq:1-photon_vertex}) with the vector potential and obtain the following relation,
\begin{widetext}
\begin{align}
  \frac{\delta^{2} \mathcal{G}(\tau_{1}, \tau_{2})}{\delta A^{\alpha}(\tau)\delta A^{\beta}(\tau^{\prime})} 
  =& \int d\bar{\tau}_{1}d\bar{\tau}_{2}\mathcal{G}(\tau_{1}, \bar{\tau}_{1})\Gamma^{\alpha}(\bar{\tau}_{1};\tau)\mathcal{G}(\bar{\tau}_{1}, \bar{\tau}_{2})\Gamma^{\beta}(\bar{\tau}_{2};\tau^{\prime})\mathcal{G}(\bar{\tau}_{2},\tau_{2}) \notag\\
  &+\int d\bar{\tau}_{1}d\bar{\tau}_{2}\mathcal{G}(\tau_{1}, \bar{\tau}_{1})\Gamma^{\beta}(\bar{\tau}_{1};\tau^{\prime})\mathcal{G}(\bar{\tau}_{1}, \bar{\tau}_{2})\Gamma^{\alpha}(\bar{\tau}_{2};\tau)\mathcal{G}(\bar{\tau}_{2},\tau_{2}) 
  -\int d\bar{\tau}_{1}\mathcal{G}(\tau_{1}, \bar{\tau}_{1})\frac{\delta\Gamma^{\alpha}(\bar{\tau}_{1};\tau)}{\delta A^{\beta}(\tau^{\prime})}\mathcal{G}(\bar{\tau}_{1}, \tau_{2}).
\end{align}
Therefore, the nonlinear response function is derived as 
\begin{align}
  &\tilde{\chi}^{\alpha;\beta\gamma}(\tau;\tau^{\prime},\tau^{\prime\prime}) \notag\\
  =&\frac{1}{4}\mathrm{Tr}\left[J^{\alpha\beta\gamma}\mathcal{G}(\tau^{-},\tau^{+})\right]\delta(\tau-\tau^{\prime})\delta(\tau-\tau^{\prime\prime}) \notag \\
  &+\frac{1}{4}\int d\tau_{1} \mathrm{Tr}\left[J^{\alpha\beta}\mathcal{G}(\tau,\tau_{1})\Gamma^{\gamma}(\tau_{1};\tau^{\prime\prime})\mathcal{G}(\tau_{1},\tau)\right]\delta(\tau - \tau^{\prime}) +\frac{1}{4}\int d\tau_{1} \mathrm{Tr}\left[J^{\alpha\gamma}\mathcal{G}(\tau,\tau_{1})\Gamma^{\beta}(\tau_{1};\tau^{\prime})\mathcal{G}(\tau_{1},\tau)\right]\delta(\tau - \tau^{\prime\prime}) \notag\\
  &+\frac{1}{4}\int d\tau_{1}d\tau_{2}\mathrm{Tr}\left[J^{\alpha}\mathcal{G}(\tau,\tau_{1})\Gamma^{\beta}(\tau_{1};\tau^{\prime})\mathcal{G}(\tau_{1},\tau_{2})\Gamma^{\gamma}(\tau_{2};\tau^{\prime\prime})\mathcal{G}(\tau_{2},\tau)\right] \notag\\
  &+\frac{1}{4}\int d\tau_{1}d\tau_{2}\mathrm{Tr}\left[J^{\alpha}\mathcal{G}(\tau,\tau_{1})\Gamma^{\gamma}(\tau_{1};\tau^{\prime\prime})\mathcal{G}(\tau_{1},\tau_{2})\Gamma^{\beta}(\tau_{2};\tau^{\prime})\mathcal{G}(\tau_{2},\tau)\right] - \frac{1}{4}\int d\tau_{1} \mathrm{Tr}\left[J^{\alpha}\mathcal{G}(\tau,\tau_{1})\frac{\delta\Gamma^{\beta}(\tau_{1};\tau^{\prime})}{\delta A^{\gamma}(\tau^{\prime\prime})}\mathcal{G}(\tau_{1},\tau)\right]. \label{eq:nonlinear_chi_1}
\end{align}
\end{widetext}
We diagrammatically show the nonlinear response function in Fig.~\ref{fig:nonlinear_chi}(a). The last term contains $\delta\Gamma^{\beta}(\tau_{1};\tau^{\prime})/\delta A^{\gamma}(\tau^{\prime\prime})$ which is known as the two-photon vertex. The self-consistent equation for the two-photon vertex is depicted in Fig.~\ref{fig:nonlinear_chi}(b), which is derived by differentiating the self-consistent equation for the one-photon vertex [Eq.~\eqref{eq:self_eq_vertex}]. 
We can write the two-photon vertex by utilizing the superoperator $Z$,
\begin{widetext}
    \begin{align}
  \frac{\delta\overleftrightarrow{\Gamma}^{\alpha}_{ab}(\bm{k},\tau_{1};\tau)}{\delta A^{\beta}(\tau^{\prime})}
  =&\overleftrightarrow{Z}_{ab}(\bm{k}, \tau_{1})\left[\int d\bar{\tau}\left\{-U(\tau_{1}, \bar{\tau})[\Gamma^{\beta}(\tau_{1};\tau^{\prime})\mathcal{G}(\tau_{1}, \bar{\tau})\Gamma^{\alpha}(\bar{\tau};\tau)] \right\}\delta(\tau_{1}-\bar{\tau})\right] \notag \\
  &+\overleftrightarrow{Z}_{ab}(\bm{k}, \tau_{1})\left[\int d\bar{\tau}\left\{-U(\tau_{1}, \bar{\tau})[\Gamma^{\alpha}(\tau_{1};\tau)\mathcal{G}(\tau_{1}, \bar{\tau})\Gamma^{\beta}(\bar{\tau};\tau^{\prime})] \right\}\delta(\tau_{1}-\bar{\tau})\right] \notag \\
  &+ \overleftrightarrow{Z}_{ab}(\bm{k}, \tau_{1})\left[-J^{\alpha\beta}\delta(\tau_{1}-\tau)\delta(\tau_{1}-\tau^{\prime})\right].
\end{align}
\end{widetext}
Thus, we obtain the self-consistent equations for the one-photon and two-photon vertices. 
By solving the self-consistent equations, we can evaluate the vertices and calculate the nonlinear response function $\tilde{\chi}^{\alpha;\beta\gamma}$ [Eq.~\eqref{eq:nonlinear_chi_1}].

\begin{figure*}[htbp]
 \includegraphics[width=0.9\linewidth]{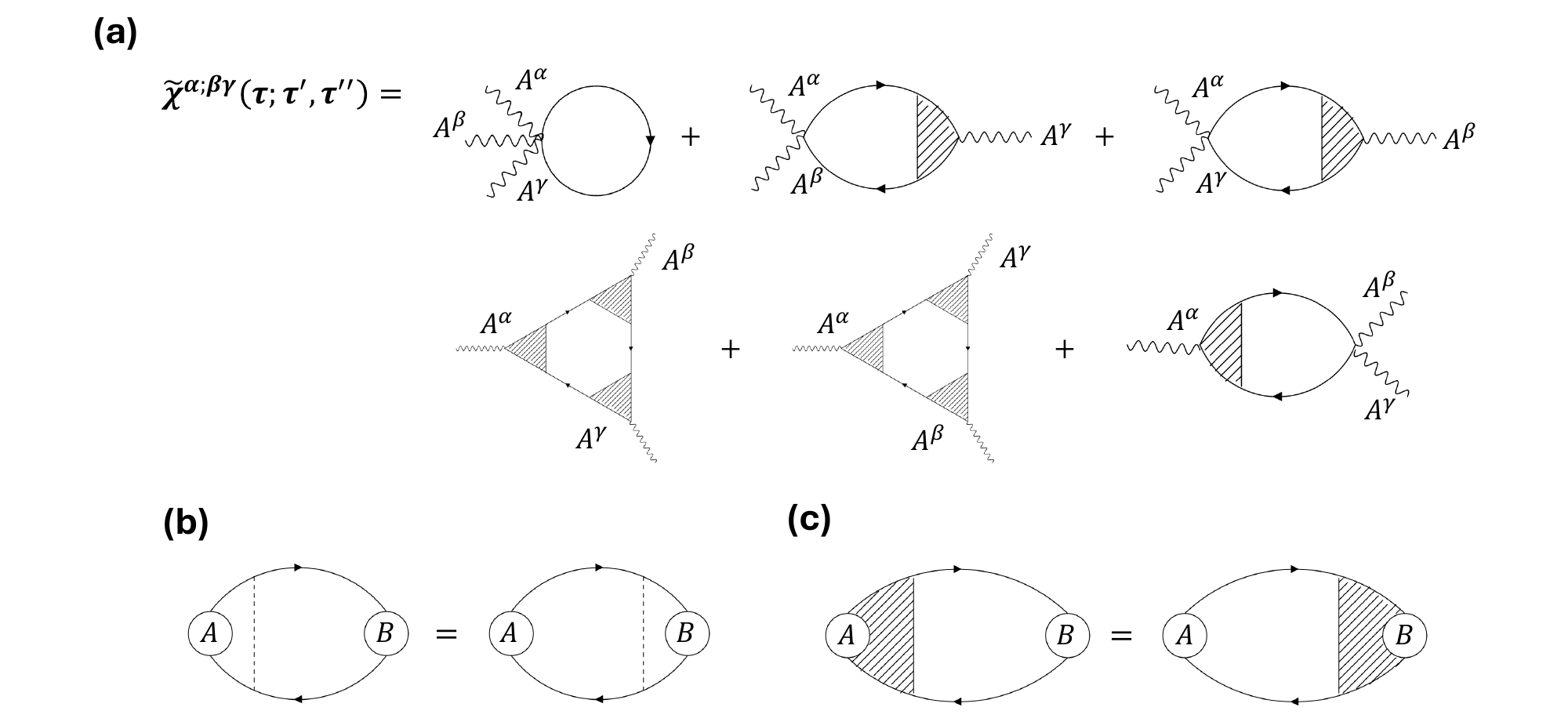}
 \caption{(a) Feynman diagrams which contribute to the second-order 
 nonlinear response function $\tilde{\chi}^{\alpha;\beta\gamma}$. (b,c) The diagrammatic representation of the properties of the superoperators (b) $U$ 
 and (c) $Z$.}
 \label{fig:nonlinear_chi2}
\end{figure*}

It is beneficial to rewrite the formula for the nonlinear current response function in a symmetric form, as illustrated in Fig.~\ref{fig:nonlinear_chi2}(a). Because the symmetric form does not contain the two-photon vertex explicitly, we do not have to solve the self-consistent equation for the two-photon vertex, and the computational task is significantly reduced. 
To derive the symmetric form, we use some useful properties of the superoperators $U$ and $Z$. 
First, the superoperators satisfy 
\begin{widetext}
\begin{align}
  &\sum\int d\tau_{1}d\tau_{2} \mathrm{Tr}\left[\overleftrightarrow{A}_{ab}(\bm{k}, \tau_{1})\overleftrightarrow{\mathcal{G}}_{bi}(\bm{k}, \tau_{1}, \tau_{2})\overleftrightarrow{U}_{ij}(\bm{k}, \tau_{2})[B]\overleftrightarrow{\mathcal{G}}_{ja}(\bm{k}, \tau_{2}, \tau_{1})\right] \notag\\
  &= \sum\int d\tau_{1}d\tau_{2}\mathrm{Tr}\left[\overleftrightarrow{U}_{ab}(\bm{k}, \tau_{1})[A]\overleftrightarrow{\mathcal{G}}_{bi}(\bm{k}, \tau_{1}, \tau_{2})\overleftrightarrow{B}_{ij}(\bm{k},\tau_{2})\overleftrightarrow{\mathcal{G}}_{ja}(\bm{k}, \tau_{2},\tau_{1})\right] \label{eq:U_relation},
\end{align}
\begin{align}
  &\sum\int d\tau_{1}d\tau_{2} \mathrm{Tr}\left[\overleftrightarrow{A}_{ab}(\bm{k}, \tau_{1})\overleftrightarrow{\mathcal{G}}_{bi}(\bm{k}, \tau_{1}, \tau_{2})\overleftrightarrow{Z}_{ij}(\bm{k}, \tau_{2})[B]\overleftrightarrow{\mathcal{G}}_{ja}(\bm{k}, \tau_{2}, \tau_{1})\right] \notag\\
  &= \sum\int d\tau_{1}d\tau_{2}\mathrm{Tr}\left[\overleftrightarrow{Z}_{ab}(\bm{k}, \tau_{1})[A]\overleftrightarrow{\mathcal{G}}_{bi}(\bm{k}, \tau_{1}, \tau_{2})\overleftrightarrow{B}_{ij}(\bm{k},\tau_{2})\overleftrightarrow{\mathcal{G}}_{ja}(\bm{k}, \tau_{2},\tau_{1})\right]. \label{eq:Z_relation}
\end{align}    
\end{widetext}
The above Eqs.~\eqref{eq:U_relation} and \eqref{eq:Z_relation} correspond to the diagrams in Figs.~\ref{fig:nonlinear_chi2}(b) and \ref{fig:nonlinear_chi2}(c), respectively. 
Second, the superoperators satisfy the linearity, 
\begin{align}
 \overleftrightarrow{U}_{ab}(\bm{k}, \tau)[X] + \overleftrightarrow{U}_{ab}(\bm{k}, \tau)[Y] = \overleftrightarrow{U}_{ab}(\bm{k}, \tau)[X + Y], \\
  \overleftrightarrow{Z}_{ab}(\bm{k},\tau)[A] + \overleftrightarrow{Z}_{ab}(\bm{k},\tau)[B] = \overleftrightarrow{Z}_{ab}(\bm{k},\tau)[A + B]. \label{eq:Z_linearity}
\end{align}
These properties of the superoperators are proved in Appendix~\ref{app:prop_superop}.

Using Eqs.~(\ref{eq:U_relation}) and (\ref{eq:Z_relation}), we can include a part of the two-photon vertex to correct the bare current operator. Then, we divide some terms and recollect them to construct the symmetric form using the linearity of $U$ and $Z$. Finally, we get the symmetric form of the second-order nonlinear response function,
\begin{align}
  &\tilde{\chi}^{\alpha;\beta\gamma}(\tau;\tau^{\prime},\tau^{\prime\prime}) = \notag \\
  &\frac{1}{2}\tilde{\chi}^{[\alpha\beta\gamma]}(\tau;\tau^{\prime},\tau^{\prime\prime})+\frac{1}{2}\tilde{\chi}^{[\alpha][\beta\gamma]}(\tau;\tau^{\prime},\tau^{\prime\prime}) \notag \\
  &+\tilde{\chi}^{[\alpha][\beta][\gamma]}(\tau;\tau^{\prime},\tau^{\prime\prime})
   + \tilde{\chi}^{[\alpha\beta][\gamma]}(\tau;\tau^{\prime},\tau^{\prime\prime}) \notag \\
   &+ \left(\beta\leftrightarrow \gamma, \quad \tau^{\prime} \leftrightarrow \tau^{\prime\prime}\right),
\end{align}
where we define the terms $\tilde{\chi}^{\cdots}(\cdots)$ as
\begin{widetext}
\begin{align}
  \tilde{\chi}^{[\alpha\beta\gamma]}(\tau;\tau^{\prime},\tau^{\prime\prime})=\frac{1}{4}\sum \mathrm{Tr}&\left[\overleftrightarrow{J}^{\alpha\beta\gamma}_{ab}(\bm{k})\overleftrightarrow{\mathcal{G}}_{ba}(\bm{k},\tau^{-},\tau^{+})\right]\delta(\tau^{\prime}-\tau^{\prime})\delta(\tau^{\prime}-\tau^{\prime\prime}),
\end{align}
\begin{align}
  \tilde{\chi}^{[\alpha][\beta\gamma]}(\tau;\tau^{\prime},\tau^{\prime\prime})=\frac{1}{4}\int d\tau_{1}\sum \mathrm{Tr}&\left[\overleftrightarrow{\Gamma}^{\alpha}_{ab}(\bm{k}, \tau_{1};\tau)\overleftrightarrow{\mathcal{G}}_{bi}(\bm{k},\tau_{1},\tau^{\prime})\overleftrightarrow{J}^{\beta\gamma}_{ij}(\bm{k})\overleftrightarrow{\mathcal{G}}_{ja}(\bm{k},\tau^{\prime},\tau_{1})\right]\delta(\tau^{\prime}-\tau^{\prime\prime}),
\end{align}
\begin{align}
  \tilde{\chi}^{[\alpha][\beta][\gamma]}(\tau;\tau^{\prime},\tau^{\prime\prime})=\frac{1}{4}\int d\tau_{1}d\tau_{2}d\tau_{3}&\sum \mathrm{Tr}\Big[\overleftrightarrow{\Gamma}^{\alpha}_{ab}(\bm{k}, \tau_{1};\tau)\overleftrightarrow{\mathcal{G}}_{bi}(\bm{k},\tau_{1},\tau_{2})\notag \\
   &\times\overleftrightarrow{\Gamma}^{\gamma}_{ij}(\bm{k},\tau_{2};\tau^{\prime\prime})\mathcal{G}_{jk}(\bm{k},\tau_{2}, \tau_{3})\overleftrightarrow{\Gamma}^{\beta}_{kl}(\bm{k},\tau_{3};\tau^{\prime})\overleftrightarrow{\mathcal{G}}_{la}(\bm{k},\tau_{3},\tau_{1})\Big], 
\end{align}
\begin{align}
  \tilde{\chi}^{[\alpha\beta][\gamma]}(\tau;\tau^{\prime},\tau^{\prime\prime})=\frac{1}{4}\int d\tau_{1}\sum \mathrm{Tr}&\left[\overleftrightarrow{J}^{\alpha\beta}_{ab}(\bm{k})\overleftrightarrow{\mathcal{G}}_{bi}(\bm{k},\tau,\tau_{1})\overleftrightarrow{\Gamma}^{\gamma}_{ij}(\bm{k},\tau_{1};\tau^{\prime\prime})\overleftrightarrow{\mathcal{G}}_{ja}(\bm{k},\tau_{1},\tau)\right]\delta(\tau-\tau^{\prime}).
\end{align}    
\end{widetext}
Note that the one-photon vertex function is obtained by solving the self-consistent equation, Eq.~\eqref{eq:self-consistent_by_Z}.
We get the nonlinear response function in the Matsubara frequency presentation $\tilde{\chi}^{\alpha;\beta\gamma}(i\omega;i\omega^{\prime},i\omega^{\prime\prime})$ after the Fourier transformation with respect to the imaginary time. Utilizing the imaginary time translation symmetry, we can simplify the nonlinear response function as,
\begin{align}
    \tilde{\chi}^{\alpha;\beta\gamma}(i\omega;i\omega^{\prime},i\omega^{\prime\prime}) = \tilde{\chi}^{\alpha;\beta\gamma}(i\omega^{\prime},i\omega^{\prime\prime})\beta\delta(\omega - \omega^{\prime} - \omega^{\prime\prime}).
\end{align}

\subsection{Analytic continuation of nonlinear response function}
To discuss the nonlinear optical responses in real time, we need the retarded nonlinear response functions. 
It is known that the real frequency representation of the retarded nonlinear response functions can be obtained by the analytic continuation of the response functions in the Matsubara frequency presentation~\cite{Michishita2021, Rostami2021}. In this subsection, we review the formalism of the analytic continuation. 

We consider the time evolution of the physical quantity $\hat{A}(t)$ under the external field $F(t)$ and assume a perturbative Hamiltonian 
\begin{align}
  \hat{V}(t) = \sum \hat{B}_{i}(t)F_{i}(t) + \sum \hat{C}_{ij}(t)F_{i}(t)F_{j}(t) + \cdots,
\end{align}
where the operators are given in the interaction representation. The expectation value of $\hat{A}(t)$ is calculated by the density matrix,
\begin{align}  
A(t)=\mathrm{Tr}\left[\hat{\rho}(t)\hat{A}(t)\right],
\end{align}
and it is expanded with respect to the external fields,
\begin{align}
  &A(t) =  \left<\hat{A}(t)\right> + \sum_{i}\int^{\infty}_{-\infty}dt_{1}\chi_{AB_{i}}(t_{1})F_{i}(t-t_{1}) \notag\\
  &+\sum_{i,j}\int^{\infty}_{-\infty}dt_{1}\int^{\infty}_{-\infty}dt_{2}\chi_{AB_{i}B_{j}}(t_{1},t_{2})F_{i}(t-t_{1})F_{j}(t-t_{2}) \notag \\
  &+ \sum_{i,j}\int^{\infty}_{-\infty}dt_{1}\chi_{AC_{ij}}(t_{1})F_{i}(t-t_{1})F_{j}(t-t_{1}) + \cdots.
\end{align}
The response functions are defined as
\begin{align}
  \chi_{AB_{i}}(t_{1})=& i\Theta(t_{1})\left<\left[\Delta\hat{B}_{i}(-t_{1}),\Delta\hat{A}(0)\right]\right>, \\
  \chi_{AB_{i}B_{j}}(t_{1},t_{2})=& \frac{i^{2}}{2!}\sum_{\mathcal{P}}\Theta(t_{2}-t_{1})\Theta(t_{1}) \notag \\
  &\times\left<\left[\Delta\hat{B}_{j}(-t_{2}),\left[\Delta\hat{B}_{i}(-t_{1}),\Delta\hat{A}(0)\right]\right]\right>, \\
  \chi_{AC_{ij}}(t_{1})=& \frac{i}{2!}\sum_{\mathcal{P}}\Theta(t_{1})\left<\left[\Delta\hat{C}_{ij}(-t_{1}),\Delta\hat{A}(0)\right]\right>.
\end{align}
The notation $\left<\cdots\right>$ describes the expectation value under the canonical distribution of an unperturbed Hamiltonian, and $\Delta\hat{X}(t) \equiv \hat{X}(t) - \left<\hat{X}(t)\right>$. The symbol $\sum_{\mathcal{P}}$ stands for summation over all permutations among the dummy variables such as 
$(B_{i}, t_{1})$ and $(B_{j}, t_{2})$. 

The linear and nonlinear response functions of the electric current are defined by 
\begin{align}
  &J^{\alpha}(t) = \left<\hat{J}^{\alpha}(t)\right> 
  + \sum_{\beta}\int^{\infty}_{-\infty}dt_{1}\chi^{\alpha\beta}(t_{1})A^{\beta}(t-t_{1}) \notag\\
  &+\sum_{\beta,\gamma}\int^{\infty}_{-\infty}dt_{1}dt_{2}\chi^{\alpha;\beta\gamma}(t_{1},t_{2})A^{\beta}(t-t_{1})A^{\gamma}(t-t_{2}) + \cdots.
\end{align}
We can calculate the current response functions $\chi^{\alpha\beta}(t_{1})$ and $\chi^{\alpha;\beta\gamma}(t_{1},t_{2})$ by replacing the operators $\hat{A}$, $\hat{B}$, and $\hat{C}$ in the previous paragraph with $\hat{J}^{\alpha}$, $\hat{J}^{\alpha\beta}$, and $\hat{J}^{\alpha\beta\gamma}$. We proceed to review the general formalism and later discuss the current responses.

Here, we consider the three-body response function $\chi_{AB_{i}B_{j}}(t_{1},t_{2})$ and discuss the relationship with the counterpart in the imaginary time representation.
First, we perform the Fourier transformation with respect to the real time $t_{i}$,
\begin{align}
  \chi_{AB_{i}B_{j}}(\omega_{1},\omega_{2}) & = \notag \\ \frac{1}{2!}\sum_{\mathcal{P}}  \sum_{\lambda_{1}\lambda_{2}\lambda_{3}} &\frac{\Delta A_{\lambda_{1}\lambda_{3}}\Delta B_{j,\lambda_{3}\lambda_{2}}\Delta B_{i,\lambda_{2}\lambda_{1}}}{\omega_{1} + \omega_{2} + E_{\lambda_{1}\lambda_{3}} + i(\eta_{1} + \eta_{2})} \notag\\
  \times & \left[\frac{P_{\lambda_{1}\lambda_{2}}}{\omega_{1} + E_{\lambda_{1}\lambda_{2}} + i\eta_{1}} - \frac{P_{\lambda_{2}\lambda_{3}}}{\omega_{2} + E_{\lambda_{2}\lambda_{1}} + i\eta_{2}} \right],\label{eq:real_frequency}
\end{align}
where $E_{\lambda_{1}\lambda_{2}} = E_{\lambda_{1}} - E_{\lambda_{2}}$, $P_{\lambda_{1}\lambda_{2}} = P_{\lambda_{1}} - P_{\lambda_{2}}$, $\Delta A_{\lambda_{1}\lambda_{3}} = \bra{\Lambda_{1}}\Delta\hat{A}(0)\ket{\Lambda_{3}}$, and $P_{\lambda_{1}} = e^{-\beta E_{\lambda_{1}}}/\sum_{\lambda}e^{-\beta E_{\lambda}}$.
The infinitesimal positive parameters $\eta_{1}$ and $\eta_{2}$ appear due to the causality of the response functions. 
Next, we derive the spectral representation of the response functions in the Matsubara frequency domain. The imaginary time evolution of the expectation value $\left<A(\tau)\right>_{\mathrm{tot}}$ is defined as Eqs.~(\ref{eq:c_tau_def}-\ref{eq:tot_expectation}), and we have
\begin{align}
  &\left<A(\tau)\right>_{\mathrm{tot}} 
  = \left<\hat{A}(\tau)\right> + \sum_{i}\int^{\beta}_{0}d\tau_{1}\tilde{\chi}_{AB_{i}}(\tau_{1})F_{i}(\tau - \tau_{1})\notag\\
  & + \sum_{i,j} \int^{\beta}_{0}d\tau_{1}\int^{\beta}_{0}d\tau_{2}\tilde{\chi}_{AB_{i}B_{j}}(\tau_{1},\tau_{2})F_{i}(\tau - \tau_{1})F_{j}(\tau - \tau_{2})  \notag \\
  & + \sum_{i,j}\int^{\beta}_{0}d\tau_{1}\tilde{\chi}_{AC_{ij}}(\tau_{1})F_{i}(\tau - \tau_{1})F_{j}(\tau - \tau_{1})+ \cdots,
\end{align}
where the imaginary time-ordered response functions are defined as 
\begin{align}
  \tilde{\chi}_{AB_{i}}(\tau_{1}) &= -\left<T\left[\Delta\hat{B}_{i}(-\tau_{1})\Delta\hat{A}(0)\right]\right>,  \\
  \tilde{\chi}_{AB_{i}B_{j}}(\tau_{1}, \tau_{2}) &= \frac{1}{2!}\left<T\left[\Delta\hat{B}_{i}(-\tau_{1})\Delta\hat{B}_{j}(-\tau_{2})\Delta\hat{A}(0)\right]\right>, \\
  \tilde{\chi}_{AC_{ij}}(\tau_{1}) & = \frac{1}{2!}\sum_{\mathcal{P}}-\left<T\left[\Delta\hat{C}_{ij}(-\tau_{1})\Delta\hat{A}(0)\right]\right>.
\end{align}
The spectral representation of $\tilde{\chi}_{AB_{i}B_{j}}$ in the Matsubara frequency domain is obtained as
\begin{align}
  &\tilde{\chi}_{AB_{i}B_{j}}(i\omega_{1}, i\omega_{2}) = \notag \\
  & \frac{1}{2!}\sum_{\mathcal{P}}\frac{A_{\lambda_{1}\lambda_{3}}B_{2,\lambda_{3}\lambda_{2}}B_{1,\lambda_{2}\lambda_{1}}}{i(\omega_{1}+\omega_{2}) + E_{\lambda_{1}\lambda_{3}}}\Bigg\{ \frac{P_{\lambda_{1}\lambda_{2}}}{i\omega_{1} + E_{\lambda_{1}\lambda_{2}}} - \frac{P_{\lambda_{2}\lambda_{3}}}{i\omega_{2} + E_{\lambda_{2}\lambda_{3}}} \Bigg\}. \label{eq:imaginary_freq}
\end{align}
Comparing Eqs.~(\ref{eq:real_frequency}) and (\ref{eq:imaginary_freq}), we see the similarity between the spectral representation of $\tilde{\chi}_{AB_{i}B_{j}}$ and the retarded response function $\chi_{AB_{i}B_{j}}$.
By performing the analytic continuation $i\omega_{i}\rightarrow \omega_{i} + i\eta_{i}$, we obtain the physical response function in the real frequency domain,
\begin{align}
  &\tilde{\chi}_{AB_{i}B_{j}}(i\omega_{1}, i\omega_{2})\notag \\
  &\rightarrow\tilde{\chi}_{AB_{i}B_{j}}(\omega_{1} + i\eta_{1}, \omega_{2} + i\eta_{2}) = \chi_{AB_{i}B_{j}}(\omega_{1}, \omega_{2}).
\end{align}
The other physical response functions such as $\chi_{AB_{i}}$ and $\chi_{AC_{ij}}$ in the real frequency domain can also be obtained by analytic continuation. 

Applying the above framework of analytic continuation to the current response functions in the Matsubara frequency domain, the linear and nonlinear current response functions are obtained as
\begin{align}
    \chi^{\alpha\beta}(\omega) = \tilde{\chi}^{\alpha\beta}(\omega + i\eta),
\end{align}
and
\begin{align}
    \quad \chi^{\alpha;\beta\gamma}(\omega_{1}, \omega_{2}) = \tilde{\chi}^{\alpha;\beta\gamma}(\omega_{1} + i\eta_{1}, \omega_{2} + i\eta_{2}),
\end{align}
respectively.
When we calculate the linear and nonlinear conductivity, we replace $\eta_{i}$ with a finite phenomenological scattering parameter $\gamma$ for numerical convergence. Replacing $\eta_{i} \rightarrow \gamma$ is reasonable to obtain physically meaningful results in the finite frequency region when $\gamma$ is sufficiently small ~\cite{Watanabe2022, Passos2018}. In the following calculation of nonlinear conductivity, we set the phenomenological parameters $\eta_{1}=\eta_{2}=\gamma$. 

Finally, we express the current response functions via the Green's functions. 
The linear current response function can be expressed 
as
\begin{widetext}
\begin{align}
    \chi^{\alpha\beta}(\omega )  
    =& \frac{1}{2}\sum_{\bm{k}}\int^{\infty}_{-\infty} \frac{d\tilde{\omega}}{2\pi i}f(\tilde{\omega})\mathrm{Tr}\Big[\overleftrightarrow{J}^{\alpha\beta}_{ab}(\bm{k})\left(\overleftrightarrow{G}^{R}_{ba}(\bm{k}, \tilde{\omega})-\overleftrightarrow{G}^{A}_{ba}(\bm{k}, \tilde{\omega})\right) \notag \\
    &+\overleftrightarrow{J}^{\alpha}_{ab}(\bm{k})\overleftrightarrow{G}^{R}_{bi}(\bm{k}, \tilde{\omega}+\omega +i\gamma)\overleftrightarrow{\Gamma}^{\beta}_{ij}(\bm{k}, \omega +i\gamma)\Big\{\overleftrightarrow{G}^{R}_{ja}(\bm{k}, \tilde{\omega}) - \overleftrightarrow{G}^{A}_{ja}(\bm{k}, \tilde{\omega})\Big\} \notag \\
    &+ \overleftrightarrow{J}^{\alpha}_{ab}(\bm{k})\Big\{\overleftrightarrow{G}^{R}_{bi}(\bm{k}, \tilde{\omega}) - \overleftrightarrow{G}^{A}_{bi}(\bm{k}, \tilde{\omega})\Big\}\overleftrightarrow{\Gamma}^{\beta}_{ij}(\bm{k}, \omega +i\gamma)\overleftrightarrow{G}^{A}_{ja}(\bm{k}, \tilde{\omega}-\omega -i\gamma)\Big],
    \label{eq:linear_response_green}
\end{align}
\end{widetext}
where $f(\omega)$ is the Fermi distribution function. The retarded and advanced Green's functions are defined as $\overleftrightarrow{G}^{R}(\bm{k}, z ) = 1/\left[z - \overleftrightarrow{\mathcal{H}} + i\delta \right]$ and $\overleftrightarrow{G}^{A}(\bm{k}, z ) = 1/\left[z - \overleftrightarrow{\mathcal{H}} - i\delta \right]$, respectively, by taking the limit $\delta \rightarrow +0$. When we transformed the Matsubara frequency summation into the real frequency integral, we used the relation $\oint_{C}\frac{d\omega}{2\pi i}f(\omega)A(\omega)=-\frac{1}{\beta}\sum_{n}A(i\omega_{n})$, where $\oint_{C}$ represents the integral path around the poles of the Fermi distribution function~\cite{Michishita2021}.

The second-order nonlinear response functions are obtained as
\begin{widetext}
\begin{align}
  \chi^{\alpha;\beta\gamma}(\omega^{\prime},\omega^{\prime\prime})
  =\sum_{\bm{k}}\Big[&\frac{1}{2}\chi^{[\alpha\beta\gamma]}(\bm{k},\omega^{\prime},\omega^{\prime\prime}) +\frac{1}{2}\chi^{[\alpha][\beta\gamma]}(\bm{k}, \omega^{\prime},\omega^{\prime\prime}) + \chi^{ [\alpha\beta][\gamma]}(\bm{k}, \omega^{\prime},\omega^{\prime\prime})  \notag \\
  &+\chi^{[\alpha][\beta][\gamma]}(\bm{k},\omega^{\prime},\omega^{\prime\prime})+ (\beta\leftrightarrow\gamma,\quad \omega^{\prime}\leftrightarrow\omega^{\prime\prime})\Big],
\end{align}
where 
we defined
\begin{align}
  \chi^{[\alpha\beta\gamma]}(\bm{k}, \omega^{\prime},\omega^{\prime\prime})= -\frac{1}{2}\int^{\infty}_{-\infty}\frac{d\tilde{\omega}}{2\pi i}f(\tilde{\omega})\mathrm{Tr}&\left[\overleftrightarrow{J}^{\alpha\beta\gamma}_{ab}(\bm{k})\left\{\overleftrightarrow{G}^{R}_{ba}(\bm{k},\tilde{\omega}) - \overleftrightarrow{G}^{A}_{ba}(\bm{k},\tilde{\omega})\right\}\right],
\end{align}
\begin{align}
  \chi^{[\alpha\beta][\gamma]}(\bm{k}, \omega^{\prime},\omega^{\prime\prime})
  =&-\frac{1}{2}\int^{\infty}_{-\infty}\frac{d\tilde{\omega}}{2\pi i}f(\tilde{\omega}) \mathrm{Tr}\Big[\overleftrightarrow{J}^{\alpha\beta}_{ab}(\bm{k})\overleftrightarrow{G}^{R}_{bi}(\bm{k},\tilde{\omega} + \omega^{\prime\prime}+i\gamma)\overleftrightarrow{\Gamma}^{\gamma}_{ij}(\bm{k},\omega^{\prime\prime}+i\gamma)\overleftrightarrow{G}^{RA}_{ja}(\bm{k},\tilde{\omega}) \notag \\
  &+\overleftrightarrow{J}^{\alpha\beta}_{ab}(\bm{k})\overleftrightarrow{G}^{RA}_{bi}(\bm{k},\tilde{\omega})\overleftrightarrow{\Gamma}^{\gamma}_{ij}(\bm{k},\omega^{\prime\prime}+i\gamma)\overleftrightarrow{G}^{A}_{ja}(\bm{k},\tilde{\omega} - \omega^{\prime\prime}-i\gamma)\Big],
\end{align}
\begin{align}
  \chi^{[\alpha][\beta][\gamma]}(\bm{k}, \omega^{\prime},\omega^{\prime\prime}) 
  &= -\frac{1}{2}\int^{\infty}_{-\infty}\frac{d\tilde{\omega}}{2\pi i}f(\tilde{\omega})\mathrm{Tr}\Big[\overleftrightarrow{\Gamma}^{\alpha}_{ab}(\bm{k}, -\omega-2i\gamma) \notag \\
  \times&\Big\{\overleftrightarrow{G}^{R}_{bi}(\bm{k},\tilde{\omega} + \omega + 2i\gamma)\overleftrightarrow{\Gamma}^{\beta}_{ij}(\bm{k},\omega^{\prime}+i\gamma)\overleftrightarrow{G}^{R}_{jk}(\bm{k},\tilde{\omega} + \omega^{\prime\prime}+i\gamma)\overleftrightarrow{\Gamma}^{\gamma}_{kl}(\bm{k},\omega^{\prime\prime}+\gamma)\overleftrightarrow{G}^{RA}_{la}(\bm{k},\tilde{\omega}) \notag \\
  &+\overleftrightarrow{G}^{R}_{bi}(\bm{k},\tilde{\omega} + \omega^{\prime} + i\gamma)\overleftrightarrow{\Gamma}^{\beta}_{ij}(\bm{k},\omega^{\prime}+i\gamma)\overleftrightarrow{G}^{RA}_{jk}(\bm{k},\tilde{\omega})\overleftrightarrow{\Gamma}^{\gamma}_{kl}(\bm{k},\omega^{\prime\prime}+i\gamma)\overleftrightarrow{G}^{A}_{la}(\bm{k},\tilde{\omega}-\omega^{\prime\prime}-i\gamma) \notag \\
    &+\overleftrightarrow{G}^{RA}_{bi}(\bm{k},\tilde{\omega})\overleftrightarrow{\Gamma}^{\beta}_{ij}(\bm{k},\omega^{\prime}+i\gamma)\overleftrightarrow{G}^{A}_{jk}(\bm{k},\tilde{\omega}-\omega^{\prime}-i\gamma)\overleftrightarrow{\Gamma}^{\gamma}_{kl}(\bm{k},\omega^{\prime\prime}+i\gamma)\overleftrightarrow{G}^{A}_{la}(\bm{k},\tilde{\omega}-\omega-2i\gamma)\Big\}\Big] ,
\end{align}
\begin{align}
  \chi^{[\alpha][\beta\gamma]}(\bm{k}, \omega^{\prime},\omega^{\prime\prime})
  =&-\frac{1}{2}\int^{\infty}_{-\infty}\frac{d\tilde{\omega}}{2\pi i}f(\tilde{\omega}) \mathrm{Tr}\Big[\overleftrightarrow{\Gamma}^{\alpha}_{ab}(\bm{k}, -\omega-2i\gamma)\overleftrightarrow{G}^{R}_{bi}(\bm{k},\tilde{\omega}+ \omega + 2i\gamma)\overleftrightarrow{J}^{\beta\gamma}_{ij}(\bm{k})\overleftrightarrow{G}^{RA}_{ja}(\bm{k},\tilde{\omega}) \notag \\
  &+\overleftrightarrow{\Gamma}^{\alpha}_{ab}(\bm{k}, -\omega-2i\gamma)\overleftrightarrow{G}^{RA}_{bi}(\bm{k},\tilde{\omega})\overleftrightarrow{J}^{\beta\gamma}_{ij}(\bm{k})\overleftrightarrow{G}^{A}_{ja}(\bm{k},\tilde{\omega}- \omega-2i\gamma)\Big].
  \label{eq:nonlinear_response_green}
\end{align}
\end{widetext}
We adopt the notation $\overleftrightarrow{G}^{RA} = \overleftrightarrow{G}^{R} - \overleftrightarrow{G}^{A}$ and $\omega = \omega^{\prime} + \omega^{\prime\prime}$ for simplicity of the formulas.
We also derive the one-photon vertex function in the real frequency domain and summarize the derivation and results in Appendix~\ref{app:ana_cont_vertex}. By using the above formulas, Eqs.~\eqref{eq:linear_response_green}-\eqref{eq:nonlinear_response_green}, the linear and nonlinear current response functions can be calculated by using the vertex function and the Green's function.

\subsection{Linear and nonlinear conductivity in the band representation} \label{sec:formulation_conductivity}

In this subsection, we show the linear and nonlinear conductivity in the band representation.
We consider the diagonalization of the BdG Hamiltonian,
\begin{align}
  \left(\overleftrightarrow{U}^{\dagger}(\bm{k})\right)_{ai}\left(\overleftrightarrow{\mathcal{H}}(\bm{k})\right)_{ij}\left(\overleftrightarrow{U}(\bm{k})\right)_{jb} = E_{a}(\bm{k})\delta_{ab},
\end{align}
where $1\leq a, b \leq 2n$ are the indices for the energy band of Bogoliubov quasiparticles, and $i,j$ mean the fermion's internal degrees of freedom in the Nambu space. 
In the following, we use the notation $\tilde{A}_{ab}$ as the band representation of $\overleftrightarrow{A}$,
\begin{align}
  \tilde{A}_{ab}(\bm{k}) = \left(\overleftrightarrow{U}^{\dagger}(\bm{k})\right)_{ai}\left(\overleftrightarrow{A}(\bm{k})\right)_{ij}\left(\overleftrightarrow{U}(\bm{k})\right)_{jb}.
\end{align}
Under the same unitary transformation, the Green's functions in the band presentation are given by
\begin{align}
  \tilde{G}^{R}_{ab}(\bm{k}, \omega)  &= \left(\omega  - E_{a}(\bm{k}) + i\delta \right)^{-1}\delta_{ab},  \\
    \tilde{G}^{A}_{ab}(\bm{k}, \omega)  &= \left(\omega  - E_{a}(\bm{k}) - i\delta \right)^{-1}\delta_{ab}.
\end{align}
Because $\overleftrightarrow{G}^{R} - \overleftrightarrow{G}^{A}$ 
in this representation is proportional to a delta function 
in the limit $\delta \rightarrow +0$, we can simplify the equations by 
\begin{align}
  &\int^{\infty}_{-\infty}\frac{d\tilde{\omega}}{2\pi i}\mathrm{Tr}\left[\overleftrightarrow{A}_{ij}(\tilde{\omega}, \bm{k})\left\{\overleftrightarrow{G}^{R}_{ji}(\bm{k}, \tilde{\omega}) - \overleftrightarrow{G}^{A}_{ji}(\bm{k}, \tilde{\omega})\right\}\right] \notag \\
  =& -\sum_{a}\tilde{A}_{aa}(\tilde{\omega}=E_{a}(\bm{k}), \bm{k}). \label{eq:clean_limit}
\end{align}

When the velocity gauge is adopted, the electric field is given by $ {\bm E}(\omega) = i \omega {\bm A}(\omega)$, and the linear conductivity $\sigma^{\alpha\beta}(\omega)$ is obtained by the linear current response function,
\begin{align}
  \sigma^{\alpha\beta}(\omega) = \frac{\chi^{\alpha\beta}(\omega)}{i\omega}.
\end{align}
Calculating Eq.~\eqref{eq:linear_response_green} by using Eq.~\eqref{eq:clean_limit}, we obtain the linear conductivity
\begin{align}
  \sigma^{\alpha\beta}(\omega) = \frac{i}{2\omega}\sum_{\bm{k}}\left[f_{a}\tilde{J}^{\alpha\beta}_{aa} + \frac{\tilde{J}^{\alpha}_{ab}\tilde{\Gamma}^{\beta}_{ba}(\omega +i\gamma)f_{ab}}{\omega - E_{ba} +i\gamma}\right], \label{eq:linear_conductivity}
\end{align}
where we define $f_{a}=f(E_{a}(\bm{k}))$, $f_{ab} = f_{a} - f_{b}$, and $E_{ab} = E_{a}(\bm{k}) - E_{b}(\bm{k})$ and omit the $\bm{k}$ description for simplicity.
We also obtain the second-order nonlinear conductivity,
\begin{widetext}
\begin{align}
  &\sigma^{\alpha;\beta\gamma}(\omega;\omega^{\prime},\omega^{\prime\prime}) \notag \\
  =& -\frac{1}{2\omega^{\prime}\omega^{\prime\prime}}\Bigg[\sum_{\bm{k}}f_{a}\tilde{J}^{\alpha\beta\gamma}_{aa}  +\sum_{\bm{k}}\frac{\tilde{J}^{\alpha\beta}_{ab}\tilde{\Gamma}^{\gamma}_{ba}(\omega^{\prime\prime}+i\gamma)f_{ab}}{\omega^{\prime\prime} - E_{ba} + i\gamma} +\frac{\tilde{J}^{\alpha\gamma}_{ab}\tilde{\Gamma}^{\beta}_{ba}(\omega^{\prime}+i\gamma)f_{ab}}{\omega^{\prime} - E_{ba} + i\gamma} 
 +\sum_{\bm{k}} \frac{\tilde{\Gamma}^{\alpha}_{ab}(-\omega-2i\gamma)\tilde{J}^{\beta\gamma}_{ba}f_{ab}}{\omega - E_{ba} + 2i\gamma} \notag \\
 &+\sum_{\bm{k}} \frac{\tilde{\Gamma}^{\alpha}_{ab}(-\omega-2i\gamma)}{\omega - E_{ba} + 2i\gamma}\Bigg\{\frac{\tilde{\Gamma}^{\beta}_{bc}(\omega^{\prime}+i\gamma)\tilde{\Gamma}^{\gamma}_{ca}(\omega^{\prime\prime}+i\gamma)f_{ac}}{\omega^{\prime\prime}-E_{ca} + i\gamma} - \frac{\tilde{\Gamma}^{\gamma}_{bc}(\omega^{\prime\prime}+i\gamma)\tilde{\Gamma}^{\beta}_{ca}(\omega^{\prime}+i\gamma)f_{cb}}{\omega^{\prime\prime} - E_{bc} + i\gamma}\Bigg\} \notag\\
 &+\sum_{\bm{k}} \frac{\tilde{\Gamma}^{\alpha}_{ab}(-\omega-2i\gamma)}{\omega - E_{ba} + 2i\gamma}\Bigg\{\frac{\tilde{\Gamma}^{\gamma}_{bc}(\omega^{\prime\prime}+i\gamma)\tilde{\Gamma}^{\beta}_{ca}(\omega^{\prime}+i\gamma)f_{ac}}{\omega^{\prime}-E_{ca} + i\gamma} - \frac{\tilde{\Gamma}^{\beta}_{bc}(\omega^{\prime}+i\gamma)\tilde{\Gamma}^{\gamma}_{ca}(\omega^{\prime\prime}+i\gamma)f_{cb}}{\omega^{\prime} - E_{bc} + i\gamma}\Bigg\}\Bigg]. \label{eq:nonlinear_conductivity}
\end{align}    
\end{widetext}

After the vertex function is determined by solving the self-consistent equation, we can calculate the linear and nonlinear conductivity by using the above formulas. In the real frequency representation, the self-consistent equation for the vertex function can be expressed as a simple linear equation. We represent the vertex function as
\begin{align}
\overleftrightarrow{\Gamma}^{\alpha}_{ab}(\bm{k}, \omega+i\gamma)&=\overleftrightarrow{J}^{\alpha}_{ab}(\bm{k})+\overleftrightarrow{\varphi}_{ab}(\bm{k}) \overleftrightarrow{\Lambda}^{\alpha}(\omega + i\gamma),
    \label{eq:vertex_rep}
\end{align}
where we define
\begin{align}
\overleftrightarrow{\varphi}_{ab}(\bm{k}) &= 
  \begin{pmatrix}
    \varphi_{ab}(\bm{k}) & 0 \\
    0 & \varphi^{\dagger}_{ab}(\bm{k})
  \end{pmatrix}, \\    
    \overleftrightarrow{\Lambda}^{\alpha}(\omega+i\gamma)&=
    \Lambda^{\alpha}_{1}(\omega+i\gamma)\tau_{x} + \Lambda^{\alpha}_{2}(\omega+i\gamma)\tau_{y},
    \label{eq:vertex_mode}
\end{align}
with the Pauli matrix in the Nambu space  $\tau_{i}$. 
When we fix the gauge so that the mean-field pair potential $\Delta$ is a real number, $\Lambda^{\alpha}_{1}$ and $\Lambda^{\alpha}_{2}$ correspond to the amplitude mode and the phase mode of collective excitations, respectively.
As we show in Appendix~\ref{sec:appendix_self-consistent_equation}, the self-consistent equation of $\overleftrightarrow{\Lambda}^{\alpha}(\omega + i\gamma)$ is written as,
\begin{align}
    \begin{pmatrix}
        \Lambda^{\alpha}_{1}(\omega + i\gamma) \\
        \Lambda^{\alpha}_{2}(\omega + i\gamma)
    \end{pmatrix}
    =
    \bm{M}(\omega + i\gamma)
    \begin{pmatrix}
        \Lambda^{\alpha}_{1}(\omega + i\gamma) \\
        \Lambda^{\alpha}_{2}(\omega + i\gamma)
    \end{pmatrix}
    +
    \begin{pmatrix}
        L^{J}_{1}(\omega+i\gamma) \\
        L^{J}_{2}(\omega+i\gamma)
    \end{pmatrix}.
\end{align}
The expressions of $\bm{M}(\omega + i\gamma)$ and $L^{J}_{i}(\omega + i\gamma)$ are shown in Appendix~\ref{sec:appendix_self-consistent_equation}. The solution of this linear equation gives the vertex function through Eq.~\eqref{eq:vertex_rep}.

\subsection{Extension to multicomponent Cooper channels}
\label{subsec:extension to multi}
So far we have implicitly assumed a single-component superconductor by adopting the pairing Hamiltonian, Eq.~\eqref{eq:pairing_Hamiltonian}.
It is straightforward to extend the self-consistent response approximation so that it can be applied to multicomponent superconductors. In multicomponent superconductors, multiple Cooper pairing channels allow exotic superconducting phases such as chiral superconductivity~\cite{Kallin2016}, nematic superconductivity~\cite{Fu2014}, and anapole superconductivity~\cite{Kanasugi2022, Kitamura2023}.
Despite the considerable interest in exotic superconductivity, there is as yet no unambiguous confirmation of any such superconductivity.
Rigorous numerical analysis of the optical response in multicomponent superconductors is required to search for strong optical probes of exotic superconductivity.
Here, we consider two-component superconductors with two Cooper pairing channels by adopting the pairing interaction, 
\begin{align}
  H_{\mathrm{pair}} =&-\frac{U_{1}}{2V}\sum_{\bm{k},\bm{k}^{\prime}}\varphi^{(1)}_{ab}(\bm{k})\varphi_{cd}^{(1)\dagger}(\bm{k}^{\prime})c_{\bm{k}a}^{\dagger}c_{-\bm{k}b}^{\dagger}c_{-\bm{k}^{\prime}c}c_{\bm{k}^{\prime}d} \notag \\
  &-\frac{U_{2}}{2V}\sum_{\bm{k},\bm{k}^{\prime}}\varphi^{(2)}_{ab}(\bm{k})\varphi_{cd}^{(2)\dagger}(\bm{k}^{\prime})c_{\bm{k}a}^{\dagger}c_{-\bm{k}b}^{\dagger}c_{-\bm{k}^{\prime}c}c_{\bm{k}^{\prime}d}.
\end{align}
The mean-field pair potential is introduced as
\begin{align}
  &\Delta^{+}_{ab}(\bm{k}, \tau; \bm{A}) = \notag\\
  & -\frac{U_{1}}{V} \varphi^{(1)\dagger}_{ab}(\bm{k})\sum \varphi^{(1)}_{dc}(\bm{k}^{\prime})\mathcal{F}^{+}_{cd}(\bm{k}^{\prime}, \tau^{+}, \tau^{-};\bm{A}) \notag \\
  & -\frac{U_{2}}{V} \varphi^{(2)\dagger}_{ab}(\bm{k})\sum \varphi^{(2)}_{dc}(\bm{k}^{\prime})\mathcal{F}^{+}_{cd}(\bm{k}^{\prime}, \tau^{+}, \tau^{-};\bm{A}),   
\end{align}
\begin{align}
  &\Delta_{ab}(\bm{k}, \tau; \bm{A}) = \notag \\
  & -\frac{U_{1}}{V} \varphi^{(1)}_{ab}(\bm{k})\sum \varphi^{(1)\dagger}_{dc}(\bm{k}^{\prime})\mathcal{F}_{cd}(\bm{k}^{\prime}, \tau^{+}, \tau^{-};\bm{A}) \notag \\
  &-\frac{U_{2}}{V} \varphi^{(2)}_{ab}(\bm{k})\sum \varphi^{(2)\dagger}_{dc}(\bm{k}^{\prime})\mathcal{F}_{cd}(\bm{k}^{\prime}, \tau^{+}, \tau^{-};\bm{A}),
\end{align}
based on the Kadanoff-Baym method explained in Sec.~\ref{sec:Kadanoff-Baym}. 

Let us verify the particle number conservation law.
The time evolution of the total particle number is given by Eq.~(\ref{eq:fourth_term}), and cancellation of the third and fourth terms in Eq.~(\ref{eq:fourth_term}) is essential for the total particle number conservation law. In the presence of the multiple Cooper pairing channels, these terms are obtained as
\begin{align}
  \sum \mathcal{F}_{aj}&(\bm{k}, \tau^{-}, \tau^{+};\bm{A})\Delta^{+}_{ja}(\bm{k}, \tau;\bm{A}) = \notag\\
  - \frac{U_{1}}{V} &\left[\sum \mathcal{F}_{aj}(\bm{k}, \tau^{-}, \tau^{+};\bm{A})\varphi^{(1)\dagger}_{ja}(\bm{k})\right] \notag \\
  &\times  \left[\sum \varphi^{(1)}_{dc}(\bm{k}^{\prime})\mathcal{F}^{+}_{cd}(\bm{k}^{\prime}, \tau^{+}, \tau^{-};\bm{A})\right] \notag \\
  - \frac{U_{2}}{V} &\left[\sum \mathcal{F}_{aj}(\bm{k}, \tau^{-}, \tau^{+};\bm{A})\varphi^{(2)\dagger}_{ja}(\bm{k})\right] \notag \\
  &\times  \left[\sum \varphi^{(2)}_{dc}(\bm{k}^{\prime})\mathcal{F}^{+}_{cd}(\bm{k}^{\prime}, \tau^{+}, \tau^{-};\bm{A})\right],
\end{align}
\begin{align}
  \sum  \Delta_{aj}&(\bm{k}, \tau;\bm{A})\mathcal{F}^{+}_{ja}(\bm{k}, \tau^{-}, \tau^{+};\bm{A}) = \notag \\
  - \frac{U_{1}}{V}&\left[\sum \varphi^{(1)\dagger}_{dc}(\bm{k}^{\prime})\mathcal{F}_{cd}(\bm{k}^{\prime}, \tau^{+}, \tau^{-};\bm{A})\right] \notag \\
  &\times  \left[\sum \varphi^{(1)}_{aj}(\bm{k})\mathcal{F}^{+}_{ja}(\bm{k}, \tau^{-}, \tau^{+};\bm{A})\right] \notag \\
     - \frac{U_{2}}{V}&\left[\sum \varphi^{(2)\dagger}_{dc}(\bm{k}^{\prime})\mathcal{F}_{cd}(\bm{k}^{\prime}, \tau^{+}, \tau^{-};\bm{A})\right] \notag \\
  &\times  \left[\sum \varphi^{(2)}_{aj}(\bm{k})\mathcal{F}^{+}_{ja}(\bm{k}, \tau^{-}, \tau^{+};\bm{A})\right].
\end{align}
Therefore, we can verify that these terms 
cancel each other, and the total particle number conservation law is satisfied. 

The formulas of the linear and second-order nonlinear conductivities are the same as Eqs.~(\ref{eq:linear_conductivity}) and (\ref{eq:nonlinear_conductivity}), respectively, while the one-photon vertex function is modified. The vertex can be represented as
\begin{align}
    \overleftrightarrow{\Gamma}^{\alpha}_{ab}(\bm{k}, \omega+i\gamma)=&\overleftrightarrow{J}^{\alpha}_{ab}(\bm{k})+
    \overleftrightarrow{\varphi^{(1)}}_{ab}(\bm{k}) \overleftrightarrow{\Lambda^{(1)}}^{\alpha}(\omega + i\gamma) \notag \\
    &+\overleftrightarrow{\varphi^{(2)}}_{ab}(\bm{k}) \overleftrightarrow{\Lambda^{(2)}}^{\alpha}(\omega + i\gamma),
    \label{eq:vertex_rep2}
\end{align}
and the self-consistent equation is written as
\begin{widetext}
\begin{align}
    \begin{pmatrix}
        \bm{\Lambda}^{(1)\alpha}(\omega + i\gamma) \\
        \bm{\Lambda}^{(2)\alpha}(\omega + i\gamma) \\
    \end{pmatrix}
    =
    \begin{pmatrix}
    \bm{M}^{(11)}(\omega + i\gamma) & \bm{M}^{(12)}(\omega + i\gamma) \\ 
    \bm{M}^{(21)}(\omega + i\gamma) & \bm{M}^{(22)}(\omega + i\gamma)
    \end{pmatrix}
    \begin{pmatrix}
        \bm{\Lambda}^{(1)\alpha}(\omega + i\gamma) \\
        \bm{\Lambda}^{(2)\alpha}(\omega + i\gamma) \\
    \end{pmatrix}
    +
    \begin{pmatrix}
        \bm{L}^{(1)J}(\omega+i\gamma) \\
        \bm{L}^{(2)J}(\omega+i\gamma) 
    \end{pmatrix},
\end{align}
\end{widetext}
where we use the notation $\bm{\Lambda}^{(i)\alpha} = \left(\Lambda^{(i)\alpha}_{1}, \Lambda^{(i)\alpha}_{2}\right)^{\top}$ and $\bm{L}^{(i)J} = \left(L^{(i)J}_{1}, L^{(i)J}_{2}\right)^{\top}$. The detailed derivation is presented in Appendix~\ref{app:extention to multi}. The off-diagonal block $\bm{M}^{(12)}$ and $\bm{M}^{(21)}$ admix the collective modes of different Cooper channels and allow the Leggett mode and the Bardasis-Schrieffer mode.   

\subsection{Summary of the self-consistent response approximation}
We have derived the formulas of linear and second-order nonlinear conductivity in the self-consistent response approximation. We applied the Kadanoff-Baym method to the equation of motion to satisfy the total particle number conservation law. Thus, the results obtained from the self-consistent response approximation are physically reasonable. 

Our formulas are applicable not only to conventional superconductors but also to various unconventional superconductors, such as spin-triplet superconductors, anisotropic superconductors, and helical superconductors. It has been shown that the amplitude mode and the phase mode contribute to the linear and nonlinear conductivities through the vertex function. In Sec.~\ref{subsec:extension to multi}, we extended the formulas to the multicomponent superconductors with multiple Cooper pairing channels. We can evaluate the contribution of the collective modes such as the Leggett mode and the Bardasis-Schrieffer mode, which can arise from multicomponent superconducting order parameters. 

\section{Numerical demonstration in topological superconductors}
\label{sec:numerical_demonstration}
In the following sections, we study the linear and second-order nonlinear conductivities in the Rashba superconductor by the self-consistent response approximation. The effects of collective modes, which are neglected in the formalism starting from the BdG Hamiltonian, are discussed. It is shown that the amplitude mode of the $s$-wave superconductivity enhances the conductivities. The interband pairing 
is essential for the enhancement. 

The Rashba superconductor has attracted interest because it is a potential platform of topological superconductivity. 
Thus, a clarification of the superconducting state by optical responses is desirable, since they could be a probe of topological superconductivity. 
In the later section, we investigate the photocurrent conductivity, namely, the photo-induced direct current, around the topological transition. 
In our previous work~\cite{Tanaka2024}, we have shown that the bare photocurrent conductivity without vertex corrections reverses the sign at the topological transition by increasing the Zeeman field. 
We show that the photocurrent conductivity obtained by the self-consistent response approximation also shows a sign reversal. 
The collective modes enhance the photocurrent conductivity through the vertex correction, but do not change qualitative behaviors at the topological transition.

\subsection{Model of Rashba $s$-wave superconductors}

\begin{figure*}[htbp]
 \includegraphics[width=0.9\linewidth]{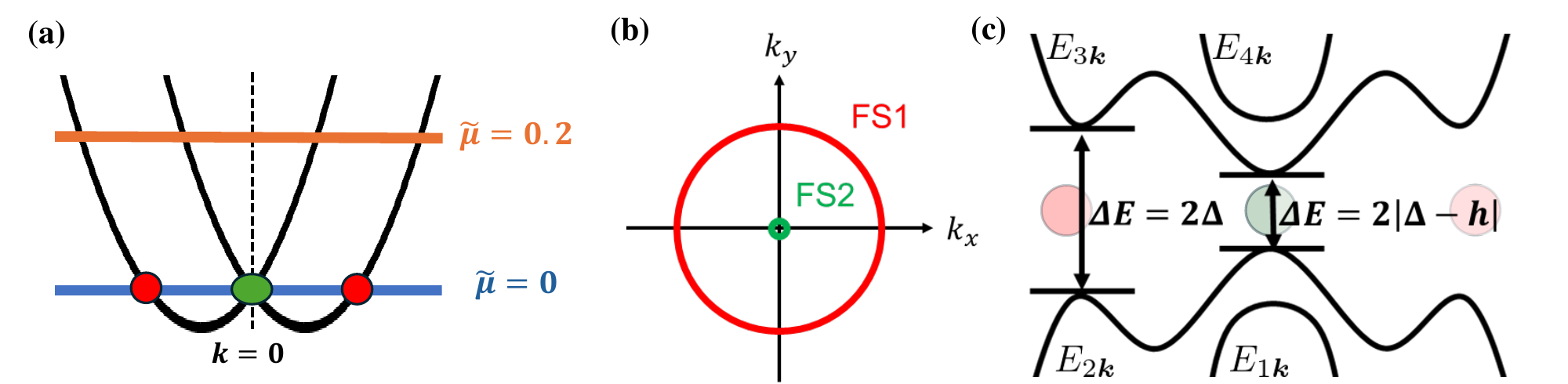}
 \caption{(a) Schematic parabolic bands around the $\Gamma$ point (Dirac point) in the normal state. The red and green points illustrate the Fermi surfaces when the Fermi level lies on the Dirac point ($\tilde{\mu}=0$). (b) The Fermi surfaces for $\tilde{\mu}=0$. The FS1 exists far from the Dirac point while the FS2 is 
 on the Dirac point. (c) Schematic illustration of the band structure in the superconducting state with the parameter $\tilde{\mu}=0$. The energy gap is obtained as $\Delta E=2\Delta$ around the FS1, while $\Delta E=2|\Delta -h|$ around the FS2.}
 \label{fig:schematic}
\end{figure*}

We consider noncentrosymmetric $s$-wave superconductors with antisymmetric spin-orbit coupling under a magnetic field. The noninteracting Hamiltonian is given by
\begin{align}
    H_{0} =& H_{\mathrm{kin}}+H_{\mathrm{ASOC}}+H_{\mathrm{Zeeman}}, \label{eq:normal_H} \\
    H_{\mathrm{kin}} =& \sum_{\bm{k},s} \xi(\bm{k})C_{\bm{k},s}^{\dagger}C_{\bm{k},s}, \\
    H_{\mathrm{ASOC}} =& \sum_{\bm{k},s,s^{\prime}}\bm{g}(\bm{k})\cdot\bm{\sigma}_{ss^{\prime}}C_{\bm{k},s}^{\dagger}C_{\bm{k},s^{\prime}}, \\
    H_{\mathrm{Zeeman}} =& \sum_{\bm{k},s,s^{\prime}}\bm{h}\cdot\bm{\sigma}_{ss^{\prime}}C_{\bm{k},s}^{\dagger}C_{\bm{k},s^{\prime}}.
\end{align}
In the following, we assume a two-dimensional crystal structure with tetragonal $\mathrm{C_{4v}}$ point group symmetry. The first term $\mathcal{H}_{\mathrm{kin}}$ is a kinetic energy measured from a chemical potential, which contains the hopping term
\begin{align}
        \xi(\bm{k}) &= -2t_{1}(\cos k_{x} + \cos k_{y}) + 4t_{2}\cos k_{x}\cos k_{y} - \mu.
\end{align}
The second term $\mathcal{H}_{\mathrm{ASOC}}$ represents a Rashba-type spin-orbit coupling and the g-vector is assumed as
\begin{align}
        \bm{g}(\bm{k}) &= \alpha (\sin k_{y}, -\sin k_{x}, 0).
\end{align}
The parameters are set as $t_{1}=1$, $t_{2}=0.2$, and $\alpha =0.4$ in the numerical calculations. The Zeeman field in $\mathcal{H}_{\mathrm{Zeeman}}$ is assumed to be
\begin{align}
    \bm{h} = (0, h\cos\theta, h\sin\theta), \quad \theta=\ang{60}, \label{eq:magnetic_field}
\end{align}
with $\theta$ being the angle between the magnetic field and the two-dimensional plane. 
We assume the magnetic field tilted from the perpendicular axis to break the $\mathrm{C_{2z}}$ rotation symmetry, which prohibits second-order nonlinear responses.

In the Rashba superconductor with an in-plane component of the magnetic field, 
the Cooper pairs have a finite total momentum in the equilibrium superconducting state. 
This state is called the helical superconducting state~\cite{Bauer2012, Smidman2017}. Thus, we introduce the Cooper pairs' momentum $2\bm{q}$ and the $s$-wave pairing interaction is introduced as
\begin{align}
  H_{\mathrm{pair}} =-\frac{U}{2V}\sum_{\bm{k},\bm{k}^{\prime}}&(i\sigma_{y})_{s_{1}s_{2}}(-i\sigma_{y})_{s_{3}s_{4}} \notag \\
  &\times C_{\bm{k}+\bm{q},s_{1}}^{\dagger}C_{-\bm{k}+\bm{q},s_{2}}^{\dagger}C_{-\bm{k}^{\prime}+\bm{q},s_{3}}C_{\bm{k}^{\prime}+\bm{q},s_{4}}.
\end{align}
The mean field is obtained as
\begin{align}
    \Delta(\bm{q})_{ab} = \Delta(\bm{q})(i\sigma_{y})_{ab}, \quad \Delta(\bm{q})\in \mathbb{R}^{+},
\end{align}
where we choose an appropriate $U(1)$ gauge.

We evaluate the condensation energy to determine the total momentum of Cooper pairs $2\bm{q}$. The free energy density is given by
\begin{align}
    \Omega({\bm q}, \Delta({\bm q})) = &\frac{1}{2V}\sum_{\bm{k}, s}\left[[\mathcal{H}_{0}]_{ss} + |\Delta({\bm q})|^{2}/2U\right] \notag \\
    & -\frac{T}{2V}\sum_{\bm{k}, n}\ln \left(1+e^{-E_{n,{\bm q}}(k)/T}\right),
\end{align}
and the condensation energy is obtained as
\begin{align}
    F({\bm q})\equiv \Omega({\bm q}, \Delta({\bm q})) - \Omega({\bm q}, 0).
\end{align}
The total momentum of Cooper pairs $2\bm{q}$ is determined so as to minimize the condensation energy,
\begin{align}
    F(\bm{q}) = \min_{\bar{\bm{q}}} F(\bar{\bm{q}}).
\end{align}
We can fix the $y$-component of the total momentum as $q_{y}=0$ because the $x$-component of the magnetic field is assumed to be zero in Eq.~(\ref{eq:magnetic_field}). The only $q_{x}$ needs to be determined by minimizing the condensation energy.


The topological superconductivity in the Rashba system is related to the Dirac electrons in the normal state.
The noninteracting Hamiltonian has a Dirac point at $\bm{k}=\bm{0}$ when the magnetic field is zero. The energy of the Dirac point $\xi_{0}$ is 
\begin{align}
    \xi_{0} = -4t_{1} + 4t_{2} = -3.2,
\end{align}
for our choice of the hopping parameters.
We define $\tilde{\mu}$ as the chemical potential from the Dirac point, $\tilde{\mu} \equiv \mu - \xi_{0}$. When we set the chemical potential as $\tilde{\mu} = 0$, the Fermi level lies on the Dirac point under zero magnetic field [see Fig.~\ref{fig:schematic}(a)]. In this case, one of the two spin-split Fermi surfaces, "Fermi surface 1" (FS1), exists far from the Dirac point, and the other "Fermi surface 2" (FS2) is the point at the Dirac point [Fig.~\ref{fig:schematic}(b)]. Although the $s$-wave superconducting state is topologically trivial at the zero magnetic field, a large Zeeman field 
can induce topological superconductivity hosting Majorana zero modes~\cite{Sato2009, Sato2010, Sau2010}.
The critical value is $h^{c}=\sqrt{{\tilde{\mu}}^{2} + \Delta^{2}}$ when the magnetic field is perpendicular to the conducting plane. 
Thus, $\tilde{\mu}$
must be small to obtain approachable $h^{c}$, and experimental searches for the condition $\mu \simeq 0$ have been performed extensively.
This is one of the reasons for which we focus on the case $\tilde{\mu}=0.0$, relevant to topological superconductivity. 
Another reason is that the Dirac point gives a strong impact on the nonlinear response phenomena. 
In our previous work~\cite{Tanaka2024}, the photocurrent conductivity calculated without vertex corrections is strongly enhanced 
around the Dirac point, and the enhancement is attributed to the quantum-geometric properties of Dirac electrons. 
In the following part, we assume $\tilde{\mu}=0.0$ unless otherwise mentioned. 


We set $t_{1} = 1.0 \times 10^{2} \mathrm{meV}$ and introduce the lattice constant $a$ for a quantitative estimate. 
The numerical calculations are conducted within the $N^2$-discretized Brillouin zone. Unless stated otherwise, $N$ is set to $1600$.
For numerical convergence, we introduce a finite temperature $T=10^{-4}$ for the Fermi-Dirac distribution function.
We evaluate the linear and nonlinear conductivities by using the self-consistent response approximation and discuss the effects of the collective modes in Secs.~\ref{sec:linear} and \ref{sec:nonlinear}.

\subsection{Intraband pairing and interband pairing}

Before showing the numerical results, in this subsection we elucidate the intraband and interband pairing in the model. In Secs.~\ref{sec:linear} and \ref{sec:nonlinear}, we show that interband pairing plays an essential role in the enhancement of the optical responses due to the collective modes.  Intraband pairing and interband pairing can be captured by the superconducting fitness~\cite{Ramires2016, Ramires2018}. We define the intraband superconducting fitness $F_{A}$ and the interband superconducting fitness $F_{C}$ at momentum $\bm{k}$ 
by
\begin{align}
    F_{A}(\bm{k})(i\sigma_{y}) =& \mathcal{H}^{\prime}_{0}(\bm{k}) [\Delta(i\sigma_{y})] + [\Delta(i\sigma_{y})]\mathcal{H}^{\prime\top}_{0}(-\bm{k}), \\
    F_{C}(\bm{k})(i\sigma_{y}) =& \mathcal{H}^{\prime}_{0}(\bm{k}) [\Delta(i\sigma_{y})] - [\Delta(i\sigma_{y})]\mathcal{H}^{\prime\top}_{0}(-\bm{k}).
\end{align}
We define 
$\mathcal{H}^{\prime}_{0}(\bm{k}) \equiv \bm{c} (\bm{k})\cdot \bm{\sigma}/|\bm{c}(\bm{k})|$ for the noninteracting Hamiltonian with the form 
$\mathcal{H}_{0}(\bm{k}) = c_{0}(\bm{k}) + \bm{c}(\bm{k})\cdot \bm{\sigma}$. In our model, $\mathcal{H}^{\prime}_{0}(\bm{k})$ is obtained as $\mathcal{H}^{\prime}_{0}(\bm{k}) = [\mathcal{H}_{0}(\bm{k}) - \xi(\bm{k})]/|\bm{g}(\bm{k}) + \bm{h}|$.
When the Cooper pairs' total momentum $2\bm{q}$ is sufficiently small and ignorable, 
the intraband and interband pairing is characterized by the fitness as
\begin{align}
    \mathrm{Tr}[F_{A}(\bm{k})^{\dagger}F_{A}(\bm{k})] =& 8\Delta^{2}g({\bm{k}})^{2}/|\bm{g}(\bm{k}) + \bm{h}|^{2},\label{eq:intrabnd_SC} \\
    \mathrm{Tr}[F_{C}(\bm{k})^{\dagger}F_{C}(\bm{k})] =& 8\Delta^{2}h^{2}/|\bm{g}(\bm{k}) + \bm{h}|^{2}. \label{eq:interband_SC}
\end{align}
These formulas tell us that the interband pairing is dominant around the Dirac point because the Rashba spin-orbit coupling vanishes at the Dirac point. On the other hand, the intraband pairing is dominant around the FS1 as usual. The energy gap in the superconducting state is also different between the FS1 and FS2 [see Fig.~\ref{fig:schematic}(c)]. The energy gap on the FS1 is obtained as $\Delta E=2\Delta$, while the gap at the FS2 is estimated as 
$\Delta E=2|\Delta -h|$~\cite{Tanaka2024}. Thus, the superconducting gap around the FS1 is robust against the Zeeman field, although it closes on the FS2 at the topological transition.

\subsection{Vertex function}
\label{sec:vertex_func}
\begin{figure*}[htbp]
 \includegraphics[width=0.9\linewidth]{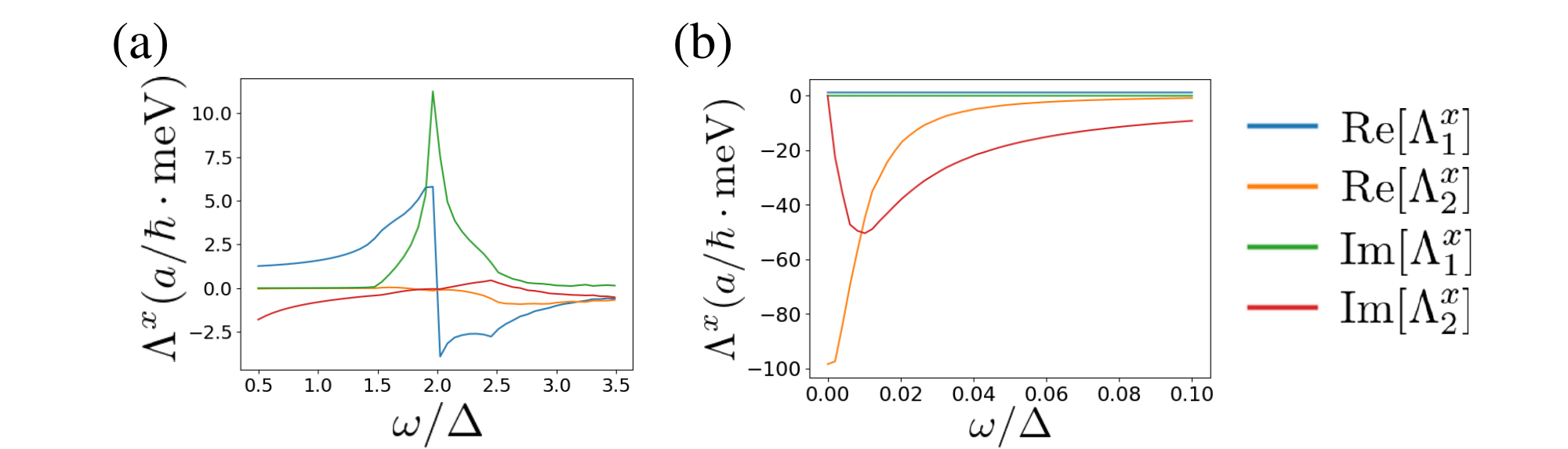}
 \caption{The frequency dependence of $\Lambda^{x}_i(\omega +i\gamma)$ [Eq.~\eqref{eq:vertex_mode}] in the vertex function, Eq.~\eqref{eq:vertex_rep}. $\Lambda^{x}_{1}$ and $\Lambda^{x}_{2}$ originate from the amplitude and phase modes, respectively. We set the parameters $\tilde{\mu}=0.0$, $U=1.15$, $h=0.02$, and  $\gamma=8\times 10^{-4}$.  Results in (a) the moderate frequency region $0.5\leq \Omega/\Delta \leq 3.5$ and (b) the low frequency region $0 \leq \Omega/\Delta \leq 0.1$.}
 \label{fig:vertex_result}
\end{figure*}

In the last part of this section, we show the self-consistent solution of the vertex function and discuss its characteristic properties. The vertex function can be represented as in Eq.~\eqref{eq:vertex_rep}. By comparing Eqs.~\eqref{eq:vertex_Delta_relation}, \eqref{eq:vertex_rep}, and \eqref{eq:vertex_mode}, we can verify the correspondence between $\Lambda^{x}_{i}$ and the amplitude and phase modes. Figure \ref{fig:vertex_result} shows the frequency dependence of $\Lambda^{x}_i(\omega +i\gamma)$. The peak of $\mathrm{Im}[\Lambda^{x}_{1}]$ around $\omega=2\Delta$ originates from the amplitude mode known as the Higgs mode. The nearly $\omega^{-1}$ divergence of $\mathrm{Im}[\Lambda^{x}_{2}]$ in the low-frequency region $\omega \ll 2\Delta$ 
may be attributed to the phase mode known as the Nambu-Goldstone mode. 
The behavior of $\Lambda^{x}_{2}$ in the low-frequency region shown in Fig.~\ref{fig:vertex_result}(b) 
can be described as
\begin{align}
    \Lambda^{x}_{2}(\omega +i\gamma) \propto \frac{i}{\omega + i\gamma},
\end{align}
by introducing the phenomenological scattering rate $\gamma$ to cut off the divergence.
Indeed, the vertex function can be numerically fitted by using a fitting function,
\begin{align}
\Lambda^{x}_{2}(\omega +i\gamma) \simeq \frac{iA}{\omega + iB}, 
\label{eq:clean_vertex}
\end{align}
with fitting parameters $A$ and $B$. We find that $B$ is almost equal to $\gamma$.

In Sec.~\ref{sec:nonlinear}, we take the clean limit for $\Lambda^{x}_{2}$ 
to obtain physically meaningful results of the photocurrent conductivity. 
According to the results discussed above, the vertex in the clean limit $\gamma\rightarrow 0$ is given by $\Lambda^{x}_{2}(\omega +i\gamma) \simeq iA/\omega$.

\section{Linear optical responses: effects of interband pairing}
\label{sec:linear}

\begin{figure*}[htbp]
 \includegraphics[width=0.9\linewidth]{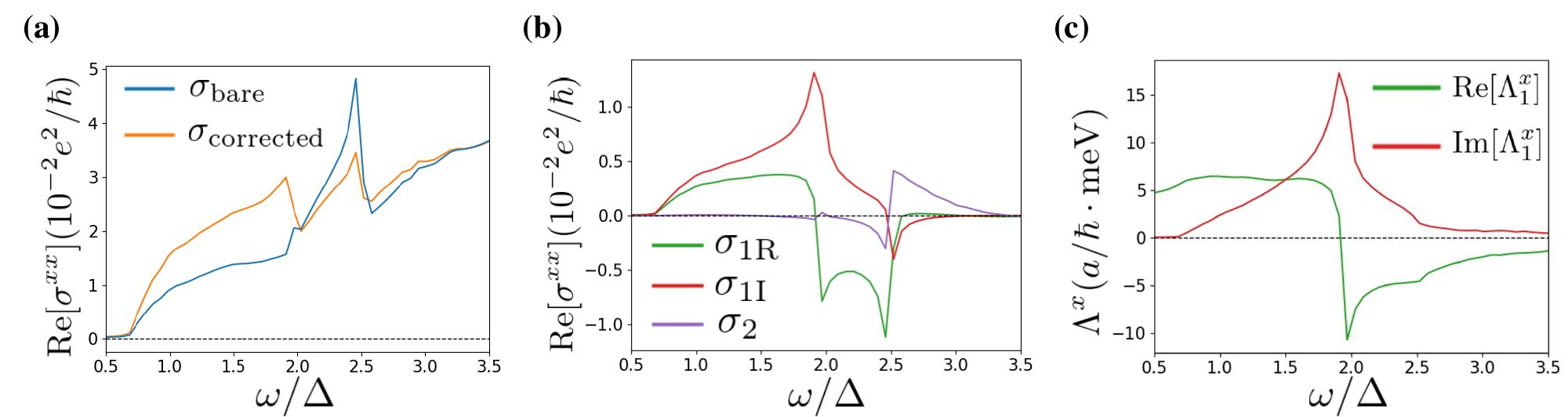}
 \caption{The frequency dependence of (a) the linear conductivity $\mathrm{Re}[\sigma^{xx}]$, (b) its components defined by Eq.~\eqref{eq:linear_conductivity_decomposition}, and (c) the vertex function $\Lambda^{x}_{1}$. (a) The bare and corrected linear conductivities are compared to clarify the role of vertex correction. (b) The contribution of the vertex correction arising from the amplitude (Higgs) mode and the phase (NG) mode to the linear conductivity. The amplitude mode gives $\sigma_{\mathrm{1R}}$ and $\sigma_{\mathrm{1I}}$, while the phase mode gives $\sigma_{\mathrm{2}}$. 
 We set the parameters as $\tilde{\mu}=0.0$ and $h = 0.05$.}
 \label{fig:linear_result}
\end{figure*}

In this section, we show the linear optical conductivity with and without the vertex correction. The roles of the collective modes are discussed. In the case of $\tilde{\mu}=0.0$, the Fermi level lies on the Dirac point, and the interband pairing is dominant around the Dirac point. We show that the amplitude mode enhances the longitudinal linear conductivity in this case. For $s$-wave superconductors, the enhancement by the amplitude mode is unique to the multiband system, because the optical conductivity is not enhanced in a single-band model as we show in Sec.~\ref{sec:multiband_effect}. In Sec.~\ref{sec:interband_pairing}, we verify that interband pairing is essential for the enhancement of linear conductivity by the amplitude mode. 

In the following, the conductivity obtained with (without) vertex correction is referred to as the corrected (bare) conductivity. 
The main result of this section is presented in Fig.~\ref{fig:linear_result}(a), which shows the bare and corrected linear conductivities, $\mathrm{Re}[\sigma^{xx}]$. 
We see that the linear conductivity is enhanced by the vertex correction in the low frequency region $\omega/\Delta \leq 2.0$. 
To identify the causes of enhancement, we decompose the contribution of the vertex using Eqs.~(\ref{eq:linear_conductivity}), (\ref{eq:vertex_rep}), and \eqref{eq:vertex_mode}. The correction to the linear conductivity can be written as
\begin{align}
     \sigma_{\mathrm{corrected}} - \sigma_{\mathrm{bare}} = 
     \sigma_{\mathrm{1}} + \sigma_{\mathrm{2}},
\label{eq:linear_conductivity_decomposition}
\end{align}
with $\sigma_{\mathrm{1}} = 
\sigma_{\mathrm{1R}} + \sigma_{\mathrm{1I}}$,
where $\sigma_{\mathrm{1R}}$, $\sigma_{\mathrm{1I}}$, and $\sigma_{\mathrm{2}}$ are contributions proportional to $\mathrm{Re}[\Lambda^{x}_{1}]$, $\mathrm{Im}[\Lambda^{x}_{1}]$, and $\Lambda^{x}_{2}$, respectively. Figure~\ref{fig:linear_result}(b) shows that $\Lambda^{x}_{1}$ makes the dominant contribution to the enhancement of the linear conductivity. Because the vertex $\Lambda^{x}_{1}$ ($\Lambda^{x}_{2}$) comes from the amplitude (phase) mode, we conclude that the amplitude mode enhances the linear conductivity.

\subsection{Multiband effect}
\label{sec:multiband_effect}

Here, we compare our Rashba superconductor model with the single-band $s$-wave superconductor model to discuss the multiband effects on the linear conductivity. In Ref.~\cite{Huang2023}, it was shown that the linear conductivity $\mathrm{Re}[\sigma^{xx}]$ is suppressed by the vertex correction in the single-band $s$-wave superconductor under supercurrent. We find that the vertex $\Lambda^{x}_{2}$ due to the phase mode gives a destructive contribution, while the effect of amplitude mode through $\Lambda^{x}_{1}$ 
is negligible in this case [Fig.~\ref{fig:single_linear_result}(b) in Appendix \ref{app:single_band}]. 

We can analytically show that the contribution of the amplitude mode 
is negligible 
in the single-band 
superconductor, 
as presented in the following with the detailed discussion in Appendix~\ref{app:single_band}.
The BdG Hamiltonian for single-band $s$-wave superconductors is given by
\begin{align}
    \mathcal{H}(\bm{k}) = 
    \begin{pmatrix}
        \xi_{1}(\bm{k}) & \psi \\
        \psi & -\xi_{2}(\bm{k})
    \end{pmatrix}, \quad \psi\in\mathbb{R}^{+},
\end{align}
where we choose an appropriate $U(1)$ gauge.
By introducing the electric current operator and the one-photon vertex function,
\begin{align}
    J^{\alpha}(\bm{k}) &= J_{0}(\bm{k}) + J_{3}(\bm{k})\sigma_{3}, \\
    \Gamma^{\alpha}(\bm{k},\omega + i\gamma) &= J^{\alpha}(\bm{k}) + \Lambda_{1}^{\alpha}(\omega + i\gamma)\sigma_{1}+ \Lambda_{2}^{\alpha}(\omega + i\gamma)\sigma_{2},
\end{align}
the resonant component of the linear optical  conductivity is obtained as
\begin{align}
    \sigma^{\alpha\beta}(\omega) = \sum_{\bm{k}}\frac{\pi\psi J^{\alpha}_{3}}{2{\Delta E}^{2}}\left\{ \frac{-\bar{\xi}\Lambda^{\beta}_{1} + J^{\beta}_{3}\psi}{\Delta E} + i\Lambda^{\beta}_{2} \right\}\delta(\omega - 2\Delta E),
\end{align}
where $\bar{\xi}(\bm{k})$ is defined as $\bar{\xi}(\bm{k})=(\xi_{1}(\bm{k}) + \xi_{2}(\bm{k}))/2$.
The interband energy difference is given by $2\Delta E = 2\sqrt{\psi^{2} + \bar{\xi}(\bm{k})^{2}}$, 
and the minimum energy gap is $2\psi$. 
The resonant component 
is enhanced around $\omega \sim 2\psi$
because of the prefactor $\Delta E^{-2}$ and the peak of the joint density of states. However, the contribution of the amplitude mode through $\Lambda^{\beta}_{1}$ is suppressed there by the prefactor $\bar{\xi}\sim 0$. Moreover, the vertex arising from the amplitude mode $\Lambda^{x}_{1}$ is relatively small compared to $\Lambda^{x}_{2}$ because $\Lambda^{x}_{1}$ originates from the weighted integral of $\bar{\xi}(\bm{k})$ around the Fermi surface~\cite{Dai2017}. 
Therefore, the effect of the amplitude mode is negligible in the single-band 
superconductors. 

However, these mechanisms do not apply to multiband systems, and indeed the vertex correction due to the amplitude mode through $\Lambda^{x}_{1}$ significantly enhances the low-energy optical conductivity in the model for Rashba superconductors, as shown in Fig.~\ref{fig:linear_result}(b).
Thus, the effects of collective modes on optical conductivity are qualitatively different between the single-band and multiband systems in the clean limit.
In our model, the amplitude mode plays a significant role, 
because the energy bands are split owing to the Rashba spin-orbit coupling and the Zeeman field. 
Therefore, the multiband effect is essential to enhance the linear conductivity by the amplitude mode in $s$-wave superconductors.

\subsection{Interband pairing}
\label{sec:interband_pairing}

To further clarify the multiband effects, we focus on the role of interband pairing. 
The resonant component of the enhanced linear conductivity in the region $\omega/\Delta <2.0$ originates from 
momentum space around the Dirac point [see Figs.~\ref{fig:schematic}(a) and (c)]. Given Eqs.~(\ref{eq:intrabnd_SC}) and (\ref{eq:interband_SC}), the interband superconducting fitness is much larger than the intraband superconducting fitness around the Dirac point. Therefore, the interband pairing is expected to play an essential role in the enhancement of linear conductivity shown in Fig.~\ref{fig:linear_result}(a). In contrast, the interband pairing is absent in single-band $s$-wave superconductors.

\begin{figure*}[htbp]
 \includegraphics[width=0.9\linewidth]{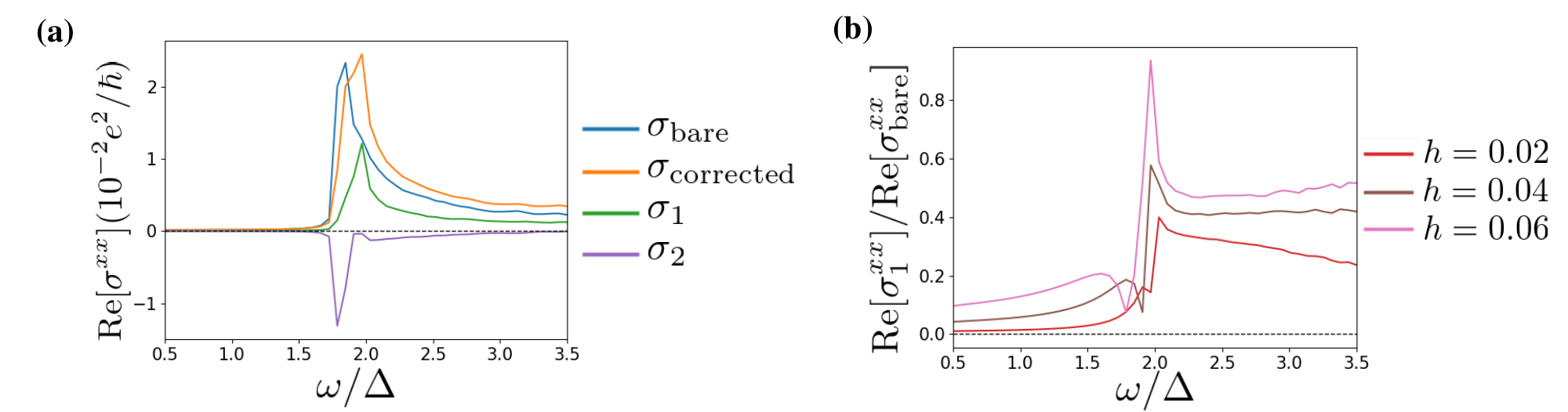}
 \caption{(a) The linear conductivity $\mathrm{Re}[\sigma^{xx}]$ for the effective chemical potential $\tilde{\mu}=0.2$ and the magnetic field $h=0.06$. (b) The ratio of $\mathrm{Re}[\sigma^{xx}_{1}]$ to $\mathrm{Re}[\sigma^{xx}_{\mathrm{bare}}]$ for various magnitudes of the magnetic field. We fix $\tilde{\mu}=0.2$. In the calculation for Fig.~\ref{fig:linear_ratio}(b), we set $N=3200$ for numerical convergence.}
 \label{fig:linear_ratio}
\end{figure*}

Here, we show the results that indicate the importance of the interband pairing for the enhancement due to the amplitude mode. 
We set $\tilde{\mu} = 0.2$ so that the Fermi surface is away from the Dirac point [Fig.~\ref{fig:schematic}(a)]. Although intraband pairing is dominant in this case, 
we can increase the interband pairing by increasing the magnetic field. Therefore, it is expected that the vertex correction arising from the amplitude mode through $\Lambda^{x}_{1}$ increases when we increase the magnetic field. Figure~\ref{fig:linear_ratio}(a) shows that the contribution $\sigma_{1} = \sigma_{\rm 1R} + \sigma_{\rm 1I}$ from the amplitude mode is cooperative to the bare conductivity $\sigma_{\mathrm{bare}}$. 
When the magnitude of the magnetic field $h$ is increased, the ratio $\mathrm{Re}[\sigma_{1}]/\mathrm{Re}[\sigma_{\mathrm{bare}}]$ increases [Fig.~\ref{fig:linear_ratio}(b)], consistent with our expectation. This indicates that the interband pairing plays an essential role for the enhancement of the linear conductivity by the amplitude mode.


In contrast to the low-frequency region discussed above, 
the vertex correction suppresses the linear conductivity in the high-frequency region $\omega/\Delta >2.0$, as shown in Fig.~\ref{fig:linear_result}(a). The suppression is also due to the amplitude mode [see Fig.~\ref{fig:linear_result}(b)]. The sign of the vertex $\mathrm{Re}[\Lambda_{1}]$ changes around $\omega/\Delta \sim 2.0$ [Fig.~\ref{fig:linear_result}(c)], and $\sigma_{\rm 1R}$ becomes negative. 
This effect of the amplitude mode does not appear in the Rashba model for $\tilde{\mu}=0.2$ and in the single-band model. Therefore, the vertex correction which suppresses the linear conductivity is also a characteristic property of 
the Dirac point.

We summarize the results of the linear conductivity in $s$-wave Rashba superconductors.
We have shown that the amplitude mode of the order parameter 
enhances the linear conductivity $\mathrm{Re}[\sigma^{xx}]$ in the low-frequency region $\omega/\Delta <2.0$. This enhancement in clean $s$-wave superconductors is unique to multiband superconductors. In particular, the interband pairing plays an essential role in the enhancement. Therefore, the linear conductivity is significantly enhanced in the model with $\tilde{\mu}=0.0$, where interband pairing is dominant around the Dirac point. The peculiar quantum-geometric properties of Dirac electrons are also important for the large optical conductivity. 
When the Fermi level lies on the Dirac point, the conductivity is expected to be large since it is related to the quantum-geometric quantities. Indeed, this is the case for our results showing that the linear conductivity 
for $\tilde{\mu}=0.0$ [Fig.~\ref{fig:linear_result}(a)] is larger than that for $\tilde{\mu}=0.2$ [Fig.~\ref{fig:linear_ratio}(a)]. Because the vertex correction due to the collective modes is expected to be of the same order as the bare conductivity, the vertex correction term $\sigma_{1}$ is also enhanced cooperatively by the interband pairing and the quantum-geometric properties both of which are enhanced around the Dirac point.

\section{Nonlinear optical responses and topological transition}
\label{sec:nonlinear}

In this section, we show the results of the nonlinear optical conductivity. In particular, we focus on the photocurrent conductivity for which a theoretical calculation without vertex correction predicted the sign reversal 
at the topological transition of superconducting states~\cite{Tanaka2024}. With the numerical calculation containing the vertex correction, we show that the sign reversal robustly occurs when the contribution from the collective modes is taken into account. 
Furthermore, the low-frequency photocurrent showing the sign reversal is enhanced by the amplitude mode. 
The sign reversal originates from the band inversion, which is directly related to the topological transition, and therefore it can be used as an indicator of topological superconductivity. 

\subsection{Calculation method of photocurrent conductivity}
The photo-induced direct current, namely, the photocurrent is characterized by the photocurrent conductivity defined as 
\begin{align}
\sigma^{\alpha;\beta\gamma}_{\mathrm{PC}}(\omega) \equiv \lim_{\delta \omega \rightarrow 0} \sigma^{\alpha;\beta\gamma}(2\delta\omega; \omega+\delta\omega, -\omega+\delta\omega),
\end{align}
where $\omega$ is a frequency of an injected monochromatic light.
Although the photocurrent conductivity has been formulated in the previous sections, we need to take care of the numerical calculation because the calculation with the vertex correction can be numerically unstable when the massless collective mode emerges.
The formula of the photocurrent conductivity, Eq.~\eqref{eq:nonlinear_conductivity}, has terms containing the vertex function $\tilde{\Gamma}^{\alpha}_{ab}(-2\delta\omega-2i\gamma)$ such as
\begin{align}
     \lim_{\delta\omega\rightarrow 0} \frac{\tilde{\Gamma}^{\alpha}_{ab}(-2\delta\omega-2i\gamma)\tilde{J}^{\beta\gamma}_{ba}f_{ab}}{4(\omega + \delta\omega)(\omega - \delta\omega)(\omega - E_{ba} + 2i\gamma)}. \label{eq:example_diverge}
\end{align}
Because a part of the vertex function $\Lambda^{\alpha}_{2}(\delta \omega +i\gamma)$ diverges in the limit $\delta \omega\rightarrow 0$ and $\gamma\rightarrow 0$ (clean limit), we should examine possible divergence of the photocurrent conductivity. However, the divergent vertex function $\Lambda^{\alpha}_{2}$ does not necessarily give the divergent photocurrent conductivity. For example, in the single-band $s$-wave model, the contribution of $\Lambda^{\alpha}_{2}$ to Eq.~(\ref{eq:example_diverge}) has a prefactor $\delta\omega$~\cite{Huang2023}. Given $\Lambda^{\alpha}_{2} \sim \delta \omega^{-1}$ in the limit $\delta\omega\rightarrow 0$ and $\gamma\rightarrow 0$, the term proportional to $\delta\omega \Lambda^{\alpha}_{2}$ gives a finite contribution to the photocurrent conductivity without divergence. 

\begin{figure*}[htbp]
 \includegraphics[width=0.9\linewidth]{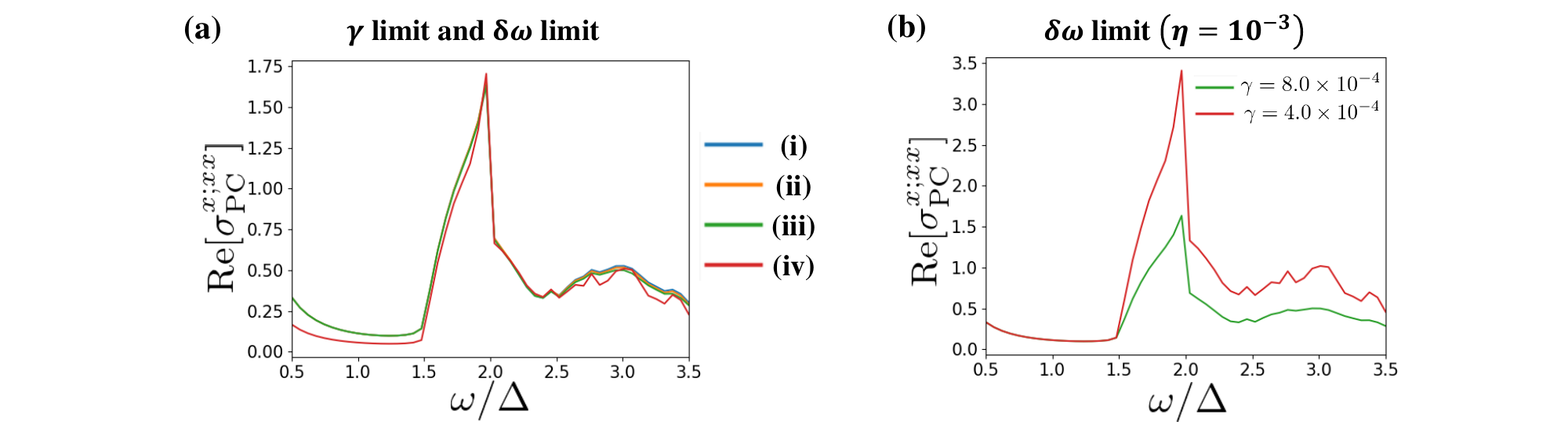}
 \caption{
 (a) Comparison of the photocurrent conductivity $\mathrm{Re}[\sigma^{x;xx}_{\mathrm{PC}}(\omega)]$ obtained in (i) the $\delta\omega$-limit and (ii)-(iv) the $\gamma$-limit. In the $\gamma$-limit, the frequency difference is set as $2\delta \omega = \eta \omega$. 
 (i) $\delta\omega$-limit with $\gamma = 8.0\times 10^{-4}$. 
 (ii) $\gamma$-limit with $\eta=10^{-4}$ and $\gamma = 8.0\times 10^{-4}$.
(iii) $\gamma$-limit with $\eta=10^{-3}$ and $\gamma = 8.0\times 10^{-4}$.
 (iv) $\mathrm{Re}[\sigma^{x;xx}_{\mathrm{PC}}(\omega)]/2$ in $\gamma$-limit with $\eta=10^{-3}$ and $\gamma = 4.0\times 10^{-4}$.
 The result is divided by 2 to show the $\gamma^{-1}$ behavior of $\mathrm{Re}[\sigma^{x;xx}_{\mathrm{PC}}(\omega)]$. (b) $\mathrm{Re}[\sigma^{x;xx}_{\mathrm{PC}}(\omega)]$ obtained in the $\delta\omega$-limit with $\gamma=8.0\times 10^{-4}$ and $\gamma=4.0\times 10^{-4}$. The resonant component appears at $\omega/\Delta >1.5$ and is proportional to $\gamma^{-1}$, indicating that the magnetic injection current is dominant. 
 The parameters of the model are set to $\tilde{\mu}=0.0$, $U=1.15$, and $h=0.02$. The unit of the vertical axis is ($e^{3}/\hbar\cdot a/\mathrm{meV}$).}
 \label{fig:photo_comp}
\end{figure*}

To avoid potential difficulty due to the divergent vertex function, we compare two methods for taking the clean and low different-frequency limit, $\gamma \rightarrow 0$ and $\delta \omega \rightarrow 0$. 
One is the $\delta\omega$-limit, where we assume a finite $\gamma$ and take the limit $\delta\omega\rightarrow 0$. Then, the vertex function is regularized as 
\begin{align}
    \Gamma^{\alpha}(-2\delta\omega-2i\gamma) \rightarrow \Gamma^{\alpha}(-2i\gamma).
\end{align}
The other is the $\gamma$-limit, where we first take the clean limit $\gamma\rightarrow 0$. For a sufficiently small frequency difference $\delta\omega$ we adopt the asymptotic form obtained 
in Sec.~\ref{sec:vertex_func},
\begin{align}
    \Gamma^{\alpha}(-2\delta\omega-2i\gamma) \rightarrow \frac{iA}{- 2 \delta \omega},
\end{align}
where $A$ is the fitting parameter in Eq.~(\ref{eq:clean_vertex}). In the numerical calculation, we introduce a sufficiently small parameter $\eta$ and the frequency difference $2\delta \omega$ is set as $\delta\omega = \eta\omega$.
The numerical results are physically meaningful if the two methods give the same results. 
Figure~\ref{fig:photo_comp}(a) compares $\mathrm{Re}[\sigma^{x;xx}_{\mathrm{PC}}(\omega)]$ obtained in the $\gamma$-limit and that obtained in the $\delta\omega$-limit. It is shown that the numerical results are almost equivalent between the two methods 
and do not depend on the choice of $\eta$ in the $\delta\omega$-limit. Thus, we can obtain physically meaningful results, which are independent of the treatment of the massless collective modes. 

Next, we discuss the dependence of the photocurrent conductivity on the phenomenological scattering rate $\gamma$ to verify that the intrinsic effect is dominant in our results. Figure~\ref{fig:photo_comp}(b) shows $\mathrm{Re}[\sigma^{x;xx}_{\mathrm{PC}}(\omega)]$ obtained in the $\delta\omega$-limit with different $\gamma$. We see that the magnitude of the resonant contribution is doubled when $\gamma$ is decreased to half. The photocurrent conductivity with $\gamma = 4.0\times 10^{-4}$ is also compared with that with $\gamma = 8.0\times 10^{-4}$ in Fig.~\ref{fig:photo_comp}(a), where the former is divided by 2. We see a coincidence of the two results at $\omega/\Delta > 1.5$. Thus, the photocurrent conductivity in this region shows the $\gamma^{-1}$ dependence 
signifying the dominance of the magnetic injection current, which is one of the intrinsic mechanisms of the photocurrent conductivity~\cite{ Zhang2019, Wang2020, Fei2020, Mu2023}. This is consistent with classification theory~\cite{Watanabe2022}, although the vertex correction due to the collective modes was not considered there. 
Note that for the parameters in Fig.~\ref{fig:photo_comp}, the photocurrent conductivity in the low-frequency region $\omega/\Delta < 1.5$ arises from the non-resonant contribution~\cite{Watanabe2022}, which is independent of the scattering rate.
From these analyses, it is verified that our formulation can adequately describe the intrinsic contribution of the photocurrent conductivity in the Rashba superconductor.

\subsection{Sign reversal of photocurrent conductivity}

\begin{figure*}[htbp]
 \includegraphics[width=0.9\linewidth]{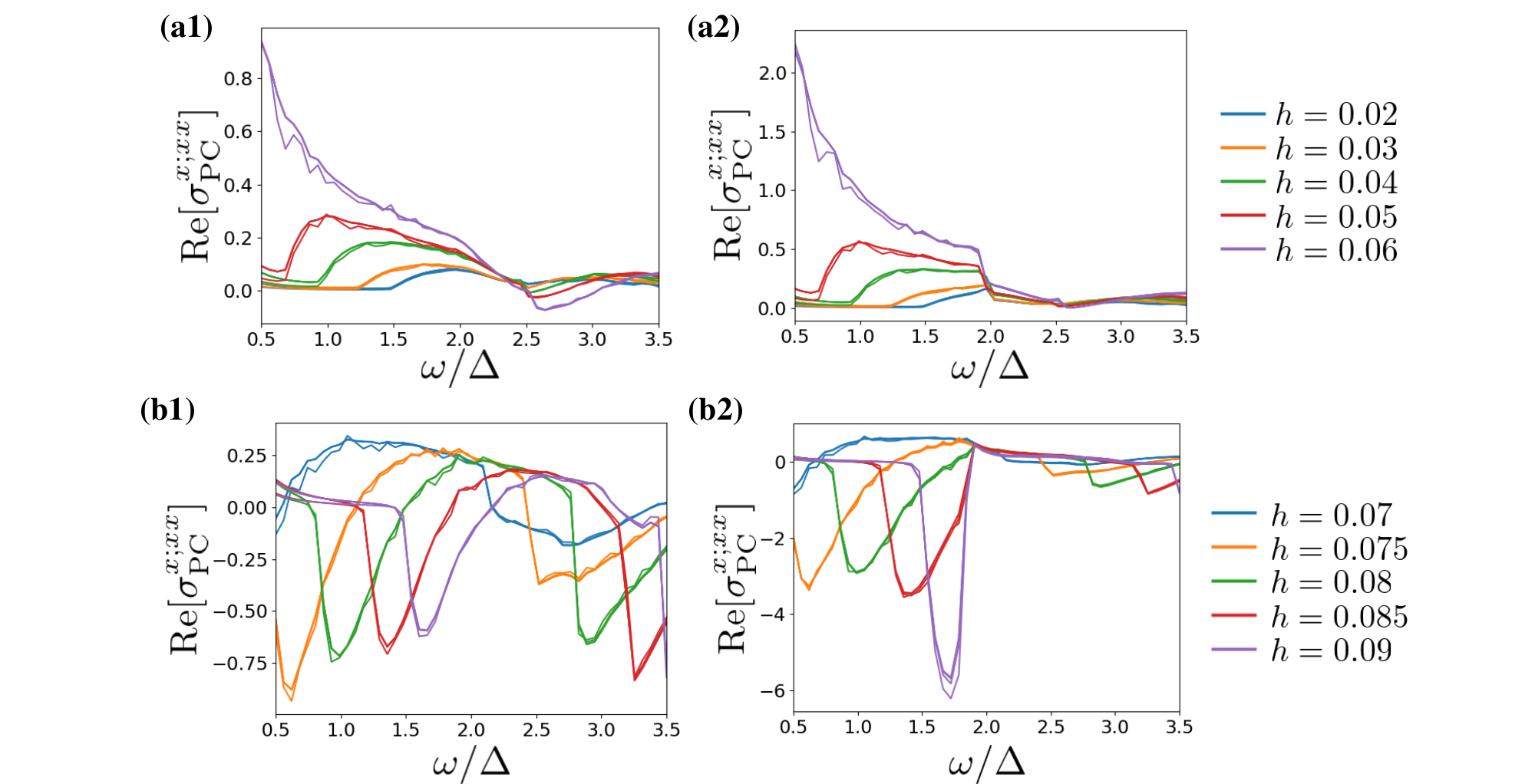}
 \caption{
 The photocurrent conductivity $\mathrm{Re}[\sigma^{x;xx}_{\mathrm{PC}}]$ in the topologically trivial states [(a1), (a2)] and the topological superconducting states [(b1), (b2)] with varying the magnitude of the magnetic field $h$. For each magnetic field, the photocurrent conductivity is calculated in the $\gamma$ limit and the $\omega$ limit methods with setting the same parameter sets $\eta$ and $\gamma$ as in Figure~\ref{fig:photo_comp}(a). We rescale the results with respect to $\gamma$ as we have done in Figure~\ref{fig:photo_comp}(a) and plot the results with the same color for each $h$. We see that the results are almost independent of details of numerical calculations in the whole parameter range.  Figures~\ref{fig:PC_result}(a1) and \ref{fig:PC_result}(b1) show the bare photocurrent conductivity without vertex correction, while Figures~\ref{fig:PC_result}(a2) and \ref{fig:PC_result}(b2) plot the corrected conductivity with vertex correction. The parameters are set as $\tilde{\mu}=0.0$ and $U=1.15$. The unit of the vertical axis is ($e^{3}/\hbar\cdot a/\mathrm{meV}$).}
 \label{fig:PC_result}
\end{figure*}

In this subsection, we carry out numerical calculations of the photocurrent conductivity around the topological transition. The results with and without the vertex correction are presented. 
We show that the sign reversal is robust against the contribution of collective modes.

When we set the parameters as $\tilde{\mu}=0.0$ and $U=1.15$, the model describes topologically trivial superconducting states in the low magnetic field region $0\leq h \leq 0.06$ but becomes topological superconducting states in the high magnetic field region $0.07 \leq h \leq 0.09$. Figure~\ref{fig:PC_result} shows the photocurrent conductivity $\mathrm{Re}[\sigma^{x;xx}_{\mathrm{PC}}]$ with varying the magnitude of the magnetic field. Figures~\ref{fig:PC_result}(a1) and \ref{fig:PC_result}(b1) plot the bare photocurrent conductivity, while the corrected one with vertex correction is shown in Figs.~\ref{fig:PC_result}(a2) and \ref{fig:PC_result}(b2).
We see that qualitative behaviors such as the shift of the resonant photocurrent in the frequency dependence are similar between the bare and corrected photocurrent conductivities. The shift of the resonant contributions reflects the suppression of the energy gap by the magnetic field.
The energy gap in the trivial state becomes smaller with $h$ being closer to the critical value $h_{c}$ and closes at the topological transition. Through the topological transition, the energy gap opens again as the magnetic field is enlarged. Upon closing of the energy gap, the resonant contribution of the photocurrent conductivity $\mathrm{Re}[\sigma^{x;xx}_{\mathrm{PC}}]$ shows the frequency shift. 


Our previous calculation without vertex correction predicted the sign reversal of the resonant contribution in the low-frequency region $\omega/\Delta < 2.0$, which is reproduced in Figs.~\ref{fig:PC_result}(a1) and \ref{fig:PC_result}(b1). The sign reversal behavior appears even when we take into account the vertex correction. As shown in Figs.~\ref{fig:PC_result}(a2) and \ref{fig:PC_result}(b2), the sign of the corrected photocurrent conductivity also reverses at the topological transition. Therefore, the sign reversal is robust against the contributions of collective modes, and thus this phenomenon is expected to be useful as an indicator of topological superconductivity.


\begin{figure}[htbp]
 \includegraphics[width=0.9\linewidth]{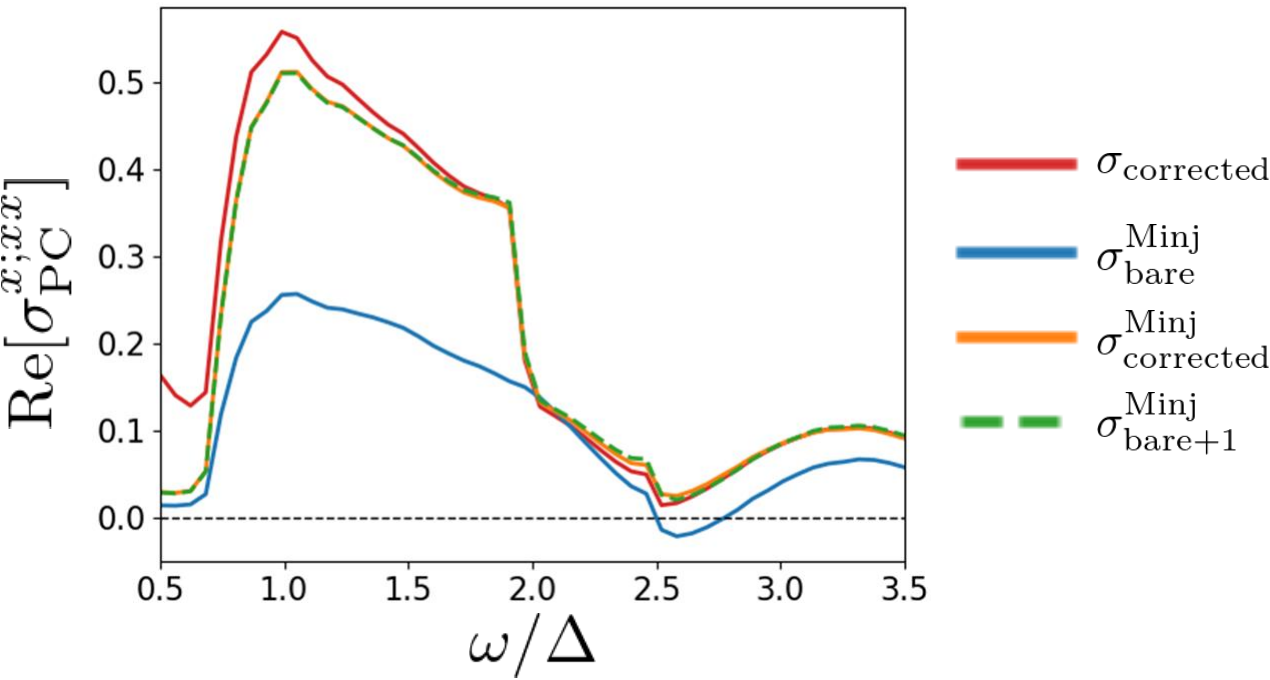}
 \caption{
 The photocurrent conductivity $\mathrm{Re}[\sigma^{x;xx}_{\mathrm{PC}}]$ and its components. The total photocurrent conductivity $\sigma_{\mathrm{corrected}}$ with vertex correction (red line) is compared with the bare and corrected magnetic injection currents $\sigma^{\mathrm{Minj}}_{\mathrm{bare}}$ (blue line) and $\sigma^{\mathrm{Minj}}_{\mathrm{corrected}}$ (orange line). We also show $\sigma^{\mathrm{Minj}}_{\mathrm{bare+1}}$ (green dashed line), which is the magnetic injection current obtained by taking into account only the vertex correction due to the amplitude mode. 
 The parameters are set as $\tilde{\mu}=0.0$, $h=0.05$ and $U=1.15$. The unit of the vertical axis is ($e^{3}/\hbar\cdot a/\mathrm{meV}$).}
 \label{fig:Minj}
\end{figure}


\subsection{Magnetic injection current}
For a step to clarify the origin of the sign reversal phenomenon, we discuss the injection current, which is one of the mechanisms of the photocurrent generation in the interband regime. The injection current grows with time in the short time pulse and saturates at the relaxation time $\tau$~\cite{Sipe2000}. Thus, the injection current is characterized by the $O(\tau)$ relaxation time dependence in the stationary state~\cite{Zhang2019}. The components corresponding to the linear and circular photogalvanic effects are called electric and magnetic injection currents, respectively. 
The bare magnetic injection current, which is calculated without vertex corrections, is formulated as
\begin{align}
\sigma^{\mathrm{Minj}}_{\mathrm{bare}}
=    -\frac{\pi}{4\gamma}\sum_{a,b, \bm{k}} \Delta\tilde{J}^{x}_{ab}\frac{\mathrm{Re}[\tilde{J}^{x}_{ba}\tilde{J}^{x}_{ab}]}{E_{ba}^{2}}F_{ab},
\label{eq:bare_Minj}
\end{align}
where we define $F_{ab}=f_{ab}\delta(\omega - E_{ba})$ and $\Delta\tilde{J}^{\alpha}_{ab}=\tilde{J}^{\alpha}_{aa} - \tilde{J}^{\alpha}_{bb}$.
It has been shown that a large magnetic injection current can be generated in some magnetic quantum materials~\cite{ Zhang2019, Wang2020, Fei2020,Mu2023}. 
In our previous study~\cite{Tanaka2024}, the magnetic injection current was shown to be dominant in the bare photocurrent conductivity in the model of Rashba superconductors. Therefore, it is also expected that the magnetic injection current is essential for the corrected photocurrent conductivity calculated with vertex corrections.

To discuss the magnetic injection current in the self-consistent approximation, we introduce the corrected magnetic injection current as
\begin{align}
\sigma^{\mathrm{Minj}}_{\mathrm{corrected}}
=    -\frac{\pi}{4\gamma}\sum_{a,b, \bm{k}} \Delta\tilde{\Gamma}^{x\mathrm{2A}}_{ab}(0)\frac{\mathrm{Re}[\tilde{\Gamma}^{x\mathrm{R}}_{ba}(\omega)\tilde{\Gamma}^{x\mathrm{R}}_{ab}(-\omega)]}{E_{ba}^{2}}F_{ab},
\label{eq:Minj_corrected}
\end{align}
where we use the following notations
\begin{align}
\Delta\tilde{\Gamma}^{x\mathrm{2A}}_{ab}(\omega) &= \tilde{\Gamma}^{x}_{aa}(\omega - 2i\gamma) - \tilde{\Gamma}^{x}_{bb}(\omega - 2i\gamma), \\ 
\tilde{\Gamma}^{x\mathrm{R}}_{ab}(\omega) &= \tilde{\Gamma}^{x}_{ab}(\omega + i\gamma).
\end{align}
This contribution is obtained from the formula for the second-order nonlinear responses [Eq.~\eqref{eq:nonlinear_conductivity}].
In the formula for the corrected magnetic injection current, the electric current operator appearing in the bare magnetic injection current is replaced with the one-photon vertex $\tilde{\Gamma}$, 
which includes the effects of the collective mode. 
Figure~\ref{fig:Minj} plots the total corrected photocurrent conductivity $\sigma_{\mathrm{corrected}}$ and the component $\sigma^{\mathrm{Minj}}_{\mathrm{corrected}}$. We see that $\sigma^{\mathrm{Minj}}_{\mathrm{corrected}}$ gives a dominant contribution to the photocurrent conductivity in this frequency region. 
This is consistent with our expectation, and thus the magnetic injection current is dominant even in the self-consistent approximation, which takes into account the vertex correction.

Comparing the bare and corrected magnetic injection current in Fig.~\ref{fig:Minj}, we find that 
the magnetic injection current is enhanced by the vertex correction. 
Let us show that the amplitude mode is essential for the enhancement.
To elucidate the importance of the amplitude mode, we calculate $\sigma^{\mathrm{Minj}}_{\mathrm{bare+1}}$ where the vertex correction to the one-photon vertex in $\sigma^{\mathrm{Minj}}_{\mathrm{corrected}}$ is evaluated by taking into account  
only $\Lambda^{x}_{1}$. The amplitude mode influences $\sigma^{\mathrm{Minj}}_{\mathrm{bare+1}}$ through the vertex $\Lambda^{x}_{1}$, while the effect of the phase mode is ignored. 
Figure~\ref{fig:Minj} shows that $\sigma^{\mathrm{Minj}}_{\mathrm{bare+1}}$ is almost equivalent to $\sigma^{\mathrm{Minj}}_{\mathrm{corrected}}$.
Therefore, we conclude that the enhancement of the photocurrent conductivity originates from the amplitude mode, whereas the contribution of the phase mode is ignorable.

\begin{figure*}[htbp]
 \includegraphics[width=0.9\linewidth]{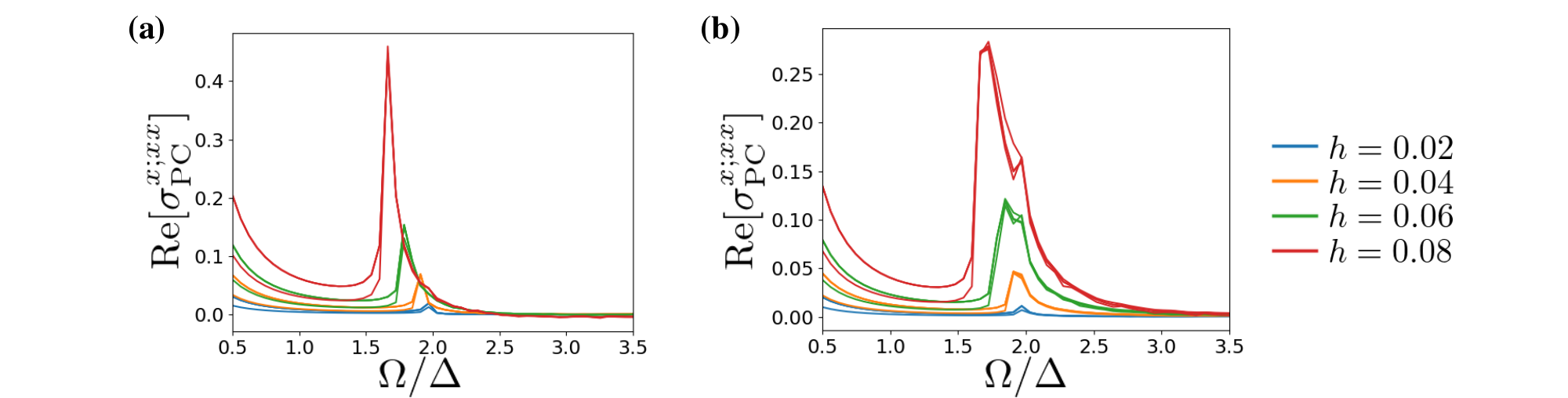}
 \caption{The photocurrent conductivity $\mathrm{Re}[\sigma^{x;xx}_{\mathrm{PC}}]$ in the trivial states with varying the magnitude of the magnetic field. Numerical results obtained by several methods are shown for each $h$, as in Fig.~\ref{fig:PC_result}. 
 (a) The bare photocurrent conductivity, and (b) the corrected photocurrent conductivity. The parameters are set as $\tilde{\mu}=0.2$ and $U=1.0$. The unit of the vertical axis is ($e^{3}/\hbar\cdot a/\mathrm{meV}$).}
 \label{fig:PC_mu_3}
\end{figure*}

We have shown that the resonant component of the photocurrent conductivity showing sign reversal is enhanced by the amplitude mode.  Here, we discuss the role of interband pairing for the enhancement. To verify the importance of interband pairing, we calculate the photocurrent conductivity in the case of $\tilde{\mu} = 0.2$. Although interband pairing is significant for $\tilde{\mu} = 0$, intraband pairing is dominant for $\tilde{\mu} = 0.2$. 
As shown in Fig.~\ref{fig:PC_mu_3}, 
for $\tilde{\mu} = 0.2$, 
the photocurrent conductivity is not enhanced but suppressed by the vertex correction in some parameter regions. 
Note that the sign reversal does not occur in this case because the critical magnetic field $h_{c}=\sqrt{{\tilde{\mu}}^{2} + \Delta^{2}}$ of the topological transition is too large to stabilize the superconducting state.
In contrast to the case of $\tilde{\mu} = 0$, sharp peaks of the photocurrent conductivity are slightly suppressed by the collective modes. 
Thus, it is indicated that the interband pairing is essential for the enhancement of the photocurrent conductivity by the amplitude mode. 

Another ingredient for the enhancement of the photocurrent conductivity is the quantum geometry of Bogoliubov quasiparticles. 
In the case of $\tilde{\mu} = 0$, the quantum-geometric properties are pronounced around the Dirac point and enhance the bare magnetic injection current~\cite{Tanaka2024}. Indeed, the bare photocurrent conductivity with $\tilde{\mu} = 0$ [Figs.~\ref{fig:PC_result}(a1) and \ref{fig:PC_result}(b1)] is larger than that with $\tilde{\mu} = 0.2$ [Fig.\ref{fig:PC_mu_3}(a)]. Because the contribution of the collective modes is expected to be of the same order as the bare photocurrent conductivity, the vertex correction is naturally enhanced by the quantum-geometric properties around the Dirac point. Further discussion of the interplay between the interband pairing and the quantum-geometric properties 
is left for future work.


\subsection{Mechanism of the sign reversal}

\begin{figure}[htbp]
 \includegraphics[width=0.9\linewidth]{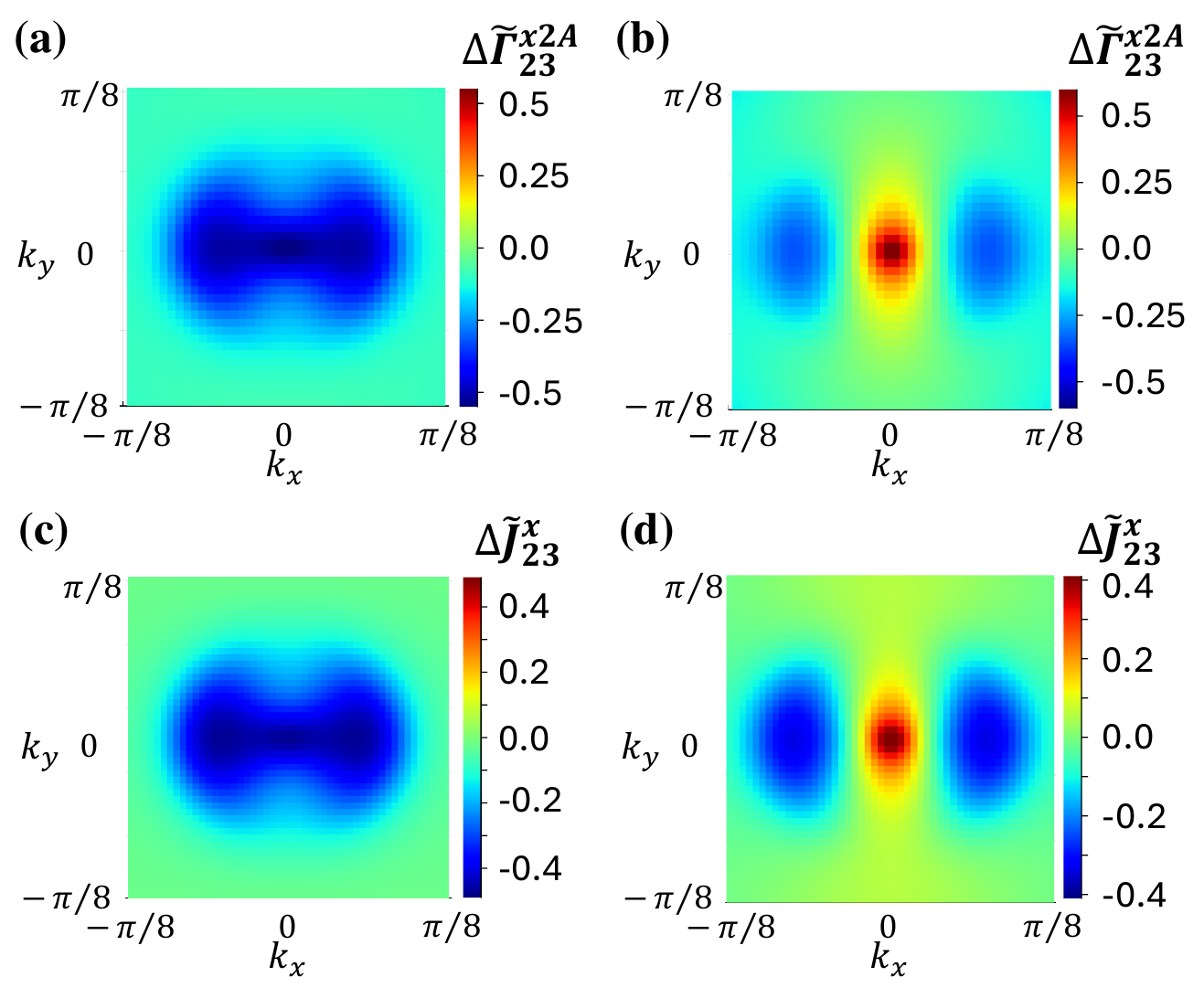}
 \caption{Distribution of (a) (b) $\Delta\tilde{\Gamma}^{x2A}_{23}$  and (c) (d) $\Delta\tilde{J}^{x}_{23}$ in the $\bm{k}$ space. We set the magnetic field (a) (c) $h=0.05$ and (b) (d) $h=0.08$ below and above the topological transition, respectively. The parameters are set as $\tilde{\mu}=0.0$ and $U=1.15$.}
 \label{fig:gp_dist}
\end{figure}

Finally, we discuss the mechanism of sign reversal of the photocurrent conductivity associated with the topological transition. 
The resonant photocurrent conductivity in the low-frequency region $\omega/\Delta < 2.0$ originates from the low-energy quasiparticle excitation~\cite{Tanaka2024}, whose energies are $E_{2\bm{k}}$ and $E_{3\bm{k}}$ in our model. Note that we have defined the Bogoliubov quasiparticle's bands as $E_{1\bm{k}}\leq E_{2\bm{k}} \leq 0 \leq E_{3\bm{k}} \leq E_{4\bm{k}}$ [see Figure \ref{fig:schematic}(c)]. 
Therefore, we focus on the interband transition between the $E_{2\bm{k}}$ and $E_{3\bm{k}}$ bands, which is called the B2 transition in this paper. 
In the low-frequency region $\omega/\Delta < 2.0$, the magnetic injection current that includes the vertex correction is dominantly given by the B2 transition because the factor $F_{ab}=f_{ab}\delta(\omega - E_{ba})$ in Eq.~\eqref{eq:Minj_corrected} prohibits the contribution originating from the other interband transition.
In the zero temperature limit, the magnetic injection current originating from the B2 transition $\sigma^{\mathrm{Minj}}_{\mathrm{corrected, B2}}$ is given by
\begin{align}
&\sigma^{\mathrm{Minj}}_{\mathrm{corrected, B2}}(\omega) \notag \\
&=    -\frac{\pi}{4\gamma}\sum_{\bm{k}} \Delta\tilde{\Gamma}^{x\mathrm{2A}}_{23}(0)\frac{\mathrm{Re}[\tilde{\Gamma}^{x\mathrm{R}}_{32}(\omega)\tilde{\Gamma}^{x\mathrm{R}}_{23}(-\omega)]}{E_{32}^{2}}\delta(\omega - E_{32}).
\end{align}
We can show that the factor $\mathrm{Re}[\tilde{\Gamma}^{x\mathrm{R}}_{32}(\omega)\tilde{\Gamma}^{x\mathrm{R}}_{23}(-\omega)]$ must be positive [see Appendix \ref{app:hermite_conj}]. Therefore, the sign of $\sigma^{\mathrm{Minj}}_{\mathrm{corrected, B2}}$ is determined by the difference in the one-photon vertex $\Delta\tilde{\Gamma}^{x\mathrm{2A}}_{23}(0)$. 
Because the resonant component of $\mathrm{Re}[\sigma^{x;xx}_{\mathrm{PC}}]$ in the low-frequency region arises from the momentum space around $\bm{k}=\bm{0}$, 
the sign reversal of $\mathrm{Re}[\sigma^{x;xx}_{\mathrm{PC}}]$ is caused by $\Delta\tilde{\Gamma}^{x\mathrm{2A}}_{23}(0)$ around $\bm{k}=\bm{0}$, which is shown in Fig.~\ref{fig:gp_dist}. 
We find that the sign of $\Delta\tilde{\Gamma}^{x\mathrm{2A}}_{23}(0)$ around $\bm{k}=\bm{0}$ is opposite between the trivial state [Fig.~\ref{fig:gp_dist}(a)] and the topological superconducting state [Fig.~\ref{fig:gp_dist}(b)]. 

Next, we discuss the reason why the sign of $\Delta\tilde{\Gamma}^{x\mathrm{2A}}_{23}(0)$ around $\bm{k}=\bm{0}$ changes at the topological transition.
The sign change is associated with the band inversion at the topological transition.
The eigenenergies of Bogoliubov quasiparticles $E_{2\bm{k}}$ and $E_{3\bm{k}}$ are obtained at $\bm{k}=\bm{0}$
\begin{align}
    (E_{2\bm{k}}, E_{3\bm{k}}) = 
    \begin{cases}
        (E^{(-)}, E^{(+)}) & (\sqrt{\tilde{\mu}^{2} + \Delta^{2}} > h) \\
        (E^{(+)}, E^{(-)}) & (\sqrt{\tilde{\mu}^{2} + \Delta^{2}} < h)
    \end{cases}, \label{eq:energy_pm}
\end{align}
where $E^{(-)}$ and $E^{(+)}$ are defined as
\begin{align}
    E^{(-)} = - \sqrt{\tilde{\mu}^{2} + \Delta^{2}} + h, \quad E^{(+)} =  \sqrt{\tilde{\mu}^{2} + \Delta^{2}} - h.
\end{align}
The band inversion between the $E_{2\bm{k}}$ and $E_{3\bm{k}}$ bands occurs 
at the critical magnetic field $h_{c}=\sqrt{\tilde{\mu}^{2} + \Delta^{2}}$; $E^{(-)}$ and $E^{(+)}$ are exchanged.
%
Here, we analytically calculate $\Delta\tilde{\Gamma}^{x\mathrm{2A}}_{23}(0)$.
When we adopt the notation $\tilde{\Gamma}^{\alpha2A}_{aa} = \tilde{\Gamma}^{\alpha}_{aa}(-2i\gamma)$ and $\Lambda^{\alpha2A}_{1} = \Lambda^{\alpha}_{1}(-2i\gamma)$, $\tilde{\Gamma}^{\alpha2A}_{aa}$ is obtained as,
\begin{align}
    \tilde{\Gamma}^{\alpha2A}_{aa} &= \partial_{\alpha}\xi+\partial_{\alpha}\bm{g}\cdot \hat{\bm{h}} - \frac{\Delta}{\sqrt{\tilde{\mu}^{2} + \Delta^{2}}}\Lambda^{\alpha2A}_{1} \quad ({\rm for} \,\, E_{a}=E^{(-)}), \\
    \tilde{\Gamma}^{\alpha2A}_{aa} &= \partial_{\alpha}\xi -\partial_{\alpha}\bm{g}\cdot \hat{\bm{h}} + \frac{\Delta}{\sqrt{\tilde{\mu}^{2} + \Delta^{2}}}\Lambda^{\alpha2A}_{1} \quad ({\rm for} \,\, E_{a}=E^{(+)}).
\end{align}
We can verify that $\tilde{\Gamma}^{\alpha2A}_{aa}$ at $\bm{k}=\bm{0}$ is exactly independent of $\Lambda^{\alpha}_{2}(-2i\gamma)$, the vertex correction due to the phase mode.
Thus, $\Delta\tilde{\Gamma}^{\alpha2A}_{23}(0)$ is given by 
\begin{align}
    \Delta\tilde{\Gamma}^{\alpha2A}_{23}(0) =
    \begin{cases}
        2(\partial_{x}\bm{g}\cdot \hat{\bm{h}}) - \frac{2\Delta}{\sqrt{\tilde{\mu}^{2} + \Delta^{2}}}\Lambda^{\alpha2A}_{1} & \left(h_{c} > h \right) \\
        -2(\partial_{x}\bm{g}\cdot \hat{\bm{h}}) + \frac{2\Delta}{\sqrt{\tilde{\mu}^{2} + \Delta^{2}}}\Lambda^{\alpha2A}_{1} & \left(h_{c} < h \right)
    \end{cases}.
\end{align}
The sign of $\Delta\tilde{\Gamma}^{\alpha2A}_{23}(0)$ is shown to change at the topological transition.

A similar sign change has been shown for the group velocity difference $\Delta\tilde{J}^{\alpha}_{23}$~\cite{Tanaka2024}, and it is the origin of the sign reversal of the bare photocurrent conductivity. 
Note that $\Delta\tilde{J}^{\alpha}_{23}$, which we call the "group velocity difference", is not exactly equal to $\partial_{k_{\alpha}}E_{2} - \partial_{k_{\alpha}}E_{3}$ due to the mixing of electron and hole in the Bogoliubov quasiparticle.
However, it is reasonable to interpret $\Delta\tilde{J}^{\alpha}_{23}$ as the group velocity difference between the Bogoliubov quasiparticles, since $\tilde{J}^{\alpha}_{a}$ can be described as the weighted sum of the group velocities of the electron and hole bands~\cite{Tanaka2024}.

To discuss the contribution of vertex correction $\Lambda^{\alpha2A}_{1}$ to $\Delta\tilde{\Gamma}^{\alpha2A}_{23}(0)$, we compare the 
group velocity difference $\Delta\tilde{J}^{\alpha}_{23}$ [Figs.~\ref{fig:gp_dist}(c) and (d)] with $\Delta\tilde{\Gamma}^{\alpha2A}_{23}(0)$ [Figs.~\ref{fig:gp_dist}(a) and (b)].
We see that $\Delta\tilde{\Gamma}^{\alpha2A}_{23}(0)$ 
and $\Delta\tilde{J}^{\alpha}_{23}$ show qualitatively the same behaviors, while the magnitude of $\Delta\tilde{\Gamma}^{\alpha2A}_{23}(0)$ is larger than $\Delta\tilde{J}^{\alpha}_{23}$. 
Thus, the vertex correction is cooperative with the group velocity difference $\Delta\tilde{J}^{\alpha}_{23}$. 
Therefore, the sign reversal of $\Delta\tilde{\Gamma}^{\alpha2A}_{23}(0)$ at the topological transition 
is robust against the contribution of the vertex corrections,
and the sign reversal behavior of the photocurrent conductivity 
is expected to appear regardless of the details of the models.

Now we arrive at the physical interpretation of the sign reversal of the magnetic injection current. The injection current is a contribution induced by the group velocity difference between the bands~\cite{Orenstein2021}. 
Indeed, the group velocity difference $\Delta\tilde{J}^{\alpha}_{ab}$ is an essential ingredient of the bare magnetic injection current in Eq.~\eqref{eq:bare_Minj}.
Therefore, the magnetic injection current is generally sensitive to the band inversion leading to the topological transition. This interpretation holds when the vertex correction due to the collective modes is taken into account, 
because the difference of the one-photon vertex $\Delta\tilde{\Gamma}^{x\mathrm{2A}}_{23}(0)$ shows qualitatively the same behaviors as the group velocity difference $\Delta\tilde{J}^{\alpha}_{23}$.

In this section, we have shown the sign reversal of the photocurrent conductivity $\mathrm{Re}[\sigma^{x;xx}_{\mathrm{PC}}]$ in the self-consistent response approximation, signaling the transition between the trivial and topological superconducting states. 
The magnetic injection current is dominant in the clean limit $\gamma \rightarrow 0$ and sensitive to band inversion at the topological transition because the injection current originates the difference in velocities between bands. Moreover, the magnetic injection current is significantly enhanced by the amplitude mode. Therefore, the photocurrent conductivity can be useful for detecting the topological transition. 
Dirac electrons are an intriguing platform for the interplay of collective modes, quantum-geometric properties, and interband pairing, leading to enhanced photocurrent generation in superconductors. 

\section{Discussion}
\label{sec:discussion}
We have demonstrated the enhancement of the linear and nonlinear optical responses due to the collective modes in the superconductors with Rashba spin-orbit coupling in the absence of the long-range Coulomb interaction. In realistic superconductors, the massless phase mode is lifted to a high energy in the scale of the plasma frequency as a consequence of the Anderson-Higgs mechanism~\cite{Anderson1958_1, Anderson1958_2, Nambu1960, Anderson1963}. This change in the phase mode due to the long-range Coulomb interaction potentially affects the optical responses 
through the vertex correction. However, the vertex correction in our model mainly arises from the amplitude (Higgs) mode, and the effect of the phase mode is negligible. Therefore, the characteristic behavior of the optical responses in our model is expected to remain unchanged even with the long-range Coulomb interaction. 
The optical responses are enhanced by the amplitude mode, which is insusceptible to the Coulomb interaction. 

We comment on a difference between the formalism in Ref.~\cite{Huang2023} and ours.  Ref.~\cite{Huang2023} deals with single-band superconductors, and the formula contains the $\tau_{3}$ components of the vertex function, which originate from the normal self-energy. The $\tau_{3}$ terms can take into account the contribution of the plasmon mode and play an important role in the Anderson-Higgs mechanism~\cite{Nambu1960}. In this study, we have investigated the multiband system, 
where the vertex correction originating from the normal self-energy can be complicated. Thus, we ignored the plasmon contributions and the Anderson-Higgs mechanism for simplicity. 
Because 
the phase mode is totally unimportant in our results, the simplification is justified. 
It was also shown that the effect of the $\tau_{3}$ components is not significant in the linear responses in the single-band $d$-wave superconductors~\cite{watanabeS2024,WatanabeS2025}. 

\subsection{Material platform and experiment}

The sign reversal of the photocurrent conductivity is a characteristic behavior associated with the topological transition. This phenomenon can be used to verify several candidates for topological superconductors. The two-dimensional $s$-wave superconductor in the presence of the Rashba spin-orbit coupling and the Zeeman field, which is the setup in this study, is a platform of class D topological superconductivity~\cite{Sato2009, Sato2010, Sau2010}. Artificial heterostructures of an $s$-wave superconductor and a ferromagnetic insulator are proposed for an experimental platform~\cite{Sau2010, Lee2012, Nagai2016}. Recent experiments also reported class D topological superconductivity in van der Waals heterostructures that combine a monolayer ferromagnet with a superconducting transition metal dichalcogenide~\cite{Kezilebieke2020, Kezilebieke2022}. In a theoretical study~\cite{Raj2024}, a model of class DIII topological superconductivity also shows a sign change in the photocurrent conductivity. 
The mechanism of sign reversal must be different between the class D and class DIII topological superconductors because the magnetic injection current is absent in the latter. 
Although a comprehensive study is needed, sign reversal of the photocurrent conductivity may be a ubiquitous phenomenon in topological superconductors.
%
In the experiment, the sign of the injection current has been detected by direct current measurements~\cite{Ma2017, Ma2019} and all-optical techniques~\cite{Ree2020, Sirica2019, Gao2020}. Therefore, the sign reversal behavior at the topological transition can be accessed. 

In Secs.~\ref{sec:linear} and \ref{sec:nonlinear}, we have shown the giant linear and nonlinear optical responses that appear when the Fermi level lies near the Dirac point. The combination of quantum geometry and collective modes plays an essential role in the enhancement. These features are expected to be realized in various quantum materials, as is exemplified in the following.
The interface of superconductors and topological insulators is an ideal platform, where superconductivity is induced in the surface Dirac/Weyl states~\cite{Meng2012, Meng2012(E), Yasuda2019}. Atomically thin layered materials are another potential platform for quantum geometry and topological nature. Monolayer $\mathrm{WTe_{2}}$ is a centrosymmetric topological insulator~\cite{Wu2018, Fei2017} and can be gate-tuned to the superconducting state~\cite{Fatemi2018,Sajadi2018}. 
In few-layer $\mathrm{WTe_{2}}$, the Berry curvature dipole was detected by the nonlinear Hall effect~\cite{Kang2019, MaQ2019}. 
The nonlinear Hall measurement in magnetic systems can also detect indication for the quantum metric, and it was observed in even-layer $\mathrm{MnBi_{2}Te_{4}}$~\cite{MnBi2Te4Science2023}. 
Superconductivity in $\mathrm{MnBi_{2}Te_{4}}$ can be realized in the heterostructure~\cite{Yuan2024}. 
An iron-based superconductor FeSe thin film 
is also an intriguing platform of nonlinear optical responses in superconductors, 
because the quantum geometry of normal electrons significantly affects superconductivity~\cite{Kitamura2021, KitamuraPRB2022, KitamuraPRR2022}.

Our results imply that the amplitude mode enhances the optical responses through interband pairing. 
In general, intraband pairing is expected to be favored, because the interband pairing does not contribute to the thermodynamic stability of superconductivity in the weak coupling limit. 
One of the methods for 
controlling interband pairing is inducing the Zeeman field into noncentrosymmetric spin-singlet superconductors. 
For this purpose, two-dimensional transition metal dichalcogenides are a promising platform because the superconductivity is robust against in-plane external magnetic fields and the orbital effect is largely avoided~\cite{Saito2016, Lu2015, Xi2016, delaBarrera2018, Liu2020, Cui2019, Rhodes2021, Zhang2023}. 
Another mechanism for a significant interband pairing was proposed in noncentrosymmetric superconductors where the spin-orbit coupling has topological defects in the momentum space, such as the Weyl points~\cite{Bauer2012, Yanase2008}.
When we go beyond the weak coupling theory, strong coupling theories for superconductivity generally predict a finite interband pairing~\cite{Bauer2012, Yanase2008}.

\section{Summary}
\label{sec:summary}
We have studied linear and nonlinear optical responses in superconductors 
based on the self-consistent response approximation, which is derived from the equation of motion satisfying the total particle number conservation.
Our formulas can be applied to the effective Hamiltonian describing various classes of unconventional superconductors including anisotropic and helical superconductors. 
The self-consistent response approximation contains the effects of the collective modes such as the amplitude and phase modes through the vertex corrections. In Sec.~\ref{sec:linear} and \ref{sec:nonlinear}, we discussed the effects of the collective modes on the linear and nonlinear conductivity. Furthermore, we extended the self-consistent response approximation to superconductors with multi-component order parameters, such as the spin-triplet superconductors and chiral superconductors.
The collective Leggett mode and Bardasis-Schriffer mode arising from multiple Cooper channels can be appropriately taken into account in the response functions. 

We applied the formulas of optical responses to a model of topological Rashba superconductors. The linear and nonlinear optical responses are strongly enhanced due to the interplay of the amplitude mode and quantum geometry when the Fermi level lies on the Dirac point. The enhancement by the amplitude mode disappears in the single-band $s$-wave superconductor, and the multiband character in the bands split by the Rashba spin-orbit coupling and the Zeeman field is essential. In particular, interband pairing is essential for the enhancement. Because we analyzed the model with a single-component order parameter, our results are not related to the Bardasis-Schriffer mode, which originates from the multiple Cooper channels. The interplay of collective modes and quantum geometry potentially gives significantly large optical responses in topological materials, and further studies on various classes of superconductors are desired. 

We have demonstrated the sign reversal of the photocurrent conductivity between the topologically trivial and nontrivial superconducting states. In our previous work, the bare photocurrent conductivity shows the sign reversal, 
and it is attributed to the magnetic injection current. 
Because the velocity difference between bands induces the injection current, the magnetic injection current is sensitive to the band inversion accompanied by the topological transition and causes a sign reversal of the photocurrent conductivity. In this study, we have extended the formula of the magnetic injection current to contain the vertex correction due to the collective modes. This contribution is dominant in the resonant photocurrent conductivity and shows qualitatively the same behavior as the bare magnetic injection current. 
Consequently, the photocurrent conductivity that takes into account the vertex correction also shows the sign reversal at the topological transition. 
Since the self-consistent response approximation satisfies the total particle number conservation law and includes the effects of collective modes, the results in this study are more accurate compared with the previous ones. 
Therefore, we expect that the nonlinear superconducting optics would provide a bulk probe for the detection of topological superconductivity in Rashba systems.
The enhancement of the photocurrent conductivity by the amplitude mode is also an advantage for detecting the sign reversal behavior.

\begin{acknowledgments}
We thank Haruki Watanabe, Sena Watanabe, Hikaru Watanabe, Koki Shinada, and Shin Kaneshiro for useful discussion. 
This work is supported by JSPS KAKENHI (Grant Numbers JP22H01181, JP22H04933, JP23K17353, JP23K22452, JP24K21530, JP24H00007).
H.T. is supported by JSPS KAKENHI (Grants Number JP23KJ1344).
\end{acknowledgments}

\appendix
\onecolumngrid

\section{Definition of the superoperator}
\label{App:def_of_superop}
We define the superoperator $\overleftrightarrow{U}_{ab}(\bm{k}, \tau_{1})[X]$ to represent the self-consistent equation of the vertex function shortly. The superoperator $\overleftrightarrow{U}_{ab}(\bm{k}, \tau_{1})[X]$ acting on $X\equiv \overleftrightarrow{X}_{ab}(\bm{k},\tau)$ is expressed as
\begin{align}
  \overleftrightarrow{U}_{ab}(\bm{k}, \tau_{1})[X] 
  \equiv -&\frac{U}{V}\varphi_{ab}(\bm{k})\sum \varphi^{\dagger}_{dc}(\bm{k}^{\prime})\overleftrightarrow{\mathrm{P}_{+}}
  \int d\bar{\tau}_{1}\overleftrightarrow{\mathcal{G}}_{ci}(\bm{k}^{\prime}, \tau_{1}, \bar{\tau}_{1})\overleftrightarrow{X}_{ij}(\bm{k}^{\prime},\bar{\tau}_{1})\overleftrightarrow{\mathcal{G}}_{jd}(\bm{k}^{\prime}, \bar{\tau}_{1},\tau_{1})\overleftrightarrow{\mathrm{P}_{-}} \notag \\
  &-\frac{U}{V}\varphi^{\dagger}_{ab}(\bm{k})\sum \varphi_{dc}(\bm{k}^{\prime})\overleftrightarrow{\mathrm{P}_{-}}
  \int d\bar{\tau}_{1}\overleftrightarrow{\mathcal{G}}_{ci}(\bm{k}^{\prime}, \tau_{1}, \bar{\tau}_{1})\overleftrightarrow{X}_{ij}(\bm{k}^{\prime},\bar{\tau}_{1})\overleftrightarrow{\mathcal{G}}_{jd}(\bm{k}^{\prime}, \bar{\tau}_{1},\tau_{1})\overleftrightarrow{\mathrm{P}_{+}}.
\end{align}

\section{Properties of the superopeartors}
\label{app:prop_superop}
In this Appendix, we verify the properties of the superoperators [Eqs.~(\ref{eq:U_relation})-(\ref{eq:Z_linearity})], which are used to rewrite the nonlinear response function into the symmetric form in the main text. 

First, we can show the following relation of the superoperator $U$,
\begin{align}
  \sum_{\bm{k}}\int d\tau_{1}d\tau_{2} \mathrm{Tr}[\overleftrightarrow{A}_{ab}(\bm{k}, \tau_{1})&\overleftrightarrow{\mathcal{G}}_{bi}(\bm{k}, \tau_{1}, \tau_{2})\overleftrightarrow{U}_{ij}(\bm{k}, \tau_{2})[B]\overleftrightarrow{\mathcal{G}}_{ja}(\bm{k}, \tau_{2}, \tau_{1})] \notag \\
  =-\frac{U}{V}\sum_{\bm{k},\bm{k}^{\prime}}\int d\tau_{1}d\tau_{2}d\bar{\tau}_{1}&\mathrm{Tr}[\overleftrightarrow{A}_{ab}(\bm{k}, \tau_{1})\overleftrightarrow{\mathcal{G}}_{bi}(\bm{k}, \tau_{1}, \tau_{2})\varphi_{ij}(\bm{k}) \varphi^{\dagger}_{dc}(\bm{k}^{\prime}) \notag \\
  &\times\overleftrightarrow{\mathrm{P}_{+}}\overleftrightarrow{\mathcal{G}}_{ck}(\bm{k}^{\prime}, \tau_{2}, \bar{\tau}_{1})\overleftrightarrow{B}_{kl}(\bm{k}^{\prime},\bar{\tau}_{1})\overleftrightarrow{\mathcal{G}}_{ld}(\bm{k}^{\prime}, \bar{\tau}_{1},\tau_{2})\overleftrightarrow{\mathrm{P}_{-}}\overleftrightarrow{\mathcal{G}}_{ja}(\bm{k}, \tau_{2}, \tau_{1})] \notag \\
    -\frac{U}{V}\sum_{\bm{k},\bm{k}^{\prime}}\int d\tau_{1}d\tau_{2}&d\bar{\tau}_{1}\mathrm{Tr}[\overleftrightarrow{A}_{ab}(\bm{k}, \tau_{1})\overleftrightarrow{\mathcal{G}}_{bi}(\bm{k}, \tau_{1}, \tau_{2})\varphi^{\dagger}_{ij}(\bm{k}) \varphi_{dc}(\bm{k}^{\prime}) \notag \\
  &\times\overleftrightarrow{\mathrm{P}_{-}}\overleftrightarrow{\mathcal{G}}_{ck}(\bm{k}^{\prime}, \tau_{2}, \bar{\tau}_{1})\overleftrightarrow{B}_{kl}(\bm{k}^{\prime},\bar{\tau}_{1})\overleftrightarrow{\mathcal{G}}_{ld}(\bm{k}^{\prime}, \bar{\tau}_{1},\tau_{2})\overleftrightarrow{\mathrm{P}_{+}}\overleftrightarrow{\mathcal{G}}_{ja}(\bm{k}, \tau_{2}, \tau_{1})] \\
  =-\frac{U}{V}\sum_{\bm{k},\bm{k}^{\prime}}\int d\tau_{1}d\tau_{2}d\bar{\tau}_{1}&\mathrm{Tr}[\varphi_{ij}(\bm{k}) \varphi^{\dagger}_{dc}(\bm{k}^{\prime})\overleftrightarrow{\mathrm{P}_{-}}\overleftrightarrow{\mathcal{G}}_{ja}(\bm{k}, \tau_{2}, \tau_{1})\overleftrightarrow{A}_{ab}(\bm{k}, \tau_{1})\overleftrightarrow{\mathcal{G}}_{bi}(\bm{k}, \tau_{1}, \tau_{2}) \overleftrightarrow{\mathrm{P}_{+}} \notag \\
  &\times\overleftrightarrow{\mathcal{G}}_{ck}(\bm{k}^{\prime}, \tau_{2}, \bar{\tau}_{1})\overleftrightarrow{B}_{kl}(\bm{k}^{\prime},\bar{\tau}_{1})\overleftrightarrow{\mathcal{G}}_{ld}(\bm{k}^{\prime}, \bar{\tau}_{1},\tau_{2})]  \notag \\
  -\frac{U}{V}\sum_{\bm{k},\bm{k}^{\prime}}\int d\tau_{1}d\tau_{2}&d\bar{\tau}_{1}\mathrm{Tr}[\varphi^{\dagger}_{ij}(\bm{k}) \varphi_{dc}(\bm{k}^{\prime})\overleftrightarrow{\mathrm{P}_{+}}\overleftrightarrow{\mathcal{G}}_{ja}(\bm{k}, \tau_{2}, \tau_{1})\overleftrightarrow{A}_{ab}(\bm{k}, \tau_{1})\overleftrightarrow{\mathcal{G}}_{bi}(\bm{k}, \tau_{1}, \tau_{2}) \overleftrightarrow{\mathrm{P}_{-}} \notag \\
  &\times\overleftrightarrow{\mathcal{G}}_{ck}(\bm{k}^{\prime}, \tau_{2}, \bar{\tau}_{1})\overleftrightarrow{B}_{kl}(\bm{k}^{\prime},\bar{\tau}_{1})\overleftrightarrow{\mathcal{G}}_{ld}(\bm{k}^{\prime}, \bar{\tau}_{1},\tau_{2})] \\
  =\sum_{\bm{k}^{\prime}}\int d\tau_{2}d\bar{\tau}_{1}\mathrm{Tr}&\left[\overleftrightarrow{U}_{dc}(\bm{k}^{\prime}, \tau_{2})[A]\overleftrightarrow{\mathcal{G}}_{ck}(\bm{k}^{\prime}, \tau_{2}, \bar{\tau}_{1})\overleftrightarrow{B}_{kl}(\bm{k}^{\prime},\bar{\tau}_{1})\overleftrightarrow{\mathcal{G}}_{ld}(\bm{k}^{\prime}, \bar{\tau}_{1},\tau_{2})\right],
\end{align}
where $\mathrm{Tr}$ represents the trace for the Numbu space, and $\varphi_{ab}(\bm{k})$ and $\varphi^{\dagger}_{ab}(\bm{k})$ act as identity operators in the Numbu space. Thus, the relation of $U$ [Eq.~(\ref{eq:U_relation})] is verified,
\begin{align}
  &\sum\int d\tau_{1}d\tau_{2} \mathrm{Tr}\left[\overleftrightarrow{A}_{ab}(\bm{k}, \tau_{1})\overleftrightarrow{\mathcal{G}}_{bi}(\bm{k}, \tau_{1}, \tau_{2})\overleftrightarrow{U}_{ij}(\bm{k}, \tau_{2})[B]\overleftrightarrow{\mathcal{G}}_{ja}(\bm{k}, \tau_{2}, \tau_{1})\right] \notag\\
  &= \sum\int d\tau_{1}d\tau_{2}\mathrm{Tr}\left[\overleftrightarrow{U}_{ab}(\bm{k}, \tau_{1})[A]\overleftrightarrow{\mathcal{G}}_{bi}(\bm{k}, \tau_{1}, \tau_{2})\overleftrightarrow{B}_{ij}(\bm{k},\tau_{2})\overleftrightarrow{\mathcal{G}}_{ja}(\bm{k}, \tau_{2},\tau_{1})\right]. \tag{\ref{eq:U_relation}}\label{eq:U_eq_app}
\end{align}

Next, we derive the relation of the superoperator $Z$,
\begin{align}
  &\sum\int d\tau_{1}d\tau_{2} \mathrm{Tr}\left[\overleftrightarrow{A}_{ab}(\bm{k}, \tau_{1})\overleftrightarrow{\mathcal{G}}_{bi}(\bm{k}, \tau_{1}, \tau_{2})\overleftrightarrow{Z}_{ij}(\bm{k}, \tau_{2})[B]\overleftrightarrow{\mathcal{G}}_{ja}(\bm{k}, \tau_{2}, \tau_{1})\right] \notag\\
  &= \sum\int d\tau_{1}d\tau_{2}\mathrm{Tr}\left[\overleftrightarrow{Z}_{ab}(\bm{k}, \tau_{1})[A]\overleftrightarrow{\mathcal{G}}_{bi}(\bm{k}, \tau_{1}, \tau_{2})\overleftrightarrow{B}_{ij}(\bm{k},\tau_{2})\overleftrightarrow{\mathcal{G}}_{ja}(\bm{k}, \tau_{2},\tau_{1})\right]. \tag{\ref{eq:Z_relation}}\label{eq:Z_eq_app} 
\end{align}    
Here, we consider expansion of $Z$ as a series of the superoperator $U$,
\begin{align}
  \overleftrightarrow{Z}_{ab}(\bm{k}, \tau_{1})[A] = \overleftrightarrow{A}_{ab}(\bm{k}, \tau_{1}) + \overleftrightarrow{Z}^{(1)}_{ab}(\bm{k}, \tau_{1})[A]+\overleftrightarrow{Z}^{(2)}_{ab}(\bm{k}, \tau_{1})[A] + \cdots,
\end{align}
where $\overleftrightarrow{Z}^{(i)}_{ab}(\bm{k}, \tau_{1})[A]$ is defined as
\begin{align}
\overleftrightarrow{Z}^{(0)}_{ab}(\bm{k}, \tau_{1})[A] =  \overleftrightarrow{A}_{ab}(\bm{k}, \tau_{1}), \quad 
  \overleftrightarrow{Z}^{(n+1)}_{ab}(\bm{k}, \tau_{1})[A] = \overleftrightarrow{U}_{ab}(\bm{k}, \tau_{1})[Z^{(n)}[A]].
\end{align}
The finite $(n+1)$th-order term $\overleftrightarrow{Z}^{(n+1)}_{ab}(\bm{k}, \tau_{1})[A]$ can be written as
\begin{align}
    \overleftrightarrow{Z}^{(n+1)}_{ab}(\bm{k}, \tau_{1})[A] = \overleftrightarrow{U}_{ab}(\bm{k}, \tau_{1})[U[U[\cdots U[A]\cdots ]]] = \overleftrightarrow{Z}^{(n)}_{ab}(\bm{k}, \tau_{1})[U[A]].
\end{align}
We can verify that any $n$-th order term of Z satisfies the relation as,
\begin{align}
  &\sum\int d\tau_{1}d\tau_{2} \mathrm{Tr}\left[\overleftrightarrow{A}_{ab}(\bm{k}, \tau_{1})\overleftrightarrow{\mathcal{G}}_{bi}(\bm{k}, \tau_{1}, \tau_{2})\overleftrightarrow{Z}^{(n)}_{ij}(\bm{k}, \tau_{2})[B]\overleftrightarrow{\mathcal{G}}_{ja}(\bm{k}, \tau_{2}, \tau_{1})\right] \notag \\
  &= \sum\int d\tau_{1}d\tau_{2}\mathrm{Tr}\left[\overleftrightarrow{Z}^{(n)}_{ab}(\bm{k}, \tau_{1})[A]\overleftrightarrow{\mathcal{G}}_{bi}(\bm{k}, \tau_{1}, \tau_{2})\overleftrightarrow{B}_{ij}(\bm{k},\tau_{2})\overleftrightarrow{\mathcal{G}}_{ja}(\bm{k}, \tau_{2},\tau_{1})\right], \label{eq:Z_nth_order}
\end{align}
which is proved 
by mathematical induction in the following way. 
For zeroth order, Eq.~(\ref{eq:Z_nth_order}) is the same as Eq.~(\ref{eq:U_relation}), and thus it is true. 
Now we assume that Eq.~(\ref{eq:Z_nth_order}) is true for $n$-th order.
Let us check for the next $(n+1)$-th order.
We can rewrite the left hand side of Eq.~(\ref{eq:Z_nth_order}) for $(n+1)$-th order as,
\begin{align}
  &\sum\int d\tau_{1}d\tau_{2} \mathrm{Tr}\left[\overleftrightarrow{A}_{ab}(\bm{k}, \tau_{1})\overleftrightarrow{\mathcal{G}}_{bi}(\bm{k}, \tau_{1}, \tau_{2})\overleftrightarrow{Z}^{(n+1)}_{ij}(\bm{k}, \tau_{2})[B]\overleftrightarrow{\mathcal{G}}_{ja}(\bm{k}, \tau_{2}, \tau_{1})\right] \notag \\
  =&\sum\int d\tau_{1}d\tau_{2} \mathrm{Tr}\left[\overleftrightarrow{A}_{ab}(\bm{k}, \tau_{1})\overleftrightarrow{\mathcal{G}}_{bi}(\bm{k}, \tau_{1}, \tau_{2})\overleftrightarrow{U}_{ij}(\bm{k}, \tau_{2})[Z^{(n)}[B]]\overleftrightarrow{\mathcal{G}}_{ja}(\bm{k}, \tau_{2}, \tau_{1})\right] \\
  =&\sum\int d\tau_{1}d\tau_{2} \mathrm{Tr}\left[\overleftrightarrow{U}_{ab}(\bm{k}, \tau_{1})[A]\overleftrightarrow{\mathcal{G}}_{bi}(\bm{k}, \tau_{1}, \tau_{2})\overleftrightarrow{Z}^{(n)}_{ij}(\bm{k}, \tau_{2})[B]\overleftrightarrow{\mathcal{G}}_{ja}(\bm{k}, \tau_{2}, \tau_{1})\right] \\
  =&\sum\int d\tau_{1}d\tau_{2} \mathrm{Tr}\left[\overleftrightarrow{Z}^{(n)}_{ab}(\bm{k}, \tau_{1})[U[A]]\overleftrightarrow{\mathcal{G}}_{bi}(\bm{k}, \tau_{1}, \tau_{2})\overleftrightarrow{B}_{ij}(\bm{k}, \tau_{2})\overleftrightarrow{\mathcal{G}}_{ja}(\bm{k}, \tau_{2}, \tau_{1})\right] \\
  =&\sum\int d\tau_{1}d\tau_{2} \mathrm{Tr}\left[\overleftrightarrow{Z}^{(n+1)}_{ab}(\bm{k}, \tau_{1})[A]\overleftrightarrow{\mathcal{G}}_{bi}(\bm{k}, \tau_{1}, \tau_{2})\overleftrightarrow{B}_{ij}(\bm{k}, \tau_{2})\overleftrightarrow{\mathcal{G}}_{ja}(\bm{k}, \tau_{2}, \tau_{1})\right].
  \end{align}
By mathematical induction, Eq.~\eqref{eq:Z_nth_order} is proved for arbitrary order. Therefore, the relation of the superoperator $Z$ [Eq.~\eqref{eq:Z_eq_app}] is true.

Next, we verify the linearity of $Z$,
\begin{align}
    \overleftrightarrow{Z}_{ab}(\bm{k},\tau)[A] + \overleftrightarrow{Z}_{ab}(\bm{k},\tau)[B] = \overleftrightarrow{Z}_{ab}(\bm{k},\tau)[A + B]. \tag{\ref{eq:Z_linearity}}
\end{align}
We consider the self-consistent equation for some opearators $\overleftrightarrow{A}_{ab}(\bm{k},\tau)$ and $\overleftrightarrow{B}_{ab}(\bm{k},\tau)$, by which
$\overleftrightarrow{Z}_{ab}(\bm{k},\tau)[A]$ and $\overleftrightarrow{Z}_{ab}(\bm{k},\tau)[B]$ are determined by 
\begin{align}
    \overleftrightarrow{Z}_{ab}(\bm{k},\tau)[A] &= \overleftrightarrow{A}_{ab}(\bm{k}, \tau) +  \overleftrightarrow{U}_{ab}(\bm{k}, \tau)[Z[A]], \\
    \overleftrightarrow{Z}_{ab}(\bm{k},\tau)[B] &= \overleftrightarrow{B}_{ab}(\bm{k}, \tau) +  \overleftrightarrow{U}_{ab}(\bm{k}, \tau)[Z[B]].
\end{align}
On the other hand, $\overleftrightarrow{Z}_{ab}(\bm{k},\tau)[A+B]$ is given by
\begin{align}
    \overleftrightarrow{Z}_{ab}(\bm{k},\tau)[A+B] = \overleftrightarrow{A}_{ab}(\bm{k}, \tau) + \overleftrightarrow{B}_{ab}(\bm{k}, \tau) +  \overleftrightarrow{U}_{ab}(\bm{k}, \tau)[Z[A+B]]. \label{eq:Z_A_B}
\end{align}
Using the linearity of the superoperator $U$,
\begin{align}
 \overleftrightarrow{U}_{ab}(\bm{k}, \tau)[X] + \overleftrightarrow{U}_{ab}(\bm{k}, \tau)[Y] = \overleftrightarrow{U}_{ab}(\bm{k}, \tau)[X + Y],
\end{align}
we can rewrite the sum of $Z[A]$ and $Z[B]$ as
\begin{align}
    \overleftrightarrow{Z}_{ab}(\bm{k},\tau)[A] + \overleftrightarrow{Z}_{ab}(\bm{k},\tau)[B] = \overleftrightarrow{A}_{ab}(\bm{k}, \tau) + \overleftrightarrow{B}_{ab}(\bm{k}, \tau) +  \overleftrightarrow{U}_{ab}(\bm{k}, \tau)[Z[A] + Z[B]]. \label{eq:Z_A_Z_B}
\end{align}
Comparing Eqs.~(\ref{eq:Z_A_B}) and (\ref{eq:Z_A_Z_B}), the linearity of Z [Eq~.(\ref{eq:Z_linearity})] is proved.

\section{Analytic continuation of the vertex function}
\label{app:ana_cont_vertex}
In this Appendix, we explain the analytic continuation of the vertex function. The vertex function in the Matsubara frequency representation is obtained by
\begin{align}
  \overleftrightarrow{\Gamma}^{\alpha}_{ab}(\bm{k}, i\omega)=\overleftrightarrow{J}^{\alpha}_{ab}(\bm{k})+\overleftrightarrow{\varphi}_{ab}(\bm{k})\overleftrightarrow{U}(i\omega)[\Gamma^{\alpha}(i\omega)],
\end{align}
where $\overleftrightarrow{U}(i\omega)$ and $\overleftrightarrow{\varphi}$ are defined as
\begin{align}
  \overleftrightarrow{U}(i\omega)[\Gamma^{\alpha}(i\omega)]
  =&\frac{U}{V}k_{\mathrm{B}}T\sum \varphi^{\dagger}_{dc}(\bm{k}^{\prime})\overleftrightarrow{\mathrm{P}_{+}}
  \overleftrightarrow{\mathcal{G}}_{ci}(\bm{k}^{\prime}, i\omega_{1})\overleftrightarrow{\Gamma}^{\alpha}_{ij}(\bm{k}^{\prime},i\omega)\overleftrightarrow{\mathcal{G}}_{jd}(\bm{k}^{\prime}, i(\omega_{1}-\omega))\overleftrightarrow{\mathrm{P}_{-}}\notag \\
  &+\frac{U}{V}k_{\mathrm{B}}T\sum \varphi_{dc}(\bm{k}^{\prime})\overleftrightarrow{\mathrm{P}_{-}}
  \overleftrightarrow{\mathcal{G}}_{ci}(\bm{k}^{\prime}, i\omega_{1})\overleftrightarrow{\Gamma}^{\alpha}_{ij}(\bm{k}^{\prime},i\omega)\overleftrightarrow{\mathcal{G}}_{jd}(\bm{k}^{\prime}, i(\omega_{1}-\omega))\overleftrightarrow{\mathrm{P}_{+}},
\end{align}
\begin{align}
  \overleftrightarrow{\varphi}_{ab}(\bm{k}) = 
  \begin{pmatrix}
    \varphi_{ab}(\bm{k}) & 0 \\
    0 & \varphi^{\dagger}_{ab}(\bm{k})
  \end{pmatrix}.
\end{align}
We introduce $\Lambda$ and represent the vertex function as
\begin{align}
  \overleftrightarrow{\Gamma}^{\alpha}_{ab}(\bm{k}, i\omega)=\overleftrightarrow{J}^{\alpha}_{ab}(\bm{k})+\overleftrightarrow{\varphi}_{ab}(\bm{k}) \overleftrightarrow{\Lambda}^{\alpha}( i\omega).
\end{align}
Then, we rewrite the self-consistent equation of the vertex function as
\begin{align}
  \overleftrightarrow{\Lambda}^{\alpha}( i\omega) = \overleftrightarrow{U}(i\omega)[J^{\alpha}] + \overleftrightarrow{U}(i\omega)[\varphi\Lambda^{\alpha}( i\omega)].
\end{align}
Here, we introduce the following notations to represent the self-consistent equation shortly,
\begin{align}
  \overleftrightarrow{\bar{u}}_{cd}(\bm{k}, i\omega)[A] &\equiv k_{\mathrm{B}}T\sum_{\omega_{1}}
  \overleftrightarrow{\mathcal{G}}_{ci}(\bm{k}, i\omega_{1})\overleftrightarrow{\varphi}_{ij}(\bm{k})\overleftrightarrow{A}\overleftrightarrow{\mathcal{G}}_{jd}(\bm{k}, i(\omega_{1}-\omega)),\\
    \overleftrightarrow{u}_{cd}(\bm{k}, i\omega)[A] &\equiv k_{\mathrm{B}}T\sum_{\omega_{1}}
  \overleftrightarrow{\mathcal{G}}_{ci}(\bm{k}, i\omega_{1})\overleftrightarrow{A}_{ij}(\bm{k})\overleftrightarrow{\mathcal{G}}_{jd}(\bm{k}, i(\omega_{1}-\omega)),\\
  \overleftrightarrow{P}[A] &\equiv \overleftrightarrow{P}_{+}\overleftrightarrow{A}\overleftrightarrow{P}_{-} + \overleftrightarrow{P}_{-}\overleftrightarrow{A}\overleftrightarrow{P}_{+}.
\end{align}
We can rewrite the first and second terms of the self-consistent equation as
\begin{align}
    \overleftrightarrow{U}(i\omega)[J^{\alpha}]
  &  =\frac{U}{V}\sum \overleftrightarrow{\varphi}^{\dagger}_{dc}(\bm{k}^{\prime})\overleftrightarrow{P}\left[\overleftrightarrow{u}_{cd}(\bm{k}^{\prime}, i\omega)\left[J^{\alpha}\right]\right], \label{eq:U_J} \\
  \overleftrightarrow{U}(i\omega)[\varphi\Lambda^{\alpha}(i\omega)]
  &=\frac{U}{V}\sum \overleftrightarrow{\varphi}^{\dagger}_{dc}(\bm{k}^{\prime})\overleftrightarrow{P}\left[\overleftrightarrow{\bar{u}}_{cd}(\bm{k}^{\prime}, i\omega)\left[\Lambda^{\alpha}(i\omega)\right]\right],
\end{align}
where $\overleftrightarrow{\varphi}^{\dagger}$ is defined as
\begin{align}
  \overleftrightarrow{\varphi}^{\dagger}_{dc}(\bm{k}) = 
  \begin{pmatrix}
    \varphi^{\dagger}_{dc}(\bm{k}) & 0 \\
    0 & \varphi_{dc}(\bm{k})
  \end{pmatrix}.
\end{align}
Using the above notations, we can obtain the vertex function in the real frequency space from the analytic continuation of $\overleftrightarrow{u}_{cd}(\bm{k}, i\omega)\left[J^{\alpha}\right]$ and $\overleftrightarrow{\bar{u}}_{cd}(\bm{k}, i\omega)\left[\Lambda^{\alpha}(i\omega)\right]$. 

We rewrite $\overleftrightarrow{\bar{u}}_{cd}(\bm{k}^{\prime}, i\omega)\left[\Lambda^{\alpha}(i\omega)\right]$ by using the relation $\oint_{C}\frac{d\omega}{2\pi i}f(\omega)A(\omega)=-\frac{1}{\beta}\sum_{n}A(i\omega_{n})$, where $\oint_{C}$ denotes the integral path around the poles of the Fermi distribution function~\cite{Michishita2021}. In the upper half-plane $\mathrm{Im}[i\omega]>0$, the function is rewritten as
\begin{align}
  \overleftrightarrow{\bar{u}}_{cd}(\bm{k}, i\omega)\left[\Lambda^{\alpha}(i\omega)\right] = -\int^{\infty}_{-\infty} \frac{d\tilde{\omega}}{2\pi i}f(\tilde{\omega})\mathrm{Tr}\Big[&\overleftrightarrow{G}^{R}_{ci}(\bm{k}, \tilde{\omega}+i\omega)\overleftrightarrow{\varphi}_{ij}(\bm{k})\overleftrightarrow{\Lambda}^{\alpha}( i\omega)\Big\{\overleftrightarrow{G}^{R}_{jd}(\bm{k}, \tilde{\omega}) - \overleftrightarrow{G}^{A}_{jd}(\bm{k}, \tilde{\omega})\Big\} \notag \\
  &+ \Big\{\overleftrightarrow{G}^{R}_{ci}(\bm{k}, \tilde{\omega}) - \overleftrightarrow{G}^{A}_{ci}(\bm{k}, \tilde{\omega})\Big\}\overleftrightarrow{\varphi}_{ij}(\bm{k}^{\prime})\overleftrightarrow{\Lambda}^{\alpha}( i\omega)\overleftrightarrow{G}^{A}_{jd}(\bm{k}, \tilde{\omega}-i\omega)\Big].
\end{align}
In the lower half-plane $\mathrm{Im}[i\omega]<0$, the function is expressed by
\begin{align}
  \overleftrightarrow{\bar{u}}_{cd}(\bm{k}, i\omega)\left[\Lambda^{\alpha}(i\omega)\right] 
    = -\int^{\infty}_{-\infty} \frac{d\tilde{\omega}}{2\pi i}f(\tilde{\omega})\mathrm{Tr}\Big[&\overleftrightarrow{G}^{A}_{ci}(\bm{k}, \tilde{\omega}+i\omega)\overleftrightarrow{\varphi}_{ij}(\bm{k})\overleftrightarrow{\Lambda}^{\alpha}( i\omega)\Big\{\overleftrightarrow{G}^{R}_{jd}(\bm{k}, \tilde{\omega}) - \overleftrightarrow{G}^{A}_{jd}(\bm{k}, \tilde{\omega})\Big\} \notag \\
  &+ \Big\{\overleftrightarrow{G}^{R}_{ci}(\bm{k}, \tilde{\omega}) - \overleftrightarrow{G}^{A}_{ci}(\bm{k}, \tilde{\omega})\Big\}\overleftrightarrow{\varphi}_{ij}(\bm{k})\overleftrightarrow{\Lambda}^{\alpha}( i\omega)\overleftrightarrow{G}^{R}_{jd}(\bm{k}, \tilde{\omega}-i\omega)\Big]. 
\end{align}
The analytic continuation $i\omega \rightarrow \omega + i\gamma$ in the upper half-plane leads to the real frequency representation,
\begin{align}
  &\overleftrightarrow{\bar{u}}_{cd}(\bm{k}, \omega + i\gamma)\left[\Lambda^{\alpha}(\omega + i\gamma)\right] \notag \\
    =& -\int^{\infty}_{-\infty} \frac{d\tilde{\omega}}{2\pi i}f(\tilde{\omega})\mathrm{Tr}\Big[\overleftrightarrow{G}^{R}_{ci}(\bm{k}, \tilde{\omega}+\omega + i\gamma)\overleftrightarrow{\varphi}_{ij}(\bm{k})\overleftrightarrow{\Lambda}^{\alpha}( \omega + i\gamma)\Big\{\overleftrightarrow{G}^{R}_{jd}(\bm{k}, \tilde{\omega}) - \overleftrightarrow{G}^{A}_{jd}(\bm{k}, \tilde{\omega})\Big\} \notag \\
  &+ \Big\{\overleftrightarrow{G}^{R}_{ci}(\bm{k}, \tilde{\omega}) - \overleftrightarrow{G}^{A}_{ci}(\bm{k}, \tilde{\omega})\Big\}\overleftrightarrow{\varphi}_{ij}(\bm{k})\overleftrightarrow{\Lambda}^{\alpha}( \omega + i\gamma)\overleftrightarrow{G}^{A}_{jd}(\bm{k}, \tilde{\omega}-\omega - i\gamma)\Big].
\end{align}
When we perform the analytic continuation $i\omega \rightarrow -\omega - i\gamma$ in the lower half-plane, the real frequency representation is obtained as
\begin{align}
  &\overleftrightarrow{\bar{u}}_{cd}(\bm{k}, -\omega - i\gamma)\left[\Lambda^{\alpha}(-\omega - i\gamma)\right] \notag \\
    =& -\int^{\infty}_{-\infty} \frac{d\tilde{\omega}}{2\pi i}f(\tilde{\omega})\mathrm{Tr}\Big[\overleftrightarrow{G}^{A}_{ci}(\bm{k}, \tilde{\omega}-\omega - i\gamma)\overleftrightarrow{\varphi}_{ij}(\bm{k})\overleftrightarrow{\Lambda}^{\alpha}( -\omega - i\gamma)\Big\{\overleftrightarrow{G}^{R}_{jd}(\bm{k}, \tilde{\omega}) - \overleftrightarrow{G}^{A}_{jd}(\bm{k}, \tilde{\omega})\Big\} \notag \\
  &+ \Big\{\overleftrightarrow{G}^{R}_{ci}(\bm{k}, \tilde{\omega}) - \overleftrightarrow{G}^{A}_{ci}(\bm{k}, \tilde{\omega})\Big\}\overleftrightarrow{\varphi}_{ij}(\bm{k})\overleftrightarrow{\Lambda}^{\alpha}( -\omega - i\gamma)\overleftrightarrow{G}^{R}_{jd}(\bm{k}, \tilde{\omega}+\omega + i\gamma)\Big] .
\end{align}
We also obtain the real frequency representation of $\overleftrightarrow{u}_{cd}(\bm{k}, i\omega)\left[J^{\alpha}\right]$ by the analytic continuation, 
\begin{align}
  \overleftrightarrow{u}_{cd}(\bm{k}, \omega + i\gamma)\left[J^{\alpha}\right] 
    = -\int^{\infty}_{-\infty} \frac{d\tilde{\omega}}{2\pi i}f(\tilde{\omega})\mathrm{Tr}\Big[&\overleftrightarrow{G}^{R}_{ci}(\bm{k}, \tilde{\omega}+\omega + i\gamma)\overleftrightarrow{J}^{\alpha}_{ij}(\bm{k})\Big\{\overleftrightarrow{G}^{R}_{jd}(\bm{k}, \tilde{\omega}) - \overleftrightarrow{G}^{A}_{jd}(\bm{k}, \tilde{\omega})\Big\} \notag \\
  &+ \Big\{\overleftrightarrow{G}^{R}_{ci}(\bm{k}, \tilde{\omega}) - \overleftrightarrow{G}^{A}_{ci}(\bm{k}, \tilde{\omega})\Big\}\overleftrightarrow{J}^{\alpha}_{ij}(\bm{k})\overleftrightarrow{G}^{A}_{jd}(\bm{k}, \tilde{\omega}-\omega - i\gamma)\Big],
\end{align}
\begin{align}
 \overleftrightarrow{u}_{cd}(\bm{k}, -\omega - i\gamma)\left[J^{\alpha}\right]
    = -\int^{\infty}_{-\infty} \frac{d\tilde{\omega}}{2\pi i}f(\tilde{\omega})\mathrm{Tr}\Big[&\overleftrightarrow{G}^{A}_{ci}(\bm{k}, \tilde{\omega}-\omega - i\gamma)\overleftrightarrow{J}^{\alpha}_{ij}(\bm{k})\Big\{\overleftrightarrow{G}^{R}_{jd}(\bm{k}, \tilde{\omega}) - \overleftrightarrow{G}^{A}_{jd}(\bm{k}, \tilde{\omega})\Big\} \notag \\
  &+ \Big\{\overleftrightarrow{G}^{R}_{ci}(\bm{k}, \tilde{\omega}) - \overleftrightarrow{G}^{A}_{ci}(\bm{k}, \tilde{\omega})\Big\}\overleftrightarrow{J}^{\alpha}_{ij}(\bm{k})\overleftrightarrow{G}^{R}_{jd}(\bm{k}, \tilde{\omega}+\omega + i\gamma)\Big].
\end{align}
By using these expressions, the real frequency representation of $\Lambda^{\alpha}(\omega + i\gamma)$ and $\Lambda^{\alpha}(-\omega - i\gamma)$ are given by the self-consistent equations,
\begin{align}
  \overleftrightarrow{\Lambda}^{\alpha}(\omega +i\gamma) &= \overleftrightarrow{\tilde{U}}(\omega +i\gamma)[J^{\alpha}] + \overleftrightarrow{\tilde{U}}(\omega +i\gamma)[\varphi\Lambda^{\alpha}( \omega +i\gamma)], \\
  \overleftrightarrow{\Lambda}^{\alpha}(-\omega -i\gamma) &= \overleftrightarrow{\tilde{U}}(-\omega -i\gamma)[J^{\alpha}] + \overleftrightarrow{\tilde{U}}(-\omega -i\gamma)[\varphi\Lambda^{\alpha}( -\omega -i\gamma)].
\end{align}

\section{Self-consistent equation of the vertex function in the clean limit}
\label{sec:appendix_self-consistent_equation}
In this Appendix, we derive the self-consistent equation of the vertex function in the clean limit. First, we consider the retarded vertex function $\overleftrightarrow{\Gamma}^{\alpha}_{ab}(\bm{k}, \omega+i\gamma)$, which is given by  
\begin{align}
  \overleftrightarrow{\Gamma}^{\alpha}_{ab}(\bm{k}, \omega+i\gamma)&=\overleftrightarrow{J}^{\alpha}_{ab}(\bm{k})+\overleftrightarrow{\varphi}_{ab}(\bm{k}) \overleftrightarrow{\Lambda}^{\alpha}(\omega + i\gamma),\\
  \overleftrightarrow{\Lambda}^{\alpha}(\omega + i\gamma) &= \frac{U}{V}\sum \overleftrightarrow{\varphi}^{\dagger}_{dc}(\bm{k}^{\prime})\overleftrightarrow{P}\left[\overleftrightarrow{\bar{u}}_{cd}(\bm{k}^{\prime}, \omega + i\gamma)\left[\Lambda^{\alpha}(\omega + i\gamma)\right] + \overleftrightarrow{u}_{cd}(\bm{k}^{\prime}, \omega + i\gamma)\left[J^{\alpha}\right]\right]. \label{eq:app_self_consistent}
\end{align}
When we use the indices for the 
fermion's internal degrees of freedom and the Nambu space, we can rewrite the functions as
\begin{align}
  \bar{u}_{mn}(\bm{k}, \omega + i\gamma)\left[\Lambda^{\alpha}(\omega + i\gamma)\right]
    = -\int^{\infty}_{-\infty} \frac{d\tilde{\omega}}{2\pi i}f(\tilde{\omega})\Big[&G^{R}_{mo}(\bm{k}, \tilde{\omega}+\omega + i\gamma)\varphi_{op}(\bm{k})\Lambda^{\alpha}_{pq}( \omega + i\gamma)\Big\{G^{R}_{qn}(\bm{k}, \tilde{\omega}) - G^{A}_{qn}(\bm{k}, \tilde{\omega})\Big\} \notag \\
  &+ \Big\{G^{R}_{mo}(\bm{k}, \tilde{\omega}) - G^{A}_{mo}(\bm{k}, \tilde{\omega})\Big\}\varphi_{op}(\bm{k})\Lambda^{\alpha}_{pq}( \omega + i\gamma)G^{A}_{qn}(\bm{k}, \tilde{\omega}-\omega - i\gamma)\Big], \\
  u_{mn}(\bm{k}, \omega + i\gamma)\left[J^{\alpha}\right] = -\int^{\infty}_{-\infty} \frac{d\tilde{\omega}}{2\pi i}f(\tilde{\omega})\Big[&G^{R}_{mo}(\bm{k}, \tilde{\omega}+\omega + i\gamma)J^{\alpha}_{op}(\bm{k})\Big\{G^{R}_{pn}(\bm{k}, \tilde{\omega}) - G^{A}_{pn}(\bm{k}, \tilde{\omega})\Big\}\notag \\
  &+ \Big\{G^{R}_{mo}(\bm{k}, \tilde{\omega}) - G^{A}_{mo}(\bm{k}, \tilde{\omega})\Big\}J^{\alpha}_{op}(\bm{k})G^{A}_{pn}(\bm{k}, \tilde{\omega}-\omega - i\gamma)\Big].
\end{align}
Here, we define 
\begin{align}
  X^{\alpha}_{mn}(\bm{k}^{\prime}, \omega+i\gamma) = \tilde{u}_{mn}(\bm{k}^{\prime}, \omega + i\gamma)\left[\Lambda^{\alpha}(\omega + i\gamma)\right] + u_{mn}(\bm{k}^{\prime}, \omega + i\gamma)\left[J^{\alpha}\right]. 
\end{align}
Using Eq.~\eqref{eq:clean_limit}, we can rewrite $X^{\alpha}$ as
\begin{align}
  \tilde{X}^{\alpha}_{mn}(\bm{k}^{\prime}, \omega+i\gamma)=\frac{\{\tilde{\varphi}_{mp}(\bm{k^{\prime}})\tilde{\Lambda}^{\alpha}_{pn}(\bm{k}^{\prime},\omega+i\gamma) + \tilde{J}^{\alpha}_{mn}(\bm{k}^{\prime})\}f_{nm}}{\omega - E_{mn} +i\gamma}.
\end{align}
Equation~\eqref{eq:app_self_consistent} leads to the relation
\begin{align}
   \Lambda^{\alpha}_{eh}(\omega + i\gamma) = \frac{U}{nV}\sum_{\bm{k}^{\prime}}\mathrm{Tr}[Y^{eh}(\bm{k}^{\prime})X^{\alpha}(\bm{k}^{\prime},\omega + i\gamma)], \quad
      \Lambda^{\alpha}_{he}(\omega + i\gamma) = \frac{U}{nV}\sum_{\bm{k}^{\prime}}\mathrm{Tr}[Y^{he}(\bm{k}^{\prime})X^{\alpha}(\bm{k}^{\prime},\omega + i\gamma)], 
\end{align}
where we use the following notations,
\begin{align}
    \overleftrightarrow{\Lambda}^{\alpha}(\omega + i\gamma) = 
  \begin{pmatrix}
      0 & \Lambda^{\alpha}_{eh}(\omega + i\gamma) \\
      \Lambda^{\alpha}_{he}(\omega + i\gamma) & 0
  \end{pmatrix}, \quad
  [Y^{eh}(\bm{k}^{\prime})]_{(i, a)(j, b)} =
  \begin{pmatrix}
    0 & 0\\
    \varphi^{\dagger}_{ij}(\bm{k}^{\prime}) & 0
  \end{pmatrix}, \quad
    [Y^{he}(\bm{k}^{\prime})]_{(i, a)(j, b)} =
  \begin{pmatrix}
    0 & \varphi_{ij}(\bm{k}^{\prime})\\
    0 & 0
  \end{pmatrix}.
\end{align}
Here, we introduce the band representation of $\overleftrightarrow{A}(\bm{k})$ as
\begin{align}
  \tilde{A}_{ab}(\bm{k}) = \left(\overleftrightarrow{U}^{\dagger}(\bm{k})\right)_{ai}\left(\overleftrightarrow{A}(\bm{k})\right)_{ij}\left(\overleftrightarrow{U}(\bm{k})\right)_{jb},
\end{align}
where the unitary transformation is performed by the unitary matrix $\overleftrightarrow{U}(\bm{k})$ for diagonalization of the BdG Hamiltonian,
\begin{align}
  \left(\overleftrightarrow{U}^{\dagger}(\bm{k})\right)_{ai}\left(\overleftrightarrow{\mathcal{H}}(\bm{k})\right)_{ij}\left(\overleftrightarrow{U}(\bm{k})\right)_{jb} = E_{a}(\bm{k})\delta_{ab}.
\end{align}
$\Lambda^{\alpha}_{eh}(\omega + i\gamma)$ can be expressed as
\begin{align}
     \Lambda^{\alpha}_{eh}(\omega + i\gamma) = \frac{U}{nV}\sum_{\bm{k}^{\prime}}\tilde{Y}^{eh}_{ab}(\bm{k}^{\prime})\tilde{X}^{\alpha}_{ba}(\bm{k}^{\prime},\omega + i\gamma)
     = \sum_{i,j} F^{eh}_{ji}(\omega+i\gamma)\Lambda^{\alpha}_{ij}(\omega+i\gamma) + G^{eh}(\omega+i\gamma), 
\end{align}
where we defined $F^{eh}_{ji}(\omega+i\gamma)$ and $G^{eh}(\omega+i\gamma)$ as
\begin{align}
  F^{eh}_{ji}(\omega+i\gamma)= \frac{U}{nV}\sum_{\bm{k}^{\prime}, a, b, c}\frac{U_{ja}(\bm{k}^{\prime})\tilde{Y}^{eh}_{ab}(\bm{k}^{\prime})\tilde{\varphi}_{bc}(\bm{k^{\prime}})U^{\dagger}_{ci}(\bm{k}^{\prime})f_{ab}}{\omega - E_{ba} +i\gamma}, \quad 
  G^{eh}(\omega+i\gamma)= \frac{U}{nV}\sum_{\bm{k}^{\prime}, a, b}\frac{\tilde{Y}^{eh}_{ab}(\bm{k}^{\prime})\tilde{J}^{\alpha}_{ba}(\bm{k}^{\prime})f_{ab}}{\omega - E_{ba} +i\gamma}.
\end{align}
The indices $i, j$ stand for the complex of internal degrees of freedom of the fermions and the Nambu space, while the indices $a,b,c$ denote the band indices. In the same way, the function $\Lambda^{\alpha}_{he}(\omega + i\gamma)$ is obtained as
\begin{align}
       \Lambda^{\alpha}_{he}(\omega + i\gamma) &= F^{he}_{ji}(\omega+i\gamma)\Lambda^{\alpha}_{ij}(\omega+i\gamma) + G^{eh}(\omega+i\gamma),\\
         F^{he}_{ji}(\omega+i\gamma)&\equiv\frac{U}{nV}\sum_{\bm{k}^{\prime},a,b,c}\frac{U_{ja}(\bm{k}^{\prime})\tilde{Y}^{he}_{ab}(\bm{k}^{\prime})\tilde{\varphi}_{bc}(\bm{k^{\prime}})U^{\dagger}_{ci}(\bm{k}^{\prime})f_{ab}}{\omega - E_{ba} +i\gamma},  \quad 
  G^{he}(\omega+i\gamma)= \frac{U}{nV}\sum_{\bm{k}^{\prime},a ,b}\frac{\tilde{Y}^{he}_{ab}(\bm{k}^{\prime})\tilde{J}^{\alpha}_{ba}(\bm{k}^{\prime})f_{ab}}{\omega - E_{ba} +i\gamma}.
\end{align}
We can rewrite the self-consistent equation as
\begin{align}
    \begin{pmatrix}
        \Lambda^{\alpha}_{eh}(\omega + i\gamma) \\
        \Lambda^{\alpha}_{he}(\omega + i\gamma)
    \end{pmatrix}
    = 
    \begin{pmatrix}
        \hat{F}^{eh}_{eh}(\omega + i\gamma) & \hat{F}^{eh}_{he}(\omega + i\gamma) \\
        \hat{F}^{he}_{eh}(\omega + i\gamma) & \hat{F}^{he}_{he}(\omega + i\gamma)
    \end{pmatrix}
        \begin{pmatrix}
        \Lambda^{\alpha}_{eh}(\omega + i\gamma) \\
        \Lambda^{\alpha}_{he}(\omega + i\gamma)
    \end{pmatrix}
    +
    \begin{pmatrix}
        G_{eh}(\omega + i\gamma) \\
        G_{he}(\omega + i\gamma)
    \end{pmatrix}.
\end{align}
We express the functions $\hat{F}^{\cdots}_{\cdots}(\omega + i\gamma)$ by the summation for the fermion's internal degrees of freedom. 
For example, $\tilde{F}^{eh}_{eh}(\omega + i\gamma)$ is given by the summation,
\begin{align}
    \hat{F}^{eh}_{eh}(\omega + i\gamma) = \sum_{k, l}F^{eh}_{(e, k)(h, l)}(\omega+i\gamma), \label{eq:internal_sum}
\end{align}
where the indices $k, l$ stand for the fermion's internal degrees of freedom. 
As a result of the unitary transformation of $\Lambda^{\alpha}$, $\Lambda^{\alpha}_{1} = (\Lambda^{\alpha}_{eh} + \Lambda^{\alpha}_{he})/2$ and $\Lambda^{\alpha}_{2} = i(\Lambda^{\alpha}_{eh} - \Lambda^{\alpha}_{he})/2$, the self-consistent equation is rewritten as
\begin{align}
    \begin{pmatrix}
        \Lambda^{\alpha}_{1}(\omega + i\gamma) \\
        \Lambda^{\alpha}_{2}(\omega + i\gamma)
    \end{pmatrix}
    =
    \bm{M}(\omega + i\gamma)
    \begin{pmatrix}
        \Lambda^{\alpha}_{1}(\omega + i\gamma) \\
        \Lambda^{\alpha}_{2}(\omega + i\gamma)
    \end{pmatrix}
    +
    \begin{pmatrix}
        L^{J}_{1}(\omega+i\gamma) \\
        L^{J}_{2}(\omega+i\gamma)
    \end{pmatrix}.
\end{align}
The components of $\bm{M}$ and $L^{J}$ are obtained as
\begin{align}
    M_{11} = \frac{1}{2}\left(\hat{F}^{eh}_{eh} + \hat{F}^{eh}_{he} + \hat{F}^{he}_{eh} + \hat{F}^{he}_{he}\right),& \quad
    M_{12} = -\frac{i}{2}\left(\hat{F}^{eh}_{eh} - \hat{F}^{eh}_{he} + \hat{F}^{he}_{eh} - \hat{F}^{he}_{he}\right), \\
    M_{21} = \frac{i}{2}\left(\hat{F}^{eh}_{eh} + \hat{F}^{eh}_{he} - \hat{F}^{he}_{eh} - \hat{F}^{he}_{he}\right),& \quad
    M_{22} = \frac{1}{2}\left(\hat{F}^{eh}_{eh} - \hat{F}^{eh}_{he} - \hat{F}^{he}_{eh} + \hat{F}^{he}_{he}\right), \\
    L^{J}_{1} = \frac{1}{2}\left(G_{eh} + G_{he}\right),& \quad L^{J}_{2} = \frac{i}{2}\left(G_{eh} - G_{he}\right).
\end{align}

\section{Extension to multicomponent Cooper channels}
\label{app:extention to multi}
In this Appendix, we derive the self-consistent equation for the vertex function in multicomponent superconductors. 
Here, we consider two-component superconductors with two Cooper pairing channels by adopting the pairing interaction, 
\begin{align}
  H_{\mathrm{pair}} =-\frac{U_{1}}{2V}\sum_{\bm{k},\bm{k}^{\prime}}\varphi^{(1)}_{ab}(\bm{k})\varphi_{cd}^{(1)\dagger}(\bm{k}^{\prime})c_{\bm{k}a}^{\dagger}c_{-\bm{k}b}^{\dagger}c_{-\bm{k}^{\prime}c}c_{\bm{k}^{\prime}d} -\frac{U_{2}}{2V}\sum_{\bm{k},\bm{k}^{\prime}}\varphi^{(2)}_{ab}(\bm{k})\varphi_{cd}^{(2)\dagger}(\bm{k}^{\prime})c_{\bm{k}a}^{\dagger}c_{-\bm{k}b}^{\dagger}c_{-\bm{k}^{\prime}c}c_{\bm{k}^{\prime}d}.
\end{align}
The mean-field pair potential is introduced as
\begin{align}
  \Delta^{+}_{ab}(\bm{k}, \tau; \bm{A}) = 
   -\frac{U_{1}}{V} \varphi^{(1)\dagger}_{ab}(\bm{k})\sum \varphi^{(1)}_{dc}(\bm{k}^{\prime})\mathcal{F}^{+}_{cd}(\bm{k}^{\prime}, \tau^{+}, \tau^{-};\bm{A}) -\frac{U_{2}}{V} \varphi^{(2)\dagger}_{ab}(\bm{k})\sum \varphi^{(2)}_{dc}(\bm{k}^{\prime})\mathcal{F}^{+}_{cd}(\bm{k}^{\prime}, \tau^{+}, \tau^{-};\bm{A}),   
\end{align}
\begin{align}
  \Delta_{ab}(\bm{k}, \tau; \bm{A}) =  -\frac{U_{1}}{V} \varphi^{(1)}_{ab}(\bm{k})\sum \varphi^{(1)\dagger}_{dc}(\bm{k}^{\prime})\mathcal{F}_{cd}(\bm{k}^{\prime}, \tau^{+}, \tau^{-};\bm{A}) -\frac{U_{2}}{V} \varphi^{(2)}_{ab}(\bm{k})\sum \varphi^{(2)\dagger}_{dc}(\bm{k}^{\prime})\mathcal{F}_{cd}(\bm{k}^{\prime}, \tau^{+}, \tau^{-};\bm{A}),
\end{align}
based on the Kadanoff-Baym method explained in Sec.~\ref{sec:Kadanoff-Baym}. 
The vertex function can be represented as
\begin{align}
    \overleftrightarrow{\Gamma^{\alpha}}_{ab}(\bm{k}, \omega+i\gamma)=\overleftrightarrow{J^{\alpha}}_{ab}(\bm{k})+
    \overleftrightarrow{\varphi^{(1)}}_{ab}(\bm{k}) \overleftrightarrow{\Lambda^{(1)\alpha}}(\omega + i\gamma) 
    +\overleftrightarrow{\varphi^{(2)}}_{ab}(\bm{k}) \overleftrightarrow{\Lambda^{(2)\alpha}}(\omega + i\gamma),
\end{align}
and the self-consistent equation is obtained as
\begin{align}
    \overleftrightarrow{\Lambda^{\alpha(i)}}(\omega + i\gamma) = \frac{U}{V}\sum \overleftrightarrow{\varphi}^{(i)\dagger}_{dc}(\bm{k}^{\prime})\overleftrightarrow{P}\Big[&\overleftrightarrow{\bar{u}^{(1)}}_{cd}(\bm{k}^{\prime}, \omega + i\gamma)\left[\Lambda^{(1)\alpha}(\omega + i\gamma)\right] \notag \\
    &+ \overleftrightarrow{\bar{u}^{(2)}}_{cd}(\bm{k}^{\prime}, \omega + i\gamma)\left[\Lambda^{(2)\alpha}(\omega + i\gamma)\right] + \overleftrightarrow{u}_{cd}(\bm{k}^{\prime}, \omega + i\gamma)\left[J^{\alpha}\right]\Big].
\end{align}
Here, $\overleftrightarrow{\bar{u}^{(i)}}$ are defined as
\begin{align}
  &\overleftrightarrow{\bar{u}^{(i)}}_{cd}(\bm{k}, \omega + i\gamma)\left[\Lambda^{(j)\alpha}(\omega + i\gamma)\right] \notag \\
    =& -\int^{\infty}_{-\infty} \frac{d\tilde{\omega}}{2\pi i}f(\tilde{\omega})\mathrm{Tr}\Big[\overleftrightarrow{G}^{R}_{ci}(\bm{k}, \tilde{\omega}+\omega + i\gamma)\overleftrightarrow{\varphi^{(i)}}_{ij}(\bm{k})\overleftrightarrow{\Lambda}^{(j)\alpha}( \omega + i\gamma)\Big\{\overleftrightarrow{G}^{R}_{jd}(\bm{k}, \tilde{\omega}) - \overleftrightarrow{G}^{A}_{jd}(\bm{k}, \tilde{\omega})\Big\} \notag \\
  &+ \Big\{\overleftrightarrow{G}^{R}_{ci}(\bm{k}, \tilde{\omega}) - \overleftrightarrow{G}^{A}_{ci}(\bm{k}, \tilde{\omega})\Big\}\overleftrightarrow{\varphi^{(i)}}_{ij}(\bm{k})\overleftrightarrow{\Lambda}^{(j)\alpha}( \omega + i\gamma)\overleftrightarrow{G}^{A}_{jd}(\bm{k}, \tilde{\omega}-\omega - i\gamma)\Big],
\end{align}
and
\begin{align}
  \overleftrightarrow{u}_{cd}(\bm{k}, \omega + i\gamma)\left[J^{\alpha}\right] 
    = -\int^{\infty}_{-\infty} \frac{d\tilde{\omega}}{2\pi i}f(\tilde{\omega})\mathrm{Tr}\Big[&\overleftrightarrow{G}^{R}_{ci}(\bm{k}, \tilde{\omega}+\omega + i\gamma)\overleftrightarrow{J}^{\alpha}_{ij}(\bm{k})\Big\{\overleftrightarrow{G}^{R}_{jd}(\bm{k}, \tilde{\omega}) - \overleftrightarrow{G}^{A}_{jd}(\bm{k}, \tilde{\omega})\Big\} \notag \\
  &+ \Big\{\overleftrightarrow{G}^{R}_{ci}(\bm{k}, \tilde{\omega}) - \overleftrightarrow{G}^{A}_{ci}(\bm{k}, \tilde{\omega})\Big\}\overleftrightarrow{J}^{\alpha}_{ij}(\bm{k})\overleftrightarrow{G}^{A}_{jd}(\bm{k}, \tilde{\omega}-\omega - i\gamma)\Big].
\end{align}
We define $F^{\cdots}_{\cdots}$ and $G_{\cdots}$ by
\begin{align}
  F^{(k)eh}_{(l)ji}(\omega+i\gamma)= \frac{U}{nV}\sum_{\bm{k}^{\prime}}\frac{U_{ja}(\bm{k}^{\prime})\tilde{Y}^{(k)eh}_{ab}(\bm{k}^{\prime})\tilde{\varphi}^{(l)}_{bc}(\bm{k^{\prime}})U^{\dagger}_{ci}(\bm{k}^{\prime})f_{ab}}{\omega - E_{ba} +i\gamma}, \quad 
  G_{(k)eh}(\omega+i\gamma)= \frac{U}{nV}\sum_{\bm{k}^{\prime}}\frac{\tilde{Y}^{(k)eh}_{ab}(\bm{k}^{\prime})\tilde{J}^{\alpha}_{ba}(\bm{k}^{\prime})f_{ab}}{\omega - E_{ba} +i\gamma},
\end{align}
\begin{align}
         F^{(k)he}_{(l)ji}(\omega+i\gamma)&=\frac{U}{nV}\sum_{\bm{k}^{\prime}}\frac{U_{ja}(\bm{k}^{\prime})\tilde{Y}^{(k)he}_{ab}(\bm{k}^{\prime})\tilde{\varphi}^{(l)}_{bc}(\bm{k^{\prime}})U^{\dagger}_{ci}(\bm{k}^{\prime})f_{ab}}{\omega - E_{ba} +i\gamma},  \quad 
  G_{(k)he}(\omega+i\gamma)= \frac{U}{nV}\sum_{\bm{k}^{\prime}}\frac{\tilde{Y}^{(k)he}_{ab}(\bm{k}^{\prime})\tilde{J}^{\alpha}_{ba}(\bm{k}^{\prime})f_{ab}}{\omega - E_{ba} +i\gamma}.
\end{align}
Then, we express $\hat{F}^{\cdots}_{\cdots}$ by the summation for the fermion's internal degrees of freedom like in Eq.~(\ref{eq:internal_sum}). For example, $\hat{F}^{(i)eh}_{(j)eh}$ is given by the summation,
\begin{align}
    \hat{F}^{(i)eh}_{(j)eh} = \sum_{k, l} F^{(i)eh}_{(j)(e,k)(h,l)}(\omega+i\gamma).
\end{align}
The self-consistent equation of $\Lambda^{(i)\alpha}$ is obtained as follows,
\begin{align}
    \begin{pmatrix}
        \Lambda^{(1)\alpha}_{eh} \\
        \Lambda^{(1)\alpha}_{he} \\
        \Lambda^{(2)\alpha}_{eh} \\
        \Lambda^{(2)\alpha}_{he} 
    \end{pmatrix}
    =
    \begin{pmatrix}
        \hat{F}^{(1)eh}_{(1)eh} & \hat{F}^{(1)eh}_{(1)he} & \hat{F}^{(1)eh}_{(2)eh} & \hat{F}^{(1)eh}_{(2)he} \\
        \hat{F}^{(1)he}_{(1)eh} & \hat{F}^{(1)he}_{(1)he} & \hat{F}^{(1)he}_{(2)eh} & \hat{F}^{(1)he}_{(2)he} \\
        \hat{F}^{(2)eh}_{(1)eh} & \hat{F}^{(2)eh}_{(1)he} & \hat{F}^{(2)eh}_{(2)eh} & \hat{F}^{(2)eh}_{(2)he} \\
        \hat{F}^{(2)he}_{(1)eh} & \hat{F}^{(2)he}_{(1)he} & \hat{F}^{(2)he}_{(2)eh} & \hat{F}^{(2)he}_{(2)he}
    \end{pmatrix}
    \begin{pmatrix}
        \Lambda^{(1)\alpha}_{eh} \\
        \Lambda^{(1)\alpha}_{he} \\
        \Lambda^{(2)\alpha}_{eh} \\
        \Lambda^{(2)\alpha}_{he} 
    \end{pmatrix}
    +
    \begin{pmatrix}
        G_{(1)eh}(\omega + i\gamma) \\
        G_{(1)he}(\omega + i\gamma) \\
        G_{(2)eh}(\omega + i\gamma) \\
        G_{(2)he}(\omega + i\gamma)
    \end{pmatrix}.
\end{align}
When we rewrite $\Lambda^{(i)\alpha}$ as $\Lambda^{(i)\alpha}_{1} = (\Lambda^{(i)\alpha}_{eh} + \Lambda^{(i)\alpha}_{he})/2$, $\Lambda^{(i)\alpha}_{2} = i(\Lambda^{(i)\alpha}_{eh} - \Lambda^{(i)\alpha}_{he})/2$, the self-consistent equation is expressed as
\begin{align}
    \begin{pmatrix}
        \bm{\Lambda}^{(1)\alpha}(\omega + i\gamma) \\
        \bm{\Lambda}^{(2)\alpha}(\omega + i\gamma) \\
    \end{pmatrix}
    =
    \begin{pmatrix}
    \bm{M}^{(11)}(\omega + i\gamma) & \bm{M}^{(12)}(\omega + i\gamma) \\ 
    \bm{M}^{(21)}(\omega + i\gamma) & \bm{M}^{(22)}(\omega + i\gamma)
    \end{pmatrix}
    \begin{pmatrix}
        \bm{\Lambda}^{(1)\alpha}(\omega + i\gamma) \\
        \bm{\Lambda}^{(2)\alpha}(\omega + i\gamma) \\
    \end{pmatrix}
    +
    \begin{pmatrix}
        \bm{L}^{(1)J}(\omega+i\gamma) \\
        \bm{L}^{(2)J}(\omega+i\gamma) 
    \end{pmatrix}.
\end{align}
We adopt the notation $\bm{\Lambda}^{(i)\alpha} = \left(\Lambda^{(i)\alpha}_{1}, \Lambda^{(i)\alpha}_{2}\right)^{\top}$, $\bm{L}^{(i)J} = \left(L^{(i)J}_{1}, L^{(i)J}_{2}\right)^{\top}$, and the components of $\bm{M}^{(ij)}$ and $\bm{L}^{(i)J}$ are obtained as
\begin{align}
    M^{(ij)}_{11} = \frac{1}{2}\left(\hat{F}^{(i)eh}_{(j)eh} + \hat{F}^{(i)eh}_{(j)he} + \hat{F}^{(i)he}_{(j)eh} + \hat{F}^{(i)he}_{(j)he}\right),& \quad
    M^{(ij)}_{12} = -\frac{i}{2}\left(\hat{F}^{(i)eh}_{(j)eh} - \hat{F}^{(i)eh}_{(j)he} + \hat{F}^{(i)he}_{(j)eh} - \hat{F}^{(i)he}_{(j)he}\right), \\
    M^{(ij)}_{21} = \frac{i}{2}\left(\hat{F}^{(i)eh}_{(j)eh} + \hat{F}^{(i)eh}_{(j)he} - \hat{F}^{(i)he}_{(j)eh} - \hat{F}^{(i)he}_{(j)he}\right),& \quad
    M^{(ij)}_{22} = \frac{1}{2}\left(\hat{F}^{(i)eh}_{(j)eh} - \hat{F}^{(i)eh}_{(j)he} - \hat{F}^{(i)he}_{(j)eh} + \hat{F}^{(i)he}_{(j)he}\right), \\
    L^{(i)J}_{1} = \frac{1}{2}\left(G_{(i)eh} + G_{(i)he}\right),& \quad L^{(i)J}_{2} = \frac{i}{2}\left(G_{(i)eh} - G_{(i)he}\right).
\end{align}

\section{Bare linear optical conductivity}
In this Appendix, we review the bare linear optical conductivity in the superconducting state and discuss the relation between the linear optical conductivity and quantum geometry. 
First, we introduce the quantum geometric quantities using the vector potential parameterization~\cite{Watanabe2022}.  Here, we consider the adiabatic parameter $\lambda$ that plays the same role as the vector potential. The Hamiltonian with the parameter $\bm{\lambda}$ is introduced as
\begin{align}
H_{\bm{\lambda}}\equiv\overleftrightarrow{\mathcal{H}}_{ab}(\bm{k};\bm{A} = \bm{\lambda}) =
  \begin{pmatrix}
    \mathcal{H}_{ab}(\bm{k}+ \bm{q} - \bm{\lambda})) & \Delta_{ab}(\bm{k}) \\
    \Delta^{+}_{ab}(\bm{k}) & -\mathcal{H}^{\top}_{ab}(-\bm{k}+ \bm{q} - \bm{\lambda})
  \end{pmatrix}.  
\end{align}
The Hellman-Feynman relation for the vector potential is obtained as
\begin{align}
    \Braket{a_{\bm{\lambda}}|\frac{\partial H_{\bm{\lambda}}}{\partial \lambda_{\alpha}}|b_{\bm{\lambda}}} =\frac{\partial E_{a}(\bm{\lambda})}{\partial \lambda_{\alpha}}\delta_{ab} - i[\xi^{\lambda_{\alpha}}, H_{\bm{\lambda}}]_{ab},
\end{align}
where we define the connection $\xi^{\lambda_{\alpha}}$,
\begin{align}
    \xi^{\lambda_{\alpha}}_{ab} \equiv i\Braket{a_{\bm{\lambda}}|\frac{\partial b_{\bm{\lambda}}}{\partial \lambda_{\alpha}}},
\end{align}
with the eigenstates $\ket{a_{\bm{\lambda}}}$, $\ket{b_{\bm{\lambda}}}$ and eigenenergy $E_{a}(\bm{\lambda})$ of  the Hamiltonian $H_{\bm{\lambda}}$. We consider the band-resolved quantum metric and Berry curvature, which are defined as
\begin{align}
    g^{\alpha\beta}_{ab} = \mathrm{Re}[\xi^{\lambda_{\alpha}}_{ab}\xi^{\lambda_{\beta}}_{ba}], \quad \Omega^{\alpha\beta}_{ab} = -2\mathrm{Im}[\xi^{\lambda_{\alpha}}_{ab}\xi^{\lambda_{\beta}}_{ba}].
\end{align}
These quantum geometric quantities are related to the paramagnetic electric current in the band representation, 
\begin{align}
    \tilde{J}^{\alpha}_{ab} = 
    - \lim_{\bm{\lambda}\rightarrow\bm{0}}\Braket{a_{\bm{\lambda}}|\frac{\partial H_{\bm{\lambda}}}{\partial \lambda_{\alpha}}|b_{\bm{\lambda}}}.
\end{align}
Actually, the band-resolved quantum metric and Berry curvature can be expressed as
\begin{align}
    g^{\alpha\beta}_{ab} = \frac{1}{E^{2}_{ab}}\mathrm{Re}[J^{\alpha}_{ab}J^{\beta}_{ba}], \quad \Omega^{\alpha\beta}_{ab} = -\frac{2}{E_{ab}^{2}}\mathrm{Im}[J^{\alpha}_{ab}J^{\beta}_{ba}].
    \label{eq:quantum_geo}
\end{align}
By using Eq.~\eqref{eq:quantum_geo}, the bare linear optical conductivity is obtained as 
\begin{align}
    \sigma^{\alpha\beta}(\omega) = \frac{1}{2i\omega}\sum_{a,b}\left( \frac{-J^{\alpha}_{ab}J^{\beta}_{ba}f_{ab}}{\omega + i\gamma - E_{ba}} - J^{\alpha\beta}_{ab}f_{a}\delta_{ab}\right)
    = \frac{1}{2i\omega}\sum_{a,b}\left( \frac{-E_{ba}^{2}f_{ab}}{\omega + i\gamma - E_{ba}}\left(g^{\alpha\beta}_{ab} - \frac{i}{2}\Omega^{\alpha\beta}_{ab}\right) - J^{\alpha\beta}_{ab}f_{a}\delta_{ab}\right).
\end{align}
In particular, the resonant component of the linear optical conductivity is given by
\begin{align}
    \sigma^{\alpha\beta}(\omega) =& \pi \sum_{a,b} E_{ba}\left(g^{\alpha\beta}_{ab} - \frac{i}{2}\Omega^{\alpha\beta}_{ab}\right)f_{ab}.
\end{align}
Therefore, the bare linear conductivity is determined by the band dispersion and the quantum geometry in the vector potential space.

Figure~\ref{fig:linear_result}(a) plots the linear optical conductivity when the Fermi level lies on the Dirac point ($\tilde{\mu}=0.0$), while Fig.~\ref{fig:linear_ratio}(a) shows the results when the Fermi level is far from the Dirac point ($\tilde{\mu}=0.2$). Comparing these figures, we find that the Dirac point close to the Fermi level enhances the linear conductivity.
The enhancement of the linear conductivity can be attributed to the quantum metric. In the normal state, the quantum metric diverges at the Dirac point due to the energy degeneracy. Although the superconducting gap prevents the divergence of the quantum metric, a large quantum metric arises when the Fermi level lies near the Dirac point~\cite{Tanaka2024}. In the main text, we show that the peculiar geometric properties of the Dirac point enhance not only the bare linear conductivity but also the corrected linear conductivity with vertex correction due to the collective modes.

\section{Linear response in the single-band model}
In this Appendix, we calculate the linear optical conductivity in single-band $s$-wave superconductors, and discuss effects of the collective modes. The BdG Hamiltonian of single-band superconductors is written as
\begin{align}
    \mathcal{H}(\bm{k}) = 
    \begin{pmatrix}
        \xi_{1}(\bm{k}) & \psi \\
        \psi & -\xi_{2}(\bm{k})
    \end{pmatrix},
\end{align}
where we use the Nambu spinor notation~\cite{Nambu1960}.
The eigenenergies are obtained as $E_{\pm}(\bm{k})=\Delta\xi(\bm{k})\pm \Delta E(\bm{k})$, where 
$\bar{\xi}(\bm{k})=(\xi_{1}(\bm{k}) + \xi_{2}(\bm{k}))/2$, $\Delta\xi(\bm{k})=(\xi_{1}(\bm{k}) - \xi_{2}(\bm{k}))/2$, and $\Delta E(\bm{k})=\sqrt{{\bar{\xi}(\bm{k})}^{2} + \psi^{2}}$.

The electric current operator and the one-photon vertex function are expressed as follows, 
\begin{align}
    J^{\alpha}(\bm{k}) = J_{0}(\bm{k}) + J_{3}(\bm{k})\sigma_{3}, \quad \Gamma^{\alpha}(\bm{k}, \omega+i\gamma) = J_{0}(\bm{k}) + J_{3}(\bm{k})\sigma_{3} + \Lambda_{1}(\omega+i\gamma)\sigma_{1} + \Lambda_{2}(\omega+i\gamma)\sigma_{2}. 
\end{align}
Here, we ignore the diamagnetic current term because we focus on the effect of collective modes.
In the zero temperature limit, the paramagnetic linear conductivity is obtained as
\begin{align}
    \sigma^{\alpha\beta}(\omega) = \sum_{\bm{k}} \frac{i\psi J^{\alpha}_{3}}{\omega\Delta E}\cdot \frac{-4\bar{\xi}\Lambda^{\beta}_{1} + 4\psi J^{\beta}_{3} + 2i(\omega + i\gamma)\Lambda^{\beta}_{2}}{(\omega + i\gamma)^{2} - 4 {\Delta E}^{2}},
\end{align}
where we omit variables $\bm{k}$ and $\omega + i\gamma$ in the right hand side of the equation for simplicity. In particular, the resonant component of linear conductivity is given by
\begin{align}
    \sigma^{\alpha\beta}(\omega) = \sum_{\bm{k}}\frac{\pi\psi J^{\alpha}_{3}}{2{\Delta E}^{2}}\left\{ \frac{-\bar{\xi}\Lambda^{\beta}_{1} + J^{\beta}_{3}\psi}{\Delta E} + i\Lambda^{\beta}_{2} \right\}\delta(\omega - 2\Delta E).
\end{align}
The large resonant component of linear conductivity can arise from the momentum space with small energy gap, 
where $\Delta E(\bm{k})$ is small, since the resonant components are proportional to ${\Delta E}^{-2}$. The energy gap becomes the smallest, $2\Delta E(\bm{k})=2\psi$, around the Fermi surface where $\bar{\xi}(\bm{k})=0$. Moreover, the joint density of states has a peak at $\omega = 2\psi$. 

\begin{figure*}[htbp]
 \includegraphics[width=0.9\linewidth]{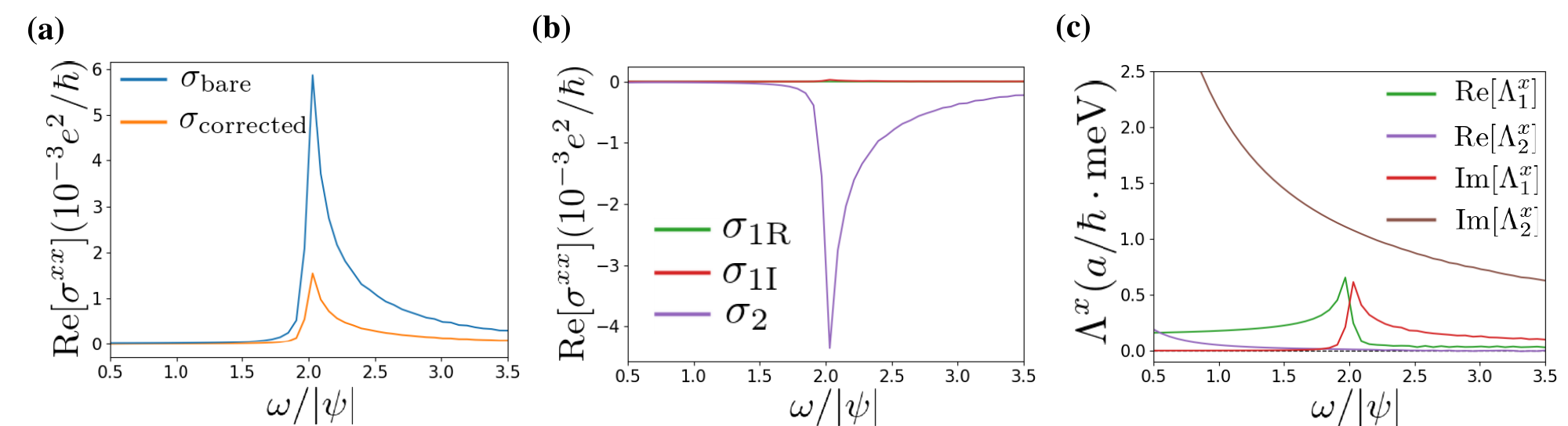}
 \caption{(a) (b) The linear conductivity $\mathrm{Re}[\sigma^{xx}]$ and (c) the vertex function $\Lambda^{x}_{1}$ in the single-band model introduced in Appendix~\ref{app:single_band}. (a) The bare and corrected linear conductivity. (b) The contributions of the vertex function in the linear conductivity, $\sigma_{\mathrm{1R}}$, $\sigma_{\mathrm{1I}}$, and $\sigma_{\mathrm{2}}$, which arise from the vertices $\mathrm{Re}[\Lambda^{x}_{1}]$, $\mathrm{Im}[\Lambda^{x}_{1}]$, and $\Lambda^{x}_{2}$, respectively. 
 }
 \label{fig:single_linear_result}
\end{figure*}

In the main text, we denote the contributions in linear conductivity proportional to $\Lambda^{\beta}_{1}$ and $\Lambda^{\beta}_{2}$ by $\sigma_{\mathrm{1}}$ and $\sigma_{\mathrm{2}}$, respectively [Eq.~\eqref{eq:linear_conductivity_decomposition}]. 
In the single-band models, we obtain large $\sigma_{\mathrm{bare}}$ and $\sigma_{\mathrm{2}}$ around $\omega = 2\psi$ due to the prefactor ${\Delta E}^{-2}$. However, $\sigma_{1}$ is suppressed by the prefactor $\bar{\xi}$ because $\bar{\xi}$ is close to zero when $\Delta E \sim \psi$.
Here, we numerically demonstrate vertex correction of the linear conductivity in the single-band model. We consider a single-band spin-degenerate $s$-wave superconductor in the presence of a DC supercurrent,
\begin{align}
\mathcal{H}(\bm{k})=
    \begin{pmatrix}
        \epsilon(\bm{k}+q\hat{x}) & \psi \\
        \psi & -\epsilon(-\bm{k}+q\hat{x})
    \end{pmatrix},
\end{align}
where $\epsilon(\bm{k})=t(2 - \cos k_{x} - \cos k_{y}) - \mu$.
We set $t=1.0\times 10^{2}\mathrm{meV}$, $q=0.02/a$, and $U=0.5\times 10^{2}\mathrm{meV}$.
Figure \ref{fig:single_linear_result}(a) shows the frequency dependence of the bare and corrected linear conductivity $\mathrm{Re}[\sigma^{xx}]$. The corrected conductivity with vertex correction is suppressed by the contribution due to the phase mode $\sigma_{2}$, while the contribution due to the amplitude mode $\sigma_{1}$ is almost zero [see Figs.~\ref{fig:single_linear_result}(a) and (b)] as expected by the above discussion. In Fig.~\ref{fig:single_linear_result}(c), it is shown that the vertex due to the amplitude mode $\Lambda^{x}_{1}$ itself is relatively small compared with $\Lambda^{x}_{2}$ due to the phase mode. This is probably because $\Lambda^{x}_{1}$ originates from the weighted integral of $\bar{\xi}(\bm{k})$ around the Fermi surface~\cite{Dai2017}.
Note that these results are expected to be ubiquitous in single-band models including those for $d$-wave superconductors. However, multiband systems can show qualitatively different results as demonstrated in the main text.
\label{app:single_band}

\section{Hermitian conjugate of the vertex function}
\label{app:hermite_conj}
We define the Hermitian conjugate of the vertex function $\overleftrightarrow{\Gamma^{\dagger}}^{\alpha}_{ab}$ as
\begin{align}
    \overleftrightarrow{\Gamma^{\dagger}}^{\alpha}_{ab}(\bm{k}, \omega +i\gamma) = \left[\overleftrightarrow{\Gamma}^{\alpha}_{ba}(\bm{k}, \omega +i\gamma)\right]^{\dagger}.
\end{align}
We can rewrite $\overleftrightarrow{\Gamma^{\dagger}}^{\alpha}_{ab}$ as
\begin{align}
    \left[\Gamma^{\dagger\alpha}(\bm{k}, \omega +i\gamma)\right]_{mn} = \left[\Gamma^{\alpha}(\bm{k}, \omega +i\gamma)\right]^{*}_{nm},
\end{align}
where the indices $m,n$ 
include the fermion's internal degrees of freedom and the Nambu space. The vertex function is given by the self-consisten equation,
\begin{align}
 \overleftrightarrow{\Gamma}^{\alpha}_{ab}(\bm{k}, \omega + i\gamma)&=\overleftrightarrow{J}^{\alpha}_{ab}(\bm{k})+\overleftrightarrow{U}_{ab}(\bm{k}, \omega + i\gamma)[\Gamma^{\alpha}(\omega + i\gamma)],
\end{align}
and therefore, the Hermitian conjugate of the vertex function is determined by the equation
\begin{align}
 \overleftrightarrow{\Gamma^{\dagger}}^{\alpha}_{ab}(\bm{k}, \omega + i\gamma)&=\overleftrightarrow{J^{\dagger}}^{\alpha}_{ab}(\bm{k})+\left[\overleftrightarrow{U}_{ab}(\bm{k}, \omega + i\gamma)[\Gamma^{\alpha}(\omega + i\gamma)] \right]^{\dagger},
\end{align}
where $U^{\dagger}$ is obtained as
\begin{align}
  \left[\overleftrightarrow{U}_{ab}(\bm{k}, \omega + i\gamma)[\Gamma^{\alpha}(\omega + i\gamma)]\right]^{\dagger}& \notag \\
    =\frac{U}{V}\int\frac{d\tilde{\omega}}{2\pi i}\sum_{\bm{k}^{\prime}}f(\tilde{\omega})\Big[&\varphi^{\dagger}_{ab}(\bm{k})\varphi^{\top}_{dc}(\bm{k}^{\prime})\overleftrightarrow{\mathrm{P}_{-}}\Big\{ \overleftrightarrow{G}^{AR}_{dj}(\bm{k}^{\prime}, \tilde{\omega})\overleftrightarrow{\Gamma^{\dagger}}^{\alpha}_{ji}(\bm{k}^{\prime},\omega + i\gamma)\overleftrightarrow{G}^{A}_{ic}(\bm{k}^{\prime},\tilde{\omega} + \omega - i\gamma) \notag \\
  &+\overleftrightarrow{G}^{R}_{dj}(\bm{k}^{\prime}, \tilde{\omega}-\omega +i\gamma)\overleftrightarrow{\Gamma^{\dagger}}^{\alpha}_{ji}(\bm{k}^{\prime},\omega+i\gamma)\overleftrightarrow{G}^{AR}_{ic}(\bm{k}^{\prime}, \tilde{\omega})\Big\}\overleftrightarrow{\mathrm{P}_{+}} \notag \\
  &+\varphi_{ab}(\bm{k})\varphi^{*}_{dc}(\bm{k}^{\prime})\overleftrightarrow{\mathrm{P}_{+}}\Big\{ \overleftrightarrow{G}^{AR}_{dj}(\bm{k}^{\prime}, \tilde{\omega})\overleftrightarrow{\Gamma^{\dagger}}^{\alpha}_{ji}(\bm{k}^{\prime},\omega + i\gamma)\overleftrightarrow{G}^{A}_{ic}(\bm{k}^{\prime}, \tilde{\omega}+\omega - i\gamma ) \notag \\
  &\overleftrightarrow{G}^{R}_{dj}(\bm{k}^{\prime}, \tilde{\omega}-\omega + i\gamma)\overleftrightarrow{\Gamma^{\dagger}}^{\alpha}_{ji}(\bm{k}^{\prime},\omega+i\gamma)\overleftrightarrow{G}^{AR}_{ic}(\bm{k}^{\prime}, \tilde{\omega})\Big\}\overleftrightarrow{\mathrm{P}_{-}}\Big]\\
    =-\frac{U}{V}\int\frac{d\tilde{\omega}}{2\pi i}\sum_{\bm{k}^{\prime}}f(\tilde{\omega})\Big[&\varphi^{\dagger}_{ab}(\bm{k})\varphi_{dc}(\bm{k}^{\prime})\overleftrightarrow{\mathrm{P}_{-}}\Big\{ \overleftrightarrow{G}^{RA}_{ci}(\bm{k}^{\prime}, \tilde{\omega})\overleftrightarrow{\Gamma^{\dagger}}^{\alpha}_{ij}(\bm{k}^{\prime},\omega + i\gamma)\overleftrightarrow{G}^{A}_{jd}(\bm{k}^{\prime},\tilde{\omega} + \omega - i\gamma) \notag \\
  &+\overleftrightarrow{G}^{R}_{ci}(\bm{k}^{\prime}, \tilde{\omega}-\omega + i\gamma)\overleftrightarrow{\Gamma^{\dagger}}^{\alpha}_{ij}(\bm{k}^{\prime},\omega+i\gamma)\overleftrightarrow{G}^{RA}_{jd}(\bm{k}^{\prime}, \tilde{\omega})\Big\}\overleftrightarrow{\mathrm{P}_{+}} \notag \\
  &+\varphi_{ab}(\bm{k})\varphi^{\dagger}_{dc}(\bm{k}^{\prime})\overleftrightarrow{\mathrm{P}_{+}}\Big\{ \overleftrightarrow{G}^{RA}_{ci}(\bm{k}^{\prime}, \tilde{\omega})\overleftrightarrow{\Gamma^{\dagger}}^{\alpha}_{ij}(\bm{k}^{\prime},\omega + i\gamma)\overleftrightarrow{G}^{A}_{jd}(\bm{k}^{\prime}, \tilde{\omega}+\omega -i\gamma ) \notag \\
  &\overleftrightarrow{G}^{R}_{ci}(\bm{k}^{\prime}, \tilde{\omega}-\omega + i\gamma)\overleftrightarrow{\Gamma^{\dagger}}^{\alpha}_{ij}(\bm{k}^{\prime},\omega+i\gamma)\overleftrightarrow{G}^{RA}_{jd}(\bm{k}^{\prime}, \tilde{\omega})\Big\}\overleftrightarrow{\mathrm{P}_{-}}\Big]\\
  =\overleftrightarrow{U}_{ab}(\bm{k}, -\omega + i\gamma)&[\Gamma^{\dagger\alpha}(\omega + i\gamma)].
\end{align}
In the above calculation, we used the relation $\left(G^{R}(\omega + i\gamma)\right)^{\dagger}=G^{A}(\omega -i\gamma)$. As a result, the Hermitian conjugate of the vertex function is obtained from the self-consistent equation
\begin{align}
 \overleftrightarrow{\Gamma^{\dagger}}^{\alpha}_{ab}(\bm{k}, \omega + i\gamma)
 =\overleftrightarrow{J^{\dagger}}^{\alpha}_{ab}(\bm{k})+\overleftrightarrow{U}_{ab}(\bm{k}, -\omega + i\gamma)\left[\Gamma^{\dagger\alpha}(\omega + i\gamma)\right].
\end{align}
Thus, when we assume that the solution of the self-consistent equation is uniquely determined, the Hermitian conjugate of the vertex function is given by
\begin{align}
    \overleftrightarrow{\Gamma^{\dagger}}^{\alpha}_{ab}(\bm{k}, \omega + i\gamma) = \overleftrightarrow{\Gamma}^{\alpha}_{ab}(\bm{k}, -\omega + i\gamma).
\end{align}
We also obtain the relation of the vertex function $\tilde{\Gamma}^{x}_{ab}(\omega + i\gamma)$ in the band representation,
\begin{align}
    \tilde{\Gamma}^{\dagger\alpha}_{ab}(\bm{k}, \omega + i\gamma) = \tilde{\Gamma}^{\alpha}_{ab}(\bm{k}, -\omega + i\gamma).
\end{align}
The magnetic injection current contains the factor $\mathrm{Re}[\tilde{\Gamma}^{\alpha}_{ba}(\bm{k}, \omega + i\gamma)\tilde{\Gamma}^{\alpha}_{ab}(\bm{k}, -\omega + i\gamma)]$. The sign of the factor is shown to be positive, because
\begin{align}
    \mathrm{Re}[\tilde{\Gamma}^{\alpha}_{ba}(\bm{k}, \omega + i\gamma)\tilde{\Gamma}^{\alpha}_{ab}(\bm{k}, -\omega + i\gamma)] = \mathrm{Re}\left[\tilde{\Gamma}^{\alpha}_{ba}(\bm{k}, \omega + i\gamma)\left(\tilde{\Gamma}^{\alpha}_{ba}(\bm{k}, \omega + i\gamma)\right)^{*}\right] \geq 0.
\end{align}

\bibliographystyle{apsrev4-1}
\bibliography{reference}

\begin{thebibliography}{100}%
\makeatletter
\providecommand \@ifxundefined [1]{%
 \@ifx{#1\undefined}
}%
\providecommand \@ifnum [1]{%
 \ifnum #1\expandafter \@firstoftwo
 \else \expandafter \@secondoftwo
 \fi
}%
\providecommand \@ifx [1]{%
 \ifx #1\expandafter \@firstoftwo
 \else \expandafter \@secondoftwo
 \fi
}%
\providecommand \natexlab [1]{#1}%
\providecommand \enquote  [1]{``#1''}%
\providecommand \bibnamefont  [1]{#1}%
\providecommand \bibfnamefont [1]{#1}%
\providecommand \citenamefont [1]{#1}%
\providecommand \href@noop [0]{\@secondoftwo}%
\providecommand \href [0]{\begingroup \@sanitize@url \@href}%
\providecommand \@href[1]{\@@startlink{#1}\@@href}%
\providecommand \@@href[1]{\endgroup#1\@@endlink}%
\providecommand \@sanitize@url [0]{\catcode `\\12\catcode `\$12\catcode `\&12\catcode `\#12\catcode `\^12\catcode `\_12\catcode `\%12\relax}%
\providecommand \@@startlink[1]{}%
\providecommand \@@endlink[0]{}%
\providecommand \url  [0]{\begingroup\@sanitize@url \@url }%
\providecommand \@url [1]{\endgroup\@href {#1}{\urlprefix }}%
\providecommand \urlprefix  [0]{URL }%
\providecommand \Eprint [0]{\href }%
\providecommand \doibase [0]{http://dx.doi.org/}%
\providecommand \selectlanguage [0]{\@gobble}%
\providecommand \bibinfo  [0]{\@secondoftwo}%
\providecommand \bibfield  [0]{\@secondoftwo}%
\providecommand \translation [1]{[#1]}%
\providecommand \BibitemOpen [0]{}%
\providecommand \bibitemStop [0]{}%
\providecommand \bibitemNoStop [0]{.\EOS\space}%
\providecommand \EOS [0]{\spacefactor3000\relax}%
\providecommand \BibitemShut  [1]{\csname bibitem#1\endcsname}%
\let\auto@bib@innerbib\@empty
\bibitem [{\citenamefont {Degiorgi}\ \emph {et~al.}(1994)\citenamefont {Degiorgi}, \citenamefont {Briceno}, \citenamefont {Fuhrer}, \citenamefont {Zettl},\ and\ \citenamefont {Wachter}}]{Degiorgi1994}%
  \BibitemOpen
  \bibfield  {author} {\bibinfo {author} {\bibfnamefont {L.}~\bibnamefont {Degiorgi}}, \bibinfo {author} {\bibfnamefont {G.}~\bibnamefont {Briceno}}, \bibinfo {author} {\bibfnamefont {M.~S.}\ \bibnamefont {Fuhrer}}, \bibinfo {author} {\bibfnamefont {A.}~\bibnamefont {Zettl}}, \ and\ \bibinfo {author} {\bibfnamefont {P.}~\bibnamefont {Wachter}},\ }\href {\doibase 10.1038/369541a0} {\bibfield  {journal} {\bibinfo  {journal} {Nature}\ }\textbf {\bibinfo {volume} {369}},\ \bibinfo {pages} {541} (\bibinfo {year} {1994})}\BibitemShut {NoStop}%
\bibitem [{\citenamefont {Molegraaf}\ \emph {et~al.}(2002)\citenamefont {Molegraaf}, \citenamefont {Presura}, \citenamefont {van~der Marel}, \citenamefont {Kes},\ and\ \citenamefont {Li}}]{Molegraaf2002}%
  \BibitemOpen
  \bibfield  {author} {\bibinfo {author} {\bibfnamefont {H.~J.~A.}\ \bibnamefont {Molegraaf}}, \bibinfo {author} {\bibfnamefont {C.}~\bibnamefont {Presura}}, \bibinfo {author} {\bibfnamefont {D.}~\bibnamefont {van~der Marel}}, \bibinfo {author} {\bibfnamefont {P.~H.}\ \bibnamefont {Kes}}, \ and\ \bibinfo {author} {\bibfnamefont {M.}~\bibnamefont {Li}},\ }\href {\doibase 10.1126/science.1069947} {\bibfield  {journal} {\bibinfo  {journal} {Science}\ }\textbf {\bibinfo {volume} {295}},\ \bibinfo {pages} {2239} (\bibinfo {year} {2002})},\ \Eprint {http://arxiv.org/abs/https://www.science.org/doi/pdf/10.1126/science.1069947} {https://www.science.org/doi/pdf/10.1126/science.1069947} \BibitemShut {NoStop}%
\bibitem [{\citenamefont {Wang}\ \emph {et~al.}(1998)\citenamefont {Wang}, \citenamefont {Tajima}, \citenamefont {Rykov},\ and\ \citenamefont {Tomimoto}}]{Wang1998}%
  \BibitemOpen
  \bibfield  {author} {\bibinfo {author} {\bibfnamefont {N.~L.}\ \bibnamefont {Wang}}, \bibinfo {author} {\bibfnamefont {S.}~\bibnamefont {Tajima}}, \bibinfo {author} {\bibfnamefont {A.~I.}\ \bibnamefont {Rykov}}, \ and\ \bibinfo {author} {\bibfnamefont {K.}~\bibnamefont {Tomimoto}},\ }\href {\doibase 10.1103/PhysRevB.57.R11081} {\bibfield  {journal} {\bibinfo  {journal} {Phys. Rev. B}\ }\textbf {\bibinfo {volume} {57}},\ \bibinfo {pages} {R11081} (\bibinfo {year} {1998})}\BibitemShut {NoStop}%
\bibitem [{\citenamefont {Mattis}\ and\ \citenamefont {Bardeen}(1958)}]{Mattis1958}%
  \BibitemOpen
  \bibfield  {author} {\bibinfo {author} {\bibfnamefont {D.~C.}\ \bibnamefont {Mattis}}\ and\ \bibinfo {author} {\bibfnamefont {J.}~\bibnamefont {Bardeen}},\ }\href {\doibase 10.1103/PhysRev.111.412} {\bibfield  {journal} {\bibinfo  {journal} {Phys. Rev.}\ }\textbf {\bibinfo {volume} {111}},\ \bibinfo {pages} {412} (\bibinfo {year} {1958})}\BibitemShut {NoStop}%
\bibitem [{\citenamefont {Crowley}\ and\ \citenamefont {Fu}(2022)}]{Crowley2022}%
  \BibitemOpen
  \bibfield  {author} {\bibinfo {author} {\bibfnamefont {P.~J.~D.}\ \bibnamefont {Crowley}}\ and\ \bibinfo {author} {\bibfnamefont {L.}~\bibnamefont {Fu}},\ }\href {\doibase 10.1103/PhysRevB.106.214526} {\bibfield  {journal} {\bibinfo  {journal} {Phys. Rev. B}\ }\textbf {\bibinfo {volume} {106}},\ \bibinfo {pages} {214526} (\bibinfo {year} {2022})}\BibitemShut {NoStop}%
\bibitem [{\citenamefont {Papaj}\ and\ \citenamefont {Moore}(2022)}]{Papaj2022}%
  \BibitemOpen
  \bibfield  {author} {\bibinfo {author} {\bibfnamefont {M.}~\bibnamefont {Papaj}}\ and\ \bibinfo {author} {\bibfnamefont {J.~E.}\ \bibnamefont {Moore}},\ }\href {\doibase 10.1103/PhysRevB.106.L220504} {\bibfield  {journal} {\bibinfo  {journal} {Phys. Rev. B}\ }\textbf {\bibinfo {volume} {106}},\ \bibinfo {pages} {L220504} (\bibinfo {year} {2022})}\BibitemShut {NoStop}%
\bibitem [{\citenamefont {Moor}\ \emph {et~al.}(2017)\citenamefont {Moor}, \citenamefont {Volkov},\ and\ \citenamefont {Efetov}}]{Moor2017}%
  \BibitemOpen
  \bibfield  {author} {\bibinfo {author} {\bibfnamefont {A.}~\bibnamefont {Moor}}, \bibinfo {author} {\bibfnamefont {A.~F.}\ \bibnamefont {Volkov}}, \ and\ \bibinfo {author} {\bibfnamefont {K.~B.}\ \bibnamefont {Efetov}},\ }\href {\doibase 10.1103/PhysRevLett.118.047001} {\bibfield  {journal} {\bibinfo  {journal} {Phys. Rev. Lett.}\ }\textbf {\bibinfo {volume} {118}},\ \bibinfo {pages} {047001} (\bibinfo {year} {2017})}\BibitemShut {NoStop}%
\bibitem [{\citenamefont {Leggett}(1966)}]{Leggett1966}%
  \BibitemOpen
  \bibfield  {author} {\bibinfo {author} {\bibfnamefont {A.~J.}\ \bibnamefont {Leggett}},\ }\href {\doibase 10.1143/PTP.36.901} {\bibfield  {journal} {\bibinfo  {journal} {Progress of Theoretical Physics}\ }\textbf {\bibinfo {volume} {36}},\ \bibinfo {pages} {901} (\bibinfo {year} {1966})},\ \Eprint {http://arxiv.org/abs/https://academic.oup.com/ptp/article-pdf/36/5/901/5256693/36-5-901.pdf} {https://academic.oup.com/ptp/article-pdf/36/5/901/5256693/36-5-901.pdf} \BibitemShut {NoStop}%
\bibitem [{\citenamefont {Kamatani}\ \emph {et~al.}(2022)\citenamefont {Kamatani}, \citenamefont {Kitamura}, \citenamefont {Tsuji}, \citenamefont {Shimano},\ and\ \citenamefont {Morimoto}}]{Kamatani2022}%
  \BibitemOpen
  \bibfield  {author} {\bibinfo {author} {\bibfnamefont {T.}~\bibnamefont {Kamatani}}, \bibinfo {author} {\bibfnamefont {S.}~\bibnamefont {Kitamura}}, \bibinfo {author} {\bibfnamefont {N.}~\bibnamefont {Tsuji}}, \bibinfo {author} {\bibfnamefont {R.}~\bibnamefont {Shimano}}, \ and\ \bibinfo {author} {\bibfnamefont {T.}~\bibnamefont {Morimoto}},\ }\href {\doibase 10.1103/PhysRevB.105.094520} {\bibfield  {journal} {\bibinfo  {journal} {Phys. Rev. B}\ }\textbf {\bibinfo {volume} {105}},\ \bibinfo {pages} {094520} (\bibinfo {year} {2022})}\BibitemShut {NoStop}%
\bibitem [{\citenamefont {Nagashima}\ \emph {et~al.}(2024)\citenamefont {Nagashima}, \citenamefont {Tian}, \citenamefont {Haenel}, \citenamefont {Tsuji},\ and\ \citenamefont {Manske}}]{Nagashima2024}%
  \BibitemOpen
  \bibfield  {author} {\bibinfo {author} {\bibfnamefont {R.}~\bibnamefont {Nagashima}}, \bibinfo {author} {\bibfnamefont {S.}~\bibnamefont {Tian}}, \bibinfo {author} {\bibfnamefont {R.}~\bibnamefont {Haenel}}, \bibinfo {author} {\bibfnamefont {N.}~\bibnamefont {Tsuji}}, \ and\ \bibinfo {author} {\bibfnamefont {D.}~\bibnamefont {Manske}},\ }\href {\doibase 10.1103/PhysRevResearch.6.013120} {\bibfield  {journal} {\bibinfo  {journal} {Phys. Rev. Res.}\ }\textbf {\bibinfo {volume} {6}},\ \bibinfo {pages} {013120} (\bibinfo {year} {2024})}\BibitemShut {NoStop}%
\bibitem [{\citenamefont {Bardasis}\ and\ \citenamefont {Schrieffer}(1961)}]{Bardasis1961}%
  \BibitemOpen
  \bibfield  {author} {\bibinfo {author} {\bibfnamefont {A.}~\bibnamefont {Bardasis}}\ and\ \bibinfo {author} {\bibfnamefont {J.~R.}\ \bibnamefont {Schrieffer}},\ }\href {\doibase 10.1103/PhysRev.121.1050} {\bibfield  {journal} {\bibinfo  {journal} {Phys. Rev.}\ }\textbf {\bibinfo {volume} {121}},\ \bibinfo {pages} {1050} (\bibinfo {year} {1961})}\BibitemShut {NoStop}%
\bibitem [{\citenamefont {Lee}\ and\ \citenamefont {Chung}(2023)}]{Lee2023}%
  \BibitemOpen
  \bibfield  {author} {\bibinfo {author} {\bibfnamefont {C.}~\bibnamefont {Lee}}\ and\ \bibinfo {author} {\bibfnamefont {S.~B.}\ \bibnamefont {Chung}},\ }\href {\doibase 10.1038/s42005-023-01421-8} {\bibfield  {journal} {\bibinfo  {journal} {Communications Physics}\ }\textbf {\bibinfo {volume} {6}},\ \bibinfo {pages} {307} (\bibinfo {year} {2023})}\BibitemShut {NoStop}%
\bibitem [{\citenamefont {Tsuji}\ and\ \citenamefont {Aoki}(2015)}]{Tsuji2015}%
  \BibitemOpen
  \bibfield  {author} {\bibinfo {author} {\bibfnamefont {N.}~\bibnamefont {Tsuji}}\ and\ \bibinfo {author} {\bibfnamefont {H.}~\bibnamefont {Aoki}},\ }\href {\doibase 10.1103/PhysRevB.92.064508} {\bibfield  {journal} {\bibinfo  {journal} {Phys. Rev. B}\ }\textbf {\bibinfo {volume} {92}},\ \bibinfo {pages} {064508} (\bibinfo {year} {2015})}\BibitemShut {NoStop}%
\bibitem [{\citenamefont {Anderson}(1958{\natexlab{a}})}]{Anderson1958_2}%
  \BibitemOpen
  \bibfield  {author} {\bibinfo {author} {\bibfnamefont {P.~W.}\ \bibnamefont {Anderson}},\ }\href {\doibase 10.1103/PhysRev.112.1900} {\bibfield  {journal} {\bibinfo  {journal} {Phys. Rev.}\ }\textbf {\bibinfo {volume} {112}},\ \bibinfo {pages} {1900} (\bibinfo {year} {1958}{\natexlab{a}})}\BibitemShut {NoStop}%
\bibitem [{\citenamefont {Matsunaga}\ \emph {et~al.}(2014)\citenamefont {Matsunaga}, \citenamefont {Tsuji}, \citenamefont {Fujita}, \citenamefont {Sugioka}, \citenamefont {Makise}, \citenamefont {Uzawa}, \citenamefont {Terai}, \citenamefont {Wang}, \citenamefont {Aoki},\ and\ \citenamefont {Shimano}}]{Matsunaga2014}%
  \BibitemOpen
  \bibfield  {author} {\bibinfo {author} {\bibfnamefont {R.}~\bibnamefont {Matsunaga}}, \bibinfo {author} {\bibfnamefont {N.}~\bibnamefont {Tsuji}}, \bibinfo {author} {\bibfnamefont {H.}~\bibnamefont {Fujita}}, \bibinfo {author} {\bibfnamefont {A.}~\bibnamefont {Sugioka}}, \bibinfo {author} {\bibfnamefont {K.}~\bibnamefont {Makise}}, \bibinfo {author} {\bibfnamefont {Y.}~\bibnamefont {Uzawa}}, \bibinfo {author} {\bibfnamefont {H.}~\bibnamefont {Terai}}, \bibinfo {author} {\bibfnamefont {Z.}~\bibnamefont {Wang}}, \bibinfo {author} {\bibfnamefont {H.}~\bibnamefont {Aoki}}, \ and\ \bibinfo {author} {\bibfnamefont {R.}~\bibnamefont {Shimano}},\ }\href {\doibase 10.1126/science.1254697} {\bibfield  {journal} {\bibinfo  {journal} {Science}\ }\textbf {\bibinfo {volume} {345}},\ \bibinfo {pages} {1145} (\bibinfo {year} {2014})},\ \Eprint {http://arxiv.org/abs/https://www.science.org/doi/pdf/10.1126/science.1254697} {https://www.science.org/doi/pdf/10.1126/science.1254697} \BibitemShut {NoStop}%
\bibitem [{\citenamefont {Shimano}\ and\ \citenamefont {Tsuji}(2020)}]{Shimano2020}%
  \BibitemOpen
  \bibfield  {author} {\bibinfo {author} {\bibfnamefont {R.}~\bibnamefont {Shimano}}\ and\ \bibinfo {author} {\bibfnamefont {N.}~\bibnamefont {Tsuji}},\ }\href {\doibase https://doi.org/10.1146/annurev-conmatphys-031119-050813} {\bibfield  {journal} {\bibinfo  {journal} {Annual Review of Condensed Matter Physics}\ }\textbf {\bibinfo {volume} {11}},\ \bibinfo {pages} {103} (\bibinfo {year} {2020})}\BibitemShut {NoStop}%
\bibitem [{\citenamefont {Nagaosa}\ and\ \citenamefont {Yanase}(2024)}]{Nagaosa2024}%
  \BibitemOpen
  \bibfield  {author} {\bibinfo {author} {\bibfnamefont {N.}~\bibnamefont {Nagaosa}}\ and\ \bibinfo {author} {\bibfnamefont {Y.}~\bibnamefont {Yanase}},\ }\href {\doibase https://doi.org/10.1146/annurev-conmatphys-032822-033734} {\bibfield  {journal} {\bibinfo  {journal} {Annual Review of Condensed Matter Physics}\ }\textbf {\bibinfo {volume} {15}},\ \bibinfo {pages} {63} (\bibinfo {year} {2024})}\BibitemShut {NoStop}%
\bibitem [{\citenamefont {Watanabe}\ \emph {et~al.}(2022{\natexlab{a}})\citenamefont {Watanabe}, \citenamefont {Daido},\ and\ \citenamefont {Yanase}}]{Watanabe2022_2}%
  \BibitemOpen
  \bibfield  {author} {\bibinfo {author} {\bibfnamefont {H.}~\bibnamefont {Watanabe}}, \bibinfo {author} {\bibfnamefont {A.}~\bibnamefont {Daido}}, \ and\ \bibinfo {author} {\bibfnamefont {Y.}~\bibnamefont {Yanase}},\ }\href {\doibase 10.1103/PhysRevB.105.L100504} {\bibfield  {journal} {\bibinfo  {journal} {Phys. Rev. B}\ }\textbf {\bibinfo {volume} {105}},\ \bibinfo {pages} {L100504} (\bibinfo {year} {2022}{\natexlab{a}})}\BibitemShut {NoStop}%
\bibitem [{\citenamefont {Watanabe}\ \emph {et~al.}(2022{\natexlab{b}})\citenamefont {Watanabe}, \citenamefont {Daido},\ and\ \citenamefont {Yanase}}]{Watanabe2022}%
  \BibitemOpen
  \bibfield  {author} {\bibinfo {author} {\bibfnamefont {H.}~\bibnamefont {Watanabe}}, \bibinfo {author} {\bibfnamefont {A.}~\bibnamefont {Daido}}, \ and\ \bibinfo {author} {\bibfnamefont {Y.}~\bibnamefont {Yanase}},\ }\href {\doibase 10.1103/PhysRevB.105.024308} {\bibfield  {journal} {\bibinfo  {journal} {Phys. Rev. B}\ }\textbf {\bibinfo {volume} {105}},\ \bibinfo {pages} {024308} (\bibinfo {year} {2022}{\natexlab{b}})}\BibitemShut {NoStop}%
\bibitem [{\citenamefont {Orenstein}\ \emph {et~al.}(2021)\citenamefont {Orenstein}, \citenamefont {Moore}, \citenamefont {Morimoto}, \citenamefont {Torchinsky}, \citenamefont {Harter},\ and\ \citenamefont {Hsieh}}]{Orenstein2021}%
  \BibitemOpen
  \bibfield  {author} {\bibinfo {author} {\bibfnamefont {J.}~\bibnamefont {Orenstein}}, \bibinfo {author} {\bibfnamefont {J.}~\bibnamefont {Moore}}, \bibinfo {author} {\bibfnamefont {T.}~\bibnamefont {Morimoto}}, \bibinfo {author} {\bibfnamefont {D.}~\bibnamefont {Torchinsky}}, \bibinfo {author} {\bibfnamefont {J.}~\bibnamefont {Harter}}, \ and\ \bibinfo {author} {\bibfnamefont {D.}~\bibnamefont {Hsieh}},\ }\href {\doibase https://doi.org/10.1146/annurev-conmatphys-031218-013712} {\bibfield  {journal} {\bibinfo  {journal} {Annual Review of Condensed Matter Physics}\ }\textbf {\bibinfo {volume} {12}},\ \bibinfo {pages} {247} (\bibinfo {year} {2021})}\BibitemShut {NoStop}%
\bibitem [{\citenamefont {Xu}\ \emph {et~al.}(2019)\citenamefont {Xu}, \citenamefont {Morimoto},\ and\ \citenamefont {Moore}}]{Xu2019}%
  \BibitemOpen
  \bibfield  {author} {\bibinfo {author} {\bibfnamefont {T.}~\bibnamefont {Xu}}, \bibinfo {author} {\bibfnamefont {T.}~\bibnamefont {Morimoto}}, \ and\ \bibinfo {author} {\bibfnamefont {J.~E.}\ \bibnamefont {Moore}},\ }\href {\doibase 10.1103/PhysRevB.100.220501} {\bibfield  {journal} {\bibinfo  {journal} {Phys. Rev. B}\ }\textbf {\bibinfo {volume} {100}},\ \bibinfo {pages} {220501} (\bibinfo {year} {2019})}\BibitemShut {NoStop}%
\bibitem [{\citenamefont {Tanaka}\ \emph {et~al.}(2023)\citenamefont {Tanaka}, \citenamefont {Watanabe},\ and\ \citenamefont {Yanase}}]{Tanaka2023}%
  \BibitemOpen
  \bibfield  {author} {\bibinfo {author} {\bibfnamefont {H.}~\bibnamefont {Tanaka}}, \bibinfo {author} {\bibfnamefont {H.}~\bibnamefont {Watanabe}}, \ and\ \bibinfo {author} {\bibfnamefont {Y.}~\bibnamefont {Yanase}},\ }\href {\doibase 10.1103/PhysRevB.107.024513} {\bibfield  {journal} {\bibinfo  {journal} {Phys. Rev. B}\ }\textbf {\bibinfo {volume} {107}},\ \bibinfo {pages} {024513} (\bibinfo {year} {2023})}\BibitemShut {NoStop}%
\bibitem [{\citenamefont {Huang}\ and\ \citenamefont {Wang}(2023)}]{Huang2023}%
  \BibitemOpen
  \bibfield  {author} {\bibinfo {author} {\bibfnamefont {L.}~\bibnamefont {Huang}}\ and\ \bibinfo {author} {\bibfnamefont {J.}~\bibnamefont {Wang}},\ }\href {\doibase 10.1103/PhysRevB.108.224516} {\bibfield  {journal} {\bibinfo  {journal} {Phys. Rev. B}\ }\textbf {\bibinfo {volume} {108}},\ \bibinfo {pages} {224516} (\bibinfo {year} {2023})}\BibitemShut {NoStop}%
\bibitem [{\citenamefont {Tanaka}\ \emph {et~al.}(2024)\citenamefont {Tanaka}, \citenamefont {Watanabe},\ and\ \citenamefont {Yanase}}]{Tanaka2024}%
  \BibitemOpen
  \bibfield  {author} {\bibinfo {author} {\bibfnamefont {H.}~\bibnamefont {Tanaka}}, \bibinfo {author} {\bibfnamefont {H.}~\bibnamefont {Watanabe}}, \ and\ \bibinfo {author} {\bibfnamefont {Y.}~\bibnamefont {Yanase}},\ }\href {\doibase 10.1103/PhysRevB.110.014520} {\bibfield  {journal} {\bibinfo  {journal} {Phys. Rev. B}\ }\textbf {\bibinfo {volume} {110}},\ \bibinfo {pages} {014520} (\bibinfo {year} {2024})}\BibitemShut {NoStop}%
\bibitem [{\citenamefont {Raj}\ \emph {et~al.}(2024)\citenamefont {Raj}, \citenamefont {Postlewaite}, \citenamefont {Chaudhary},\ and\ \citenamefont {Fiete}}]{Raj2024}%
  \BibitemOpen
  \bibfield  {author} {\bibinfo {author} {\bibfnamefont {A.}~\bibnamefont {Raj}}, \bibinfo {author} {\bibfnamefont {A.}~\bibnamefont {Postlewaite}}, \bibinfo {author} {\bibfnamefont {S.}~\bibnamefont {Chaudhary}}, \ and\ \bibinfo {author} {\bibfnamefont {G.~A.}\ \bibnamefont {Fiete}},\ }\href {\doibase 10.1103/PhysRevB.109.184514} {\bibfield  {journal} {\bibinfo  {journal} {Phys. Rev. B}\ }\textbf {\bibinfo {volume} {109}},\ \bibinfo {pages} {184514} (\bibinfo {year} {2024})}\BibitemShut {NoStop}%
\bibitem [{\citenamefont {Watanabe}\ and\ \citenamefont {Watanabe}(2024)}]{watanabeS2024}%
  \BibitemOpen
  \bibfield  {author} {\bibinfo {author} {\bibfnamefont {S.}~\bibnamefont {Watanabe}}\ and\ \bibinfo {author} {\bibfnamefont {H.}~\bibnamefont {Watanabe}},\ }\href {https://arxiv.org/abs/2410.18679} {\enquote {\bibinfo {title} {A gauge-invariant formulation of optical responses in superconductors},}\ } (\bibinfo {year} {2024}),\ \Eprint {http://arxiv.org/abs/2410.18679} {arXiv:2410.18679 [cond-mat.supr-con]} \BibitemShut {NoStop}%
\bibitem [{\citenamefont {Watanabe}\ and\ \citenamefont {Watanabe}(2025)}]{WatanabeS2025}%
  \BibitemOpen
  \bibfield  {author} {\bibinfo {author} {\bibfnamefont {S.}~\bibnamefont {Watanabe}}\ and\ \bibinfo {author} {\bibfnamefont {H.}~\bibnamefont {Watanabe}},\ }\href {https://arxiv.org/abs/2501.13722} {\enquote {\bibinfo {title} {Gauge-invariant electromagnetic responses in superconductors},}\ } (\bibinfo {year} {2025}),\ \Eprint {http://arxiv.org/abs/2501.13722} {arXiv:2501.13722 [cond-mat.supr-con]} \BibitemShut {NoStop}%
\bibitem [{\citenamefont {Schrieffer}(2018)}]{schrieffer2018}%
  \BibitemOpen
  \bibfield  {author} {\bibinfo {author} {\bibfnamefont {J.~R.}\ \bibnamefont {Schrieffer}},\ }\href@noop {} {\emph {\bibinfo {title} {Theory of superconductivity}}}\ (\bibinfo  {publisher} {CRC press},\ \bibinfo {year} {2018})\BibitemShut {NoStop}%
\bibitem [{\citenamefont {Nambu}(1960)}]{Nambu1960}%
  \BibitemOpen
  \bibfield  {author} {\bibinfo {author} {\bibfnamefont {Y.}~\bibnamefont {Nambu}},\ }\href {\doibase 10.1103/PhysRev.117.648} {\bibfield  {journal} {\bibinfo  {journal} {Phys. Rev.}\ }\textbf {\bibinfo {volume} {117}},\ \bibinfo {pages} {648} (\bibinfo {year} {1960})}\BibitemShut {NoStop}%
\bibitem [{\citenamefont {Dai}\ and\ \citenamefont {Lee}(2017)}]{Dai2017}%
  \BibitemOpen
  \bibfield  {author} {\bibinfo {author} {\bibfnamefont {Z.}~\bibnamefont {Dai}}\ and\ \bibinfo {author} {\bibfnamefont {P.~A.}\ \bibnamefont {Lee}},\ }\href {\doibase 10.1103/PhysRevB.95.014506} {\bibfield  {journal} {\bibinfo  {journal} {Phys. Rev. B}\ }\textbf {\bibinfo {volume} {95}},\ \bibinfo {pages} {014506} (\bibinfo {year} {2017})}\BibitemShut {NoStop}%
\bibitem [{\citenamefont {Oh}\ and\ \citenamefont {Watanabe}(2024)}]{Oh2024}%
  \BibitemOpen
  \bibfield  {author} {\bibinfo {author} {\bibfnamefont {C.-g.}\ \bibnamefont {Oh}}\ and\ \bibinfo {author} {\bibfnamefont {H.}~\bibnamefont {Watanabe}},\ }\href {\doibase 10.1103/PhysRevResearch.6.013058} {\bibfield  {journal} {\bibinfo  {journal} {Phys. Rev. Res.}\ }\textbf {\bibinfo {volume} {6}},\ \bibinfo {pages} {013058} (\bibinfo {year} {2024})}\BibitemShut {NoStop}%
\bibitem [{\citenamefont {He}\ \emph {et~al.}(2017)\citenamefont {He}, \citenamefont {Wang},\ and\ \citenamefont {Guo}}]{He2017}%
  \BibitemOpen
  \bibfield  {author} {\bibinfo {author} {\bibfnamefont {Y.}~\bibnamefont {He}}, \bibinfo {author} {\bibfnamefont {Y.-X.}\ \bibnamefont {Wang}}, \ and\ \bibinfo {author} {\bibfnamefont {H.}~\bibnamefont {Guo}},\ }\href {\doibase https://doi.org/10.1016/j.physleta.2017.03.010} {\bibfield  {journal} {\bibinfo  {journal} {Physics Letters A}\ }\textbf {\bibinfo {volume} {381}},\ \bibinfo {pages} {1603} (\bibinfo {year} {2017})}\BibitemShut {NoStop}%
\bibitem [{\citenamefont {Guo}\ \emph {et~al.}(2013)\citenamefont {Guo}, \citenamefont {Chien},\ and\ \citenamefont {He}}]{Guo2013}%
  \BibitemOpen
  \bibfield  {author} {\bibinfo {author} {\bibfnamefont {H.}~\bibnamefont {Guo}}, \bibinfo {author} {\bibfnamefont {C.-C.}\ \bibnamefont {Chien}}, \ and\ \bibinfo {author} {\bibfnamefont {Y.}~\bibnamefont {He}},\ }\href {\doibase 10.1007/s10909-012-0853-7} {\bibfield  {journal} {\bibinfo  {journal} {Journal of Low Temperature Physics}\ }\textbf {\bibinfo {volume} {172}},\ \bibinfo {pages} {5} (\bibinfo {year} {2013})}\BibitemShut {NoStop}%
\bibitem [{\citenamefont {Kadanoff}\ and\ \citenamefont {Martin}(1961)}]{Kadanoff1961}%
  \BibitemOpen
  \bibfield  {author} {\bibinfo {author} {\bibfnamefont {L.~P.}\ \bibnamefont {Kadanoff}}\ and\ \bibinfo {author} {\bibfnamefont {P.~C.}\ \bibnamefont {Martin}},\ }\href {\doibase 10.1103/PhysRev.124.670} {\bibfield  {journal} {\bibinfo  {journal} {Phys. Rev.}\ }\textbf {\bibinfo {volume} {124}},\ \bibinfo {pages} {670} (\bibinfo {year} {1961})}\BibitemShut {NoStop}%
\bibitem [{\citenamefont {Kulik}\ \emph {et~al.}(1981)\citenamefont {Kulik}, \citenamefont {Entin-Wohlman},\ and\ \citenamefont {Orbach}}]{Kulik1981}%
  \BibitemOpen
  \bibfield  {author} {\bibinfo {author} {\bibfnamefont {I.~O.}\ \bibnamefont {Kulik}}, \bibinfo {author} {\bibfnamefont {O.}~\bibnamefont {Entin-Wohlman}}, \ and\ \bibinfo {author} {\bibfnamefont {R.}~\bibnamefont {Orbach}},\ }\href {\doibase 10.1007/BF00115617} {\bibfield  {journal} {\bibinfo  {journal} {Journal of Low Temperature Physics}\ }\textbf {\bibinfo {volume} {43}},\ \bibinfo {pages} {591} (\bibinfo {year} {1981})}\BibitemShut {NoStop}%
\bibitem [{\citenamefont {Zha}\ \emph {et~al.}(1995)\citenamefont {Zha}, \citenamefont {Levin},\ and\ \citenamefont {Liu}}]{Zha1995}%
  \BibitemOpen
  \bibfield  {author} {\bibinfo {author} {\bibfnamefont {Y.}~\bibnamefont {Zha}}, \bibinfo {author} {\bibfnamefont {K.}~\bibnamefont {Levin}}, \ and\ \bibinfo {author} {\bibfnamefont {D.~Z.}\ \bibnamefont {Liu}},\ }\href {\doibase 10.1103/PhysRevB.51.6602} {\bibfield  {journal} {\bibinfo  {journal} {Phys. Rev. B}\ }\textbf {\bibinfo {volume} {51}},\ \bibinfo {pages} {6602} (\bibinfo {year} {1995})}\BibitemShut {NoStop}%
\bibitem [{\citenamefont {Kosztin}\ \emph {et~al.}(2000)\citenamefont {Kosztin}, \citenamefont {Chen}, \citenamefont {Kao},\ and\ \citenamefont {Levin}}]{Ioan2000}%
  \BibitemOpen
  \bibfield  {author} {\bibinfo {author} {\bibfnamefont {I.}~\bibnamefont {Kosztin}}, \bibinfo {author} {\bibfnamefont {Q.}~\bibnamefont {Chen}}, \bibinfo {author} {\bibfnamefont {Y.-J.}\ \bibnamefont {Kao}}, \ and\ \bibinfo {author} {\bibfnamefont {K.}~\bibnamefont {Levin}},\ }\href {\doibase 10.1103/PhysRevB.61.11662} {\bibfield  {journal} {\bibinfo  {journal} {Phys. Rev. B}\ }\textbf {\bibinfo {volume} {61}},\ \bibinfo {pages} {11662} (\bibinfo {year} {2000})}\BibitemShut {NoStop}%
\bibitem [{\citenamefont {Anderson}\ \emph {et~al.}(2016)\citenamefont {Anderson}, \citenamefont {Boyack}, \citenamefont {Wu},\ and\ \citenamefont {Levin}}]{Anderson2016}%
  \BibitemOpen
  \bibfield  {author} {\bibinfo {author} {\bibfnamefont {B.~M.}\ \bibnamefont {Anderson}}, \bibinfo {author} {\bibfnamefont {R.}~\bibnamefont {Boyack}}, \bibinfo {author} {\bibfnamefont {C.-T.}\ \bibnamefont {Wu}}, \ and\ \bibinfo {author} {\bibfnamefont {K.}~\bibnamefont {Levin}},\ }\href {\doibase 10.1103/PhysRevB.93.180504} {\bibfield  {journal} {\bibinfo  {journal} {Phys. Rev. B}\ }\textbf {\bibinfo {volume} {93}},\ \bibinfo {pages} {180504} (\bibinfo {year} {2016})}\BibitemShut {NoStop}%
\bibitem [{\citenamefont {Boyack}\ \emph {et~al.}(2016)\citenamefont {Boyack}, \citenamefont {Anderson}, \citenamefont {Wu},\ and\ \citenamefont {Levin}}]{Boyack2016}%
  \BibitemOpen
  \bibfield  {author} {\bibinfo {author} {\bibfnamefont {R.}~\bibnamefont {Boyack}}, \bibinfo {author} {\bibfnamefont {B.~M.}\ \bibnamefont {Anderson}}, \bibinfo {author} {\bibfnamefont {C.-T.}\ \bibnamefont {Wu}}, \ and\ \bibinfo {author} {\bibfnamefont {K.}~\bibnamefont {Levin}},\ }\href {\doibase 10.1103/PhysRevB.94.094508} {\bibfield  {journal} {\bibinfo  {journal} {Phys. Rev. B}\ }\textbf {\bibinfo {volume} {94}},\ \bibinfo {pages} {094508} (\bibinfo {year} {2016})}\BibitemShut {NoStop}%
\bibitem [{\citenamefont {Lutchyn}\ \emph {et~al.}(2008)\citenamefont {Lutchyn}, \citenamefont {Nagornykh},\ and\ \citenamefont {Yakovenko}}]{Lutchyn2008}%
  \BibitemOpen
  \bibfield  {author} {\bibinfo {author} {\bibfnamefont {R.~M.}\ \bibnamefont {Lutchyn}}, \bibinfo {author} {\bibfnamefont {P.}~\bibnamefont {Nagornykh}}, \ and\ \bibinfo {author} {\bibfnamefont {V.~M.}\ \bibnamefont {Yakovenko}},\ }\href {\doibase 10.1103/PhysRevB.77.144516} {\bibfield  {journal} {\bibinfo  {journal} {Phys. Rev. B}\ }\textbf {\bibinfo {volume} {77}},\ \bibinfo {pages} {144516} (\bibinfo {year} {2008})}\BibitemShut {NoStop}%
\bibitem [{\citenamefont {Baym}\ and\ \citenamefont {Kadanoff}(1961)}]{Baym1961}%
  \BibitemOpen
  \bibfield  {author} {\bibinfo {author} {\bibfnamefont {G.}~\bibnamefont {Baym}}\ and\ \bibinfo {author} {\bibfnamefont {L.~P.}\ \bibnamefont {Kadanoff}},\ }\href {\doibase 10.1103/PhysRev.124.287} {\bibfield  {journal} {\bibinfo  {journal} {Phys. Rev.}\ }\textbf {\bibinfo {volume} {124}},\ \bibinfo {pages} {287} (\bibinfo {year} {1961})}\BibitemShut {NoStop}%
\bibitem [{\citenamefont {Kadanoff}\ and\ \citenamefont {Baym}(1962)}]{kadanoff1962}%
  \BibitemOpen
  \bibfield  {author} {\bibinfo {author} {\bibfnamefont {L.}~\bibnamefont {Kadanoff}}\ and\ \bibinfo {author} {\bibfnamefont {G.}~\bibnamefont {Baym}},\ }\href {https://books.google.co.jp/books?id=1-FEAAAAIAAJ} {\emph {\bibinfo {title} {Quantum Statistical Mechanics: Green's Function Methods in Equilibrium and Nonequilibrium Problems}}},\ Frontiers in Physics. A Lecture Note and Reprint Series\ (\bibinfo  {publisher} {W.A. Benjamin},\ \bibinfo {year} {1962})\BibitemShut {NoStop}%
\bibitem [{\citenamefont {Sato}\ \emph {et~al.}(2009)\citenamefont {Sato}, \citenamefont {Takahashi},\ and\ \citenamefont {Fujimoto}}]{Sato2009}%
  \BibitemOpen
  \bibfield  {author} {\bibinfo {author} {\bibfnamefont {M.}~\bibnamefont {Sato}}, \bibinfo {author} {\bibfnamefont {Y.}~\bibnamefont {Takahashi}}, \ and\ \bibinfo {author} {\bibfnamefont {S.}~\bibnamefont {Fujimoto}},\ }\href {\doibase 10.1103/PhysRevLett.103.020401} {\bibfield  {journal} {\bibinfo  {journal} {Phys. Rev. Lett.}\ }\textbf {\bibinfo {volume} {103}},\ \bibinfo {pages} {020401} (\bibinfo {year} {2009})}\BibitemShut {NoStop}%
\bibitem [{\citenamefont {Sato}\ \emph {et~al.}(2010)\citenamefont {Sato}, \citenamefont {Takahashi},\ and\ \citenamefont {Fujimoto}}]{Sato2010}%
  \BibitemOpen
  \bibfield  {author} {\bibinfo {author} {\bibfnamefont {M.}~\bibnamefont {Sato}}, \bibinfo {author} {\bibfnamefont {Y.}~\bibnamefont {Takahashi}}, \ and\ \bibinfo {author} {\bibfnamefont {S.}~\bibnamefont {Fujimoto}},\ }\href {\doibase 10.1103/PhysRevB.82.134521} {\bibfield  {journal} {\bibinfo  {journal} {Phys. Rev. B}\ }\textbf {\bibinfo {volume} {82}},\ \bibinfo {pages} {134521} (\bibinfo {year} {2010})}\BibitemShut {NoStop}%
\bibitem [{\citenamefont {Sau}\ \emph {et~al.}(2010)\citenamefont {Sau}, \citenamefont {Lutchyn}, \citenamefont {Tewari},\ and\ \citenamefont {Das~Sarma}}]{Sau2010}%
  \BibitemOpen
  \bibfield  {author} {\bibinfo {author} {\bibfnamefont {J.~D.}\ \bibnamefont {Sau}}, \bibinfo {author} {\bibfnamefont {R.~M.}\ \bibnamefont {Lutchyn}}, \bibinfo {author} {\bibfnamefont {S.}~\bibnamefont {Tewari}}, \ and\ \bibinfo {author} {\bibfnamefont {S.}~\bibnamefont {Das~Sarma}},\ }\href {\doibase 10.1103/PhysRevLett.104.040502} {\bibfield  {journal} {\bibinfo  {journal} {Phys. Rev. Lett.}\ }\textbf {\bibinfo {volume} {104}},\ \bibinfo {pages} {040502} (\bibinfo {year} {2010})}\BibitemShut {NoStop}%
\bibitem [{\citenamefont {Sigrist}\ and\ \citenamefont {Ueda}(1991)}]{Sigrist1991}%
  \BibitemOpen
  \bibfield  {author} {\bibinfo {author} {\bibfnamefont {M.}~\bibnamefont {Sigrist}}\ and\ \bibinfo {author} {\bibfnamefont {K.}~\bibnamefont {Ueda}},\ }\href {\doibase 10.1103/RevModPhys.63.239} {\bibfield  {journal} {\bibinfo  {journal} {Rev. Mod. Phys.}\ }\textbf {\bibinfo {volume} {63}},\ \bibinfo {pages} {239} (\bibinfo {year} {1991})}\BibitemShut {NoStop}%
\bibitem [{\citenamefont {Jujo}(2005)}]{Jujo2005}%
  \BibitemOpen
  \bibfield  {author} {\bibinfo {author} {\bibfnamefont {T.}~\bibnamefont {Jujo}},\ }\href {\doibase 10.1143/JPSJ.74.1111} {\bibfield  {journal} {\bibinfo  {journal} {Journal of the Physical Society of Japan}\ }\textbf {\bibinfo {volume} {74}},\ \bibinfo {pages} {1111} (\bibinfo {year} {2005})},\ \Eprint {http://arxiv.org/abs/https://doi.org/10.1143/JPSJ.74.1111} {https://doi.org/10.1143/JPSJ.74.1111} \BibitemShut {NoStop}%
\bibitem [{\citenamefont {Rostami}\ and\ \citenamefont {Cappelluti}(2021)}]{Rostami2021npj}%
  \BibitemOpen
  \bibfield  {author} {\bibinfo {author} {\bibfnamefont {H.}~\bibnamefont {Rostami}}\ and\ \bibinfo {author} {\bibfnamefont {E.}~\bibnamefont {Cappelluti}},\ }\href {\doibase 10.1038/s41699-021-00217-0} {\bibfield  {journal} {\bibinfo  {journal} {npj 2D Materials and Applications}\ }\textbf {\bibinfo {volume} {5}},\ \bibinfo {pages} {50} (\bibinfo {year} {2021})}\BibitemShut {NoStop}%
\bibitem [{\citenamefont {Baym}(1962)}]{Baym1962}%
  \BibitemOpen
  \bibfield  {author} {\bibinfo {author} {\bibfnamefont {G.}~\bibnamefont {Baym}},\ }\href {\doibase 10.1103/PhysRev.127.1391} {\bibfield  {journal} {\bibinfo  {journal} {Phys. Rev.}\ }\textbf {\bibinfo {volume} {127}},\ \bibinfo {pages} {1391} (\bibinfo {year} {1962})}\BibitemShut {NoStop}%
\bibitem [{\citenamefont {Luttinger}\ and\ \citenamefont {Ward}(1960)}]{Luttinger1960}%
  \BibitemOpen
  \bibfield  {author} {\bibinfo {author} {\bibfnamefont {J.~M.}\ \bibnamefont {Luttinger}}\ and\ \bibinfo {author} {\bibfnamefont {J.~C.}\ \bibnamefont {Ward}},\ }\href {\doibase 10.1103/PhysRev.118.1417} {\bibfield  {journal} {\bibinfo  {journal} {Phys. Rev.}\ }\textbf {\bibinfo {volume} {118}},\ \bibinfo {pages} {1417} (\bibinfo {year} {1960})}\BibitemShut {NoStop}%
\bibitem [{\citenamefont {Michishita}\ and\ \citenamefont {Peters}(2021)}]{Michishita2021}%
  \BibitemOpen
  \bibfield  {author} {\bibinfo {author} {\bibfnamefont {Y.}~\bibnamefont {Michishita}}\ and\ \bibinfo {author} {\bibfnamefont {R.}~\bibnamefont {Peters}},\ }\href {\doibase 10.1103/PhysRevB.103.195133} {\bibfield  {journal} {\bibinfo  {journal} {Phys. Rev. B}\ }\textbf {\bibinfo {volume} {103}},\ \bibinfo {pages} {195133} (\bibinfo {year} {2021})}\BibitemShut {NoStop}%
\bibitem [{\citenamefont {Rostami}\ \emph {et~al.}(2021)\citenamefont {Rostami}, \citenamefont {Katsnelson}, \citenamefont {Vignale},\ and\ \citenamefont {Polini}}]{Rostami2021}%
  \BibitemOpen
  \bibfield  {author} {\bibinfo {author} {\bibfnamefont {H.}~\bibnamefont {Rostami}}, \bibinfo {author} {\bibfnamefont {M.~I.}\ \bibnamefont {Katsnelson}}, \bibinfo {author} {\bibfnamefont {G.}~\bibnamefont {Vignale}}, \ and\ \bibinfo {author} {\bibfnamefont {M.}~\bibnamefont {Polini}},\ }\href {\doibase https://doi.org/10.1016/j.aop.2021.168523} {\bibfield  {journal} {\bibinfo  {journal} {Annals of Physics}\ }\textbf {\bibinfo {volume} {431}},\ \bibinfo {pages} {168523} (\bibinfo {year} {2021})}\BibitemShut {NoStop}%
\bibitem [{\citenamefont {Passos}\ \emph {et~al.}(2018)\citenamefont {Passos}, \citenamefont {Ventura}, \citenamefont {Lopes}, \citenamefont {Santos},\ and\ \citenamefont {Peres}}]{Passos2018}%
  \BibitemOpen
  \bibfield  {author} {\bibinfo {author} {\bibfnamefont {D.~J.}\ \bibnamefont {Passos}}, \bibinfo {author} {\bibfnamefont {G.~B.}\ \bibnamefont {Ventura}}, \bibinfo {author} {\bibfnamefont {J.~M. V.~P.}\ \bibnamefont {Lopes}}, \bibinfo {author} {\bibfnamefont {J.~M. B. L.~d.}\ \bibnamefont {Santos}}, \ and\ \bibinfo {author} {\bibfnamefont {N.~M.~R.}\ \bibnamefont {Peres}},\ }\href {\doibase 10.1103/PhysRevB.97.235446} {\bibfield  {journal} {\bibinfo  {journal} {Phys. Rev. B}\ }\textbf {\bibinfo {volume} {97}},\ \bibinfo {pages} {235446} (\bibinfo {year} {2018})}\BibitemShut {NoStop}%
\bibitem [{\citenamefont {Kallin}\ and\ \citenamefont {Berlinsky}(2016)}]{Kallin2016}%
  \BibitemOpen
  \bibfield  {author} {\bibinfo {author} {\bibfnamefont {C.}~\bibnamefont {Kallin}}\ and\ \bibinfo {author} {\bibfnamefont {J.}~\bibnamefont {Berlinsky}},\ }\href {\doibase 10.1088/0034-4885/79/5/054502} {\bibfield  {journal} {\bibinfo  {journal} {Reports on Progress in Physics}\ }\textbf {\bibinfo {volume} {79}},\ \bibinfo {pages} {054502} (\bibinfo {year} {2016})}\BibitemShut {NoStop}%
\bibitem [{\citenamefont {Fu}(2014)}]{Fu2014}%
  \BibitemOpen
  \bibfield  {author} {\bibinfo {author} {\bibfnamefont {L.}~\bibnamefont {Fu}},\ }\href {\doibase 10.1103/PhysRevB.90.100509} {\bibfield  {journal} {\bibinfo  {journal} {Phys. Rev. B}\ }\textbf {\bibinfo {volume} {90}},\ \bibinfo {pages} {100509} (\bibinfo {year} {2014})}\BibitemShut {NoStop}%
\bibitem [{\citenamefont {Kanasugi}\ and\ \citenamefont {Yanase}(2022)}]{Kanasugi2022}%
  \BibitemOpen
  \bibfield  {author} {\bibinfo {author} {\bibfnamefont {S.}~\bibnamefont {Kanasugi}}\ and\ \bibinfo {author} {\bibfnamefont {Y.}~\bibnamefont {Yanase}},\ }\href {\doibase 10.1038/s42005-022-00804-7} {\bibfield  {journal} {\bibinfo  {journal} {Communications Physics}\ }\textbf {\bibinfo {volume} {5}},\ \bibinfo {pages} {39} (\bibinfo {year} {2022})}\BibitemShut {NoStop}%
\bibitem [{\citenamefont {Kitamura}\ \emph {et~al.}(2023)\citenamefont {Kitamura}, \citenamefont {Kanasugi}, \citenamefont {Chazono},\ and\ \citenamefont {Yanase}}]{Kitamura2023}%
  \BibitemOpen
  \bibfield  {author} {\bibinfo {author} {\bibfnamefont {T.}~\bibnamefont {Kitamura}}, \bibinfo {author} {\bibfnamefont {S.}~\bibnamefont {Kanasugi}}, \bibinfo {author} {\bibfnamefont {M.}~\bibnamefont {Chazono}}, \ and\ \bibinfo {author} {\bibfnamefont {Y.}~\bibnamefont {Yanase}},\ }\href {\doibase 10.1103/PhysRevB.107.214513} {\bibfield  {journal} {\bibinfo  {journal} {Phys. Rev. B}\ }\textbf {\bibinfo {volume} {107}},\ \bibinfo {pages} {214513} (\bibinfo {year} {2023})}\BibitemShut {NoStop}%
\bibitem [{\citenamefont {Bauer}\ and\ \citenamefont {Sigrist}(2012)}]{Bauer2012}%
  \BibitemOpen
  \bibfield  {author} {\bibinfo {author} {\bibfnamefont {E.}~\bibnamefont {Bauer}}\ and\ \bibinfo {author} {\bibfnamefont {M.}~\bibnamefont {Sigrist}},\ }\href@noop {} {\emph {\bibinfo {title} {Non-centrosymmetric superconductors: introduction and overview}}}\ (\bibinfo  {publisher} {Springer Science \& Business Media},\ \bibinfo {year} {2012})\BibitemShut {NoStop}%
\bibitem [{\citenamefont {Smidman}\ \emph {et~al.}(2017)\citenamefont {Smidman}, \citenamefont {Salamon}, \citenamefont {Yuan},\ and\ \citenamefont {Agterberg}}]{Smidman2017}%
  \BibitemOpen
  \bibfield  {author} {\bibinfo {author} {\bibfnamefont {M.}~\bibnamefont {Smidman}}, \bibinfo {author} {\bibfnamefont {M.~B.}\ \bibnamefont {Salamon}}, \bibinfo {author} {\bibfnamefont {H.~Q.}\ \bibnamefont {Yuan}}, \ and\ \bibinfo {author} {\bibfnamefont {D.~F.}\ \bibnamefont {Agterberg}},\ }\href {\doibase 10.1088/1361-6633/80/3/036501} {\bibfield  {journal} {\bibinfo  {journal} {Reports on Progress in Physics}\ }\textbf {\bibinfo {volume} {80}},\ \bibinfo {pages} {036501} (\bibinfo {year} {2017})}\BibitemShut {NoStop}%
\bibitem [{\citenamefont {Ramires}\ and\ \citenamefont {Sigrist}(2016)}]{Ramires2016}%
  \BibitemOpen
  \bibfield  {author} {\bibinfo {author} {\bibfnamefont {A.}~\bibnamefont {Ramires}}\ and\ \bibinfo {author} {\bibfnamefont {M.}~\bibnamefont {Sigrist}},\ }\href {\doibase 10.1103/PhysRevB.94.104501} {\bibfield  {journal} {\bibinfo  {journal} {Phys. Rev. B}\ }\textbf {\bibinfo {volume} {94}},\ \bibinfo {pages} {104501} (\bibinfo {year} {2016})}\BibitemShut {NoStop}%
\bibitem [{\citenamefont {Ramires}\ \emph {et~al.}(2018)\citenamefont {Ramires}, \citenamefont {Agterberg},\ and\ \citenamefont {Sigrist}}]{Ramires2018}%
  \BibitemOpen
  \bibfield  {author} {\bibinfo {author} {\bibfnamefont {A.}~\bibnamefont {Ramires}}, \bibinfo {author} {\bibfnamefont {D.~F.}\ \bibnamefont {Agterberg}}, \ and\ \bibinfo {author} {\bibfnamefont {M.}~\bibnamefont {Sigrist}},\ }\href {\doibase 10.1103/PhysRevB.98.024501} {\bibfield  {journal} {\bibinfo  {journal} {Phys. Rev. B}\ }\textbf {\bibinfo {volume} {98}},\ \bibinfo {pages} {024501} (\bibinfo {year} {2018})}\BibitemShut {NoStop}%
\bibitem [{\citenamefont {Zhang}\ \emph {et~al.}(2019)\citenamefont {Zhang}, \citenamefont {Holder}, \citenamefont {Ishizuka}, \citenamefont {de~Juan}, \citenamefont {Nagaosa}, \citenamefont {Felser},\ and\ \citenamefont {Yan}}]{Zhang2019}%
  \BibitemOpen
  \bibfield  {author} {\bibinfo {author} {\bibfnamefont {Y.}~\bibnamefont {Zhang}}, \bibinfo {author} {\bibfnamefont {T.}~\bibnamefont {Holder}}, \bibinfo {author} {\bibfnamefont {H.}~\bibnamefont {Ishizuka}}, \bibinfo {author} {\bibfnamefont {F.}~\bibnamefont {de~Juan}}, \bibinfo {author} {\bibfnamefont {N.}~\bibnamefont {Nagaosa}}, \bibinfo {author} {\bibfnamefont {C.}~\bibnamefont {Felser}}, \ and\ \bibinfo {author} {\bibfnamefont {B.}~\bibnamefont {Yan}},\ }\href {\doibase 10.1038/s41467-019-11832-3} {\bibfield  {journal} {\bibinfo  {journal} {Nature Communications}\ }\textbf {\bibinfo {volume} {10}},\ \bibinfo {pages} {3783} (\bibinfo {year} {2019})}\BibitemShut {NoStop}%
\bibitem [{\citenamefont {Wang}\ and\ \citenamefont {Qian}(2020)}]{Wang2020}%
  \BibitemOpen
  \bibfield  {author} {\bibinfo {author} {\bibfnamefont {H.}~\bibnamefont {Wang}}\ and\ \bibinfo {author} {\bibfnamefont {X.}~\bibnamefont {Qian}},\ }\href {\doibase 10.1038/s41524-020-00462-9} {\bibfield  {journal} {\bibinfo  {journal} {npj Computational Materials}\ }\textbf {\bibinfo {volume} {6}},\ \bibinfo {pages} {199} (\bibinfo {year} {2020})}\BibitemShut {NoStop}%
\bibitem [{\citenamefont {Fei}\ \emph {et~al.}(2020)\citenamefont {Fei}, \citenamefont {Song},\ and\ \citenamefont {Yang}}]{Fei2020}%
  \BibitemOpen
  \bibfield  {author} {\bibinfo {author} {\bibfnamefont {R.}~\bibnamefont {Fei}}, \bibinfo {author} {\bibfnamefont {W.}~\bibnamefont {Song}}, \ and\ \bibinfo {author} {\bibfnamefont {L.}~\bibnamefont {Yang}},\ }\href {\doibase 10.1103/PhysRevB.102.035440} {\bibfield  {journal} {\bibinfo  {journal} {Phys. Rev. B}\ }\textbf {\bibinfo {volume} {102}},\ \bibinfo {pages} {035440} (\bibinfo {year} {2020})}\BibitemShut {NoStop}%
\bibitem [{\citenamefont {Mu}\ \emph {et~al.}(2023)\citenamefont {Mu}, \citenamefont {Xue}, \citenamefont {Sun},\ and\ \citenamefont {Zhou}}]{Mu2023}%
  \BibitemOpen
  \bibfield  {author} {\bibinfo {author} {\bibfnamefont {X.}~\bibnamefont {Mu}}, \bibinfo {author} {\bibfnamefont {Q.}~\bibnamefont {Xue}}, \bibinfo {author} {\bibfnamefont {Y.}~\bibnamefont {Sun}}, \ and\ \bibinfo {author} {\bibfnamefont {J.}~\bibnamefont {Zhou}},\ }\href {\doibase 10.1103/PhysRevResearch.5.013001} {\bibfield  {journal} {\bibinfo  {journal} {Phys. Rev. Res.}\ }\textbf {\bibinfo {volume} {5}},\ \bibinfo {pages} {013001} (\bibinfo {year} {2023})}\BibitemShut {NoStop}%
\bibitem [{\citenamefont {Sipe}\ and\ \citenamefont {Shkrebtii}(2000)}]{Sipe2000}%
  \BibitemOpen
  \bibfield  {author} {\bibinfo {author} {\bibfnamefont {J.~E.}\ \bibnamefont {Sipe}}\ and\ \bibinfo {author} {\bibfnamefont {A.~I.}\ \bibnamefont {Shkrebtii}},\ }\href {\doibase 10.1103/PhysRevB.61.5337} {\bibfield  {journal} {\bibinfo  {journal} {Phys. Rev. B}\ }\textbf {\bibinfo {volume} {61}},\ \bibinfo {pages} {5337} (\bibinfo {year} {2000})}\BibitemShut {NoStop}%
\bibitem [{\citenamefont {Anderson}(1958{\natexlab{b}})}]{Anderson1958_1}%
  \BibitemOpen
  \bibfield  {author} {\bibinfo {author} {\bibfnamefont {P.~W.}\ \bibnamefont {Anderson}},\ }\href {\doibase 10.1103/PhysRev.110.827} {\bibfield  {journal} {\bibinfo  {journal} {Phys. Rev.}\ }\textbf {\bibinfo {volume} {110}},\ \bibinfo {pages} {827} (\bibinfo {year} {1958}{\natexlab{b}})}\BibitemShut {NoStop}%
\bibitem [{\citenamefont {Anderson}(1963)}]{Anderson1963}%
  \BibitemOpen
  \bibfield  {author} {\bibinfo {author} {\bibfnamefont {P.~W.}\ \bibnamefont {Anderson}},\ }\href {\doibase 10.1103/PhysRev.130.439} {\bibfield  {journal} {\bibinfo  {journal} {Phys. Rev.}\ }\textbf {\bibinfo {volume} {130}},\ \bibinfo {pages} {439} (\bibinfo {year} {1963})}\BibitemShut {NoStop}%
\bibitem [{\citenamefont {Lee}\ \emph {et~al.}(2012)\citenamefont {Lee}, \citenamefont {Alicea},\ and\ \citenamefont {Refael}}]{Lee2012}%
  \BibitemOpen
  \bibfield  {author} {\bibinfo {author} {\bibfnamefont {S.-P.}\ \bibnamefont {Lee}}, \bibinfo {author} {\bibfnamefont {J.}~\bibnamefont {Alicea}}, \ and\ \bibinfo {author} {\bibfnamefont {G.}~\bibnamefont {Refael}},\ }\href {\doibase 10.1103/PhysRevLett.109.126403} {\bibfield  {journal} {\bibinfo  {journal} {Phys. Rev. Lett.}\ }\textbf {\bibinfo {volume} {109}},\ \bibinfo {pages} {126403} (\bibinfo {year} {2012})}\BibitemShut {NoStop}%
\bibitem [{\citenamefont {Nagai}\ \emph {et~al.}(2016)\citenamefont {Nagai}, \citenamefont {Hoshino},\ and\ \citenamefont {Ota}}]{Nagai2016}%
  \BibitemOpen
  \bibfield  {author} {\bibinfo {author} {\bibfnamefont {Y.}~\bibnamefont {Nagai}}, \bibinfo {author} {\bibfnamefont {S.}~\bibnamefont {Hoshino}}, \ and\ \bibinfo {author} {\bibfnamefont {Y.}~\bibnamefont {Ota}},\ }\href {\doibase 10.1103/PhysRevB.93.220505} {\bibfield  {journal} {\bibinfo  {journal} {Phys. Rev. B}\ }\textbf {\bibinfo {volume} {93}},\ \bibinfo {pages} {220505} (\bibinfo {year} {2016})}\BibitemShut {NoStop}%
\bibitem [{\citenamefont {Kezilebieke}\ \emph {et~al.}(2020)\citenamefont {Kezilebieke}, \citenamefont {Huda}, \citenamefont {Va{\v{n}}o}, \citenamefont {Aapro}, \citenamefont {Ganguli}, \citenamefont {Silveira}, \citenamefont {G{\l}odzik}, \citenamefont {Foster}, \citenamefont {Ojanen},\ and\ \citenamefont {Liljeroth}}]{Kezilebieke2020}%
  \BibitemOpen
  \bibfield  {author} {\bibinfo {author} {\bibfnamefont {S.}~\bibnamefont {Kezilebieke}}, \bibinfo {author} {\bibfnamefont {M.~N.}\ \bibnamefont {Huda}}, \bibinfo {author} {\bibfnamefont {V.}~\bibnamefont {Va{\v{n}}o}}, \bibinfo {author} {\bibfnamefont {M.}~\bibnamefont {Aapro}}, \bibinfo {author} {\bibfnamefont {S.~C.}\ \bibnamefont {Ganguli}}, \bibinfo {author} {\bibfnamefont {O.~J.}\ \bibnamefont {Silveira}}, \bibinfo {author} {\bibfnamefont {S.}~\bibnamefont {G{\l}odzik}}, \bibinfo {author} {\bibfnamefont {A.~S.}\ \bibnamefont {Foster}}, \bibinfo {author} {\bibfnamefont {T.}~\bibnamefont {Ojanen}}, \ and\ \bibinfo {author} {\bibfnamefont {P.}~\bibnamefont {Liljeroth}},\ }\href {\doibase 10.1038/s41586-020-2989-y} {\bibfield  {journal} {\bibinfo  {journal} {Nature}\ }\textbf {\bibinfo {volume} {588}},\ \bibinfo {pages} {424} (\bibinfo {year} {2020})}\BibitemShut {NoStop}%
\bibitem [{\citenamefont {Kezilebieke}\ \emph {et~al.}(2022)\citenamefont {Kezilebieke}, \citenamefont {Va{\v{n}}o}, \citenamefont {Huda}, \citenamefont {Aapro}, \citenamefont {Ganguli}, \citenamefont {Liljeroth},\ and\ \citenamefont {Lado}}]{Kezilebieke2022}%
  \BibitemOpen
  \bibfield  {author} {\bibinfo {author} {\bibfnamefont {S.}~\bibnamefont {Kezilebieke}}, \bibinfo {author} {\bibfnamefont {V.}~\bibnamefont {Va{\v{n}}o}}, \bibinfo {author} {\bibfnamefont {M.~N.}\ \bibnamefont {Huda}}, \bibinfo {author} {\bibfnamefont {M.}~\bibnamefont {Aapro}}, \bibinfo {author} {\bibfnamefont {S.~C.}\ \bibnamefont {Ganguli}}, \bibinfo {author} {\bibfnamefont {P.}~\bibnamefont {Liljeroth}}, \ and\ \bibinfo {author} {\bibfnamefont {J.~L.}\ \bibnamefont {Lado}},\ }\href {\doibase 10.1021/acs.nanolett.1c03856} {\bibfield  {journal} {\bibinfo  {journal} {Nano Letters}\ }\textbf {\bibinfo {volume} {22}},\ \bibinfo {pages} {328} (\bibinfo {year} {2022})}\BibitemShut {NoStop}%
\bibitem [{\citenamefont {Ma}\ \emph {et~al.}(2017)\citenamefont {Ma}, \citenamefont {Xu}, \citenamefont {Chan}, \citenamefont {Zhang}, \citenamefont {Chang}, \citenamefont {Lin}, \citenamefont {Xie}, \citenamefont {Palacios}, \citenamefont {Lin}, \citenamefont {Jia}, \citenamefont {Lee}, \citenamefont {Jarillo-Herrero},\ and\ \citenamefont {Gedik}}]{Ma2017}%
  \BibitemOpen
  \bibfield  {author} {\bibinfo {author} {\bibfnamefont {Q.}~\bibnamefont {Ma}}, \bibinfo {author} {\bibfnamefont {S.-Y.}\ \bibnamefont {Xu}}, \bibinfo {author} {\bibfnamefont {C.-K.}\ \bibnamefont {Chan}}, \bibinfo {author} {\bibfnamefont {C.-L.}\ \bibnamefont {Zhang}}, \bibinfo {author} {\bibfnamefont {G.}~\bibnamefont {Chang}}, \bibinfo {author} {\bibfnamefont {Y.}~\bibnamefont {Lin}}, \bibinfo {author} {\bibfnamefont {W.}~\bibnamefont {Xie}}, \bibinfo {author} {\bibfnamefont {T.}~\bibnamefont {Palacios}}, \bibinfo {author} {\bibfnamefont {H.}~\bibnamefont {Lin}}, \bibinfo {author} {\bibfnamefont {S.}~\bibnamefont {Jia}}, \bibinfo {author} {\bibfnamefont {P.~A.}\ \bibnamefont {Lee}}, \bibinfo {author} {\bibfnamefont {P.}~\bibnamefont {Jarillo-Herrero}}, \ and\ \bibinfo {author} {\bibfnamefont {N.}~\bibnamefont {Gedik}},\ }\href {\doibase 10.1038/nphys4146} {\bibfield  {journal} {\bibinfo  {journal} {Nature Physics}\ }\textbf {\bibinfo {volume} {13}},\ \bibinfo {pages} {842} (\bibinfo {year} {2017})}\BibitemShut {NoStop}%
\bibitem [{\citenamefont {Ma}\ \emph {et~al.}(2019{\natexlab{a}})\citenamefont {Ma}, \citenamefont {Gu}, \citenamefont {Liu}, \citenamefont {Lai}, \citenamefont {Yu}, \citenamefont {Zhuo}, \citenamefont {Liu}, \citenamefont {Chen}, \citenamefont {Feng},\ and\ \citenamefont {Sun}}]{Ma2019}%
  \BibitemOpen
  \bibfield  {author} {\bibinfo {author} {\bibfnamefont {J.}~\bibnamefont {Ma}}, \bibinfo {author} {\bibfnamefont {Q.}~\bibnamefont {Gu}}, \bibinfo {author} {\bibfnamefont {Y.}~\bibnamefont {Liu}}, \bibinfo {author} {\bibfnamefont {J.}~\bibnamefont {Lai}}, \bibinfo {author} {\bibfnamefont {P.}~\bibnamefont {Yu}}, \bibinfo {author} {\bibfnamefont {X.}~\bibnamefont {Zhuo}}, \bibinfo {author} {\bibfnamefont {Z.}~\bibnamefont {Liu}}, \bibinfo {author} {\bibfnamefont {J.-H.}\ \bibnamefont {Chen}}, \bibinfo {author} {\bibfnamefont {J.}~\bibnamefont {Feng}}, \ and\ \bibinfo {author} {\bibfnamefont {D.}~\bibnamefont {Sun}},\ }\href {\doibase 10.1038/s41563-019-0296-5} {\bibfield  {journal} {\bibinfo  {journal} {Nature Materials}\ }\textbf {\bibinfo {volume} {18}},\ \bibinfo {pages} {476} (\bibinfo {year} {2019}{\natexlab{a}})}\BibitemShut {NoStop}%
\bibitem [{\citenamefont {Rees}\ \emph {et~al.}(2020)\citenamefont {Rees}, \citenamefont {Manna}, \citenamefont {Lu}, \citenamefont {Morimoto}, \citenamefont {Borrmann}, \citenamefont {Felser}, \citenamefont {Moore}, \citenamefont {Torchinsky},\ and\ \citenamefont {Orenstein}}]{Ree2020}%
  \BibitemOpen
  \bibfield  {author} {\bibinfo {author} {\bibfnamefont {D.}~\bibnamefont {Rees}}, \bibinfo {author} {\bibfnamefont {K.}~\bibnamefont {Manna}}, \bibinfo {author} {\bibfnamefont {B.}~\bibnamefont {Lu}}, \bibinfo {author} {\bibfnamefont {T.}~\bibnamefont {Morimoto}}, \bibinfo {author} {\bibfnamefont {H.}~\bibnamefont {Borrmann}}, \bibinfo {author} {\bibfnamefont {C.}~\bibnamefont {Felser}}, \bibinfo {author} {\bibfnamefont {J.~E.}\ \bibnamefont {Moore}}, \bibinfo {author} {\bibfnamefont {D.~H.}\ \bibnamefont {Torchinsky}}, \ and\ \bibinfo {author} {\bibfnamefont {J.}~\bibnamefont {Orenstein}},\ }\href {\doibase 10.1126/sciadv.aba0509} {\bibfield  {journal} {\bibinfo  {journal} {Science Advances}\ }\textbf {\bibinfo {volume} {6}},\ \bibinfo {pages} {eaba0509} (\bibinfo {year} {2020})},\ \Eprint {http://arxiv.org/abs/https://www.science.org/doi/pdf/10.1126/sciadv.aba0509} {https://www.science.org/doi/pdf/10.1126/sciadv.aba0509} \BibitemShut {NoStop}%
\bibitem [{\citenamefont {Sirica}\ \emph {et~al.}(2019)\citenamefont {Sirica}, \citenamefont {Tobey}, \citenamefont {Zhao}, \citenamefont {Chen}, \citenamefont {Xu}, \citenamefont {Yang}, \citenamefont {Shen}, \citenamefont {Yarotski}, \citenamefont {Bowlan}, \citenamefont {Trugman}, \citenamefont {Zhu}, \citenamefont {Dai}, \citenamefont {Azad}, \citenamefont {Ni}, \citenamefont {Qiu}, \citenamefont {Taylor},\ and\ \citenamefont {Prasankumar}}]{Sirica2019}%
  \BibitemOpen
  \bibfield  {author} {\bibinfo {author} {\bibfnamefont {N.}~\bibnamefont {Sirica}}, \bibinfo {author} {\bibfnamefont {R.~I.}\ \bibnamefont {Tobey}}, \bibinfo {author} {\bibfnamefont {L.~X.}\ \bibnamefont {Zhao}}, \bibinfo {author} {\bibfnamefont {G.~F.}\ \bibnamefont {Chen}}, \bibinfo {author} {\bibfnamefont {B.}~\bibnamefont {Xu}}, \bibinfo {author} {\bibfnamefont {R.}~\bibnamefont {Yang}}, \bibinfo {author} {\bibfnamefont {B.}~\bibnamefont {Shen}}, \bibinfo {author} {\bibfnamefont {D.~A.}\ \bibnamefont {Yarotski}}, \bibinfo {author} {\bibfnamefont {P.}~\bibnamefont {Bowlan}}, \bibinfo {author} {\bibfnamefont {S.~A.}\ \bibnamefont {Trugman}}, \bibinfo {author} {\bibfnamefont {J.-X.}\ \bibnamefont {Zhu}}, \bibinfo {author} {\bibfnamefont {Y.~M.}\ \bibnamefont {Dai}}, \bibinfo {author} {\bibfnamefont {A.~K.}\ \bibnamefont {Azad}}, \bibinfo {author} {\bibfnamefont {N.}~\bibnamefont {Ni}}, \bibinfo {author} {\bibfnamefont {X.~G.}\ \bibnamefont {Qiu}}, \bibinfo {author} {\bibfnamefont {A.~J.}\ \bibnamefont {Taylor}}, \ and\ \bibinfo {author} {\bibfnamefont {R.~P.}\ \bibnamefont {Prasankumar}},\ }\href {\doibase 10.1103/PhysRevLett.122.197401} {\bibfield  {journal} {\bibinfo  {journal} {Phys. Rev. Lett.}\ }\textbf {\bibinfo {volume} {122}},\ \bibinfo {pages} {197401} (\bibinfo {year} {2019})}\BibitemShut {NoStop}%
\bibitem [{\citenamefont {Gao}\ \emph {et~al.}(2020)\citenamefont {Gao}, \citenamefont {Kaushik}, \citenamefont {Philip}, \citenamefont {Li}, \citenamefont {Qin}, \citenamefont {Liu}, \citenamefont {Zhang}, \citenamefont {Su}, \citenamefont {Chen}, \citenamefont {Weng}, \citenamefont {Kharzeev}, \citenamefont {Liu},\ and\ \citenamefont {Qi}}]{Gao2020}%
  \BibitemOpen
  \bibfield  {author} {\bibinfo {author} {\bibfnamefont {Y.}~\bibnamefont {Gao}}, \bibinfo {author} {\bibfnamefont {S.}~\bibnamefont {Kaushik}}, \bibinfo {author} {\bibfnamefont {E.~J.}\ \bibnamefont {Philip}}, \bibinfo {author} {\bibfnamefont {Z.}~\bibnamefont {Li}}, \bibinfo {author} {\bibfnamefont {Y.}~\bibnamefont {Qin}}, \bibinfo {author} {\bibfnamefont {Y.~P.}\ \bibnamefont {Liu}}, \bibinfo {author} {\bibfnamefont {W.~L.}\ \bibnamefont {Zhang}}, \bibinfo {author} {\bibfnamefont {Y.~L.}\ \bibnamefont {Su}}, \bibinfo {author} {\bibfnamefont {X.}~\bibnamefont {Chen}}, \bibinfo {author} {\bibfnamefont {H.}~\bibnamefont {Weng}}, \bibinfo {author} {\bibfnamefont {D.~E.}\ \bibnamefont {Kharzeev}}, \bibinfo {author} {\bibfnamefont {M.~K.}\ \bibnamefont {Liu}}, \ and\ \bibinfo {author} {\bibfnamefont {J.}~\bibnamefont {Qi}},\ }\href {\doibase 10.1038/s41467-020-14463-1} {\bibfield  {journal} {\bibinfo  {journal} {Nature Communications}\ }\textbf {\bibinfo {volume} {11}},\ \bibinfo {pages} {720} (\bibinfo {year} {2020})}\BibitemShut {NoStop}%
\bibitem [{\citenamefont {Meng}\ and\ \citenamefont {Balents}(2012)}]{Meng2012}%
  \BibitemOpen
  \bibfield  {author} {\bibinfo {author} {\bibfnamefont {T.}~\bibnamefont {Meng}}\ and\ \bibinfo {author} {\bibfnamefont {L.}~\bibnamefont {Balents}},\ }\href {\doibase 10.1103/PhysRevB.86.054504} {\bibfield  {journal} {\bibinfo  {journal} {Phys. Rev. B}\ }\textbf {\bibinfo {volume} {86}},\ \bibinfo {pages} {054504} (\bibinfo {year} {2012})}\BibitemShut {NoStop}%
\bibitem [{\citenamefont {Meng}\ and\ \citenamefont {Balents}(2017)}]{Meng2012(E)}%
  \BibitemOpen
  \bibfield  {author} {\bibinfo {author} {\bibfnamefont {T.}~\bibnamefont {Meng}}\ and\ \bibinfo {author} {\bibfnamefont {L.}~\bibnamefont {Balents}},\ }\href {\doibase 10.1103/PhysRevB.96.019901} {\bibfield  {journal} {\bibinfo  {journal} {Phys. Rev. B}\ }\textbf {\bibinfo {volume} {96}},\ \bibinfo {pages} {019901} (\bibinfo {year} {2017})}\BibitemShut {NoStop}%
\bibitem [{\citenamefont {Yasuda}\ \emph {et~al.}(2019)\citenamefont {Yasuda}, \citenamefont {Yasuda}, \citenamefont {Liang}, \citenamefont {Yoshimi}, \citenamefont {Tsukazaki}, \citenamefont {Takahashi}, \citenamefont {Nagaosa}, \citenamefont {Kawasaki},\ and\ \citenamefont {Tokura}}]{Yasuda2019}%
  \BibitemOpen
  \bibfield  {author} {\bibinfo {author} {\bibfnamefont {K.}~\bibnamefont {Yasuda}}, \bibinfo {author} {\bibfnamefont {H.}~\bibnamefont {Yasuda}}, \bibinfo {author} {\bibfnamefont {T.}~\bibnamefont {Liang}}, \bibinfo {author} {\bibfnamefont {R.}~\bibnamefont {Yoshimi}}, \bibinfo {author} {\bibfnamefont {A.}~\bibnamefont {Tsukazaki}}, \bibinfo {author} {\bibfnamefont {K.~S.}\ \bibnamefont {Takahashi}}, \bibinfo {author} {\bibfnamefont {N.}~\bibnamefont {Nagaosa}}, \bibinfo {author} {\bibfnamefont {M.}~\bibnamefont {Kawasaki}}, \ and\ \bibinfo {author} {\bibfnamefont {Y.}~\bibnamefont {Tokura}},\ }\href {\doibase 10.1038/s41467-019-10658-3} {\bibfield  {journal} {\bibinfo  {journal} {Nature Communications}\ }\textbf {\bibinfo {volume} {10}},\ \bibinfo {pages} {2734} (\bibinfo {year} {2019})}\BibitemShut {NoStop}%
\bibitem [{\citenamefont {Wu}\ \emph {et~al.}(2018)\citenamefont {Wu}, \citenamefont {Fatemi}, \citenamefont {Gibson}, \citenamefont {Watanabe}, \citenamefont {Taniguchi}, \citenamefont {Cava},\ and\ \citenamefont {Jarillo-Herrero}}]{Wu2018}%
  \BibitemOpen
  \bibfield  {author} {\bibinfo {author} {\bibfnamefont {S.}~\bibnamefont {Wu}}, \bibinfo {author} {\bibfnamefont {V.}~\bibnamefont {Fatemi}}, \bibinfo {author} {\bibfnamefont {Q.~D.}\ \bibnamefont {Gibson}}, \bibinfo {author} {\bibfnamefont {K.}~\bibnamefont {Watanabe}}, \bibinfo {author} {\bibfnamefont {T.}~\bibnamefont {Taniguchi}}, \bibinfo {author} {\bibfnamefont {R.~J.}\ \bibnamefont {Cava}}, \ and\ \bibinfo {author} {\bibfnamefont {P.}~\bibnamefont {Jarillo-Herrero}},\ }\href {\doibase 10.1126/science.aan6003} {\bibfield  {journal} {\bibinfo  {journal} {Science}\ }\textbf {\bibinfo {volume} {359}},\ \bibinfo {pages} {76} (\bibinfo {year} {2018})},\ \Eprint {http://arxiv.org/abs/https://www.science.org/doi/pdf/10.1126/science.aan6003} {https://www.science.org/doi/pdf/10.1126/science.aan6003} \BibitemShut {NoStop}%
\bibitem [{\citenamefont {Fei}\ \emph {et~al.}(2017)\citenamefont {Fei}, \citenamefont {Palomaki}, \citenamefont {Wu}, \citenamefont {Zhao}, \citenamefont {Cai}, \citenamefont {Sun}, \citenamefont {Nguyen}, \citenamefont {Finney}, \citenamefont {Xu},\ and\ \citenamefont {Cobden}}]{Fei2017}%
  \BibitemOpen
  \bibfield  {author} {\bibinfo {author} {\bibfnamefont {Z.}~\bibnamefont {Fei}}, \bibinfo {author} {\bibfnamefont {T.}~\bibnamefont {Palomaki}}, \bibinfo {author} {\bibfnamefont {S.}~\bibnamefont {Wu}}, \bibinfo {author} {\bibfnamefont {W.}~\bibnamefont {Zhao}}, \bibinfo {author} {\bibfnamefont {X.}~\bibnamefont {Cai}}, \bibinfo {author} {\bibfnamefont {B.}~\bibnamefont {Sun}}, \bibinfo {author} {\bibfnamefont {P.}~\bibnamefont {Nguyen}}, \bibinfo {author} {\bibfnamefont {J.}~\bibnamefont {Finney}}, \bibinfo {author} {\bibfnamefont {X.}~\bibnamefont {Xu}}, \ and\ \bibinfo {author} {\bibfnamefont {D.~H.}\ \bibnamefont {Cobden}},\ }\href {\doibase 10.1038/nphys4091} {\bibfield  {journal} {\bibinfo  {journal} {Nature Physics}\ }\textbf {\bibinfo {volume} {13}},\ \bibinfo {pages} {677} (\bibinfo {year} {2017})}\BibitemShut {NoStop}%
\bibitem [{\citenamefont {Fatemi}\ \emph {et~al.}(2018)\citenamefont {Fatemi}, \citenamefont {Wu}, \citenamefont {Cao}, \citenamefont {Bretheau}, \citenamefont {Gibson}, \citenamefont {Watanabe}, \citenamefont {Taniguchi}, \citenamefont {Cava},\ and\ \citenamefont {Jarillo-Herrero}}]{Fatemi2018}%
  \BibitemOpen
  \bibfield  {author} {\bibinfo {author} {\bibfnamefont {V.}~\bibnamefont {Fatemi}}, \bibinfo {author} {\bibfnamefont {S.}~\bibnamefont {Wu}}, \bibinfo {author} {\bibfnamefont {Y.}~\bibnamefont {Cao}}, \bibinfo {author} {\bibfnamefont {L.}~\bibnamefont {Bretheau}}, \bibinfo {author} {\bibfnamefont {Q.~D.}\ \bibnamefont {Gibson}}, \bibinfo {author} {\bibfnamefont {K.}~\bibnamefont {Watanabe}}, \bibinfo {author} {\bibfnamefont {T.}~\bibnamefont {Taniguchi}}, \bibinfo {author} {\bibfnamefont {R.~J.}\ \bibnamefont {Cava}}, \ and\ \bibinfo {author} {\bibfnamefont {P.}~\bibnamefont {Jarillo-Herrero}},\ }\href {\doibase 10.1126/science.aar4642} {\bibfield  {journal} {\bibinfo  {journal} {Science}\ }\textbf {\bibinfo {volume} {362}},\ \bibinfo {pages} {926} (\bibinfo {year} {2018})},\ \Eprint {http://arxiv.org/abs/https://www.science.org/doi/pdf/10.1126/science.aar4642} {https://www.science.org/doi/pdf/10.1126/science.aar4642} \BibitemShut {NoStop}%
\bibitem [{\citenamefont {Sajadi}\ \emph {et~al.}(2018)\citenamefont {Sajadi}, \citenamefont {Palomaki}, \citenamefont {Fei}, \citenamefont {Zhao}, \citenamefont {Bement}, \citenamefont {Olsen}, \citenamefont {Luescher}, \citenamefont {Xu}, \citenamefont {Folk},\ and\ \citenamefont {Cobden}}]{Sajadi2018}%
  \BibitemOpen
  \bibfield  {author} {\bibinfo {author} {\bibfnamefont {E.}~\bibnamefont {Sajadi}}, \bibinfo {author} {\bibfnamefont {T.}~\bibnamefont {Palomaki}}, \bibinfo {author} {\bibfnamefont {Z.}~\bibnamefont {Fei}}, \bibinfo {author} {\bibfnamefont {W.}~\bibnamefont {Zhao}}, \bibinfo {author} {\bibfnamefont {P.}~\bibnamefont {Bement}}, \bibinfo {author} {\bibfnamefont {C.}~\bibnamefont {Olsen}}, \bibinfo {author} {\bibfnamefont {S.}~\bibnamefont {Luescher}}, \bibinfo {author} {\bibfnamefont {X.}~\bibnamefont {Xu}}, \bibinfo {author} {\bibfnamefont {J.~A.}\ \bibnamefont {Folk}}, \ and\ \bibinfo {author} {\bibfnamefont {D.~H.}\ \bibnamefont {Cobden}},\ }\href {\doibase 10.1126/science.aar4426} {\bibfield  {journal} {\bibinfo  {journal} {Science}\ }\textbf {\bibinfo {volume} {362}},\ \bibinfo {pages} {922} (\bibinfo {year} {2018})},\ \Eprint {http://arxiv.org/abs/https://www.science.org/doi/pdf/10.1126/science.aar4426} {https://www.science.org/doi/pdf/10.1126/science.aar4426} \BibitemShut {NoStop}%
\bibitem [{\citenamefont {Kang}\ \emph {et~al.}(2019)\citenamefont {Kang}, \citenamefont {Li}, \citenamefont {Sohn}, \citenamefont {Shan},\ and\ \citenamefont {Mak}}]{Kang2019}%
  \BibitemOpen
  \bibfield  {author} {\bibinfo {author} {\bibfnamefont {K.}~\bibnamefont {Kang}}, \bibinfo {author} {\bibfnamefont {T.}~\bibnamefont {Li}}, \bibinfo {author} {\bibfnamefont {E.}~\bibnamefont {Sohn}}, \bibinfo {author} {\bibfnamefont {J.}~\bibnamefont {Shan}}, \ and\ \bibinfo {author} {\bibfnamefont {K.~F.}\ \bibnamefont {Mak}},\ }\href {\doibase 10.1038/s41563-019-0294-7} {\bibfield  {journal} {\bibinfo  {journal} {Nature Materials}\ }\textbf {\bibinfo {volume} {18}},\ \bibinfo {pages} {324} (\bibinfo {year} {2019})}\BibitemShut {NoStop}%
\bibitem [{\citenamefont {Ma}\ \emph {et~al.}(2019{\natexlab{b}})\citenamefont {Ma}, \citenamefont {Xu}, \citenamefont {Shen}, \citenamefont {MacNeill}, \citenamefont {Fatemi}, \citenamefont {Chang}, \citenamefont {Mier~Valdivia}, \citenamefont {Wu}, \citenamefont {Du}, \citenamefont {Hsu}, \citenamefont {Fang}, \citenamefont {Gibson}, \citenamefont {Watanabe}, \citenamefont {Taniguchi}, \citenamefont {Cava}, \citenamefont {Kaxiras}, \citenamefont {Lu}, \citenamefont {Lin}, \citenamefont {Fu}, \citenamefont {Gedik},\ and\ \citenamefont {Jarillo-Herrero}}]{MaQ2019}%
  \BibitemOpen
  \bibfield  {author} {\bibinfo {author} {\bibfnamefont {Q.}~\bibnamefont {Ma}}, \bibinfo {author} {\bibfnamefont {S.-Y.}\ \bibnamefont {Xu}}, \bibinfo {author} {\bibfnamefont {H.}~\bibnamefont {Shen}}, \bibinfo {author} {\bibfnamefont {D.}~\bibnamefont {MacNeill}}, \bibinfo {author} {\bibfnamefont {V.}~\bibnamefont {Fatemi}}, \bibinfo {author} {\bibfnamefont {T.-R.}\ \bibnamefont {Chang}}, \bibinfo {author} {\bibfnamefont {A.~M.}\ \bibnamefont {Mier~Valdivia}}, \bibinfo {author} {\bibfnamefont {S.}~\bibnamefont {Wu}}, \bibinfo {author} {\bibfnamefont {Z.}~\bibnamefont {Du}}, \bibinfo {author} {\bibfnamefont {C.-H.}\ \bibnamefont {Hsu}}, \bibinfo {author} {\bibfnamefont {S.}~\bibnamefont {Fang}}, \bibinfo {author} {\bibfnamefont {Q.~D.}\ \bibnamefont {Gibson}}, \bibinfo {author} {\bibfnamefont {K.}~\bibnamefont {Watanabe}}, \bibinfo {author} {\bibfnamefont {T.}~\bibnamefont {Taniguchi}}, \bibinfo {author} {\bibfnamefont {R.~J.}\ \bibnamefont {Cava}}, \bibinfo {author} {\bibfnamefont {E.}~\bibnamefont {Kaxiras}}, \bibinfo {author} {\bibfnamefont {H.-Z.}\ \bibnamefont {Lu}}, \bibinfo {author} {\bibfnamefont {H.}~\bibnamefont {Lin}}, \bibinfo {author} {\bibfnamefont {L.}~\bibnamefont {Fu}}, \bibinfo {author} {\bibfnamefont {N.}~\bibnamefont {Gedik}}, \ and\ \bibinfo {author} {\bibfnamefont {P.}~\bibnamefont {Jarillo-Herrero}},\ }\href {\doibase 10.1038/s41586-018-0807-6} {\bibfield  {journal} {\bibinfo  {journal} {Nature}\ }\textbf {\bibinfo {volume} {565}},\ \bibinfo {pages} {337} (\bibinfo {year} {2019}{\natexlab{b}})}\BibitemShut {NoStop}%
\bibitem [{\citenamefont {Gao}\ \emph {et~al.}(2023)\citenamefont {Gao}, \citenamefont {Liu}, \citenamefont {Qiu}, \citenamefont {Ghosh}, \citenamefont {Trevisan}, \citenamefont {Onishi}, \citenamefont {Hu}, \citenamefont {Qian}, \citenamefont {Tien}, \citenamefont {Chen}, \citenamefont {Huang}, \citenamefont {Bérubé}, \citenamefont {Li}, \citenamefont {Tzschaschel}, \citenamefont {Dinh}, \citenamefont {Sun}, \citenamefont {Ho}, \citenamefont {Lien}, \citenamefont {Singh}, \citenamefont {Watanabe}, \citenamefont {Taniguchi}, \citenamefont {Bell}, \citenamefont {Lin}, \citenamefont {Chang}, \citenamefont {Du}, \citenamefont {Bansil}, \citenamefont {Fu}, \citenamefont {Ni}, \citenamefont {Orth}, \citenamefont {Ma},\ and\ \citenamefont {Xu}}]{MnBi2Te4Science2023}%
  \BibitemOpen
  \bibfield  {author} {\bibinfo {author} {\bibfnamefont {A.}~\bibnamefont {Gao}}, \bibinfo {author} {\bibfnamefont {Y.-F.}\ \bibnamefont {Liu}}, \bibinfo {author} {\bibfnamefont {J.-X.}\ \bibnamefont {Qiu}}, \bibinfo {author} {\bibfnamefont {B.}~\bibnamefont {Ghosh}}, \bibinfo {author} {\bibfnamefont {T.~V.}\ \bibnamefont {Trevisan}}, \bibinfo {author} {\bibfnamefont {Y.}~\bibnamefont {Onishi}}, \bibinfo {author} {\bibfnamefont {C.}~\bibnamefont {Hu}}, \bibinfo {author} {\bibfnamefont {T.}~\bibnamefont {Qian}}, \bibinfo {author} {\bibfnamefont {H.-J.}\ \bibnamefont {Tien}}, \bibinfo {author} {\bibfnamefont {S.-W.}\ \bibnamefont {Chen}}, \bibinfo {author} {\bibfnamefont {M.}~\bibnamefont {Huang}}, \bibinfo {author} {\bibfnamefont {D.}~\bibnamefont {Bérubé}}, \bibinfo {author} {\bibfnamefont {H.}~\bibnamefont {Li}}, \bibinfo {author} {\bibfnamefont {C.}~\bibnamefont {Tzschaschel}}, \bibinfo {author} {\bibfnamefont {T.}~\bibnamefont {Dinh}}, \bibinfo {author} {\bibfnamefont {Z.}~\bibnamefont {Sun}}, \bibinfo {author} {\bibfnamefont {S.-C.}\ \bibnamefont {Ho}}, \bibinfo {author} {\bibfnamefont {S.-W.}\ \bibnamefont {Lien}}, \bibinfo {author} {\bibfnamefont {B.}~\bibnamefont {Singh}}, \bibinfo {author} {\bibfnamefont {K.}~\bibnamefont {Watanabe}}, \bibinfo {author} {\bibfnamefont {T.}~\bibnamefont {Taniguchi}}, \bibinfo {author} {\bibfnamefont {D.~C.}\ \bibnamefont {Bell}}, \bibinfo {author} {\bibfnamefont {H.}~\bibnamefont {Lin}}, \bibinfo {author} {\bibfnamefont {T.-R.}\ \bibnamefont {Chang}}, \bibinfo {author} {\bibfnamefont {C.~R.}\ \bibnamefont {Du}}, \bibinfo {author} {\bibfnamefont {A.}~\bibnamefont {Bansil}}, \bibinfo {author} {\bibfnamefont {L.}~\bibnamefont {Fu}}, \bibinfo {author} {\bibfnamefont {N.}~\bibnamefont {Ni}}, \bibinfo {author} {\bibfnamefont {P.~P.}\ \bibnamefont {Orth}}, \bibinfo {author} {\bibfnamefont {Q.}~\bibnamefont {Ma}}, \ and\ \bibinfo {author} {\bibfnamefont {S.-Y.}\ \bibnamefont {Xu}},\ }\href {\doibase 10.1126/science.adf1506} {\bibfield  {journal} {\bibinfo  {journal} {Science}\ }\textbf {\bibinfo {volume} {381}},\ \bibinfo {pages} {181} (\bibinfo {year} {2023})},\ \Eprint {http://arxiv.org/abs/https://www.science.org/doi/pdf/10.1126/science.adf1506} {https://www.science.org/doi/pdf/10.1126/science.adf1506} \BibitemShut {NoStop}%
\bibitem [{\citenamefont {Yuan}\ \emph {et~al.}(2024)\citenamefont {Yuan}, \citenamefont {Yan}, \citenamefont {Yi}, \citenamefont {Wang}, \citenamefont {Paolini}, \citenamefont {Zhao}, \citenamefont {Zhou}, \citenamefont {Wang}, \citenamefont {Wang}, \citenamefont {Prokscha}, \citenamefont {Salman}, \citenamefont {Suter}, \citenamefont {Balakrishnan}, \citenamefont {Grutter}, \citenamefont {Winter}, \citenamefont {Singleton}, \citenamefont {Chan},\ and\ \citenamefont {Chang}}]{Yuan2024}%
  \BibitemOpen
  \bibfield  {author} {\bibinfo {author} {\bibfnamefont {W.}~\bibnamefont {Yuan}}, \bibinfo {author} {\bibfnamefont {Z.-J.}\ \bibnamefont {Yan}}, \bibinfo {author} {\bibfnamefont {H.}~\bibnamefont {Yi}}, \bibinfo {author} {\bibfnamefont {Z.}~\bibnamefont {Wang}}, \bibinfo {author} {\bibfnamefont {S.}~\bibnamefont {Paolini}}, \bibinfo {author} {\bibfnamefont {Y.-F.}\ \bibnamefont {Zhao}}, \bibinfo {author} {\bibfnamefont {L.}~\bibnamefont {Zhou}}, \bibinfo {author} {\bibfnamefont {A.~G.}\ \bibnamefont {Wang}}, \bibinfo {author} {\bibfnamefont {K.}~\bibnamefont {Wang}}, \bibinfo {author} {\bibfnamefont {T.}~\bibnamefont {Prokscha}}, \bibinfo {author} {\bibfnamefont {Z.}~\bibnamefont {Salman}}, \bibinfo {author} {\bibfnamefont {A.}~\bibnamefont {Suter}}, \bibinfo {author} {\bibfnamefont {P.~P.}\ \bibnamefont {Balakrishnan}}, \bibinfo {author} {\bibfnamefont {A.~J.}\ \bibnamefont {Grutter}}, \bibinfo {author} {\bibfnamefont {L.~E.}\ \bibnamefont {Winter}}, \bibinfo {author} {\bibfnamefont {J.}~\bibnamefont {Singleton}}, \bibinfo {author} {\bibfnamefont {M.~H.~W.}\ \bibnamefont {Chan}}, \ and\ \bibinfo {author} {\bibfnamefont {C.-Z.}\ \bibnamefont {Chang}},\ }\href {\doibase 10.1021/acs.nanolett.4c01407} {\bibfield  {journal} {\bibinfo  {journal} {Nano Letters}\ }\textbf {\bibinfo {volume} {24}},\ \bibinfo {pages} {7962} (\bibinfo {year} {2024})}\BibitemShut {NoStop}%
\bibitem [{\citenamefont {Kitamura}\ \emph {et~al.}(2021)\citenamefont {Kitamura}, \citenamefont {Ishizuka}, \citenamefont {Daido},\ and\ \citenamefont {Yanase}}]{Kitamura2021}%
  \BibitemOpen
  \bibfield  {author} {\bibinfo {author} {\bibfnamefont {T.}~\bibnamefont {Kitamura}}, \bibinfo {author} {\bibfnamefont {J.}~\bibnamefont {Ishizuka}}, \bibinfo {author} {\bibfnamefont {A.}~\bibnamefont {Daido}}, \ and\ \bibinfo {author} {\bibfnamefont {Y.}~\bibnamefont {Yanase}},\ }\href {\doibase 10.1103/PhysRevB.103.245114} {\bibfield  {journal} {\bibinfo  {journal} {Phys. Rev. B}\ }\textbf {\bibinfo {volume} {103}},\ \bibinfo {pages} {245114} (\bibinfo {year} {2021})}\BibitemShut {NoStop}%
\bibitem [{\citenamefont {Kitamura}\ \emph {et~al.}(2022{\natexlab{a}})\citenamefont {Kitamura}, \citenamefont {Daido},\ and\ \citenamefont {Yanase}}]{KitamuraPRB2022}%
  \BibitemOpen
  \bibfield  {author} {\bibinfo {author} {\bibfnamefont {T.}~\bibnamefont {Kitamura}}, \bibinfo {author} {\bibfnamefont {A.}~\bibnamefont {Daido}}, \ and\ \bibinfo {author} {\bibfnamefont {Y.}~\bibnamefont {Yanase}},\ }\href {\doibase 10.1103/PhysRevB.106.184507} {\bibfield  {journal} {\bibinfo  {journal} {Phys. Rev. B}\ }\textbf {\bibinfo {volume} {106}},\ \bibinfo {pages} {184507} (\bibinfo {year} {2022}{\natexlab{a}})}\BibitemShut {NoStop}%
\bibitem [{\citenamefont {Kitamura}\ \emph {et~al.}(2022{\natexlab{b}})\citenamefont {Kitamura}, \citenamefont {Yamashita}, \citenamefont {Ishizuka}, \citenamefont {Daido},\ and\ \citenamefont {Yanase}}]{KitamuraPRR2022}%
  \BibitemOpen
  \bibfield  {author} {\bibinfo {author} {\bibfnamefont {T.}~\bibnamefont {Kitamura}}, \bibinfo {author} {\bibfnamefont {T.}~\bibnamefont {Yamashita}}, \bibinfo {author} {\bibfnamefont {J.}~\bibnamefont {Ishizuka}}, \bibinfo {author} {\bibfnamefont {A.}~\bibnamefont {Daido}}, \ and\ \bibinfo {author} {\bibfnamefont {Y.}~\bibnamefont {Yanase}},\ }\href {\doibase 10.1103/PhysRevResearch.4.023232} {\bibfield  {journal} {\bibinfo  {journal} {Phys. Rev. Res.}\ }\textbf {\bibinfo {volume} {4}},\ \bibinfo {pages} {023232} (\bibinfo {year} {2022}{\natexlab{b}})}\BibitemShut {NoStop}%
\bibitem [{\citenamefont {Saito}\ \emph {et~al.}(2016)\citenamefont {Saito}, \citenamefont {Nakamura}, \citenamefont {Bahramy}, \citenamefont {Kohama}, \citenamefont {Ye}, \citenamefont {Kasahara}, \citenamefont {Nakagawa}, \citenamefont {Onga}, \citenamefont {Tokunaga}, \citenamefont {Nojima}, \citenamefont {Yanase},\ and\ \citenamefont {Iwasa}}]{Saito2016}%
  \BibitemOpen
  \bibfield  {author} {\bibinfo {author} {\bibfnamefont {Y.}~\bibnamefont {Saito}}, \bibinfo {author} {\bibfnamefont {Y.}~\bibnamefont {Nakamura}}, \bibinfo {author} {\bibfnamefont {M.~S.}\ \bibnamefont {Bahramy}}, \bibinfo {author} {\bibfnamefont {Y.}~\bibnamefont {Kohama}}, \bibinfo {author} {\bibfnamefont {J.}~\bibnamefont {Ye}}, \bibinfo {author} {\bibfnamefont {Y.}~\bibnamefont {Kasahara}}, \bibinfo {author} {\bibfnamefont {Y.}~\bibnamefont {Nakagawa}}, \bibinfo {author} {\bibfnamefont {M.}~\bibnamefont {Onga}}, \bibinfo {author} {\bibfnamefont {M.}~\bibnamefont {Tokunaga}}, \bibinfo {author} {\bibfnamefont {T.}~\bibnamefont {Nojima}}, \bibinfo {author} {\bibfnamefont {Y.}~\bibnamefont {Yanase}}, \ and\ \bibinfo {author} {\bibfnamefont {Y.}~\bibnamefont {Iwasa}},\ }\href {\doibase 10.1038/nphys3580} {\bibfield  {journal} {\bibinfo  {journal} {Nature Physics}\ }\textbf {\bibinfo {volume} {12}},\ \bibinfo {pages} {144} (\bibinfo {year} {2016})}\BibitemShut {NoStop}%
\bibitem [{\citenamefont {Lu}\ \emph {et~al.}(2015)\citenamefont {Lu}, \citenamefont {Zheliuk}, \citenamefont {Leermakers}, \citenamefont {Yuan}, \citenamefont {Zeitler}, \citenamefont {Law},\ and\ \citenamefont {Ye}}]{Lu2015}%
  \BibitemOpen
  \bibfield  {author} {\bibinfo {author} {\bibfnamefont {J.~M.}\ \bibnamefont {Lu}}, \bibinfo {author} {\bibfnamefont {O.}~\bibnamefont {Zheliuk}}, \bibinfo {author} {\bibfnamefont {I.}~\bibnamefont {Leermakers}}, \bibinfo {author} {\bibfnamefont {N.~F.~Q.}\ \bibnamefont {Yuan}}, \bibinfo {author} {\bibfnamefont {U.}~\bibnamefont {Zeitler}}, \bibinfo {author} {\bibfnamefont {K.~T.}\ \bibnamefont {Law}}, \ and\ \bibinfo {author} {\bibfnamefont {J.~T.}\ \bibnamefont {Ye}},\ }\href {\doibase 10.1126/science.aab2277} {\bibfield  {journal} {\bibinfo  {journal} {Science}\ }\textbf {\bibinfo {volume} {350}},\ \bibinfo {pages} {1353} (\bibinfo {year} {2015})},\ \Eprint {http://arxiv.org/abs/https://www.science.org/doi/pdf/10.1126/science.aab2277} {https://www.science.org/doi/pdf/10.1126/science.aab2277} \BibitemShut {NoStop}%
\bibitem [{\citenamefont {Xi}\ \emph {et~al.}(2016)\citenamefont {Xi}, \citenamefont {Wang}, \citenamefont {Zhao}, \citenamefont {Park}, \citenamefont {Law}, \citenamefont {Berger}, \citenamefont {Forr{\'o}}, \citenamefont {Shan},\ and\ \citenamefont {Mak}}]{Xi2016}%
  \BibitemOpen
  \bibfield  {author} {\bibinfo {author} {\bibfnamefont {X.}~\bibnamefont {Xi}}, \bibinfo {author} {\bibfnamefont {Z.}~\bibnamefont {Wang}}, \bibinfo {author} {\bibfnamefont {W.}~\bibnamefont {Zhao}}, \bibinfo {author} {\bibfnamefont {J.-H.}\ \bibnamefont {Park}}, \bibinfo {author} {\bibfnamefont {K.~T.}\ \bibnamefont {Law}}, \bibinfo {author} {\bibfnamefont {H.}~\bibnamefont {Berger}}, \bibinfo {author} {\bibfnamefont {L.}~\bibnamefont {Forr{\'o}}}, \bibinfo {author} {\bibfnamefont {J.}~\bibnamefont {Shan}}, \ and\ \bibinfo {author} {\bibfnamefont {K.~F.}\ \bibnamefont {Mak}},\ }\href {\doibase 10.1038/nphys3538} {\bibfield  {journal} {\bibinfo  {journal} {Nature Physics}\ }\textbf {\bibinfo {volume} {12}},\ \bibinfo {pages} {139} (\bibinfo {year} {2016})}\BibitemShut {NoStop}%
\bibitem [{\citenamefont {de~la Barrera}\ \emph {et~al.}(2018)\citenamefont {de~la Barrera}, \citenamefont {Sinko}, \citenamefont {Gopalan}, \citenamefont {Sivadas}, \citenamefont {Seyler}, \citenamefont {Watanabe}, \citenamefont {Taniguchi}, \citenamefont {Tsen}, \citenamefont {Xu}, \citenamefont {Xiao},\ and\ \citenamefont {Hunt}}]{delaBarrera2018}%
  \BibitemOpen
  \bibfield  {author} {\bibinfo {author} {\bibfnamefont {S.~C.}\ \bibnamefont {de~la Barrera}}, \bibinfo {author} {\bibfnamefont {M.~R.}\ \bibnamefont {Sinko}}, \bibinfo {author} {\bibfnamefont {D.~P.}\ \bibnamefont {Gopalan}}, \bibinfo {author} {\bibfnamefont {N.}~\bibnamefont {Sivadas}}, \bibinfo {author} {\bibfnamefont {K.~L.}\ \bibnamefont {Seyler}}, \bibinfo {author} {\bibfnamefont {K.}~\bibnamefont {Watanabe}}, \bibinfo {author} {\bibfnamefont {T.}~\bibnamefont {Taniguchi}}, \bibinfo {author} {\bibfnamefont {A.~W.}\ \bibnamefont {Tsen}}, \bibinfo {author} {\bibfnamefont {X.}~\bibnamefont {Xu}}, \bibinfo {author} {\bibfnamefont {D.}~\bibnamefont {Xiao}}, \ and\ \bibinfo {author} {\bibfnamefont {B.~M.}\ \bibnamefont {Hunt}},\ }\href {\doibase 10.1038/s41467-018-03888-4} {\bibfield  {journal} {\bibinfo  {journal} {Nature Communications}\ }\textbf {\bibinfo {volume} {9}},\ \bibinfo {pages} {1427} (\bibinfo {year} {2018})}\BibitemShut {NoStop}%
\bibitem [{\citenamefont {Liu}\ \emph {et~al.}(2020)\citenamefont {Liu}, \citenamefont {Xu}, \citenamefont {Sun}, \citenamefont {Liu}, \citenamefont {Liu}, \citenamefont {Wang}, \citenamefont {Zhang}, \citenamefont {Gu}, \citenamefont {Tang}, \citenamefont {Ding}, \citenamefont {Liu}, \citenamefont {Yao}, \citenamefont {Lin}, \citenamefont {Wang}, \citenamefont {Xue},\ and\ \citenamefont {Wang}}]{Liu2020}%
  \BibitemOpen
  \bibfield  {author} {\bibinfo {author} {\bibfnamefont {Y.}~\bibnamefont {Liu}}, \bibinfo {author} {\bibfnamefont {Y.}~\bibnamefont {Xu}}, \bibinfo {author} {\bibfnamefont {J.}~\bibnamefont {Sun}}, \bibinfo {author} {\bibfnamefont {C.}~\bibnamefont {Liu}}, \bibinfo {author} {\bibfnamefont {Y.}~\bibnamefont {Liu}}, \bibinfo {author} {\bibfnamefont {C.}~\bibnamefont {Wang}}, \bibinfo {author} {\bibfnamefont {Z.}~\bibnamefont {Zhang}}, \bibinfo {author} {\bibfnamefont {K.}~\bibnamefont {Gu}}, \bibinfo {author} {\bibfnamefont {Y.}~\bibnamefont {Tang}}, \bibinfo {author} {\bibfnamefont {C.}~\bibnamefont {Ding}}, \bibinfo {author} {\bibfnamefont {H.}~\bibnamefont {Liu}}, \bibinfo {author} {\bibfnamefont {H.}~\bibnamefont {Yao}}, \bibinfo {author} {\bibfnamefont {X.}~\bibnamefont {Lin}}, \bibinfo {author} {\bibfnamefont {L.}~\bibnamefont {Wang}}, \bibinfo {author} {\bibfnamefont {Q.-K.}\ \bibnamefont {Xue}}, \ and\ \bibinfo {author} {\bibfnamefont {J.}~\bibnamefont {Wang}},\ }\href {\doibase 10.1021/acs.nanolett.0c01356} {\bibfield  {journal} {\bibinfo  {journal} {Nano Letters}\ }\textbf {\bibinfo {volume} {20}},\ \bibinfo {pages} {5728} (\bibinfo {year} {2020})}\BibitemShut {NoStop}%
\bibitem [{\citenamefont {Cui}\ \emph {et~al.}(2019)\citenamefont {Cui}, \citenamefont {Li}, \citenamefont {Zhou}, \citenamefont {He}, \citenamefont {Huang}, \citenamefont {Yi}, \citenamefont {Fan}, \citenamefont {Ji}, \citenamefont {Jing}, \citenamefont {Qu}, \citenamefont {Cheng}, \citenamefont {Yang}, \citenamefont {Lu}, \citenamefont {Suenaga}, \citenamefont {Liu}, \citenamefont {Law}, \citenamefont {Lin}, \citenamefont {Liu},\ and\ \citenamefont {Liu}}]{Cui2019}%
  \BibitemOpen
  \bibfield  {author} {\bibinfo {author} {\bibfnamefont {J.}~\bibnamefont {Cui}}, \bibinfo {author} {\bibfnamefont {P.}~\bibnamefont {Li}}, \bibinfo {author} {\bibfnamefont {J.}~\bibnamefont {Zhou}}, \bibinfo {author} {\bibfnamefont {W.-Y.}\ \bibnamefont {He}}, \bibinfo {author} {\bibfnamefont {X.}~\bibnamefont {Huang}}, \bibinfo {author} {\bibfnamefont {J.}~\bibnamefont {Yi}}, \bibinfo {author} {\bibfnamefont {J.}~\bibnamefont {Fan}}, \bibinfo {author} {\bibfnamefont {Z.}~\bibnamefont {Ji}}, \bibinfo {author} {\bibfnamefont {X.}~\bibnamefont {Jing}}, \bibinfo {author} {\bibfnamefont {F.}~\bibnamefont {Qu}}, \bibinfo {author} {\bibfnamefont {Z.~G.}\ \bibnamefont {Cheng}}, \bibinfo {author} {\bibfnamefont {C.}~\bibnamefont {Yang}}, \bibinfo {author} {\bibfnamefont {L.}~\bibnamefont {Lu}}, \bibinfo {author} {\bibfnamefont {K.}~\bibnamefont {Suenaga}}, \bibinfo {author} {\bibfnamefont {J.}~\bibnamefont {Liu}}, \bibinfo {author} {\bibfnamefont {K.~T.}\ \bibnamefont {Law}}, \bibinfo {author} {\bibfnamefont {J.}~\bibnamefont {Lin}}, \bibinfo {author} {\bibfnamefont {Z.}~\bibnamefont {Liu}}, \ and\ \bibinfo {author} {\bibfnamefont {G.}~\bibnamefont {Liu}},\ }\href {\doibase 10.1038/s41467-019-09995-0} {\bibfield  {journal} {\bibinfo  {journal} {Nature Communications}\ }\textbf {\bibinfo {volume} {10}},\ \bibinfo {pages} {2044} (\bibinfo {year} {2019})}\BibitemShut {NoStop}%
\bibitem [{\citenamefont {Rhodes}\ \emph {et~al.}(2021)\citenamefont {Rhodes}, \citenamefont {Jindal}, \citenamefont {Yuan}, \citenamefont {Jung}, \citenamefont {Antony}, \citenamefont {Wang}, \citenamefont {Kim}, \citenamefont {Chiu}, \citenamefont {Taniguchi}, \citenamefont {Watanabe}, \citenamefont {Barmak}, \citenamefont {Balicas}, \citenamefont {Dean}, \citenamefont {Qian}, \citenamefont {Fu}, \citenamefont {Pasupathy},\ and\ \citenamefont {Hone}}]{Rhodes2021}%
  \BibitemOpen
  \bibfield  {author} {\bibinfo {author} {\bibfnamefont {D.~A.}\ \bibnamefont {Rhodes}}, \bibinfo {author} {\bibfnamefont {A.}~\bibnamefont {Jindal}}, \bibinfo {author} {\bibfnamefont {N.~F.~Q.}\ \bibnamefont {Yuan}}, \bibinfo {author} {\bibfnamefont {Y.}~\bibnamefont {Jung}}, \bibinfo {author} {\bibfnamefont {A.}~\bibnamefont {Antony}}, \bibinfo {author} {\bibfnamefont {H.}~\bibnamefont {Wang}}, \bibinfo {author} {\bibfnamefont {B.}~\bibnamefont {Kim}}, \bibinfo {author} {\bibfnamefont {Y.-c.}\ \bibnamefont {Chiu}}, \bibinfo {author} {\bibfnamefont {T.}~\bibnamefont {Taniguchi}}, \bibinfo {author} {\bibfnamefont {K.}~\bibnamefont {Watanabe}}, \bibinfo {author} {\bibfnamefont {K.}~\bibnamefont {Barmak}}, \bibinfo {author} {\bibfnamefont {L.}~\bibnamefont {Balicas}}, \bibinfo {author} {\bibfnamefont {C.~R.}\ \bibnamefont {Dean}}, \bibinfo {author} {\bibfnamefont {X.}~\bibnamefont {Qian}}, \bibinfo {author} {\bibfnamefont {L.}~\bibnamefont {Fu}}, \bibinfo {author} {\bibfnamefont {A.~N.}\ \bibnamefont {Pasupathy}}, \ and\ \bibinfo {author} {\bibfnamefont {J.}~\bibnamefont {Hone}},\ }\href {\doibase 10.1021/acs.nanolett.0c04935} {\bibfield  {journal} {\bibinfo  {journal} {Nano Letters}\ }\textbf {\bibinfo {volume} {21}},\ \bibinfo {pages} {2505} (\bibinfo {year} {2021})}\BibitemShut {NoStop}%
\bibitem [{\citenamefont {Zhang}\ \emph {et~al.}(2023)\citenamefont {Zhang}, \citenamefont {Xie}, \citenamefont {Fang}, \citenamefont {Zhang}, \citenamefont {Xu}, \citenamefont {Zou}, \citenamefont {Leng}, \citenamefont {Gao}, \citenamefont {Zhang}, \citenamefont {Ai}, \citenamefont {Zhang}, \citenamefont {Jia}, \citenamefont {Liu}, \citenamefont {Yan}, \citenamefont {Zhao}, \citenamefont {Haigh}, \citenamefont {Kou}, \citenamefont {Yang}, \citenamefont {Huang}, \citenamefont {Law}, \citenamefont {Xiu},\ and\ \citenamefont {Dong}}]{Zhang2023}%
  \BibitemOpen
  \bibfield  {author} {\bibinfo {author} {\bibfnamefont {E.}~\bibnamefont {Zhang}}, \bibinfo {author} {\bibfnamefont {Y.-M.}\ \bibnamefont {Xie}}, \bibinfo {author} {\bibfnamefont {Y.}~\bibnamefont {Fang}}, \bibinfo {author} {\bibfnamefont {J.}~\bibnamefont {Zhang}}, \bibinfo {author} {\bibfnamefont {X.}~\bibnamefont {Xu}}, \bibinfo {author} {\bibfnamefont {Y.-C.}\ \bibnamefont {Zou}}, \bibinfo {author} {\bibfnamefont {P.}~\bibnamefont {Leng}}, \bibinfo {author} {\bibfnamefont {X.-J.}\ \bibnamefont {Gao}}, \bibinfo {author} {\bibfnamefont {Y.}~\bibnamefont {Zhang}}, \bibinfo {author} {\bibfnamefont {L.}~\bibnamefont {Ai}}, \bibinfo {author} {\bibfnamefont {Y.}~\bibnamefont {Zhang}}, \bibinfo {author} {\bibfnamefont {Z.}~\bibnamefont {Jia}}, \bibinfo {author} {\bibfnamefont {S.}~\bibnamefont {Liu}}, \bibinfo {author} {\bibfnamefont {J.}~\bibnamefont {Yan}}, \bibinfo {author} {\bibfnamefont {W.}~\bibnamefont {Zhao}}, \bibinfo {author} {\bibfnamefont {S.~J.}\ \bibnamefont {Haigh}}, \bibinfo {author} {\bibfnamefont {X.}~\bibnamefont {Kou}}, \bibinfo {author} {\bibfnamefont {J.}~\bibnamefont {Yang}}, \bibinfo {author} {\bibfnamefont {F.}~\bibnamefont {Huang}}, \bibinfo {author} {\bibfnamefont {K.~T.}\ \bibnamefont {Law}}, \bibinfo {author} {\bibfnamefont {F.}~\bibnamefont {Xiu}}, \ and\ \bibinfo {author} {\bibfnamefont {S.}~\bibnamefont {Dong}},\ }\href {\doibase 10.1038/s41567-022-01812-8} {\bibfield  {journal} {\bibinfo  {journal} {Nature Physics}\ }\textbf {\bibinfo {volume} {19}},\ \bibinfo {pages} {106} (\bibinfo {year} {2023})}\BibitemShut {NoStop}%
\bibitem [{\citenamefont {Yanase}\ and\ \citenamefont {Sigrist}(2008)}]{Yanase2008}%
  \BibitemOpen
  \bibfield  {author} {\bibinfo {author} {\bibfnamefont {Y.}~\bibnamefont {Yanase}}\ and\ \bibinfo {author} {\bibfnamefont {M.}~\bibnamefont {Sigrist}},\ }\href {\doibase 10.1143/JPSJ.77.124711} {\bibfield  {journal} {\bibinfo  {journal} {Journal of the Physical Society of Japan}\ }\textbf {\bibinfo {volume} {77}},\ \bibinfo {pages} {124711} (\bibinfo {year} {2008})}\BibitemShut {NoStop}%
\end{thebibliography}%

\end{document}